 \documentclass[twoside,11pt]{Classes/CUEDthesisPSnPDF}

\ifpdf
    \pdfcatalog { /PageMode (/UseOutlines)
                  /OpenAction (fitbh)  }
\fi

\title{Phenomenology of the minimal $\boldsymbol{B-L}$\\[0.8ex]
        extension of the Standard Model\\[0.8ex]
	at the LHC}

\titlelong{PHENOMENOLOGY OF THE MINIMAL $\boldsymbol{B-L}$ EXTENSION\\[0.8ex]
         OF THE STANDARD MODEL AT THE LHC}
	
\ifpdf
  \author{\href{mailto:lb4x07@soton.ac.uk}{Lorenzo Basso}}
  \collegeordept{\href{http://www.eng.cam.ac.uk}{School of Physics and Astronomy}}
  \university{\href{http://www.soton.ac.uk}{University of Southampton}}
\else
  \author{Lorenzo Basso}
  \collegeordept{School of Physics and Astronomy}
  \university{University of Southampton}
\fi
\degree{Doctor of Philosophy}
\degreedate{Yet to be decided}

\supervisor{Prof.~Stefano~Moretti}
\authors{Lorenzo~Basso}
\date{June, 2011}
\addresses{\groupname\\\deptname\\\univname}

\usepackage{makeidx}
\makeindex

\usepackage{latexsym}
\usepackage{amsfonts}
\usepackage[latin1]{inputenc}
\usepackage{subfig}
\usepackage{graphicx}
\usepackage{mathrsfs}
\usepackage{amssymb}
\usepackage{amsmath}
\usepackage{verbatim}
\usepackage{axodraw}

\def\met{{\slash\!\!\!\!\!\:E}_T}
\def\mpt{{\slash\!\!\!\!\!\:P}_T}

\def\mptv{{\slash\!\!\!\!\!\:\vec{P}}_T}
\def\alp{{'\!\!\!\!\!\:\alpha}}

\def\blue{\color{Blue}}
\newcommand{\bH}{\blue{$H$}}
\begin{document}




\maketitle

\setcounter{secnumdepth}{4}
\setcounter{tocdepth}{3}

\frontmatter 
\pagenumbering{roman}

\begin{dedication} 

To my Love. $\Sigma\tau\eta\nu$ $\alpha\gamma$ $\alp\pi\eta$ $\mu o\upsilon$. 

\end{dedication}



\cleardoublepage


\begin{abstract}        

A well-motivated framework to naturally introduce neutrino masses is the $B-L$ model, a $U(1)$ extension of the standard model related to the baryon minus lepton gauged number. Besides three right-handed neutrinos, that are included to cancel the anomalies (thereby naturally providing neutrino masses), this model also encompasses a complex scalar for the spontaneous symmetry breaking of the extended gauge sector and to give mass to the $Z'$ boson. We present the phenomenology, the discovery potential at the LHC, and the most up-to-date experimental and theoretical limits of the new particles in this model. 
In the gauge sector, a $Z'$ boson is present. We study its properties (i.e., production cross sections, branching ratios, total width), showing that it is dominantly coupled to leptons. We also present a detailed discovery power study at the LHC and at Tevatron for the $Z'$ boson.
In the fermion sector, after implementing the see-saw mechanism, we end up with three heavy neutrinos. We show that they can be long-lived particles (therefore providing displaced vertices in the detector), and that they can induce spectacular multi-lepton decays of the $Z'$ boson. We also study the full signature $pp \to Z' \to \nu_h \nu_h$, and present a parton level and a detector level analysis for the tri-lepton decay mode of the $Z'$ boson via heavy neutrinos.
In the gauge sector, the two Higgs fields mix. We derive the unitarity bound and the constraints from the renormalisation group equations study. In the allowed region of the parameter space, we delineate the phenomenology of the Higgs bosons and we show characteristic signatures of the latter, at the LHC, involving the $Z'$ boson and the heavy neutrinos.

\end{abstract}



%
\tableofcontents
\addcontentsline{toc}{chapter}{Contents}
\listoffigures
\listoftables
\Declaration{
  I, Lorenzo Basso, declare that this thesis titled, `Phenomenology of the minimal $B-L$ extension of the Standard Model at the LHC' and the work presented in it are my own. I
  confirm that:

  \begin{itemize}
  \item This work was done wholly or mainly while in candidature for a
    research degree at this University.
  \item Where any part of this thesis has previously been submitted
    for a degree or any other qualification at this University or any
    other institution, this has been clearly stated.
  \item Where I have consulted the published work of others, this is
    always clearly attributed.
  \item Where I have quoted from the work of others, the source is
    always given. With the exception of such quotations, this thesis
    is entirely my own work.
  \item I have acknowledged all main sources of help.
  \item Where the
    thesis is based on work done by myself jointly with others, I have
    made clear exactly what was done by others and what I have
    contributed myself.
  \end{itemize}
 
  Signed:\\
 
  Date:\\
}
\clearpage 


\begin{acknowledgements}      

I would like to start by thanking my supervisor, Stefano Moretti, for inspiration and support throughout the entire project. It has been a great pleasure to work in collaboration with Alexander Belyaev, Claire H. Shepherd-Themistocleous and, of course, Giovanni Marco Pruna, without whom this work would have never been possible.

My gratitude goes also to people I had the chance to work with, to be enlightened by, or simply to discuss aspects of this work with: Carlo Carloni Calame, Luca Fedeli, Muge Karag\"oz \"Unel, Steve King, Doug Ross and Ian Tomalin. I also thank the NExT Institute for financial support, for the stimulating environment and for the unique chances offered.

My stay in Southampton would have not been so pleasant without the multitude of friends and mates I have been fortunate to meet in the past few years.
To all of them I give a warm thank you for the good moments and the several dinners spent altogether. 

Finally, I would like to thank my family, whose support has been vital.

\end{acknowledgements}



\printnomenclature  
\addcontentsline{toc}{chapter}{Nomenclature}

\mainmatter 
\chapter{Introduction}
\ifpdf
    \graphicspath{{Introduction/IntroductionFigs/PNG/}{Introduction/IntroductionFigs/PDF/}{Introduction/IntroductionFigs/}}
\else
    \graphicspath{{Introduction/IntroductionFigs/EPS/}{Introduction/IntroductionFigs/}}
\fi

A new era in particle physics has started. The Large Hadron Collider (LHC) at CERN, Geneva, is finally taking data and a huge effort from both theoretical and experimental communities is required to meet the challenge.

The standard model (SM) \footnote{For reviews, excellent textbooks exist, see Refs.~\cite{Cheng:1985bj,Peskin:1995ev}.} of electroweak (EW) and strong interactions \footnote{We do not consider here the inclusion and description of gravity at the quantum level, often kept aside of the traditional SM framework.} is once again going to be severely tested, as well as many of its extensions that have been proposed to cure its flaws. The observed pattern of neutrino masses \cite{Fogli:2008jx,Altarelli:2010fk}, the existence of dark matter \cite{Jarosik:2010iu} and the observed matter-antimatter asymmetry \cite{Jarosik:2010iu,Sakharov:1967dj} are the most severe evidences the SM fails to explain. Deep criticism can be moved against the SM also from the theoretical point of view. The
so-called `hierarchy problem' (see, e.g., \cite{Martin:1997ns} and references therein) is an example. The inadequacy of the SM is reinforced by the fact that one of its fundamental component is still missing: the Higgs boson. 

It is widely accepted that the SM ought to be extended, but no one knows if the proper way has already been explored in the literature.  A joint collaboration is therefore needed between the experimental and the theoretical communities. 


The project of this Thesis has been guided by these principles. The aim is to fill some of the gaps in the overall preparation towards real data, as well as to interact proficiently with experimentalists. 
An extension of the SM has been systematically studied, from the definition of its parameter space to the collider signatures,
leading to some novel and exciting possibilities for the LHC to shed light on.
Within the ambition of a complete study, in all its aspects, the experimental help has been fundamental to efficiently concentrate on aspects of actual interest in a way useful for both communities.

The main motivations \footnote{Other fundamental reasons to extend the SM being the inclusion of the gravity, the lack of a dark matter candidate, the dark energy, the hierarchy problem, the matter-antimatter asymmetry.} for the extension of the SM that we will describe concern the lack of a natural explanation for the observed pattern of neutrino masses, the unknown origin of a global and not anomalous accidental $U(1)$ symmetry in the SM (related to the baryon minus lepton ($B-L$) quantum numbers), and the absence of any observation of a fundamental scalar degree of freedom (the Higgs boson). 


Besides the global $U(1)_{B-L}$ symmetry of the SM, that can be thought as an accidental symmetry (not imposed, as the custodial symmetry \cite{Barbieri:2007gi}), the previous issues can be tackled separately. The neutrino masses and mixing angles are traditionally explained by means of the so-called `see-saw mechanism' \cite{Minkowski:1977sc,VanNieuwenhuizen:1979hm,Yanagida:1979as,S.L.Glashow,Mohapatra:1979ia}, simply including $3$ right-handed (RH) neutrinos and imposing that  their effective mass term is much bigger than the Dirac one, that couples the traditional left-handed (LH) neutrinos and the RH neutrinos to the Higgs field. The right scale for the masses of the $3$ SM-like neutrinos is then reproduced by an effective mass term for a RH neutrino of $\mathcal{O}(10^{14}\div10^{16})$ GeV. 

By now, the Higgs boson is the only undetected particle of the SM, and its properties are still unknown. Therefore, it is not unreasonable to think of modifications of the scalar sector that are still compatible with  
experimental constraints. The most economical way to modify the scalar sector of the SM is to include one (or more) scalar singlets, either real \cite{O'Connell:2006wi,BahatTreidel:2006kx,Barger:2006sk,Profumo:2007wc,Bhattacharyya:2007pb} or complex \cite{Barger:2008jx}, whose phenomenology at hadronic and leptonic colliders has been studied in great detail, as well as their impact on precision observables. In fact, the latter are able to constrain the viable parameter space of the extended Higgs sectors (see, e.g., \cite{Barger:2007im,Profumo:2007wc,Dawson:2009yx} and references therein).

Augmenting the scalar sector only does not provide an explanation for the observed pattern of the neutrino masses and mixing angles. 
Following a bottom-up approach, this Thesis discusses a well motivated framework that remedies at once such flaws of the SM: the $B-L$ model \cite{Jenkins:1987ue,Buchmuller:1991ce,Khalil:2006yi,BL_master_thesis}. This is 
 a {\it triply-minimal} extension of the SM. It is minimal in the gauge sector, in which a single $U(1)$ factor is added, related to the $B-L$ number, by simply promoting to local the already existing $U(1)_{B-L}$ global symmetry of the SM. It is minimal in the fermion sector, in which a SM singlet fermion per generation is added, to cure the new $U(1)_{B-L}$ related anomalies. These fermions can naturally be interpreted as the RH neutrinos. It is minimal in the scalar sector, in which a complex neutral scalar singlet is added to spontaneously break the new $U(1)$ symmetry, and at the same time to give to the new gauge boson a mass (to evade present experimental bounds).
The two latter points, once the $U(1)_{B-L}$ symmetry is spontaneously broken, naturally provide a dynamical implementation of the see-saw mechanism, explaining the neutrino masses. As we will see, the remnant degree of freedom of the new complex scalar severely impinges in the phenomenology of the scalar sector.


As mentioned, we follow a bottom-up approach, being mainly interested in studying the phenomenology of the model at colliders. In the light of this, we are not here interested in seeking an embedding of this model into grand unified theories (GUT). Direct consequence is that the gauge coupling related to the extra $U(1)$ factor, controlling the interaction of the $Z'$ boson, is a free parameter.
Even though we are not interested in studying the details of it, we are aware that extra $U(1)$ factors appear naturally in GUT scenarios, conceptually motivating our minimal extension.

The general model we introduce is a one-dimensional class of $U(1)$ extensions of the SM, in which each element is characterised by the properties of the new gauge boson associated to the extra $U(1)$ factor. In all generality, the latter couples to fermions proportionally to a linear combination  of the standard hypercharge ($Y$) and the $B-L$ number. The coefficients of this linear combination are $2$ \footnote{In extensions of the SM with $n$ $U(1)$ factors, the Abelian gauge group mix. For consistency, $\displaystyle \sum _{k=1}^{n+1}k$ gauge couplings appear. In our model, where $n=1$, we will have $2$ more gauge coupling besides the SM one, $g$ \cite{delAguila:1988jz,delAguila:1995rb,Chankowski:2006jk}. More details are given in section~\ref{sect:min_ext}.} gauge couplings, that in our bottom-up approach are free parameters. A variety of standard benchmark models is then recovered. For example, the sequential standard model (SSM), the $U(1)_R$, the $U(1)_\chi$, and the `pure' $B-L$ model, all anomaly-free with the SM particle content augmented by the RH neutrinos. The latter one is the benchmark model on which we will focus our numerical study.

The `pure' $B-L$ model is identified by the fact that the extra gauge boson, or $Z'_{B-L}$, couples to fermions proportionally to their $B-L$ number only.
On the one side, this directly implies a vanishing $Z-Z'$ mixing (at the tree-level), that is consistent with the existing tight constraints on such mixing, compatible with a negligible value \cite{Abreu:1994ria}. Moreover, the $B-L$ charge does not distinguish the chirality, i.e., the LH and the RH degrees of freedom of the same fermion have the same $B-L$ quantum numbers. Hence, Dirac fermions have just vectorial couplings to the $Z'_{B-L}$ boson:
\begin{equation}
g_{Z'}^R=g_{Z'}^L \Rightarrow \left\{
\begin{array}{ccccc}
g_{Z'}^V &=& \displaystyle \frac{g_{Z'}^R+g_{Z'}^L}{2} &=& (B-L)\; g'_1\, ,\\
g_{Z'}^A &=& \displaystyle \frac{g_{Z'}^R-g_{Z'}^L}{2} &=& 0\, .
\end{array}\right.
\end{equation}
On the contrary, the mass-eigenstates for the neutrinos, after the see-saw mechanism, are Majorana particles. Therefore, they have pure axial couplings to the $Z'_{B-L}$ boson \cite{Perez:2009mu}.

As a consequence, we decided not to study the asymmetries of the decay products stemming from the $Z'_{B-L}$ boson, given their trivial distribution at the peak. However, asymmetries can be important in the interference region, especially just before the $Z'$ boson peak, where the $Z-Z'$ interference will effectively provide an asymmetric distribution somewhat milder than the case in which there is no $Z'$ boson. This is a powerful method of discovery and identification of a $Z'$ boson and it deserves future investigations.

It is important to emphasise that this is a TeV scale extension of the SM. This means that the $U(1)_{B-L}$ breaking vacuum expectation value (VEV) is of $\mathcal{O}$(TeV). Hence, also the new particles will have masses at the TeV scale. In particular, we will consider both the $Z'$ boson and the new Higgs particle with masses in the $0.1 \div 10$ TeV range. Regarding the neutrinos, with a suitable choice of the parameters, the $3$ light neutrinos will have masses in the sub-eV range, while the $3$ heavy ones will have masses of $\mathcal{O}(100)$ GeV.

An attractive feature of the $B-L$ model is that a successful realisation of the mechanism of baryogenesis through leptogenesis (to explain the observed matter-antimatter asymmetry) might be possible in this model, as shown in Refs.~\cite{Abbas:2007ag,Iso:2010mv}.
It is very suggestive that this realisation takes place at the TeV scale. However, its correct implementation results in constraints on the neutrino masses, both on the light ones and on the heavy ones, and on their {\it CP}-violating phases.


As already stated, in this Thesis we are mainly concerned with the impact of the model at colliders, on which the above phases do not play any role. Also, to maximise and highlight the interesting patterns that can be observed, and for illustrative purposes, we decided to not include here the above constraints on the neutrino masses. To reconcile, the results that we will show can be thought to be valid for at least one generation.

Bearing this in mind, we will
consider all the heavy neutrinos as degenerate and with masses that are free parameters, varying them in the $50 \div 500$ GeV range. Hence, in most of the cases, smaller than $Z'$ boson mass. We will show that the $Z'$ boson in this model, with TeV scale heavy neutrinos, can decay into pairs of the latter. Another difference with respect to the traditional $Z'$ literature is, therefore, that the $Z'_{B-L}$ boson branching ratios (BRs) are not fixed. Neither its intrinsic width is, being the gauge coupling as well a free parameter. 

The presence of new coupled matter, the heavy neutrinos, has important phenomenological consequences. The possibility of the $Z'$ boson (and of the Higgs bosons, as we will show) to decay into pairs of them, will provide new and exciting signatures. It is worth to briefly mention that the peculiar decays of the Higgs bosons into pairs of heavy neutrinos, or into pairs of $Z'$ bosons, is a distinctive signature of this model, offering the chance to distinguish it from the plethora of the otherwise identical, concerning the scalar sector, singlet extensions of the SM in the literature.

In this Thesis we decided to study the details of the gauge and fermion sectors. Their mutual interactions are fully included in the $Z'$ decay into pairs of heavy neutrinos. Altogether, this decay provides new and spectacular multi-lepton signatures of the $Z'$ boson. One of the main results of this Thesis is the study of one of them, the tri-lepton decay mode (i.e., when the $Z'$ decays into exactly $3$ charged leptons and other particles, such as jets and/or missing energy), together with the related backgrounds. We will present a parton level strategy for reducing the latter in order to isolate the signal, that will be validated at the detector level.

It is very interesting that heavy neutrinos can be long-lived particles. For the experimental community, this model represents a test laboratory to study trigger efficiencies and detector resolutions for such a clear signal of physics beyond the SM (see Ref.~\cite{Fairbairn:2006gg} for a review). Moreover, from the simultaneous measurement of both the heavy neutrino decay length (from the displacement of its secondary vertex in the detector) and mass (we will show that this task is achievable in the tri-lepton signature) one can estimate the absolute mass of the parent light neutrino, for which at present, only limits exist. Altogether, this realises a spectacular and very peculiar link between very high energy and very low energy physics.

It shall be noticed that a first insight on the phenomenology of the pure $B-L$ model at the LHC was presented in Ref.~\cite{Emam:2007dy}.

%
%

\vspace{0.5cm}

The general structure of each chapter is to present the constraints for the new parameters in each sector and to study the phenomenology related to the new particles in the $B-L$ model, with respect to the SM. They concern the parameters in the gauge sector (i.e., the $Z'$ mass and the gauge coupling $g'_1$), in the fermion sector (i.e., the light and the heavy neutrino masses), and in the scalar sector (i.e., the scalar masses and mixing angle $\alpha$). 

Both experimental searches and theoretical arguments can give informations about the regions of the parameters that are allowed.

From the experimental side, past and existing facilities have looked for new particles, directly and indirectly, still with no success. This poses constraints on the parameters to evade limits. Tevatron, still accumulating data, is updating its constraints from direct searches, but for many cases the most stringent bounds come from the searches performed at LEP.

The LEP searches are still setting some of the most stringent bounds in all the sectors. Key factors for their success are manifold:  
the strength of the interaction between the Higgs boson and the gauge bosons, in the SM; the clean experimental environment of such a leptonic collider; the enormous amount of data collected at the SM $Z$ boson peak. They allow, in turn, to set stringent limits also in the $B-L$ model, on the scalar masses (as a function of the scalar mixing angle), on the $Z'$ mass and $g'_1$ coupling, and on the heavy neutrino masses.

The studies performed with the data collected by LEP (at the SM $Z$ boson peak and beyond) are the so-called precision tests of the SM, since certain observables are measured with per mil accuracy. These observables are usually referred to as precision observables. The level of agreement with the SM predictions is remarkable. This agreement results in tight constraints for most of the SM extensions, since they generally predict new contributions to the precision observables (usually to be evaluated at loop-level). In studying these limits, we did not pursue a complete beyond the tree-level study; we rather refer to the literature for that. In fact, we decided to concentrate on the study of the collider signatures and to leave the precision test analysis for future investigations.

Tevatron is currently collecting data and updating its analyses regarding the SM Higgs boson and $Z'$ boson masses (of a discrete choice of models) from direct searches. However, for the latter, the $Z'_{B-L}$ is not part of its traditional literature. Hence, the most up-to-date limits on the $Z'_{B-L}$ mass at  Tevatron, presented in section~\ref{sec:expbounds:Zp_Tev}, are a novel contribution (firstly in Ref.~\cite{Basso:2010pe}).

From the theoretical side, the model has to be well-defined. Several meanings apply to this concept, the main ones regarding the scalar potential to be bounded from below and with a stable vacuum solution, and the perturbative expansion to hold. The most powerful techniques require to go beyond the tree-level order, by studying the evolution of the renormalisation group equations (RGEs) of the parameters of the model.
The imposition of specific relations
results in constraints on the yet free parameters. At the same time, strong constraints, especially on the scalar masses, come from requiring the model to be unitary.

We present the analysis of the theoretical constraints on the new parameters of the model, namely the unitarity bound and those coming from the study of the RGEs. The unitarity assumption \cite{Basso:2010jt} relies on demanding that the cross sections for the scattering of all the particles in the model are unitarised, i.e., with an occurring probability less than one. The tightest bounds come from the scattering of the longitudinally-polarised gauge bosons, and they constrain mainly the scalar masses.

The RGE constraints \cite{Basso:2010jm} come from the analysis of the model at one loop, assuming the evolved parameters do not hit any Landau pole or spoil the theory in the scalar sector, e.g., the vacuum of the model. The latter request is known as the vacuum stability condition, while the former is the triviality condition.

The one-loop RGEs we collected from the literature have been completed with direct calculations of the top quark one and of those regarding the scalar sector. The complete set is collected in Appendix~\ref{App:RGE}. As introduced in section~\ref{sect:min_ext}, all the parameters of the model are running parameters, and the requirement of some to vanish at the EW scale just sets the boundary conditions of their evolution. In our case, the mixing parameter in the gauge sector, i.e., $\widetilde{g}$, is set to vanish at the EW scale, defining the `pure' $B-L$ model, argument of the numerical analysis of this Thesis. However, its evolution must be taken into account.


The numerical study of the collider phenomenology in this Thesis has been done with the \textsc{calchep} tool~\cite{Pukhov:2004ca} (where not specified otherwise). The model has been implemented therein with the \textsc{lanhep} module~\cite{Semenov:1996es}. 
The availability of the model implementation into \textsc{calchep} in both the
unitary and t'Hooft-Feynman gauges allowed us to perform powerful
cross-checks to test the consistency of the model itself.

The RGE have been solved and studied in \textsc{mathematica} $5.0$ \cite{mathematica5}.

\section{Outline}

The Thesis is organises as follows. Chapter~\ref{Ch:2} introduces the model and contains a description of its consequences, i.e., spontaneous symmetry breaking and the particle spectrum.

Chapter~\ref{Ch:3} presents the study of the gauge sector. It starts with the constraints on the free gauge coupling and on the $Z'$ mass and it studies the $Z'$ boson production cross sections and $2$-body decays BRs.
Finally, the discovery reach at Tevatron and LHC of the $Z'$ boson in Drell-Yan processes is shown.

Chapter~\ref{Ch:4} describes the study of the fermion sector. After reviewing the limits on neutrino masses, a detailed description of the decay properties of the heavy neutrinos is given. Their main production mechanism is in pair via the $Z'$ boson, 
and, bringing pieces together, we show that they can induce new spectacular decay patterns for the latter. Among them, we perform detailed parton and detector level analyses of one specific pattern, the tri-lepton decay mode. This is the main result of this Thesis.

Chapter~\ref{Ch:5} analyses the theoretical bounds, i.e., unitarity, triviality, vacuum stability, and some experimental constraints on the scalar sector, presenting also production cross sections and BRs for the Higgs bosons, as well as cross sections for some signatures involving decays of the Higgs bosons into pairs of $Z'$ bosons and heavy neutrinos.

In appendix~\ref{App:calchep} is explained the implementation of the model in the numerical tools used and some related technicalities. We also list the complete set of Feynman rules for the pure $B-L$ model. In appendix~\ref{App:RGE} are collected and described the RGEs of the model.



%

\chapter{The Model}\label{Ch:2}
\ifpdf
    \graphicspath{{Chapter2/Chapter2Figs/PNG/}{Chapter2/Chapter2Figs/PDF/}{Chapter2/Chapter2Figs/}}
\else
    \graphicspath{{Chapter2/Chapter2Figs/EPS/}{Chapter2/Chapter2Figs/}}
\fi


In this chapter we introduce the model that we will study in this Thesis. We will introduce the notation and highlight the differences with respect to the SM. For the motivations, we refer to the introduction.

The model under study is a triply-minimal extension of the SM. It is minimal in the gauge sector, where the SM gauge group is augmented by a single $U(1)$ factor, related to the baryon minus lepton ($B-L$) gauged number. It is minimal in the scalar sector, where a complex scalar singlet is required for the spontaneous breaking of the new $U(1)$ gauge factor, that we assume to take place at the TeV scale. Finally, it is minimal in the fermion sector, in which a SM singlet
RH fermion per generation is introduced. Its $B-L$ charge, in particular, is chosen to cure the new gauge and mixed $U(1)-$gravitational anomalies, related to the $U(1)_{B-L}$ gauge symmetry factor \cite{BL_master_thesis}. It is important to stress that three and only three RH fermions are introduced (one per generation, as required by the anomally cancellation conditions), that can naturally be identified with the RH neutrinos.

The extra Higgs singlet needs to be charged under $B-L$ to trigger the spontaneous symmetry breaking of the related $U(1)$ factor. Its charge is then fixed by the gauge invariance of the Yukawa term that couples it to the RH neutrinos. Altogether, this provides a natural and dynamical framework for the implementation of the type-I see-saw mechanism, contrary to its usual realisation through an effective mass term for the RH neutrinos. Hence, in this model, neutrinos are naturally massive particles and their masses are generated through the Higgs mechanism.

\nomenclature[zBL]{$B-L$}{Baryon minus lepton number}
\nomenclature[zSM]{$SM$}{Standard model}
\nomenclature[zRGE]{$RGE$}{Renormalisation group equation}
\nomenclature[zRH]{$RH$}{Right-handed}
\nomenclature[zLH]{$LH$}{Left-handed}
\nomenclature[zVEV]{$VEV$}{Vacuum expectation value}
\nomenclature[zEWSB]{$EWSB$}{Electro-weak symmetry breaking}
\nomenclature[zCM]{$CM$}{Center-of-mass}
\nomenclature[zBR]{$BR$}{Branching ratio}
\nomenclature[zLC]{$LC$}{Linear collider}
\nomenclature[zLHC]{$LHC$}{Large hadron collider}
\nomenclature[zLEP]{$LEP$}{Large electron positron (collider)}
\nomenclature[zPDF]{$PDF$}{Parton distribution function}
\nomenclature[zGUT]{$GUT$}{Grand unified theory}
\nomenclature[zNNLO]{$(N)NLO$}{(Next-to-)next-to-leading-order}

\nomenclature[gp]{$\pi$}{ $\simeq 3.1415$}                                             

\section{A minimal $U(1)$ extension}\label{sect:min_ext}

As mentioned, we present here the general model under study, a minimal $U(1)$ extension of the SM, related to the $B-L$ number. The classical gauge invariant Lagrangian, obeying the $SU(3)_C\times SU(2)_L\times U(1)_Y\times U(1)_{B-L}$
gauge symmetry, can be decomposed as:
\begin{equation}\label{L}
\mathscr{L}=\mathscr{L}_s + \mathscr{L}_{YM} + \mathscr{L}_f + \mathscr{L}_Y \, ,
\end{equation}
where the terms on the right hand side identify the scalar part, the gauge (or Yang-Mills) part, the fermion part and the Yukawa part, respectively.

\subsection{Scalar sector}
The spontaneous symmetry breaking of the new extra $U(1)$ factor in the gauge group is achieved by introducing in the particle content a further scalar field, that is required to acquire a VEV at the TeV scale. To keep the realisation of the Higgs mechanism minimal, only one extra degree of freedom (beside the one that will become a physical particle) is required, to give mass to the new neutral gauge field associated to the extra  $U(1)$ factor. Hence, the minimal choice for the new scalar is a complex singlet, that we will denote as $\chi$.

The scalar Lagrangian is
\begin{equation}\label{new-scalar_L}
\mathscr{L}_s=\left( D^{\mu} H\right) ^{\dagger} D_{\mu}H + 
\left( D^{\mu} \chi\right) ^{\dagger} D_{\mu}\chi - V(H,\chi ) \, ,
\end{equation}
with the scalar potential \index{Scalar potential} given by
\begin{eqnarray}\nonumber
V(H,\chi )&=&m^2H^{\dagger}H + \mu ^2\mid\chi\mid ^2 +\left( \begin{array}{cc} H^{\dagger}H& \mid\chi\mid ^2\end{array}\right)
					\left( \begin{array}{cc} \lambda _1 & \frac{\lambda _3}{2} \\
			 \frac{\lambda _3}{2} & \lambda _2 \\ \end{array} \right) \left( \begin{array}{c} H^{\dagger}H \\ \mid\chi\mid ^2 \\ \end{array} \right)\\
			  \nonumber \\ \label{BL-potential}
		&=&m^2H^{\dagger}H + \mu ^2\mid\chi\mid ^2 + \lambda _1 (H^{\dagger}H)^2 +\lambda _2 \mid\chi\mid ^4 + \lambda _3 H^{\dagger}H\mid\chi\mid ^2  \, ,
\end{eqnarray}
where $H$ and $\chi$ are the complex scalar Higgs doublet and singlet fields. Regarding the new $B-L$ charges for these scalar fields, we take $0$ and $+2$ for $H$ and $\chi$, respectively. Notice that the charge of the $\chi$ field has been chosen to ensure the gauge invariance of the fermion sector for the minimal model under discussion. A non-minimal matter content would lead to a different requirement for the charge of $\chi$ (see, e.g., Ref.~\cite{Khalil:2010iu}).

\subsection{Yang-Mills sector}
Moving to the $\mathscr{L}_{YM}$, the non-Abelian field strengths therein are the same as in the SM whereas the Abelian ones can be written as follows:
\begin{equation}\label{La}
\mathscr{L}^{\rm Abel}_{YM} = 
-\frac{1}{4}F^{\mu\nu}F_{\mu\nu}-\frac{1}{4}F^{\prime\mu\nu}F^\prime _{\mu\nu}\, ,
\end{equation}
where
\begin{eqnarray}\label{new-fs3}
F_{\mu\nu}		&=&	\partial _{\mu}B_{\nu} - \partial _{\nu}B_{\mu} \, , \\ \label{new-fs4}
F^\prime_{\mu\nu}	&=&	\partial _{\mu}B^\prime_{\nu} - \partial _{\nu}B^\prime_{\mu} \, .
\end{eqnarray}
We define $B_\mu$ and $B'_\mu$ as the $U(1)_Y$ and $U(1)_{B-L}$ gauge fields, respectively.
In this field basis, the covariant derivative \index{Covariant derivative} is \footnote{In all generality, Abelian field strengths tend to mix. The kinetic term can be diagonalised with a $GL(2,R)$ transformation, leading to the form of the covariant derivative in eq.~(\ref{cov_der}), where the mixing between the two $U(1)$ factors becomes once again evident \cite{delAguila:1995rb,Chankowski:2006jk}. For the curios reader, we refer to Ref.~\cite{BL_master_thesis} for a more comprehensive explanation.}:
\begin{equation}\label{cov_der}
D_{\mu}\equiv \partial _{\mu} + ig_S T^{\alpha}G_{\mu}^{\phantom{o}\alpha} 
+ igT^aW_{\mu}^{\phantom{o}a} +ig_1YB_{\mu} +i(\widetilde{g}Y + g_1'Y_{B-L})B'_{\mu}\, .
\end{equation}

In our bottom-up approach, we will not require gauge unification at some specific, yet arbitrary, energy scale \footnote{The gauge unification could in principle happen in a multi-step process, passing through intermediate gauge groups until the final single group.}, fixing the conditions for the extra gauge couplings at that scale. Therefore, in this model, the gauge couplings $\widetilde{g}$ and $g'_1$ are free parameters. To better understand their meaning, let us focus on eq.~(\ref{cov_der}).
This form of the covariant derivative can be re-written defining an effective coupling $Y^E$ and an effective charge $g_E$:
\begin{equation}\label{eff_par}
g_E Y^E \equiv \, \widetilde{g}Y + g_1'Y_{B-L}.
\end{equation}

As any other parameter in the Lagrangian, $\widetilde{g}$ and $g_1'$ are running parameters \cite{delAguila:1988jz,delAguila:1995rb,Chankowski:2006jk}, therefore their values ought to be defined at some scale. A discrete set of popular $Z'$ models (see, e.g., Refs.~\cite{Carena:2004xs,Appelquist:2002mw}) can be recovered by a suitable definition of both $\widetilde{g}$ and $g_1'$.

We will focus in our numerical analysis on the `pure' $B-L$ model, that is defined by the condition $\widetilde{g}(Q_{EW}) = 0$, i.e., we nullify it at the EW scale. This implies no mixing at the tree level between the $B-L$ $Z'$ and the SM $Z$ gauge bosons. Other benchmark models of our general parameterisation are for example the Sequential SM (SSM), defined by $Y^E = Y$ (that in our notation corresponds to the conditions $g'_1=0$ and $\widetilde{g}=g_1$ at the EW scale) and the $U(1)_R$ model, for which RH fermion charges vanish (that is recovered here by the condition $\widetilde{g}=-2g'_1$ at the EW scale).

It is important to note that none of the models described so far is orthogonal to the $U(1)_Y$ of the SM, therefore the RGE running of the fundamental parameters, $\widetilde{g}$ and $g_1'$, will modify the relations above. The only orthogonal $U(1)$ extension of the SM is the `$SO(10)$-inspired' $U(1)_\chi$ model, that in our notation reads $\widetilde{g}=-\frac{4}{5}g_1'$. Although the $\widetilde{g}$ and $g_1'$ couplings run with a different behaviour, the EW relation $\widetilde{g}/g_1' = -4/5$ is preserved (at one loop) at any scale.

\subsection{Fermion sector}
The fermion Lagrangian (where $k$ is the
generation index) is given by
\begin{eqnarray} \nonumber
\mathscr{L}_f &=& \sum _{k=1}^3 \Big( i\overline {q_{kL}} \gamma _{\mu}D^{\mu} q_{kL} + i\overline {u_{kR}}
			\gamma _{\mu}D^{\mu} u_{kR} +i\overline {d_{kR}} \gamma _{\mu}D^{\mu} d_{kR} +\\ \label{L_f}
			  && + i\overline {l_{kL}} \gamma _{\mu}D^{\mu} l_{kL} + i\overline {e_{kR}}
			\gamma _{\mu}D^{\mu} e_{kR} +i\overline {\nu _{kR}} \gamma _{\mu}D^{\mu} \nu
			_{kR} \Big)  \, ,
\end{eqnarray}
 where the fields' charges are the usual SM and $B-L$ ones (in particular, $B-L = 1/3$ for quarks and $-1$ for leptons with no distinction between generations, hence ensuring universality).
  The $B-L$ charge assignments of the fields as well as the introduction of new
  fermion RH heavy neutrinos ($\nu_R$'s, charged $-1$ under $B-L$)
  are designed to eliminate the triangular $B-L$ gauge anomalies of the theory. Regarding the new scalar Higgs field ($\chi$) the charge $+2$ under $B-L$ is chosen to ensure the gauge invariance of the model (for a detailed discussion, see Ref.~\cite{BL_master_thesis}).

  Therefore, the $B-L$  gauge extension of the SM gauge group broken at the TeV scale does necessarily require at least one new scalar field and
  three new fermion fields which are charged with respect to the $B-L$ group.

\subsection{Yukawa interactions}
Finally, the Yukawa interactions are
\begin{eqnarray}\nonumber
\mathscr{L}_Y &=& -y^d_{jk}\overline {q_{jL}} d_{kR}H 
                 -y^u_{jk}\overline {q_{jL}} u_{kR}\widetilde H 
		 -y^e_{jk}\overline {l_{jL}} e_{kR}H \\ \label{L_Yukawa}
	      & & -y^{\nu}_{jk}\overline {l_{jL}} \nu _{kR}\widetilde H 
	         -y^M_{jk}\overline {(\nu _R)^c_j} \nu _{kR}\chi +  {\rm 
h.c.}  \, ,
\end{eqnarray}
{where $\widetilde H=i\sigma^2 H^*$ and  $i,j,k$ take the values $1$ to $3$},
where the last term is the Majorana contribution and the others are the usual Dirac ones. Without loss of generality, we work on the basis in which the RH neutrino Yukawa coupling matrices, $y^M$, are diagonal, real, and positive. These are the only allowed gauge invariant terms \footnote{Notice that an effective mass term such as $M\overline {(\nu _R)^c}\nu_R$ is now forbidden by the gauge invariance.}. In particular, the last term in eq.~(\ref{L_Yukawa}) couples the neutrinos to the new scalar singlet field, $\chi$, and it allows for the dynamical generation of neutrino masses, as $\chi$ acquires a VEV through the Higgs mechanism. Due to this, in the $B-L$ model, the neutrinos couple to the scalar sector, in particular the right-handed (RH) ones couple strongly to the singlet scalar.

Neutrino mass eigenstates, obtained after applying the see-saw mechanism, will be called $\nu_l$ (with $l$ standing for light) and $\nu_h$
(with $h$ standing for heavy), where the first ones are the SM-like ones (see section~\ref{sect:neutrino_masses}). With a reasonable choice of Yukawa couplings, the heavy neutrinos can have masses $m_{\nu_h} \sim \mathcal{O}(100)$ GeV.

\section{Symmetry breaking}

We generalise here the SM discussion of spontaneous EW symmetry breaking (EWSB) to the more complicated classical potential of eq.~(\ref{BL-potential}). To determine the condition for $V(H,\chi )$ to be bounded from below, it is sufficient to study its behaviour
for large field values, controlled by the matrix in the first line of eq.~(\ref{BL-potential}). Requiring such a matrix to be 
positive definite, we obtain the conditions: \index{Scalar potential!Bounded-from-below}
\begin{eqnarray}\label{inf_limitated}
4 \lambda _1 \lambda _2 - \lambda _3^2 &>&0 \, , \\ \label{positivity}
\lambda _1, \lambda _2 &>& 0 \, .
\end{eqnarray}

If the conditions of eqs.~(\ref{inf_limitated}) and (\ref{positivity}) are satisfied, we can proceed to the minimisation of $V$ as a function of a constant VEV for the two Higgs fields. In the Feynman gauge, we can parametrise the scalar fields as \index{Higgs fields}
\begin{equation}\label{Higgs_goldstones}
H=\frac{1}{\sqrt{2}}
\left(
\begin{array}{c}
-i(w^1-iw^2) \\
v+(h+iz)
\end{array}
\right), \qquad
\chi =
\frac{1}{\sqrt{2}}
\left(x+(h'+iz')\right),
\end{equation}
where $w^{\pm}=w^1\mp iw^2$, $z$ and $z'$ are the would-be Goldstone
bosons of $W^{\pm}$, $Z$ and $Z'$, respectively.
Making use of gauge invariance, for the minimisation of the scalar potential, it is not restrictive to assume \index{Higgs fields! VEVs}
\begin{equation}\label{min}
\left< H \right> \equiv \frac{1}{\sqrt{2}} \left( \begin{array}{c} 0 \\ v \end{array} \right)\, , 
	\hspace{2cm} \left< \chi \right> \equiv \frac{x}{\sqrt{2}}\, ,
\end{equation} 
with $v$ and $x$ real and non-negative. The physically most interesting solutions to the minimisation of eq.~(\ref{BL-potential}) are obtained for $v$ and $x$ both non-vanishing:
\begin{eqnarray}\label{sol_min1}
v^2 &=& \frac{-\lambda _2 m^2 + \frac{\lambda _3}{2}\mu ^2}{\lambda _1 \lambda _2 - \frac{\lambda _3^{\phantom{o}2}}{4}} \, ,\\
\nonumber  \\ \label{sol_min2}
x^2 &=& \frac{-\lambda _1 \mu ^2 + \frac{\lambda _3}{2}m ^2}{\lambda _1 \lambda _2 - \frac{\lambda _3^{\phantom{o}2}}{4}} \, .
\end{eqnarray}

Physical solutions must have both $v^2>0$ and $x^2>0$ conditions satisfied. Eq~(\ref{inf_limitated}) implies the denominators are always positive, so we just need to require the numerators to be positive too. In terms of the parameters in the Lagrangian, this in turn means
\begin{displaymath} 
\left\{ 
\begin{array}{ccc} 
\displaystyle \lambda _2 m^2 & < & \displaystyle \frac{\lambda _3}{2} \mu ^2  \, , \\ &&\\
\displaystyle \lambda _1 \mu^2 & < &\displaystyle  \frac{\lambda _3}{2} m ^2 \, . 
\end{array} 
\right. 
\end{displaymath}
It is interesting to note that there are no solutions for both $\mu ^2,m ^2 >0$, irrespectively of $\lambda _3$. Solutions with
($\mu ^2>0$, $m^2<0$) or ($\mu ^2<0$, $m^2>0$) are allowed only for $\lambda _3 <0$. Solutions for ($\mu ^2<0$, $m^2<0$) are
allowed for both signs of $\lambda _3 $.

\subsection{Gauge eigenstates}\label{subsec:gauge-eigen}

To determine the gauge boson spectrum, we have to expand the scalar kinetic terms as for the SM. We expect
that there exists a massless gauge boson, the photon, whilst the other gauge bosons become massive. The extension we are studying
is in the Abelian sector of the SM gauge group, so that the charged gauge bosons $W^\pm$ will have masses given by their SM expressions,
being related to the $SU(2)_L$ factor only. Using the unitary-gauge parameterisation, 
 the kinetic terms in eq.
(\ref{new-scalar_L}) become:%
\begin{eqnarray}\nonumber
\left( D^{\mu} H\right) ^\dagger D_{\mu}H &=& \frac{1}{2}\partial ^{\mu} h \partial _{\mu}h + \frac{1}{8} (h+v)^2 \big(
				0\; 1 \big) \Big[ g W_a ^{\phantom{o}\mu }\sigma _a + g_1B^{\mu}+\widetilde g B'^{\mu}
				\Big] ^2 \left( \begin{array}{c} 0\\1\end{array} \right) \\ \nonumber
	&=& \frac{1}{2}\partial ^{\mu} h \partial _{\mu}h + \frac{1}{8} (h+v)^2 \left[ g^2 \left| W_1 ^{\phantom{o}\mu } -
		iW_2 ^{\phantom{o}\mu } \right| ^2  \right.\\ \label{boson_masses1}
	&&	\left. \hspace{4cm} + \left( gW_3 ^{\phantom{o}\mu } - g_1 B^{\mu} - \widetilde g B'^{\mu}\right) ^2 \right] 
\end{eqnarray}
and
\begin{eqnarray}\label{boson_masses2}
\left( D^{\mu} \chi\right) ^\dagger D_{\mu}\chi &=& \frac{1}{2}\partial ^{\mu} h' \partial _{\mu}h' + \frac{1}{2}(h'+x)^2
(g_1' 2B'^{\mu})^2\, ,
\end{eqnarray}
where we have taken $Y^{B-L}_\chi = 2$ in order to guarantee the gauge invariance of the Yukawa terms (see eq. (\ref{L_Yukawa})).
In eq.~(\ref{boson_masses1}) we can recognise immediately the SM
charged gauge bosons $W^\pm$, with $\displaystyle M_W=gv/2$ as in the SM. The other gauge boson masses are
not so simple to identify, because of mixing. In fact, in analogy with the SM, the fields of definite mass are linear combinations
of $B^\mu$, $W_3^\mu$ and $B'^\mu$. The explicit expressions are: \index{Mixing! Gauge bosons}
\begin{equation}\label{neutral_bosons}
\left( \begin{array}{c} B^{\mu} \\ W_3^{\phantom{o}\mu}\\ B'^{\mu} \end{array}\right) = \left(
		\begin{array}{ccc}
		\cos{\vartheta _w} & -\sin{\vartheta _w}\cos{\vartheta '} & \sin{\vartheta _w}\sin{\vartheta '}\\
		\sin{\vartheta _w} & \cos{\vartheta _w}\cos{\vartheta '} & -\cos{\vartheta _w}\sin{\vartheta '}\\
		0 & \sin{\vartheta '} & \cos{\vartheta '}
		\end{array} \right) \left( \begin{array}{c} A^{\mu} \\ Z^{\mu}\\ Z'^{\mu} \end{array}\right)\, ,
\end{equation}
with $-\frac{\pi}{4}\leq \vartheta '\leq \frac{\pi}{4}$, such that:
\begin{equation}\label{tan2theta_prime}
\tan{2\vartheta '}=\frac{2\widetilde{g}\sqrt{g^2+g_1^2}}{\widetilde{g}^2 + 16(\frac{x}{v})^2 g_1^{'2}-g^2-g_1^2}
\end{equation}
and
\begin{eqnarray}
M_A &=& 0\, ,\\ \label{Mzz'}
M_{Z,Z'} &=& \sqrt{g^2 + g_1^2}\cdot \frac{v}{2} \left[ \frac{1}{2}\left(  \frac{\widetilde{g}^2 + 16(\frac{x}{v})^2 g_1^{'2}}{g^2 + g_1^2}+1
	\right) \mp \frac{\widetilde{g}}{\sin{2\vartheta '}\sqrt{g^2+g_1^2}} \right] ^{\frac{1}{2}}\, ,
\end{eqnarray}
where
\begin{equation}\label{sin_thetap}
\sin{2\vartheta '}=\frac{2\widetilde{g}\sqrt{g^2+g_1^2}}{\sqrt{\left(\widetilde{g}^2 + 16(\frac{x}{v})^2 g_1^{'2}-g^2-g_1^2\right) ^2
		+(2\widetilde{g})^2(g^2+g_1^2)}}\, .
\end{equation}
The LEP experiments \cite{Abreu:1994ria} constrain $|\vartheta '| \lesssim 10^{-3}$. Present constraints on the VEV $x$ (see section~\ref{sec:expbounds:Zp}) allow a generous range of $\widetilde{g}$.

In the pure $B-L$ model, that we remind is defined by the condition $\widetilde{g} = 0$, one can easily see, from eq.~(\ref{sin_thetap}), that also
the mixing angle $\vartheta '$ vanishes, implying no mixing, at the tree level, between the SM $Z$ and the new $Z'$ gauge bosons, identically satisfying the LEP bound. The $Z$ and $Z'_{B-L}$ masses then simplify, respectively, to
\begin{eqnarray} 
M_Z &=& \sqrt{g^2 + g_1^2}\cdot \frac{v}{2} \, , \\ \label{Mz'}
M_{Z'_{B-L}} &=&2g'_1 x\, .
\end{eqnarray}

\subsection{Scalar eigenstates}

To compute the scalar masses, we must expand the potential in eq. (\ref{BL-potential}) around the minima
in eqs. (\ref{sol_min1}) and (\ref{sol_min2}).
We denote by $h_1$ and $h_2$ the scalar fields of definite masses, $m_{h_1}$ and $m_{h_2}$, respectively, and we conventionally choose $m^2_{h_1}~<~m^2_{h_2}$. After standard manipulations, the explicit expressions for the scalar mass eigenvalues and eigenvectors are \index{Mixing! Scalar bosons}
\begin{eqnarray}\label{mh1}
m^2_{h_1} &=& \lambda _1 v^2 + \lambda _2 x^2 - \sqrt{(\lambda _1 v^2 - \lambda _2 x^2)^2 + (\lambda _3 xv)^2} \, ,\\ \label{mh2}
m^2_{h_2} &=& \lambda _1 v^2 + \lambda _2 x^2 + \sqrt{(\lambda _1 v^2 - \lambda _2 x^2)^2 + (\lambda _3 xv)^2} \, ,
\end{eqnarray}
\begin{equation}\label{scalari_autostati_massa}
\left( \begin{array}{c} h_1\\h_2\end{array}\right) = \left( \begin{array}{cc} \cos{\alpha}&-\sin{\alpha}\\ \sin{\alpha}&\cos{\alpha}
	\end{array}\right) \left( \begin{array}{c} h\\h'\end{array}\right) \, ,
\end{equation}
where $-\frac{\pi}{2}\leq \alpha \leq \frac{\pi}{2}$ fulfils \footnote{In all generality, the whole interval $0\leq \alpha < 2\pi$ is halved
because an orthogonal transformation is invariant under $\alpha \rightarrow \alpha + \pi$. We could re-halve the interval by noting
that it is invariant also under $\alpha \rightarrow -\alpha$ if we permit the eigenvalues inversion, but this is forbidden by our
convention $m^2_{h_1} < m^2_{h_2}$. Thus $\alpha$ and $-\alpha$ are independent solutions.}\label{scalar_angle}
\begin{eqnarray}\label{sin2a}
\sin{2\alpha} &=& \frac{\lambda _3 xv}{\sqrt{(\lambda _1 v^2 - \lambda _2 x^2)^2 + (\lambda _3 xv)^2}} \, ,\\ \label{cos2a}
\cos{2\alpha} &=& \frac{\lambda _1 v^2 - \lambda _2 x^2}{\sqrt{(\lambda _1 v^2 - \lambda _2 x^2)^2 + (\lambda _3 xv)^2}}\, .
\end{eqnarray}

Eqs.~(\ref{BL-potential}), (\ref{scalari_autostati_massa}), and (\ref{sin2a}) describe a rather general scalar sector as in the minimally extended SM with scalar singlets, in which the light(heavy) Higgs boson couples to SM matters proportionally to $\cos{\alpha}$($\sin{\alpha}$).
Proper differences come from having new coupled matter, as we will explain in chapter~\ref{Ch:5}. Notice that the light(heavy) Higgs boson couples to the new matter content (heavy neutrinos and $Z'$ boson) with the complementary angle, $\sin{\alpha}$($\cos{\alpha}$), respectively, as the latter is directly coupled only to the scalar singlet $\chi$.

It is interesting to study the decoupling limit (when $\alpha \to 0$). In this limit, $h_1$ is purely the SM-like boson, while $h_2$ does not couple to SM particles, making its discovery impossible at the LHC. We will further comment on this in section~\ref{sect:Higgs}.
The other possible decoupling limit, 
for $\alpha \to \frac{\pi}{2}$ (where $h_2$ is the SM-like Higgs boson), is excluded by present experimental constraints \cite{Dawson:2009yx}.

For our numerical study of the extended Higgs sector, it is useful to invert eqs.~(\ref{mh1}), (\ref{mh2}) and (\ref{sin2a}), 
to extract the parameters in the Lagrangian in terms of the physical quantities $m_{h_1}$, $m_{h_2}$ and $\sin{2\alpha}$:
\begin{eqnarray}
\lambda _1 &=& \frac{m_{h_2}^2}{4v^2}(1-\cos{2\alpha}) + \frac{m_{h_1}^2}{4v^2}(1+\cos{2\alpha}),\\ \label{inversion_lam2}
\lambda _2 &=& \frac{m_{h_1}^2}{4x^2}(1-\cos{2\alpha}) + \frac{m_{h_2}^2}{4x^2}(1+\cos{2\alpha}),\\ \label{inversion}
\lambda _3 &=& \sin{2\alpha} \left( \frac{m_{h_2}^2-m_{h_1}^2}{2xv} \right).
\end{eqnarray}

\subsection{Fermion eigenstates}\label{sect:neutrino_masses}
The last line in eq.~(\ref{L_Yukawa}) contains the Dirac and the Majorana mass terms for the neutrinos, respectively. In contrast to the usual effective implementation of the see-saw mechanism, the last term of eq.~(\ref{L_Yukawa}) is a proper Yukawa interaction with the new Higgs singlet.

To extract the neutrino masses we have to diagonalise the neutrino mass matrix from eq. (\ref{L_Yukawa}):
\begin{equation}\label{nu_mass_matrix} 
{\mathscr{M}} = 
\left( \begin{array}{cc} 0 & m_D \\ 
                  m_D &  M 
 \end{array} \right)\, , 
\end{equation} 
where
\begin{equation} 
m_D = \frac{y^{\nu}}{\sqrt{2}} \, v \, , \qquad M = \sqrt{2} \, y^{M} \, x \, ,
\end{equation}
where $x$ is the VEV of the $\chi$ field. This matrix can be  diagonalised 
by a rotation about an angle $\alpha _\nu$, such that: \index{Mixing! Neutrinos}
\begin{equation}\label{nu_mix_angle} 
\tan{2 \alpha_\nu} = -\frac{2m_D}{M}\, .
\end{equation}

For simplicity we neglect the inter-generational mixing
so that neutrinos of each generation can be
diagonalised independently.
We also require that the neutrinos be mass 
degenerate. 
Thus, $\nu_{L,R}$ can be written as the following linear combination
 of Majorana mass eigenstates $\nu_{l,h}$ :
\begin{equation}\label{nu_mixing} 
\left( \begin{array}{c} \nu_L\\ \nu_R \end{array} \right) = 
\left( \begin{array}{cc} 
\cos{\alpha _\nu} & -\sin{\alpha_\nu} \\ 
\sin{\alpha _\nu} &\cos{\alpha _\nu} 
\end{array} \right) \times \left( \begin{array}{c} \nu_l\\ \nu_h \end{array} \right)\, , 
\end{equation} 
whose masses are respectively
\begin{eqnarray}
m_{\nu _l} & \cong & \frac{m_D^2}{M}\, ,\\
m_{\nu _h} & \cong & M\, .
\end{eqnarray}

We can now rewrite eq.~(\ref{nu_mix_angle}) in term of the physical masses:
\begin{equation}\label{nu_mix_angle_masses} 
\tan{2 \alpha_\nu} \cong -2 \sqrt{\frac{m_{\nu _l}}{m_{\nu _h}}}\, .
\end{equation}
For $m_{\nu _l} \sim 0.1$ eV and $m_{\nu _h} \sim 100$ GeV, $\tan{\alpha_\nu} \sim 10^{-6}$. Hence, $\nu_l$($\nu_h$) are mostly the LH(RH) neutrinos. One consequence is that the heavy neutrinos can be long-lived particles, as we will show in section~\ref{sect:nu_properties}. Also, being mostly RH, the heavy neutrinos are strongly coupled to the heavy(light) Higgs boson for small(big) scalar mixing angle $\alpha$, while the SM-like ones (being mostly LH) are not.

\section{Summary}
We summarise here the matter content of the pure $B-L$ model, argument of the numerical analysis of this Thesis. We list the new independent parameters with respect to the SM, with the used symbols.

\subsection{Gauge sector}
We remind the reader that the pure $B-L$ model is defined by $\widetilde{g}=0$.
Therefore, the only independent parameters in the gauge sector are the $g'_1$ coupling and the $Z'$ boson mass $M_{Z'}$ \footnote{Hereafter, when not confusing, we will call simply $Z'$ boson the $Z'_{B-L}$ (the gauge boson of the pure $B-L$ model), and $M_{Z'}$ its mass.}. The $B-L$ breaking VEV $x$ is then defined as
\begin{equation}
x=\frac{M_{Z'}}{2g'_1}\, .
\end{equation}

\subsection{Fermion sector}
All the neutrino masses are independent parameters of the model. For simplicity, we neglect the mixing within the generations and, furthermore, we take all the light neutrinos and all the heavy neutrinos to be degenerate in mass:
\begin{eqnarray}
m_{\nu ^1_l} = m_{\nu ^2_l} = m_{\nu ^3_l} &=& m_{\nu _l} \, , \\
m_{\nu ^1_h} = m_{\nu ^2_h} = m_{\nu ^3_h} &=& m_{\nu _h} \, .
\end{eqnarray}
The mixing angle $\alpha _\nu$ is then fixed, as in eqs.~(\ref{nu_mix_angle}) and (\ref{nu_mix_angle_masses}).

\subsection{Scalar sector}
The Higgs boson masses, $m_{h_1}$ and $m_{h_2}$, and mixing angle, $\alpha$, are the only new parameters in this sector. Notice that, by convention, $m_{h_1}~\leq~m_{h_2}$ and $-\frac{\pi}{2}~\leq~\alpha~\leq~\frac{\pi}{2}$.

Table~\ref{tab:quantum_number_assignation} summarises the charge assignments to fermion and scalar fields, while table~\ref{tab:symbols} shows the fields' labels and masses.
 
\begin{table}[!h]
\begin{center}
\begin{tabular}{|c||c|c|c||c|c|c||c|c|}\hline
$\boldsymbol {\psi} $   & $q_L$ & $u_R$ & $d_R$ & $\ell_L$ & $e_R$ & $\nu _R$ & $H$ & $\chi $ \\ \hline &&&&&&&&\\
$\boldsymbol {SU(3)_C}$ & $3$   & $3$   & $3$   &  $1$  & $1$   &  $1$     & $1$ & $1$ \\  &&&&&&&&\\
$\boldsymbol {SU(2)_L}$ & $2$   & $1$   & $1$   &  $2$  & $1$   &  $1$     & $2$ & $1$ \\  &&&&&&&&\\
$\boldsymbol {Y} $      & $\displaystyle\frac{1}{6}$ & $\displaystyle\frac{2}{3}$ & $\displaystyle -\frac{1}{3}$ &  $\displaystyle -\frac{1}{2}$ &  $-1$ &  $0$ & $\displaystyle \frac{1}{2}$ & $0$ \\  &&&&&&&&\\
$\boldsymbol {B-L} $  & $\displaystyle\frac{1}{3}$ & $\displaystyle\frac{1}{3}$ & $\displaystyle\frac{1}{3}$   & $-1$ & $-1$ & $-1$ & $0$ & $2$ \\  &&&&&&&&\\ \hline 
\end{tabular}
\end{center}\caption{\it $Y$ and $B-L$ quantum number assignation to chiral fermion and scalar fields. \label{tab:quantum_number_assignation}}
\end{table}

\begin{table}[!h]
\begin{center}
\begin{tabular}{|c||c|c||c|c||c|c|}\hline
\mbox{Name} & \mbox{Quarks} & \mbox{Leptons} & \multicolumn{2}{|c||}{Neutrinos} & \multicolumn{2}{|c|}{Higgses} \\ \hline &&&&&&\\
$\boldsymbol \psi $   	& $q$ 	& $\ell$ & $\nu _l$ & $\nu _h$ & $h_1$   & $h_2$ \\ &&&&&&\\ 
\mbox{\bf Mass}  		& $m_q$	& $m_{\ell}$   & $m_{\nu _l}$   &  $m_{\nu _h}$  & $m_{h_1}$   & $m_{h_2}$  \\ \hline
\end{tabular}
\end{center}\caption{\it Mass eigenstates.}\label{tab:symbols}
\end{table}

\chapter{Gauge sector}\label{Ch:3}
\ifpdf
    \graphicspath{{Chapter3/Chapter3Figs/PNG/}{Chapter3/Chapter3Figs/PDF/}{Chapter3/Chapter3Figs/}}
\else
    \graphicspath{{Chapter3/Chapter3Figs/EPS/}{Chapter3/Chapter3Figs/}}
\fi

In this chapter we discuss the gauge sector of the pure $B-L$ model, starting by delineating the viable parameter space for its new independent parameters, as listed at the end of chapter~\ref{Ch:2}. Both experimental searches and  theoretical arguments can give informations about the regions of the parameters that are allowed.

From the experimental side, the leading constraints come from Tevatron and LEP, in two different and complementary kinematic regions, and they are presented in section~\ref{sec:expbounds:Zp}. Tevatron, still accumulating data, is updating its constraints from direct searches, and its constraints are tighter for low $Z'$ boson masses. For values of the mass beyond its kinematic reach, the most stringent bounds come from the searches performed at LEP.

From the theoretical side, the study of the triviality bound (from a RGE analysis) of the parameters of the model gives an upper bound on the gauge coupling. Notice that the equations pertaining to the gauge sector decouple from the rest. Therefore, they can be studied independently. We will present their analysis in section~\ref{subsubsect:RGE_gauge}. 

After the viable parameter space is individuated, in section~\ref{sect:pheno_gauge} is studied the phenomenology at the LHC of the new particle in the gauge sector of the pure $B-L$ model, the $Z'$ boson. We present the results of our investigation by delineating its properties (i.e., production cross sections, intrinsic width, BRs).
The possibility of the $Z'$ boson to decay into pairs of heavy neutrinos is certainly the most interesting of its features. Also, a parton level discovery potential and exclusion power study at the LHC is presented in section~\ref{subsect:disc_power}.
Finally, we draw the conclusions for this chapter in section~\ref{ch3:concl}.

The main results in this chapter are the presentation of the experimental limits on the $Z'$ mass from Tevatron (published in \cite{Basso:2010pe}), the discussion of the theoretical bounds (published in \cite{Basso:2010jm}) and the analysis of the $Z'$ boson properties (published in \cite{Basso:2008iv}) and discovery potential at the LHC (published in \cite{Basso:2010pe}).

\section{Constraints}\label{sect:Zp}

In this section is delineated the viable parameter space for the new independent parameters in the gauge sector, i.e., $g'_1$ and $M_{Z'}$. 

First, we present the experimental constraints using the data from Tevatron, and we summarise those from LEP.
Then, we analyse the theoretical constraints, coming from the study of the RGEs.

\subsection{Experimental constraints}\label{sect:exp_constraints}\label{sec:expbounds:Zp}

A further neutral gauge boson is present in the $B-L$ extension of the SM, the so-called $Z'$, whose mass, in eq.~(\ref{neutral_bosons}), is generically not predicted. In the pure version of the model, argument of this Thesis, eq.~(\ref{neutral_bosons}) simplifies to $M_{Z'_{B-L}}=2g'_1 x$.
As specified in the introduction, we consider the gauge coupling $g'_1$ as a free parameter of the model. That is, a lower bound on the $Z'$ mass will, in general, be a function of the gauge coupling.

Generally speaking, two ways exist to constrain the mass. The most simple one is to look for the decay products of the gauge boson in direct production. If no deviation from the background is observed, lower limits can be derived. Obviously, the direct production is limited by the kinematic reach of the experimental facility. In the case of a hadronic machine, the Parton Distribution Functions (PDFs) will also lower its reach. The most stringent bounds from direct detection are coming from Tevatron. They will be discussed in section~\ref{sec:expbounds:Zp_Tev}.

A leptonic machine, being LEP its latest realisation, offers other possibilities to look for neutral gauge bosons. In all generality, the latter will mix with the SM $Z$ boson, affecting its mass. The LEP-I experiments have measured the mass and other properties of the SM $Z$ boson rather accurately, confirming the SM predicted values at the per mil level \cite{Altarelli:2004fq,:2005ema}. Therefore, the mixing of an extra neutral gauge boson to the SM $Z$ boson is tightly constrained \cite{Abreu:1994ria}. In the pure $B-L$ model, this bound is evaded by definition, as the vanishing of the mixing gauge coupling $\widetilde{g}$ (that defines the model) also implies no mixing at the tree level in the gauge sector (see also section~\ref{subsec:gauge-eigen}).

Effective contact interactions, arising from an expansion in $s/M^2_{Z'}$ (when $M^2_{Z'}>s$), are a complementary way for setting bounds when the $Z'$ mass is larger than the largest collider energy (about $209$ GeV for LEP-II). Once again, the non-observation of any deviation from the SM expectations sets lower limits on the ratio $M_{Z'}/g'_1$, rather than on the $Z'$ mass alone.
They will be summarised in section~\ref{sec:expbounds:Zp_LEP}.

\subsubsection{Tevatron bounds on the $Z'_{B-L}$ boson}\label{sec:expbounds:Zp_Tev}
The Tevatron collider has been collecting data for several years. So far, no deviations in SM Drell-Yan production have been observed. Regarding the $Z'_{B-L}$ boson of this Thesis, the decay into pairs of electrons and into pairs of muons are considered. In general, the experimental communities release the data in the form of $95\%$ C.L. excluded cross sections. Even though an explicit lower limit on the $Z'_{B-L}$ mass is not shown, it is then easy to extract it from the data. We extract it by comparing the $95\%$ C.L. excluded cross sections of Ref.~\cite{Abazov:2010ti,Aaltonen:2011gp} with our theoretical prediction for the $p\overline{p}\rightarrow Z'_{B-L} \rightarrow e^+e^-(\mu ^+\mu ^-)$ cross sections. As mentioned in the introduction, the gauge coupling $g'_1$ is a free parameter, so in all generality it is possible to identify the lower limit on the $Z'$ mass per each $g'_1$ fixed value. 

The latest available analyses are the D$\O$ analysis of Ref.~\cite{Abazov:2010ti} using $5.4\,\mbox{fb}^{-1}$ and the CDF analysis of Ref.~\cite{Aaltonen:2011gp} using $4.6\,\mbox{fb}^{-1}$ of data, respectively, for electrons and muons in the final state.
For the numerical evaluation of the $Z'$ boson signal, we used {\rm{CTEQ6L}} \cite{CTEQ_website} as default PDFs, evaluated at the scale $Q^2=M_{\ell \ell}^2$. The leading order (LO) cross sections are then multiplied by a mass independent $k-$factor of $1.3$ \cite{Carena:2004xs}, as in Refs.~\cite{Abazov:2010ti,Aaltonen:2011gp}, to get in agreement with the Next-to-Next-to-Leading-Order (NNLO) QCD corrections.
The Tevatron limits for the $Z'_{B-L}$ boson are shown in table~\ref{mzp-low_bound} (for selected masses and couplings) \cite{Basso:2010pe}. Notice that these are the most conservative limits, as they are evaluated for decoupled heavy neutrinos, i.e., with masses bigger than $M_{Z'}/2$. \index{Tevatron constraints on the $Z'$ boson}
\begin{table}[h]
\begin{center}
\begin{tabular}{|c|c||c|c|}
\hline
\multicolumn{2}{|c||}{$p\overline{p}\rightarrow e^+ e^-$} & \multicolumn{2}{|c|}{$p\overline{p}\rightarrow \mu^+ \mu^-$} \\
\hline
 $g_1'$         & $M_{Z'}$ (GeV) & $g_1'$         & $M_{Z'}$ (GeV)\\
\hline
0.0197 & 300  & 0.0179	& 300        \\  
0.0193 & 400  & 0.0189	& 400        \\  
0.0281 & 500  & 0.0456	& 500        \\ 
0.0351 & 600  & 0.0380	& 600        \\  
0.0587 & 700  & 0.0544	& 700        \\ 
0.0880 & 800  & 0.0830	& 800        \\ 
0.1350 & 900  & 0.1360	& 900        \\ 
0.2411 & 1000 & 0.2220	& 1000        \\
0.3880 & 1100 & 0.3380	& 1100        \\
\hline
\end{tabular}
\end{center}
\vskip -0.5cm
\caption{\it Lower bounds on the $Z'$ boson mass for selected $g_1'$ values in the $B-L$ model, at $95\%$ C.L.,
by comparing the collected data of Ref.~\cite{Abazov:2010ti,Aaltonen:2011gp} with our theoretical prediction for $p\overline{p}\rightarrow Z'_{B-L} \rightarrow e^+e^-(\mu ^+\mu ^-)$ at Tevatron. 
\label{mzp-low_bound}}
\end{table}

\subsubsection{LEP bounds on the $Z'_{B-L}$ boson}\label{sec:expbounds:Zp_LEP}
When the $Z'$ mass is larger than the largest collider energy (about $209$ GeV for LEP-II), an expansion in $s/M^2_{Z'}$ can be performed. This results in effective four-fermion contact interactions that have been bounded by LEP-II. Recall that the gauge coupling $g'_1$ is a free parameter. One could suppress the $Z'$ boson impact at lower energies also by reducing the coupling. Hence, an inverse law dependence from $g'_1$ is expected, and, in fact, the bound reads \cite{Carena:2004xs} \index{LEP constraints!Z$'$ boson}
\begin{equation}\label{LEP_bound_carena}
\frac{M_{Z'}}{g'_1} \geq 6\; \rm{TeV}\, ,
\end{equation}
or, in a more recent re-analysis, at $99\%$ C.L., as \cite{Cacciapaglia:2006pk}
\begin{equation}\label{LEP_bound}
\frac{M_{Z'}}{g'_1} \geq 7\; \rm{TeV}\, .
\end{equation}
In this Thesis, we adopt the bound of eq.~(\ref{LEP_bound}), that is the most conservative one.

The LEP experiments are also able to provide a lower bound for the $B-L$ breaking VEV $x$. In fact, making use of eq.~(\ref{Mz'}), the LEP 
bound on the $Z'_{B-L}$ boson mass of eq.~(\ref{LEP_bound}) can be rewritten as a lower bound for the $B-L$ breaking VEV
\begin{equation}\label{LEP_bound_vev}
x \geq 3.5\; \rm{TeV}\, .
\end{equation}


\subsection{Theoretical constraints}\label{subsubsect:RGE_gauge}\index{RGE!Gauge sector}

The RGE evolution gives us indications for the validity of the model concerning the gauge couplings. In particular, their evolution must stay perturbative up to some particular scale. In the $B-L$ model, the conditions that the free parameters in the gauge sector must fulfil are \footnote{Notice that the triviality condition, as known in the literature, reads $g'_1(Q)<k$, where $k=1$ or $\sqrt{4\pi}$. As shown in Ref.~\cite{Basso:2010hk}, this `ad-hoc' prescription could be systematically improved by combining the unitarity and RGE evolution techniques.}
\begin{equation}\label{cond_g}
g'_1 (Q') < 1 \hspace{0.5cm} \forall\; Q'\leq Q \qquad \mbox{and} \qquad \widetilde{g}(Q_{EW}) = 0\, ,
\end{equation}
where the second condition in eq.~(\ref{cond_g}) defines the pure $B-L$ model.

\begin{figure}[!h]
  \centering
  \includegraphics[angle=0,width=0.65\textwidth ]{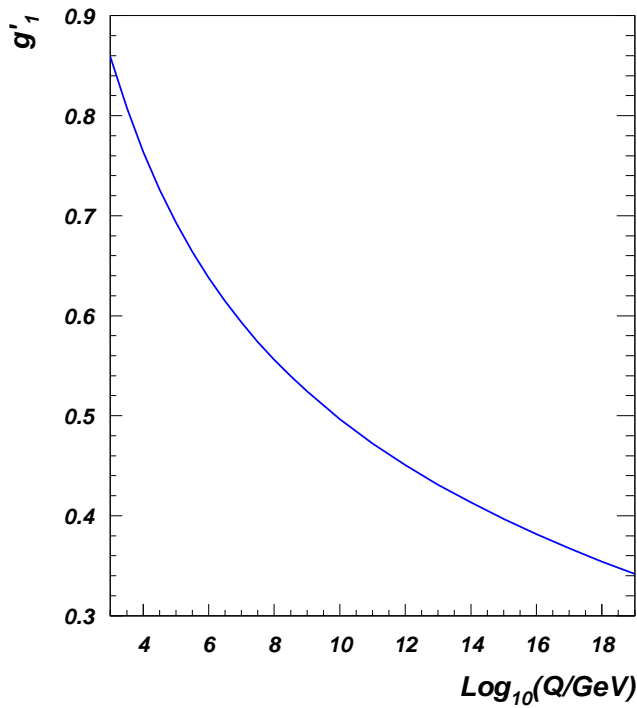}
  \vspace*{-0.5cm}
  \caption{\it Maximum allowed values by eq.~(\ref{cond_g}) for $g'_1(Q_{EW})$  in the $B-L$ model as a function of the scale $Q$.   
\label{g1p_vs_Q}}
\end{figure}

Varying the scale $Q$, the maximum scale up to which we want the model to be well-defined, we get an 
upper bound on $g'_1(Q_{EW})$ as a function of $Q$, as shown in figure~\ref{g1p_vs_Q}. 
Typical results are summarised in table~\ref{g1p-up_bound}.

\vspace*{0.5cm}
\begin{table}[h]
\begin{center}
\begin{tabular}{|c|c|c|c|c|c|c|}
\hline
$Log_{10} (Q/ \mbox{GeV}) $ & 3 & 5 & 7 & 10 & 15 & 19  \\  \hline
$g_1'(Q_{EW})$  & 0.860 & 0.693 & 0.593 & 0.497 & 0.397 & 0.342  \\ 
\hline
\end{tabular}
\end{center}
\vskip -0.5cm
\caption{\it Maximum allowed values by eq.~(\ref{cond_g}) for $g'_1(Q_{EW})$ in the $B-L$ model for selected values of the scale $Q$.
\label{g1p-up_bound}}
\end{table}

\section{Phenomenology of the gauge sector}\label{sect:pheno_gauge}

In this section we present the analysis of the new state in the gauge sector, the $Z'_{B-L}$ boson (or simply $Z'$ boson, when the meaning is clear).  

An important feature of this $Z'$ boson is the chiral structure of its couplings to fermions: since the $B-L$ charge does not distinguish between left-handed and right-handed fermions (as clear in table~\ref{tab:quantum_number_assignation}), the $B-L$ neutral current is purely vector-like, with a vanishing axial part \footnote{This is strictly true for Dirac fermions, like quarks and charged leptons. Regarding the neutrinos (both light and heavy), they are Majorana states and, hence, the coupling to the $Z'_{B-L}$ boson is just axial \cite{Perez:2009mu}.}. In fact,
\begin{equation}
g_{Z'}^A = \frac{g_{Z'}^R-g_{Z'}^L}{2}=0\, .
\end{equation}
As a consequence, we decided to not study the asymmetries of the decay products stemming from the $Z'_{B-L}$ boson, given their trivial distribution at the peak.

The $Z'_{B-L}$ boson is not always considered as a traditional benchmark for generic collider reach studies \cite{Erler:1999ub,Appelquist:2002mw,Rizzo:2006nw,Langacker:2008yv,Contino:2008xg,Gulov:2009tn,Erler:2009jh,Ball:2007zza} or data analyses \cite{Abazov:2010ti,Aaltonen:2011gp}.
Main differences with the most popular $Z'$ models in the literature (aside the vanishing axial coupling) are the fact that the gauge coupling $g'_1$ is a free parameter (as we do not consider here constraints from possible embeddings into GUTs) and the presence of new coupled matter (the $\nu _h$'s). Since their mass is as well a free parameter, they will affect the $Z'_{B-L}$ boson BRs, which will be a function of the neutrino masses.
Finally, in the pure $B-L$ model, no mixing between the $Z'_{B-L}$ boson and the SM $Z$ boson is present.

In this section we give a detailed description of the $Z'$ properties. In section~\ref{sect:Zp_properties}, the production cross sections and decay BRs are shown. A detailed investigation of the discovery potential for the LHC is presented in section~\ref{subsect:disc_power}, for the foreseen center-of-mass (CM) energies of $7$ and $14$ TeV and integrated luminosities up to $1$ fb$^{-1}$ and to $100$ fb$^{-1}$, respectively. For the former CM energy, a comparison with Tevatron is also shown, for an integrated luminosity up to $10$ fb$^{-1}$.

\subsection{Production cross sections and decay properties}\label{sect:Zp_properties}

\begin{figure}[!ht]
  \centering
  \subfloat[]{ 
  \label{Zpxs}
  \includegraphics[width=0.5\textwidth]{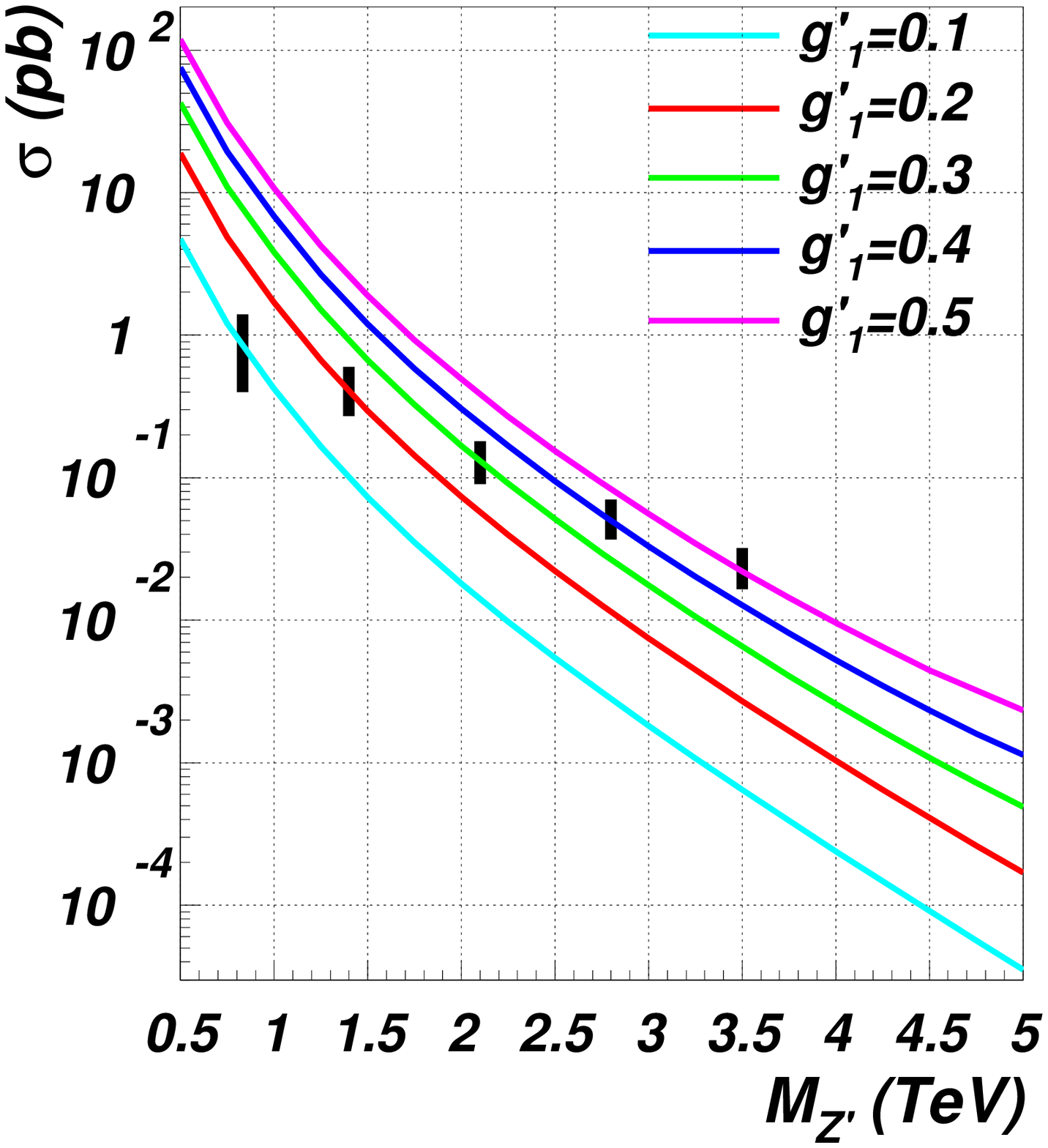}}%
  \subfloat[]{
  \label{zpxs_cont}
  \includegraphics[angle=0,width=0.5\textwidth]{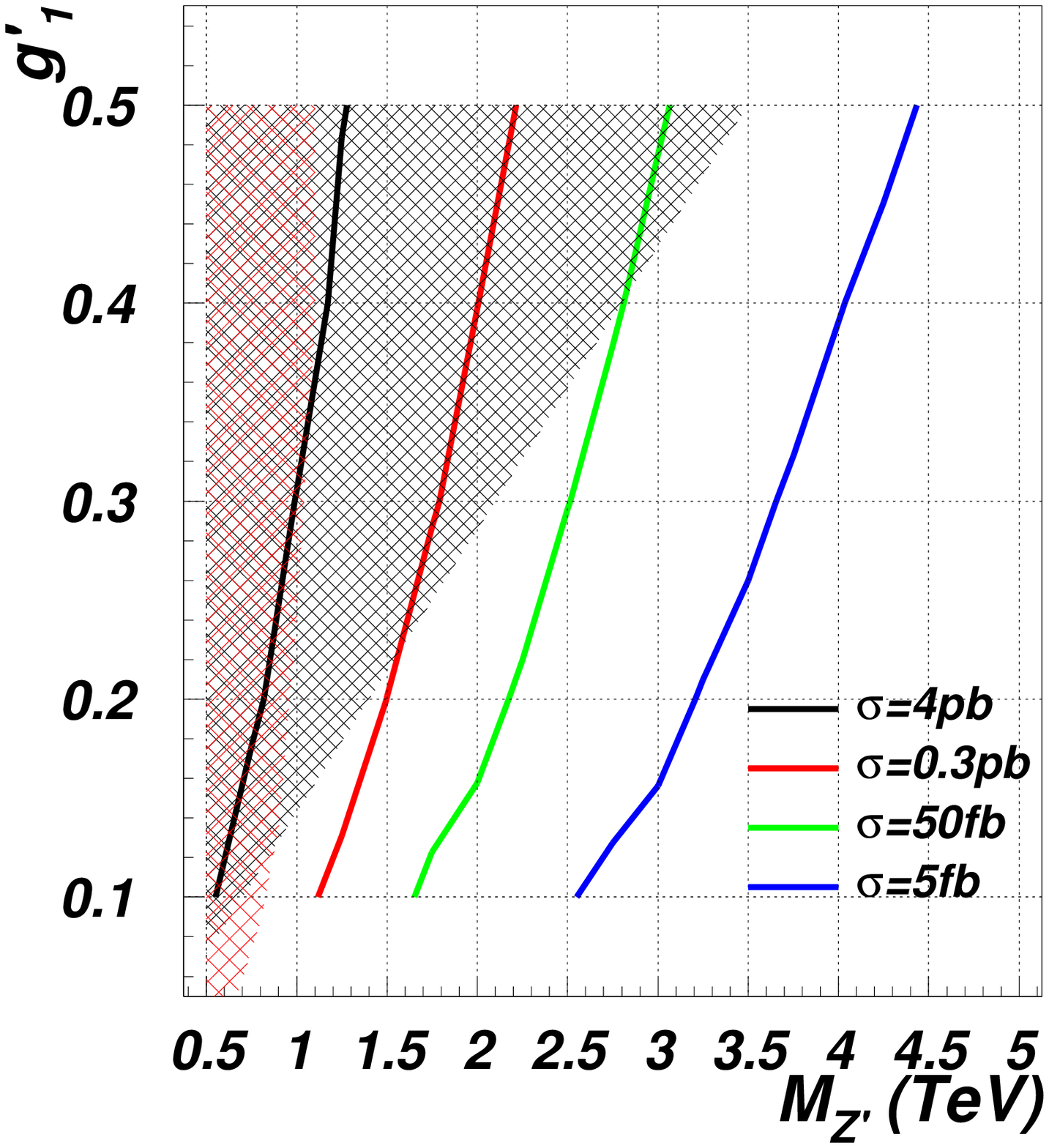}}
  \caption{\it $Z'$ hadro-production cross sections at the LHC over
 the ($M_{Z'}$, $g'_1$) plane:
(\ref{Zpxs}) as a function of  $M_{Z'}$ for various $g'_1$ values and
(\ref{zpxs_cont}) in the form of contour lines for four fixed values of
production rates. In the left frame, the vertical ticks indicate the region excluded experimentally (on the left from the ticks), in accordance with section~\ref{sec:expbounds:Zp}. In the right frame, the black shaded area is the portion excluded by eq.~(\ref{LEP_bound}) (LEP bounds). The red shaded area is the region excluded by Tevatron, in accordance with table~\ref{mzp-low_bound}.}
  \label{Zp_xs}
\end{figure}

The most efficient hadro-production process involving a
$Z'$ boson is the Drell-Yan mode
\begin{equation}\label{Z_B-L_prod}
q\bar{q} \rightarrow Z'\, ,
\end{equation}
where $q$ is either a valence quark or a sea quark in the proton. At the parton level, the $Z'$ production cross section for process (\ref{Z_B-L_prod})
depends on two main parameters: $M_{Z'}$ and $g'_1$. In figure~\ref{Zpxs} we present the  $Z'$ boson hadro-production cross section at the LHC as a function of both $M_{Z'}$ and $g'_1$, in the ranges $0.5$ TeV $< M_{Z'}< 5$ TeV and $0.1 < g'_1 < 0.5$, respectively, while figure~\ref{zpxs_cont} presents the contour levels in the ($M_{Z'}$, $g'_1$) plane for cross sections of $4$ pb, $0.3$ pb, $50$ fb and $5$ fb. \index{Production cross sections!Z$'$ boson}

\begin{figure}[!h]
\centering
  \includegraphics[angle=0,width=0.8\textwidth ]{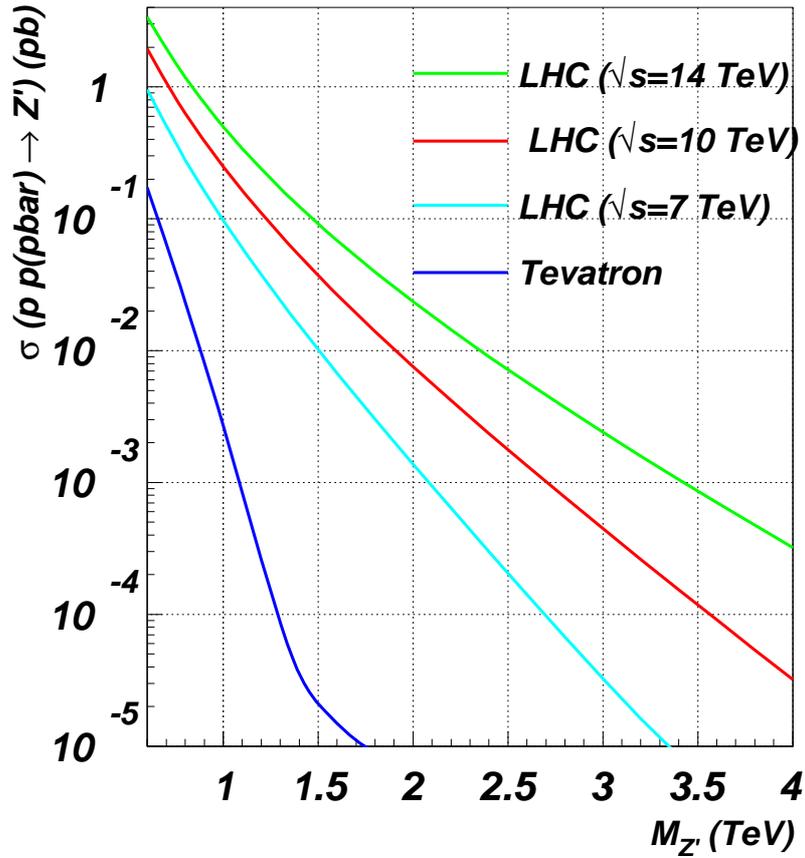}
  \caption{\it Cross sections for $pp(\overline{p}) \rightarrow Z'_{B-L}$ at Tevatron and at the LHC (for $\sqrt{s}=7,10$ and $14$ TeV) for $g'_1=0.1$.}
  \label{Zp_xs_S}
\end{figure}

For a fixed value of the coupling, we can compare the production cross sections for different hadronic machines and CM energies. For $g'_1=0.1$, the cross sections at Tevatron and the LHC (for $\sqrt{s}=7$, $10$ and $14$ TeV) are shown in figure~\ref{Zp_xs_S}.
Note that although at Tevatron the production cross section is 
smaller than at the LHC, the integrated luminosity we are 
considering here for the LHC at $7$ TeV (i.e., $1$ fb$^{-1}$) is smaller than for Tevatron (i.e., $10$ fb$^{-1}$).

\subsubsection{Decay properties}
As discussed earlier, the extra $U(1)_{B-L}$ gauge group provides an additional neutral gauge boson, $Z'$, with no mixing with the SM $Z$ boson. Therefore, the $Z'$ boson decays only to fermions at the tree level and its width is given by 
the following expression:
\begin{equation}\label{Z'_width}
\Gamma (Z'\rightarrow f\overline{f})=
\frac{M_{Z'}}{12\pi}C_f (v^f)^2 
\left[ 1 +2\frac{m_f^2}{M_{Z'}^2}\right]\sqrt{1-\frac{4m_f^2}{M_{Z'}^2}}\, ,
\end{equation}
where $m_f$ is the mass and $C_f$ the number of colours of the fermion type $f$ and $v^f = (B-L) \cdot g'_1$ is the coupling (see table~\ref{tab:quantum_number_assignation}).

\begin{figure}[!ht]
  \centering
  \includegraphics[angle=0,width=1\textwidth]{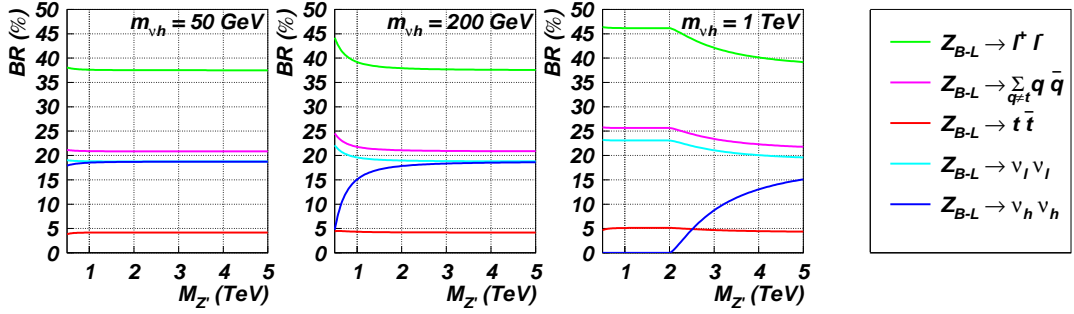}
  \caption{\it $Z'$ boson BRs as a function of $M_{Z'}$ for several heavy neutrino masses: $m_{\nu_h} = 50$, $200$ GeV, and $1$ TeV, from left to right, respectively. A summation over all lepton/neutrino flavours is implied throughout, whereas, in the case of quarks we distinguish between light flavours ($q=d,u,s,c,b$) and the top quark.}
  \label{Zp_BR}
\end{figure}

Clearly, the $Z'$ boson BRs depend strongly on the heavy neutrino mass and figure~\ref{Zp_BR} shows how they change with fixed (although arbitrary) values of $m_{\nu_h}$, for the following three cases: a heavy neutrino (i) much lighter than, (ii) lighter than, and (iii) comparable in mass to the $Z'$ boson, in the range $0.5$ TeV $<M_{Z'}<5$ TeV, after summing over the generations. \index{BRs!Z$'$ boson}

A feature of the current $B-L$ model illustrated in the previous figure is that the $Z'$ boson predominantly couples to leptons. In fact, after summing over the generations, $k=1...3$, we roughly get for leptons and quarks:
\begin{displaymath}
\sum _k BR\left( Z' \rightarrow \ell_k \overline{\ell_k} + \nu_k \nu_k \right) \sim \frac{3}{4}\, ,\qquad
\sum _k BR\left( Z' \rightarrow q_k \overline{q_k} \right) \sim \frac{1}{4}\, .
\end{displaymath}
In particular, $BR(Z'\rightarrow \ell^+ \ell^-)$ varies between $12.5\%$ and $15.5\%$ ($\ell=e,\mu$), while $BR(Z'\rightarrow q \overline{q})$ varies between $4\%$ and $5\%$.

Not surprisingly, then, for a relatively light (with respect to the $Z'$ gauge boson) heavy neutrino, the $Z'$ BR into pairs of such particles is relatively high: $\sim 18\% $ (at most, again, after summing over the generations). As we will see in section~\ref{sect:nu_properties}, the pair production from $Z'$ boson decays is an effective way to produce the heavy neutrinos at the LHC, also because of the relatively high BR. The importance of this channel is due to the fact that it allows for the $B-L$ model discrimination: as already mentioned, asymmetries of the standard $Z'$ decay modes are trivial at the peak, as true for a variety of other models. To remove this degeneracy, one should look for further consequences of our model. Among these, the decay into heavy neutrino pairs is certainly the most spectacular, allowing for unusual $Z'$ decays, such as multi-lepton and/or multi-jet decay patterns (see, for instance, Refs.~\cite{Huitu:2008gf,Basso:2008iv,Perez:2009mu}).
After introducing the heavy neutrino properties, the whole section~\ref{sect:Zp_to_nuh} will be devoted to the study of this signature, with particular emphasis on one specific decay pattern of the heavy neutrino pair, the so-called tri-lepton decay mode.

One should finally note that possible $Z'$ decays into one light and one heavy neutrino are highly suppressed by the corresponding (heavy-light) neutrino mixing and thus they can safely be neglected.

\begin{figure}[!htb]
  \begin{center}
  \vspace*{-0.2cm}
  \subfloat[]{ 
  \label{Zp_width_vs_MZp}
  \includegraphics[width=0.5\textwidth]{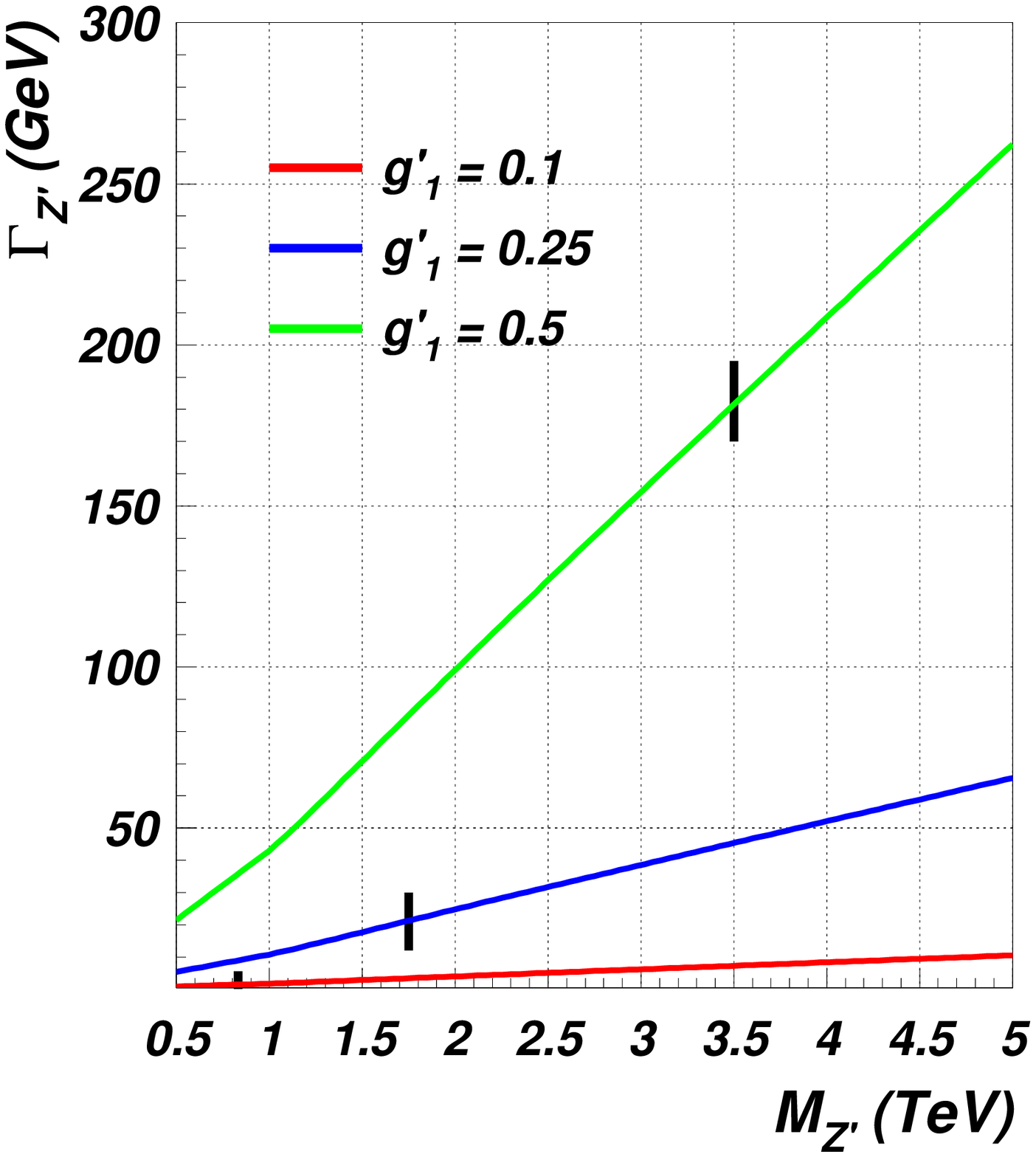}}%
  \subfloat[]{
  \label{Zp_width_vs_g1p}
  \includegraphics[width=0.5\textwidth]{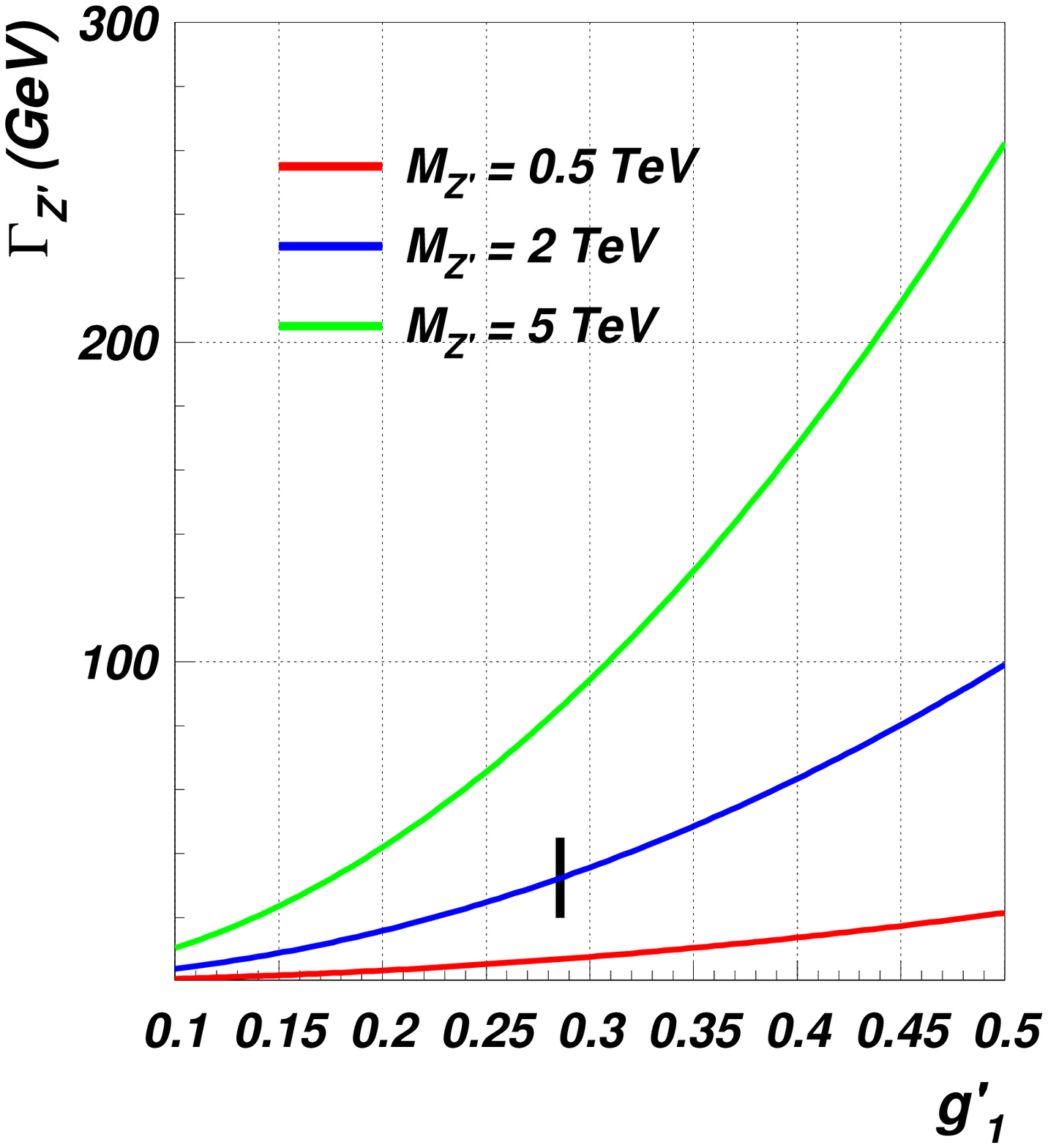}}
  \\ \vspace*{-0.3cm} \centering
  \subfloat[]{ 
  \label{Zp_width_vs_Mhn_rel}
  \includegraphics[width=0.5\textwidth]{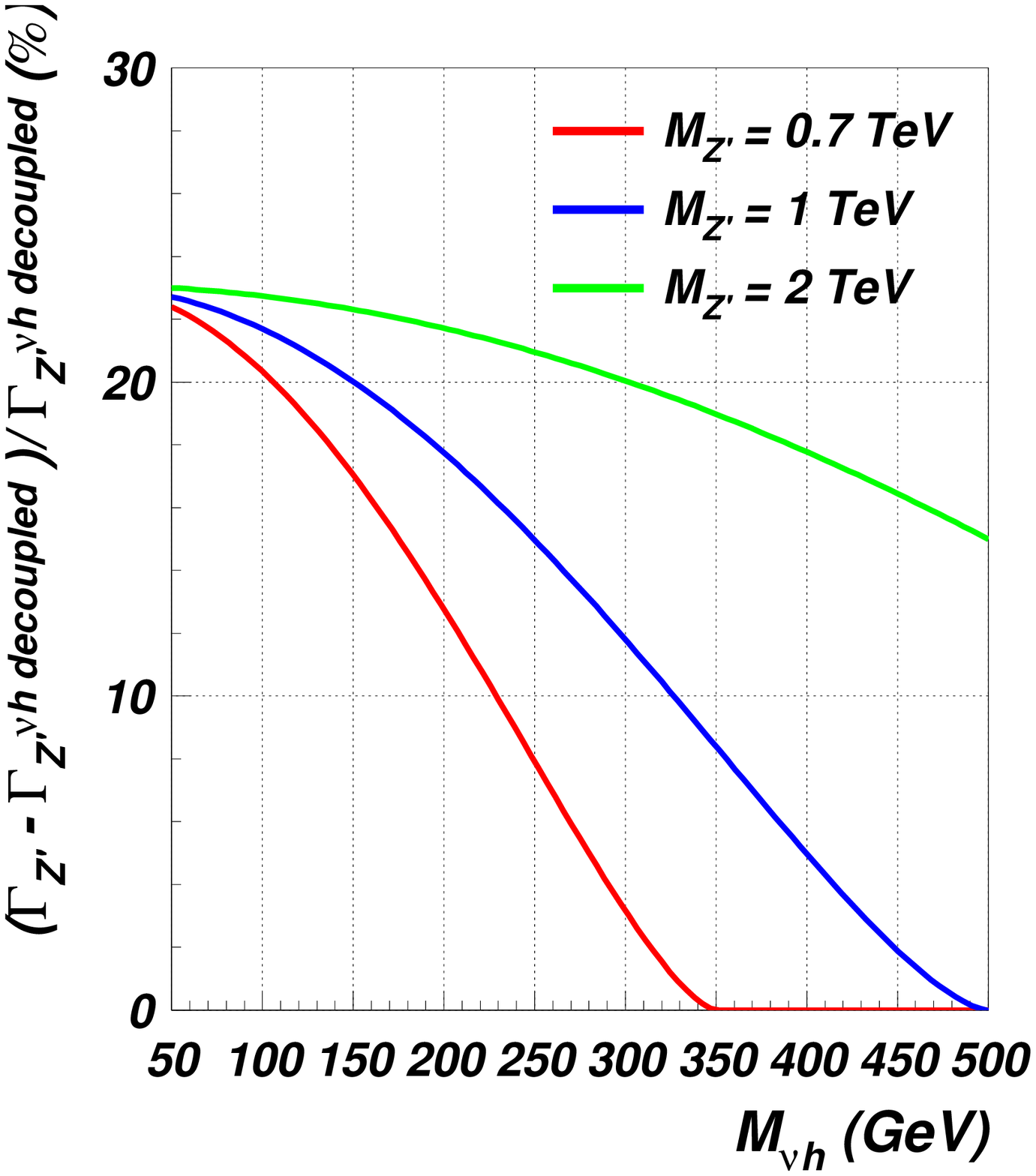}}%
  \end{center}
  \vspace*{-0.6cm} 
  \caption{\it $Z'$ total width as a function of: (\ref{Zp_width_vs_MZp}) $M_{Z'}$ (for fixed values of $g'_1$), (\ref{Zp_width_vs_g1p})
 $g'_1$ (for fixed values of $M_{Z'}$) and (\ref{Zp_width_vs_Mhn_rel}) $m_{\nu_h}$ (for fixed values of $M_{Z'}$ and $g'_1 = 0.1$).
  In plots (\ref{Zp_width_vs_MZp})[(\ref{Zp_width_vs_g1p})]
  the portion of the curves to the left[right] of the vertical ticks (when appearing) are experimentally excluded, in accordance with eq.~(\ref{LEP_bound}) and table~\ref{mzp-low_bound}. Notice finally that in plot \ref{Zp_width_vs_g1p} the curve for $M_{Z'}=0.5$ TeV is shown just for sake of comparison, as it is all excluded.}
  \label{Zp_width}
\end{figure}

In figures~\ref{Zp_width_vs_MZp}  and  \ref{Zp_width_vs_g1p} we present the total  decay width of the $Z'$ boson as a function of $M_{Z'}$  and  $g'_1$, respectively (with the other parameters held fixed to three
different values), assuming that the partial decay width into heavy neutrinos vanishes. Also, figure~\ref{Zp_width_vs_Mhn_rel}  presents the relative variation of the total width as a function of the $\nu_h$ mass for three different values of  $M_{Z'}$ and for $g'_1 = 0.1$.

From the first two plots we see that the total width of a $Z'$ gauge boson varies from a few to hundreds of GeV over a mass range of $0.5$ TeV $<M_{Z'}<5$ TeV, depending on the value of $g'_1$, while from the third plot one can gather the importance of taking into consideration the heavy neutrinos, since their relative contribution to the total $Z'$ width can be as large as $25\%$ (whenever this channel is open).

%
%


\subsection{Discovery power at the LHC}\label{subsect:disc_power}
In this section we determine the discovery potential of the LHC considering several CM energies, $7$ and $14$ TeV, using the expected integrated luminosities. For $\sqrt{s}=7$ TeV only, we also compare our results for the LHC to the expected ultimate reach at Tevatron.

For sake of comparison, a complementary discovery power study at the LHC for the $Z'_{B-L}$ boson also appears, for CM energy of $7$ TeV, in Ref.~\cite{Salvioni:2009mt} and for $14$ TeV in Ref.~\cite{Emam:2008zz} (for the $Z'\to e^+ e^-$ channel only).

\vspace*{-2cm}
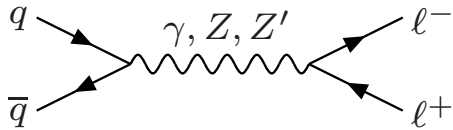
\begin{figure}[!ht] \centering \scalebox{1.75}{
\begin{picture}(95,79)(0,30)
\unitlength=1.0 pt
\SetScale{1.0}
\SetWidth{0.5}      
\scriptsize    

\Text(-14.0,70.0)[l]{$q$}
\ArrowLine(-8.0,70.0)(12.0,60.0)
\Text(-14.0,50.0)[l]{$\overline{q}$}
\ArrowLine(12.0,60.0)(-8.0,50.0)

\Photon(12.0,60.0)(50.0,60.0){2.0}{6}
\Text(19.0,68.0)[l]{$\gamma,Z,Z'$}

\Text(72.0,70.0)[l]{$\ell ^-$}
\ArrowLine(50.0,60.0)(70.0,70.0) 
\Text(72.0,50.0)[l]{$\ell ^+$}
\ArrowLine(70.0,50.0)(50.0,60.0) 
\end{picture}}
\vspace*{-1cm}
\caption{\it Feynman diagram for the Drell-Yan di-lepton production ($\ell=e,\,\mu$). \label{DY_pic}}
\end{figure}

The process we are interested in is the di-lepton production, shown in figure~\ref{DY_pic}. We define our signal as $pp\rightarrow \gamma,\,Z,\,Z'_{B-L}\rightarrow  \ell^+ \ell^-$ ($\ell=e,\,\mu$), i.e., all possible sources together with their mutual interferences, and the background as $pp\rightarrow \gamma,\,Z\rightarrow \ell^+ \ell^-$ ($\ell=e,\,\mu$), i.e., SM Drell-Yan production (including interference). No other sources of background, such as $WW$, $ZZ$, $WZ$ or $t\overline{t}$ have been taken into account. These can be suppressed or are insignificant \cite{Ball:2007zza,Abazov:2010ti}.
For both the signal and the background, we have assumed standard
acceptance cuts (for both electrons and muons) at the LHC:
\begin{equation}\label{LHC_cut}
p_T^\ell > 10~{\rm GeV},\qquad |\eta^\ell|<2.5\qquad (\ell=e,\,\mu),
\end{equation} 
and we apply 
the following requirements on the di-lepton invariant mass,  $M_{\ell \ell}$, depending on whether we are considering electrons or muons.
We distinguish two different scenarios: an `early' one (for $\sqrt{s}=7$ TeV) and an `improved' one  (for $\sqrt{s}=14$ TeV), and, in computing the signal significances, we will select a window as large as either one width of the $Z'_{B-L}$ boson or twice the di-lepton mass resolution \footnote{We take  the CMS di-electron and di-muon mass resolutions \cite{Bayatian:2006zz,Ball:2007zza} as typical for the LHC environment. ATLAS resolutions \cite{:1999fq} do not differ substantially.}, 
whichever the largest. The half windows in the invariant mass distributions respectively read, for the `early scenario': \index{Di-lepton resolution!LHC}
\begin{eqnarray}\label{LHC_ris_el}
\mbox{electrons: }\; |M_{ee}-M_{Z'}| &<& 
\mbox{max} \left( \frac{\Gamma_{Z'}}{2},\; \left( 0.02\frac{M_{Z'}}{\rm GeV} \right) {\rm GeV}\; \right),\\ \label{LHC_ris_mu}
\mbox{muons: }\; |M_{\mu\mu}-M_{Z'}| &<& 
\mbox{max} \left( \frac{\Gamma_{Z'}}{2},\; \left( 0.08\frac{M_{Z'}}{\rm GeV} \right) {\rm GeV}\; \right),
\end{eqnarray}
and for the `improved scenario':
\begin{eqnarray}\label{LHC_ris_el_imp}
\mbox{electrons: }\; |M_{ee}-M_{Z'}| &<& 
\mbox{max} \left( \frac{\Gamma_{Z'}}{2},\; \left( 0.005\frac{M_{Z'}}{\rm GeV} \right) {\rm GeV}\, \right),\\ \label{LHC_ris_mu_imp}
\mbox{muons: }\; |M_{\mu\mu}-M_{Z'}| &<& 
\mbox{max} \left( \frac{\Gamma_{Z'}}{2},\; \left( 0.04\frac{M_{Z'}}{\rm GeV} \right) {\rm GeV}\, \right).
\end{eqnarray}
Our choice reflects the fact that what we will observe is in fact the convolution 
between the Gaussian detector resolution and the Breit-Wigner 
shape of the peak, and such convolution will be dominated by the largest of the two. Our approach is to take the convolution width exactly equal to the resolution width or to the peak width, whichever is largest \footnote{In details, for resolutions below $\Gamma /2$, we take the convolution equal to the resolution width. For resolutions above $3\Gamma$, we take the convolution equal to the peak width. When the resolution $\in \left[ \Gamma /2,3\Gamma\right]$, the convolution is taken as a linear interpolation between the two regimes.}, and to count all the events within this window. Finally, only $68\%$ of signal events are considered: intrinsically, when the peak width is dominating, effectively (by rescaling the signal), otherwise.

\begin{figure}[!h]
\centering
  \includegraphics[angle=0,width=0.8\textwidth ]{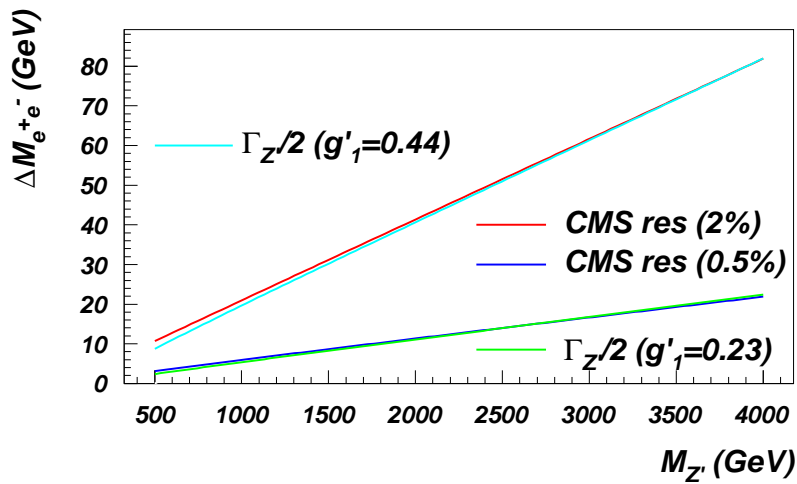}
  \caption{\it Comparison of the CMS electron resolution according to eqs.~(\ref{LHC_ris_el}) and (\ref{LHC_ris_el_imp}), with the two different constant terms as described in the text, compared to $\, \Gamma _{Z'}/2$ for $g'_1=0.23$ and $g'_1=0.44$.}
  \label{CMS_res}
\end{figure}

In figure \ref{CMS_res} we compare the LHC resolutions for electrons for the two aforementioned scenarios (eqs.~(\ref{LHC_ris_el}) and (\ref{LHC_ris_el_imp})) with $\Gamma_{Z'}/2$.
It is clear that, whichever the $Z'_{B-L}$ mass, for a value of the coupling $g'_1$ smaller than roughly $0.4$, the peak will be dominated by the early experimental resolution,
%
%
%
i.e., the half window 
will contain an amount of signal as big as the one produced with
$|M_{\ell \ell}-M_{Z'}| = \Gamma_{Z'}/2$. The region of interest in the parameter space we are going to study almost always fulfils the condition $g'_1<0.4$, as we will
see from the plots in the following section. The muon resolution is much worse and in such a plot it would be an order of magnitude higher than the other curves. Hence, for this final state, the peak is always dominated by the experimental resolution, for the values of the gauge coupling we are considering.

In the next section we will compare the LHC and Tevatron discovery reach. In the derivation of the experimental constraints, we refer to the latest publications, being the D$\O$ analysis of Ref.~\cite{Abazov:2010ti} for the electron case and the CDF analysis of Ref.~\cite{Aaltonen:2011gp} for the muon final state, discussed in section~\ref{sec:expbounds:Zp_Tev}. Hence, we have considered the typical acceptance cuts (for  electrons and muons) for the respective detector:
\begin{eqnarray}\label{Tev_cut}
p_T^e > 25~{\rm GeV}, &\qquad& |\eta^e|<1.1, \\
p_T^\mu > 18~{\rm GeV}, &\qquad& |\eta^\mu|<1,
\end{eqnarray}
and the following requirements on the di-lepton invariant mass,  $M_{\ell \ell}$, depending on whether we are considering electrons or 
muons \footnote{We take the D$\O$ di-electron \cite{D0res} and the CDF di-muon \cite{Aaltonen:2011gp} mass resolutions as a typical Tevatron environment, in accordance with the most up-to-date limits.}:\index{Di-lepton resolution!Tevatron}
\begin{eqnarray}\label{Tev_ris}
\mbox{electron: } |M_{ee}-M_{Z'}| &<& 
\mbox{max} \left( \frac{\Gamma_{Z'}}{2},\, \left( 0.16 \sqrt{\frac{M_{Z'}}{\rm GeV}}+ 0.04\frac{M_{Z'}}{\rm GeV} \right) {\rm GeV} \right),\\
\mbox{muons: } |M_{\mu\mu}-M_{Z'}| &<& 
\mbox{max} \left( \frac{\Gamma_{Z'}}{2},\, \left( 0.017\%\, \left(\frac{M_{Z'}}{\rm GeV}\right) ^2 \right) {\rm GeV} \right).
\end{eqnarray}

The selection of an invariant mass window centred at the $Z'$ boson mass is comparable to the standard experimental analysis, as in Ref.~\cite{Abazov:2010ti} (the electron channel at D$\O$), where signal and background are integrated from $M_{Z'}-10\Gamma _{Z'}$ (where $\Gamma _{Z'}$ is the $Z'$ boson width obtained by rescaling the SM $Z$ boson width by the ratio of the $Z'$ to the $Z$ boson mass) to infinity.
Since the background (in proximity of the narrow resonance) can be reasonably thought as flat, while the signal is not, the procedure we propose enhances the signal more than the background and it is expected to be more sensitive than the aforementioned one. Ref.~\cite{Aaltonen:2011gp} applies a different strategy and figure~\ref{contour7_ecal} shows that our procedure is comparable to it,
 although less involved.
A Bayesan approach is being used at the LHC~\cite{Khachatryan:2010fa}, similar to the CDF case. Hence, we present our results for a comparison `a posteriori'.

In our analysis we use a definition of signal significance $\sigma$, as follows. 
In the region where the number
of both signal ($s$) and background ($b$) 
events is `large' (here taken to be bigger than 20),
we use a definition of significance
based on Gaussian statistics:\index{Significance}
\begin{equation}\label{signif}
{\sigma} \equiv {\it s}/{\sqrt{\it b}}.
\end{equation}
Otherwise, in case of smaller statistics, we used the Bityukov algorithm \cite{Bityukov:2000tt}, which basically uses the Poisson `true' distribution instead of the `approximate' Gaussian one.

Finally, for the numerical evaluation of the cross sections, we use the same setup as in section~\ref{sec:expbounds:Zp_Tev}. In particular, the same mass independent $k-$factor of $1.3$ is also used for the LHC (as in Ref.~\cite{Khachatryan:2010fa} \footnote{Notice that in Ref.~\cite{Khachatryan:2010fa} the $k$-factor used was mass-dependent. Here we use the average value.}).

A typical detector resolution has effectively been taken into account by our procedure, that consists in counting all the events that occur within the window (in invariant mass) previously described, and by rescaling to $68\%$ the signal events when the peak is dominated by the experimental resolution.
Nonetheless, our simulation does not account for Initial State Radiation (ISR) effects. 
ISR can have two main sources: QED-like ISR (i.e., photon emission), that has the effect of shifting the peak and of creating a tail towards smaller energy, and QCD-like ISR (i.e., gluon emission), that has similar effects and might also induce trigger issues
in the intent of removing backgrounds (e.g., by cutting on final state jets). Although we are aware of such effects, we believe that their analysis goes beyond the scope of this work and it will be argument of future investigations. Altogether, we are confident that, while particular aspects of
our analysis may be sensitive to such effects, the general picture will not depend upon these substantially.
Also, the only background considered here was the irreducible SM Drell-Yan. Reducible backgrounds, photon-to-electron conversion, efficiencies in reconstructing electrons/muons, jets faking leptons etc., whose overall effect is to deplete the signal, were neglected (being $t\overline{t}$ the most important source, at the level of $ \lesssim 10\%$). However, for this analysis they are not quantitatively important  \cite{Abulencia:2005ix,Ball:2007zza,Abazov:2010ti}.
The net effect of the factors above is usually regarded as an overall reduction of the total acceptance, being the lepton identification the most important source, about $80\div 90\%$ per each lepton. We comment on this in the conclusion of the chapter.

\subsubsection{LHC at $\sqrt{s}=7$ TeV and comparison with Tevatron}
\begin{figure}[!h]
  \subfloat[]{ 
  \label{contour7_ecal}
  \includegraphics[angle=0,width=0.48\textwidth ]{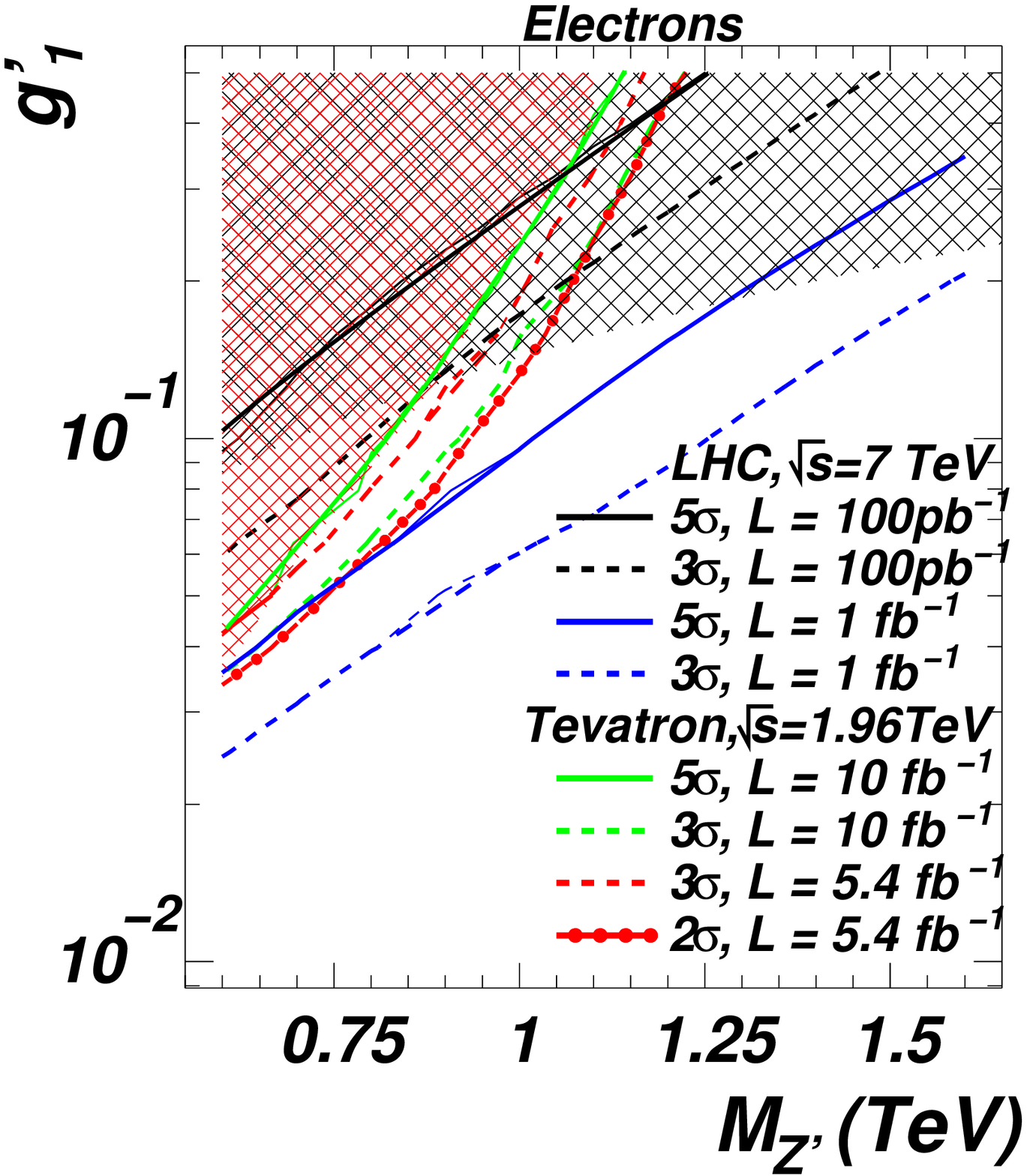}}
  \subfloat[]{
  \label{contour7_pt}
  \includegraphics[angle=0,width=0.48\textwidth ]{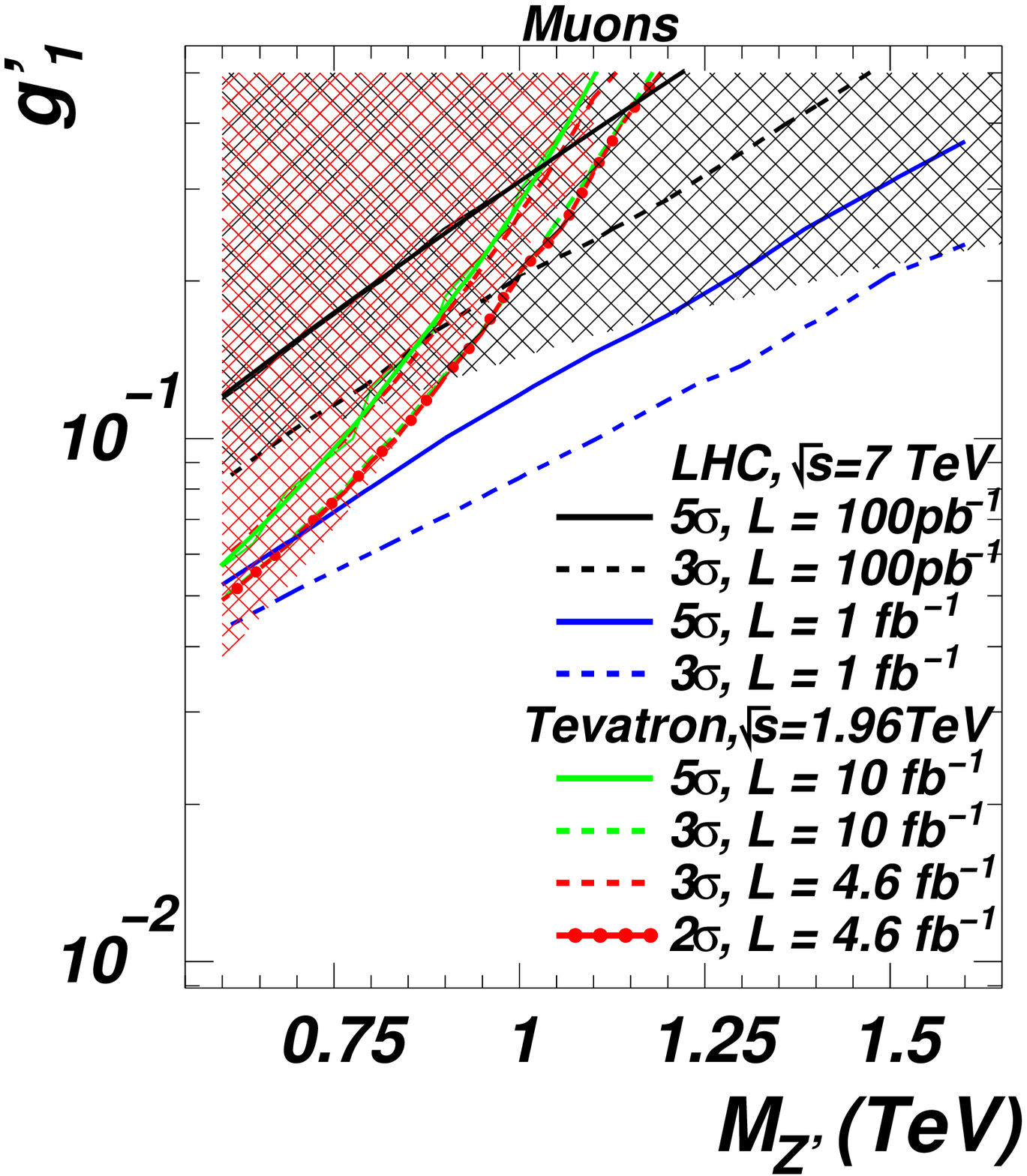}}
  \vspace*{-0.25cm}
  \caption{\it Significance contour levels plotted against $g_1'$
and $M_{Z'}$ at the LHC for $\sqrt{s}=7$ TeV for $0.1-1$ fb$^{-1}$ and at  Tevatron ($\sqrt{s}=1.96$ TeV) for (\ref{contour7_ecal}, electrons) $5.4-10\,\mbox{fb}^{-1}$ and (\ref{contour7_pt}, muons) $4.6-10\,\mbox{fb}^{-1}$ of integrated luminosity. The shaded areas correspond to the region of parameter space excluded
experimentally, in accordance with eq.~(\ref{LEP_bound}) (LEP bounds, in black) and table~\ref{mzp-low_bound} (Tevatron bounds, in red).}
  \label{contour7}
\end{figure}

The first years of the LHC work will be at a CM energy of $7$ TeV, where the total integrated luminosity is likely to be of the order of $1$ fb$^{-1}$. Figure~\ref{contour7} shows the discovery potential under these conditions, as well as the most recent limit from LEP (see eq.~\ref{LEP_bound})
and from Tevatron (as in section~\ref{sec:expbounds:Zp_Tev}).

In the same figure we also include for comparison the Tevatron discovery potential at the integrated luminosities used for the latest published analyses \cite{Abazov:2010ti,Aaltonen:2011gp} ($5.4\,\mbox{fb}^{-1}$ and $4.6\,\mbox{fb}^{-1}$ for electrons and muons, respectively) as well as the expected reaches at $\mathcal{L}=10\,\mbox{fb}^{-1}$.

Notice that the Tevatron excluded area are based on the actual data, while the dot-dashed $2\sigma$ curves are theoretical curves. Thus, if from the one side theory cannot reproduce experiments, from the other side we are comparing two methods of extracting the results.
%
%
%
%
As mentioned previously, figure~\ref{contour7} shows that the procedures used in experimental analyses for the electron channel \cite{D0,Abazov:2010ti} are not quite optimisied for maximising the signal significance. The alternative analysis described in this work has the potential to improve sensitivities and can be easily developed even further.

It is then clear that Tevatron will still be 
competitive with the LHC (for $\sqrt{s}=7$ TeV CM energy), especially in the lower mass region where the LHC requires $1$ fb$^{-1}$ to be sensitive to the same couplings as Tevatron.
The LHC will be able to probe, at $5\sigma$ level, the $Z'_{B-L}$ boson for values of the coupling
down to $3.7-5.2 \cdot 10^{-2}$ (for electrons and muons, respectively), while 
Tevatron can be sensitive down to $4.2 \cdot 10^{-2}$ with electrons. 
The kinematic reach of the two machines is different. The LHC for $1$ fb$^{-1}$ can discover the $Z'_{B-L}$ boson up to masses of $1.20-1.25$ TeV, while at Tevatron a $3\sigma$ evidence will be possible up to a value of the mass of $1$ TeV in the electron channel, for a suitable choice of the coupling. 
As clear from figure~\ref{contour7_pt}, the muon channel at Tevatron requires more than $10$ fb$^{-1}$ to start probing (at $3\sigma$) points in the $M_{Z'}-g'_1$ plane allowed by the CDF constraints, as in table~\ref{mzp-low_bound}. This total integrate luminosity appears to be more than what can be collected, due to the announced shutdown by the end of the year $2011$ \cite{Tevatron_reach,Tevatron_shutdown}.

\begin{figure}[!h]
  \subfloat[]{ 
  \label{contour7_ecal_LHC}
  \includegraphics[angle=0,width=0.48\textwidth ]{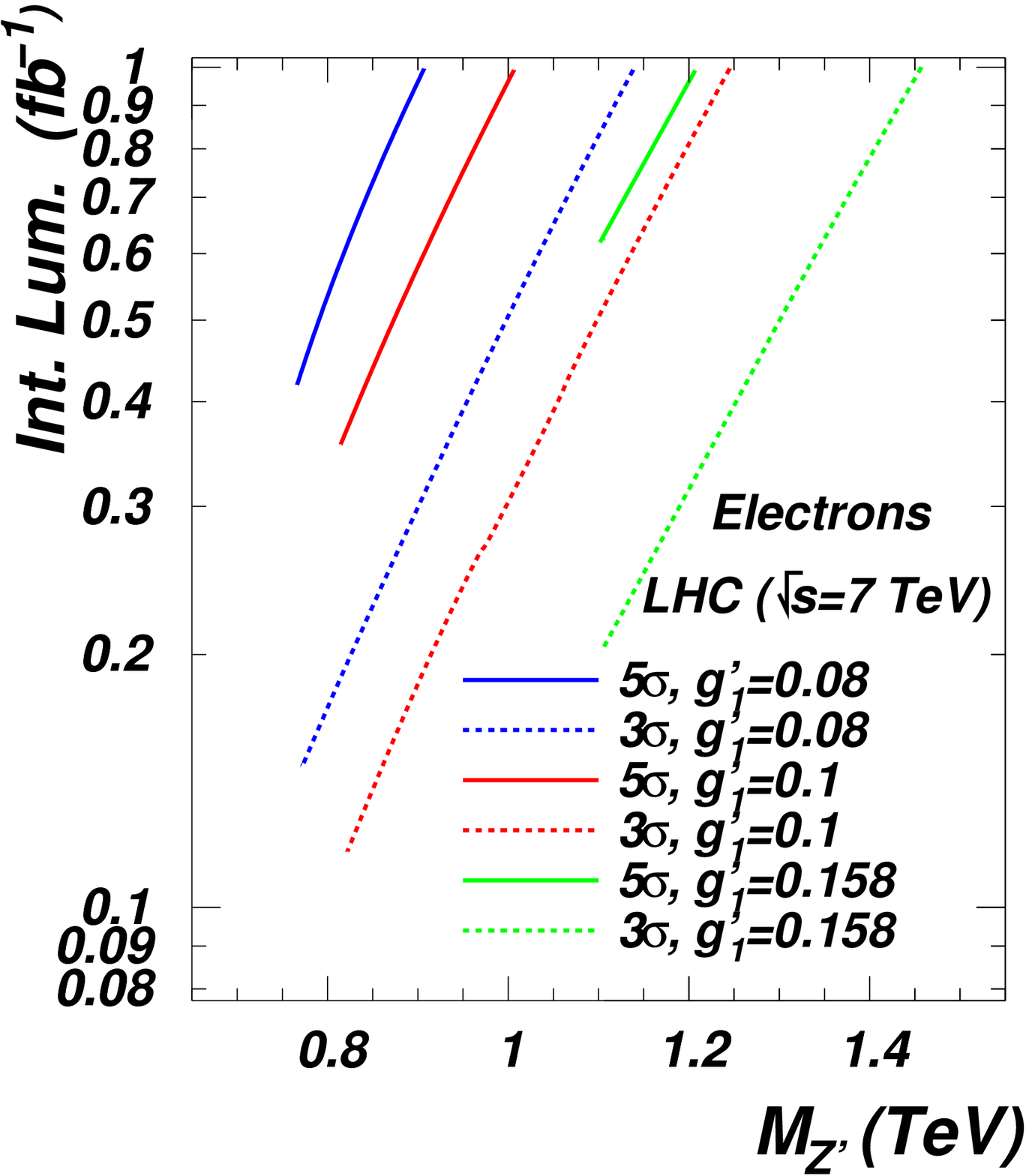}}
  \subfloat[]{
  \label{contour7_ecal_Tev}
  \includegraphics[angle=0,width=0.48\textwidth ]{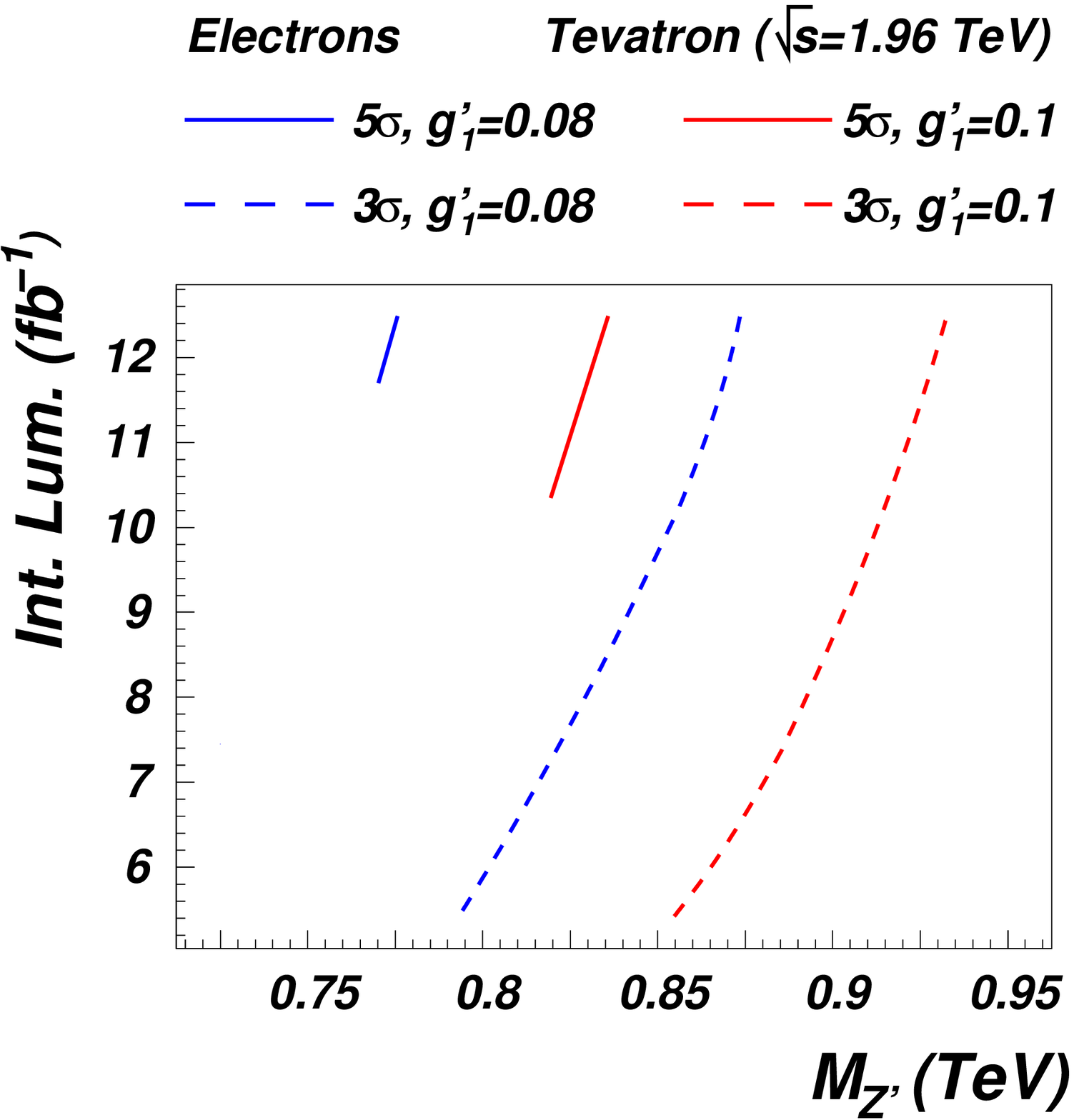}}\\
\flushleft{  
  \subfloat[]{
  \label{contour7_pt_LHC}
  \includegraphics[angle=0,width=0.48\textwidth ]{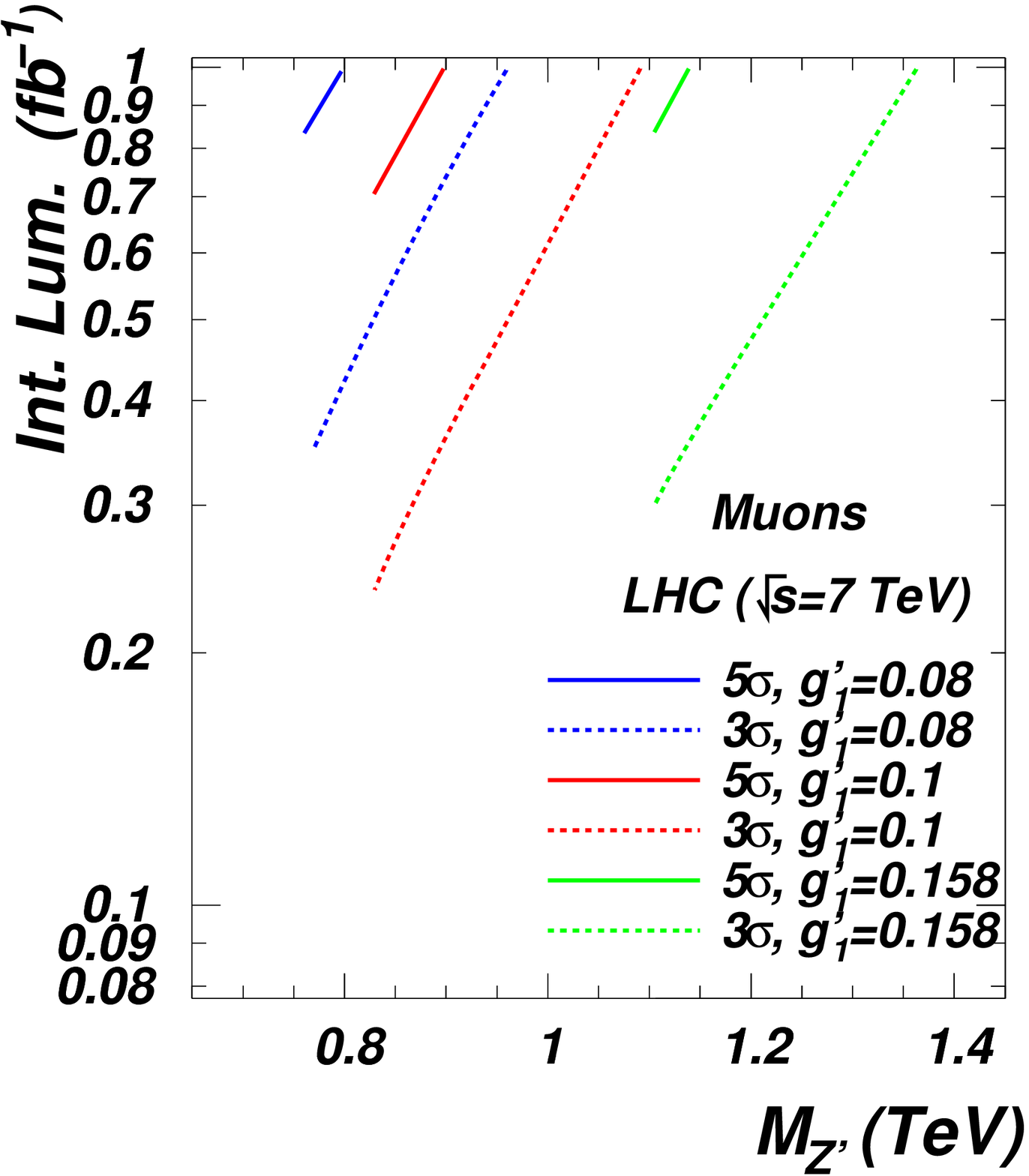}}  }
  \vspace*{-0.25cm}
  \caption{\it Integrated luminosity required for observation at $3\sigma$ and $5\sigma$ vs. $M_{Z'}$ for selected values of $g_1'$ at the LHC for $\sqrt{s}=7$ TeV for (\ref{contour7_ecal_LHC}) electrons and (\ref{contour7_pt_LHC}) muons and at Tevatron ($\sqrt{s}=1.96$ TeV) for (\ref{contour7_ecal_Tev}) electrons (the muon channel requires more than $10$ fb$^{-1}$ and is, hence, not shown).
Only allowed combinations of masses and couplings are shown.}
  \label{lumi_vs_mzp_7TeV}
\end{figure}

Figure~\ref{lumi_vs_mzp_7TeV} shows the integrated luminosity required for $3\sigma$ evidence and $5\sigma$ discovery as a function of the $Z'_{B-L}$ boson mass for selected values of the coupling for both electron and muon final states at the LHC, and for the electron channel only at Tevatron. The muon channel at Tevatron requires more than $10$ fb$^{-1}$ to start probing the $Z'_{B-L}$ boson at $3\sigma$, and, hence, we do not present it.

We now fix some values for the coupling ($g'_1 = 0.158,\, 0.1,\, 0.08$ for the LHC analysis, $g'_1=0.1,\, 0.08$ for Tevatron) and we see what luminosity is required for discovery at each machine. 
For $g'_1=0.1$ the LHC requires $0.35-0.70$ fb$^{-1}$ to be sensitive at $5\sigma$, with electrons and muons, respectively, while Tevatron requires $10$ fb$^{-1}$ with electrons. For the same value of the coupling, Tevatron can discover the $Z'_{B-L}$ boson up to $M_{Z'}=840$ GeV, with electrons in the final state, with $12$ fb$^{-1}$ of data. The LHC can extend the Tevatron reach up to $M_{Z'}=1.0(0.9)$ TeV for $g'_1=0.1$. Regarding $g'_1=0.08$, a discovery can be made chiefly with electrons, requiring $0.4(12)$ fb$^{-1}$, for masses up to $900(780)$ GeV at the LHC(Tevatron). For muons, the LHC requires $0.85$ fb$^{-1}$ for masses up to $800$ GeV. Both machines will be sensitive at $3\sigma$ with much less integrated luminosities, requiring roughly $0.12-0.15(0.25-0.35)$ fb$^{-1}$ to probe the $Z'_{B-L}$ at the LHC for electrons(muons) in the final state, for $g'_1=0.1-0.08$. At Tevatron, $5.5$ fb$^{-1}$ are required for probing at $3\sigma$ both values of the coupling, for electrons only. Finally, bigger values of the coupling, such as $g'_1=0.158$, can be probed just at the LHC, that is sensitive at $3\sigma$ to masses up $1.45(1.35)$ TeV using electrons(muons) and at $5\sigma$ to masses up to $1.2(1.15)$ GeV, requiring $0.2-0.6(0.3-0.9)$ fb$^{-1}$ at least, at $3\sigma-5\sigma$ for electrons(muons).

The $5\sigma$ discovery potential for the LHC at $\sqrt{s}=7$ TeV and for  Tevatron are summarised in the table~\ref{5sigma_at_7TeV}, for selected values of couplings and integrated luminosities.
\begin{table}[h]
\begin{center}
\begin{tabular}{|c||c|c|c||c|c|c|}
\hline
LHC & \multicolumn{3}{|c||}{$pp\rightarrow e^+ e^-$} & \multicolumn{3}{|c|}{$pp\rightarrow \mu^+ \mu^-$} \\
\hline
$ \mathcal{L}$ (fb$^{-1}$)  & $g'_1=0.08$ & $g'_1=0.1$ & $g'_1=0.158$ & $g'_1=0.08$ & $g'_1=0.1$ & $g'_1=0.158$ \\
\hline
0.2 & 820(-)   & 925(-)     & 1100(-)	 & -(-)    & -(-)     & -(-)      \\ 
0.3 & 900(-)   & 1000(800)  & 1200(-)	 & -(-)    & 850(-)   & 1100(-)   \\  
0.5 & 1000(775)& 1100(875)  & 1300(-)    & 825(-)  & 950(-)   & 1200(-)   \\ 
1   & 1130(900)& 1250(1000) & 1450(1200) & 950(800)& 1080(900)& 1360(1130)\\ 
\hline
\hline
Tevatron & \multicolumn{3}{|c||}{$pp\rightarrow e^+ e^-$} & \multicolumn{3}{|c|}{$pp\rightarrow \mu^+ \mu^-$} \\
\hline
$ \mathcal{L}$ (fb$^{-1}$)  & $g'_1=0.08$ & $g'_1=0.1$ & $g'_1=0.158$ & $g'_1=0.08$ & $g'_1=0.1$ & $g'_1=0.158$ \\
\hline
8  & 825(-)   & 895(-)& -(-)& -(-)    & -(-) & -(-)       \\  
10 & 850(-)   & 915(825)& -(-)& -(-)    & -(-) & -(-)       \\ 
12 & 870(775) & 930(830)& -(-)& -(-)    & -(-) & -(-)    \\ 
\hline
\end{tabular}
\end{center}
\vskip -0.5cm
\caption{\it Maximum $Z'_{B-L}$ boson masses (in GeV) for a $3\sigma(5\sigma)$ discovery for selected $g_1'$ and integrated luminosities in the $B-L$ model, both at the LHC (for $\sqrt{s}=7$ TeV) and at Tevatron (for $\sqrt{s}=1.96$ TeV).  No numbers are quoted for already excluded configurations.}
\label{5sigma_at_7TeV}
\end{table}

\paragraph{Exclusion power}
~

\vspace*{0.3cm}
\noindent If no evidence for a signal is found at the LHC at the considered energy and luminosity configuration, $95\%$ C.L. exclusion limits can be derived: in this subsection we present exclusion plots for the early stage of the LHC CM energy ($\sqrt{s}=7$ TeV). We also show the expected exclusions at Tevatron for $\mathcal{L}=10$ fb$^{-1}$.

\begin{figure}[!h]
\centering
  \subfloat[]{ 
  \label{contour7_excl}
  \includegraphics[angle=0,width=0.48\textwidth ]{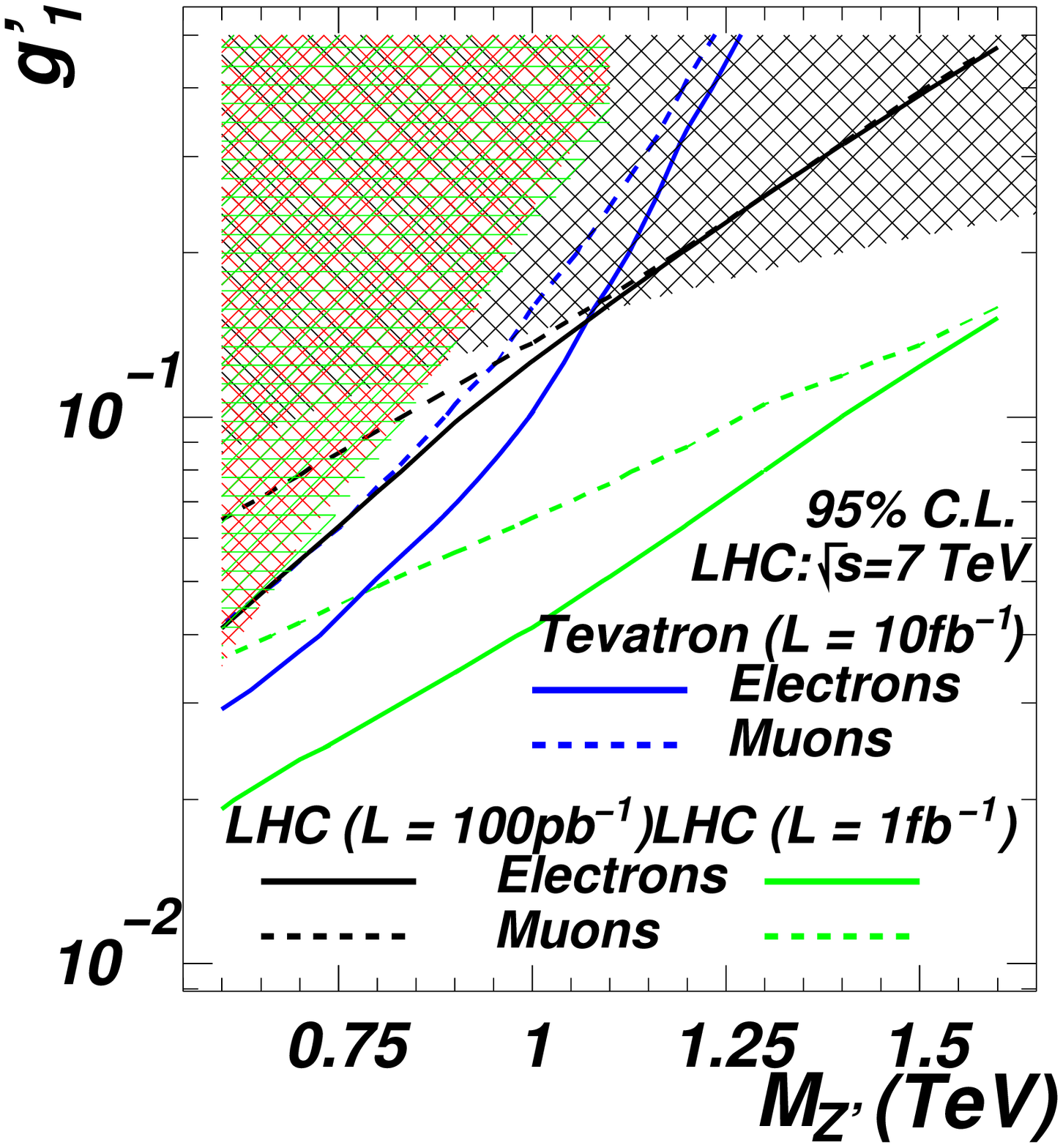}}\\
  \subfloat[]{
  \label{lumi7LHC_excl}
  \includegraphics[angle=0,width=0.48\textwidth ]{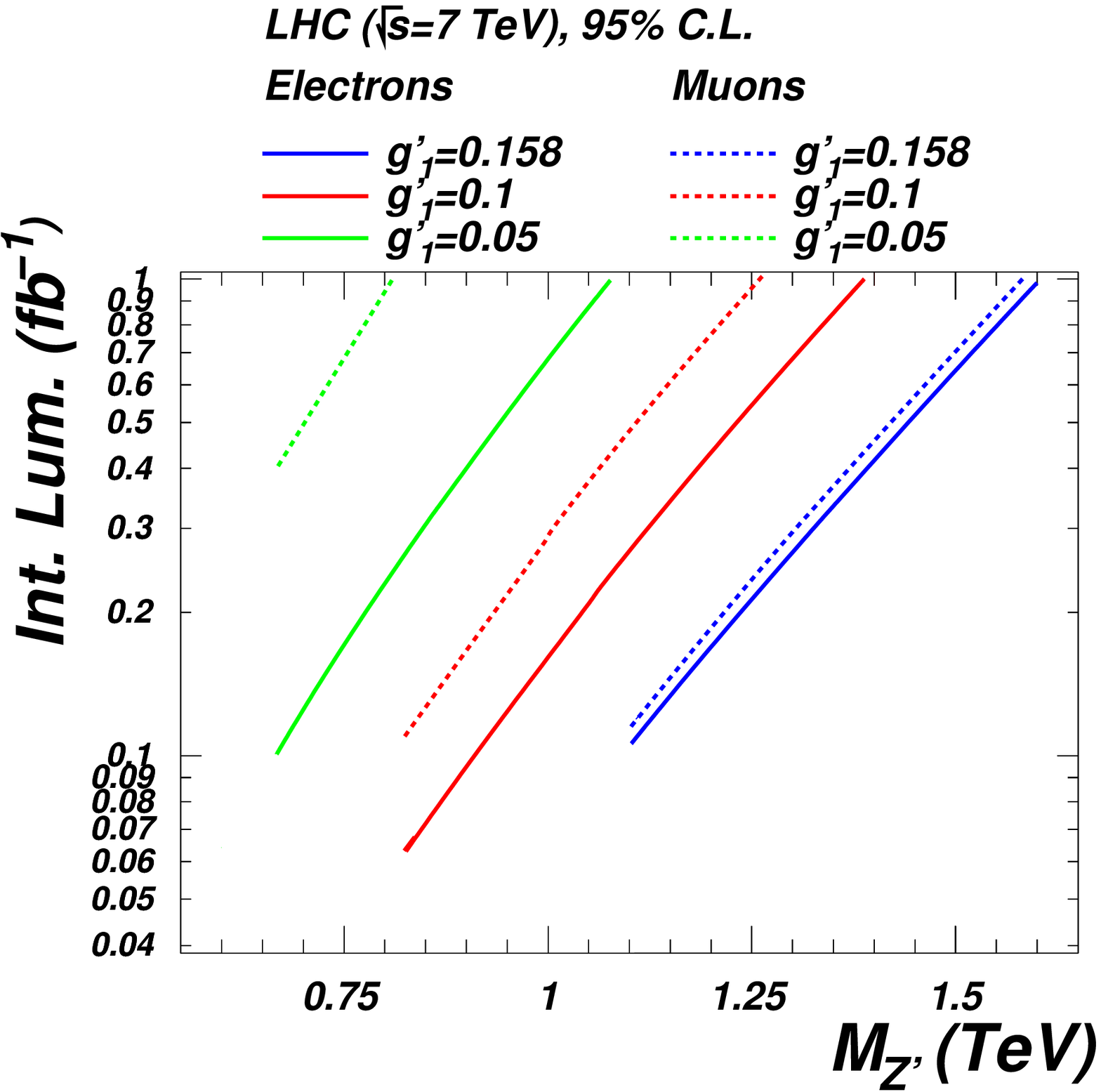}}
  \subfloat[]{
  \label{lumi7Tev_excl}
  \includegraphics[angle=0,width=0.48\textwidth ]{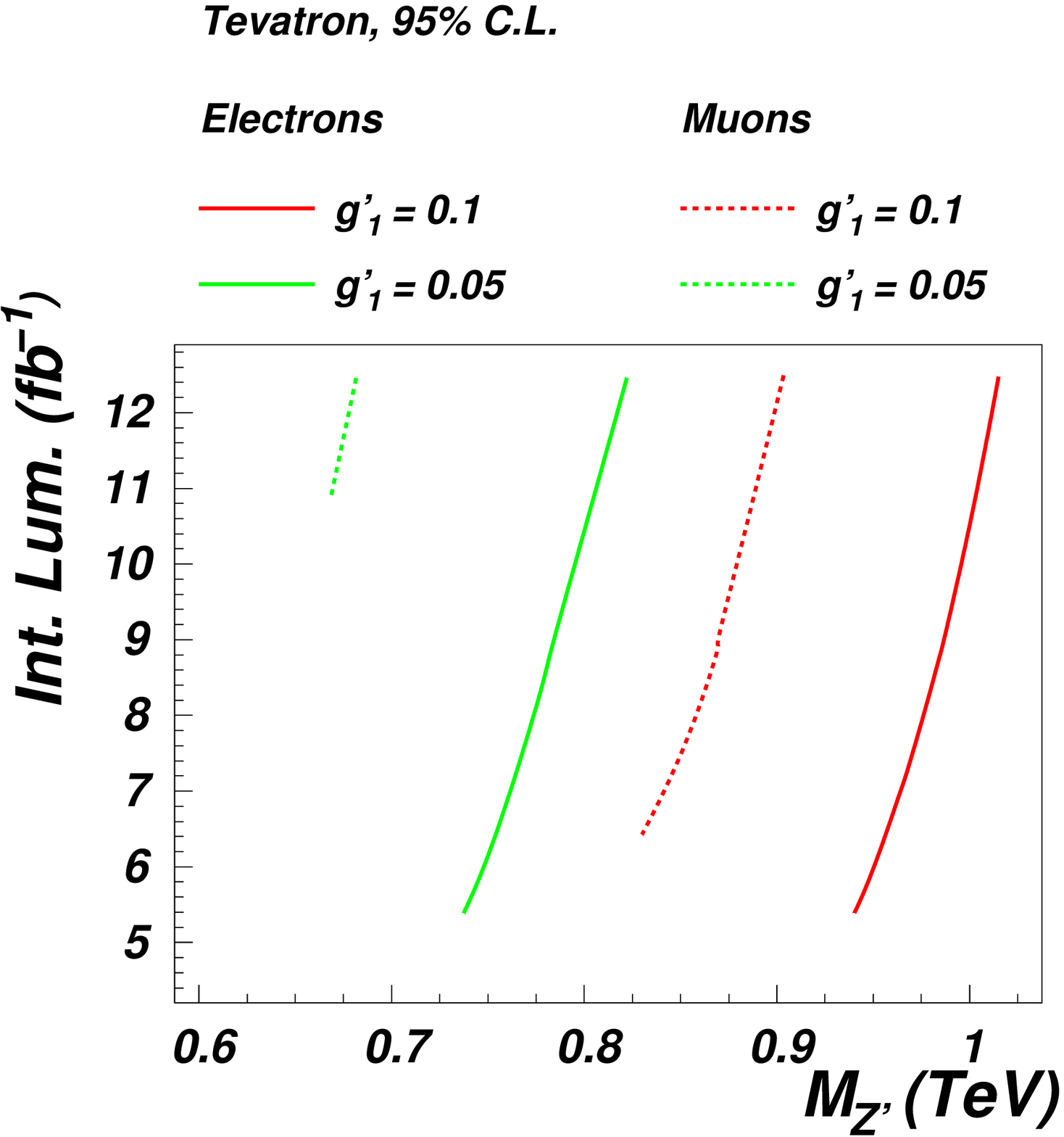}}
  \vspace*{-0.25cm}
  \caption{\it (\ref{contour7_excl}) Contour levels for $95\%$ C.L. plotted against $g_1'$ and $M_{Z'}$ at the LHC for selected integrated luminosities and integrated luminosity required for observation at $3\sigma$ and $5\sigma$ vs. $M_{Z'}$ for selected values of $g_1'$ (in which only the allowed combination of masses and couplings are shown), for (\ref{lumi7LHC_excl}) the LHC at $\sqrt{s}=7$ TeV and (\ref{lumi7Tev_excl}) at Tevatron ($\sqrt{s}=1.96$ TeV), for both electrons and muons.
The shaded areas and the allowed $(M_{Z'},g'_1)$ shown are in accordance with eq.~(\ref{LEP_bound}) (LEP bounds, in black) and table~\ref{mzp-low_bound} (Tevatron bounds, in red for electrons and in green for muons).}
  \label{excl_7}
\end{figure}

We start by looking at the $95\%$ C.L. limits presented in figure~\ref{excl_7} for Tevatron and for this stage of the LHC (for $10$ fb$^{-1}$ and $1$ fb$^{-1}$ of integrated luminosity, respectively).

One can see that the different resolutions imply that the limits derived using electrons are always more stringent than those derived using muons in excluding the $Z'_{B-L}$ boson.
As for the discovery reach, Tevatron is also competitive in setting limits, especially in the lower mass region.
In particular, using electrons and in case of no evidence at Tevatron with $10$ fb$^{-1}$, the $Z'_{B-L}$ boson can be excluded for values of the coupling down to $0.03$ ($0.04$ for muons) for $M_{Z'}=600$ GeV. For the LHC, to set the same exclusion limit for the same mass, $1$ fb$^{-1}$ of integrated luminosity is required, allowing to exclude $g'_1>0.02(0.35)$ using electron(muons) in the final state. For the same integrated luminosity, the LHC has much more scope in excluding a $Z'_{B-L}$, for $M_{Z'}>1.0$ TeV.

For a coupling of $0.1$, the $Z'_{B-L}$ boson can be excluded up to $1.40(1.25)$ TeV at the LHC considering electrons(muons) for $1$ fb$^{-1}$, and up to $1.0(0.9)$ TeV at Tevatron for $10$ fb$^{-1}$ of data. For $g'_1=0.05$, the LHC when looking at muons(electrons) will require $400(100)$ pb$^{-1}$ to start improving the current available limits, while with $100$ pb$^{-1}$ it can set limits on $g'_1=0.158$, out of the reach of Tevatron. It will ultimately be able to exclude $Z'_{B-L}$ boson up to $M_{Z'}=1.6$ TeV for $1$ fb$^{-1}$ (both with electrons and muons).

The $95\%$ C.L. exclusions for the LHC at $\sqrt{s}=7$ TeV and at Tevatron are summarised in table~\ref{2sigma_at_7TeV}, for selected values of couplings and integrated luminosities.
\begin{table}[h]
\begin{center}
\begin{tabular}{|c||c|c|c||c|c|c|}
\hline
LHC & \multicolumn{3}{|c||}{$pp\rightarrow e^+ e^-$} & \multicolumn{3}{|c|}{$pp\rightarrow \mu^+ \mu^-$} \\
\hline
$ \mathcal{L}$ (fb$^{-1}$) & $g'_1=0.05$   & $g'_1=0.1$ & $g'_1=0.158$ & $g'_1=0.05$   & $g'_1=0.1$ & $g'_1=0.158$ \\
\hline 
0.1 & 670  & 900  & 1100 & $-$  & 820  & $-$  \\
0.2 & 770  & 1050 & 1250 & $-$  & 950  & 1225  \\
0.5 & 950  & 1225 & 1450 & 700  & 1100 & 1425  \\
  1 & 1075 & 1375 & 1600 & 800  & 1250 & 1575  \\
\hline
\hline
Tevatron & \multicolumn{3}{|c||}{$pp\rightarrow e^+ e^-$} & \multicolumn{3}{|c|}{$pp\rightarrow \mu^+ \mu^-$} \\
\hline
$ \mathcal{L}$ (fb$^{-1}$) & $g'_1=0.05$   & $g'_1=0.1$ & $g'_1=0.158$ & $g'_1=0.05$   & $g'_1=0.1$ & $g'_1=0.158$ \\
\hline
 6  & 750 & 950  & $-$ & $-$  & $-$ & $-$  \\
 8  & 775 & 975  & $-$ & $-$  & 860 & $-$  \\
 10 & 800 & 1000 & $-$ & $-$  & 875 & $-$  \\
 12 & 825 & 1020 & $-$ & 680  & 900 & $-$  \\
\hline
\end{tabular}
\end{center}
\vskip -0.5cm
\caption{\it Maximum $Z'_{B-L}$ boson masses (in GeV) for a $95\%$ C.L. exclusion for selected $g_1'$ and integrated luminosities in the $B-L$ model. No numbers are quoted for already excluded configurations.}
\label{2sigma_at_7TeV}
\end{table}


\subsubsection{LHC at $\sqrt{s}=14$ TeV}
\begin{figure}[!h]
  \subfloat[]{ 
  \label{contour14_ecal}
  \includegraphics[angle=0,width=0.48\textwidth ]{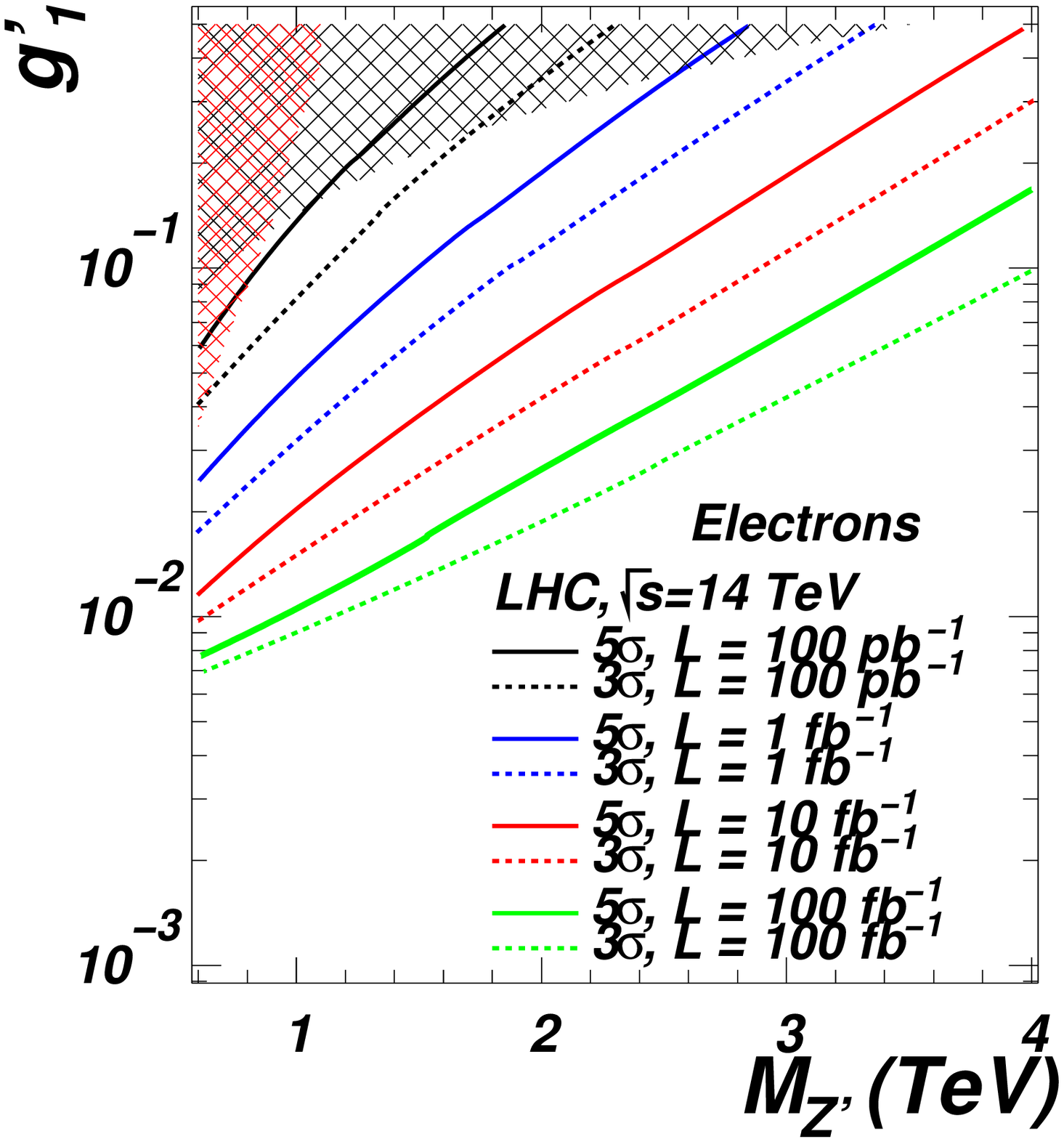}}
  \subfloat[]{
  \label{contour14_pt}
  \includegraphics[angle=0,width=0.48\textwidth ]{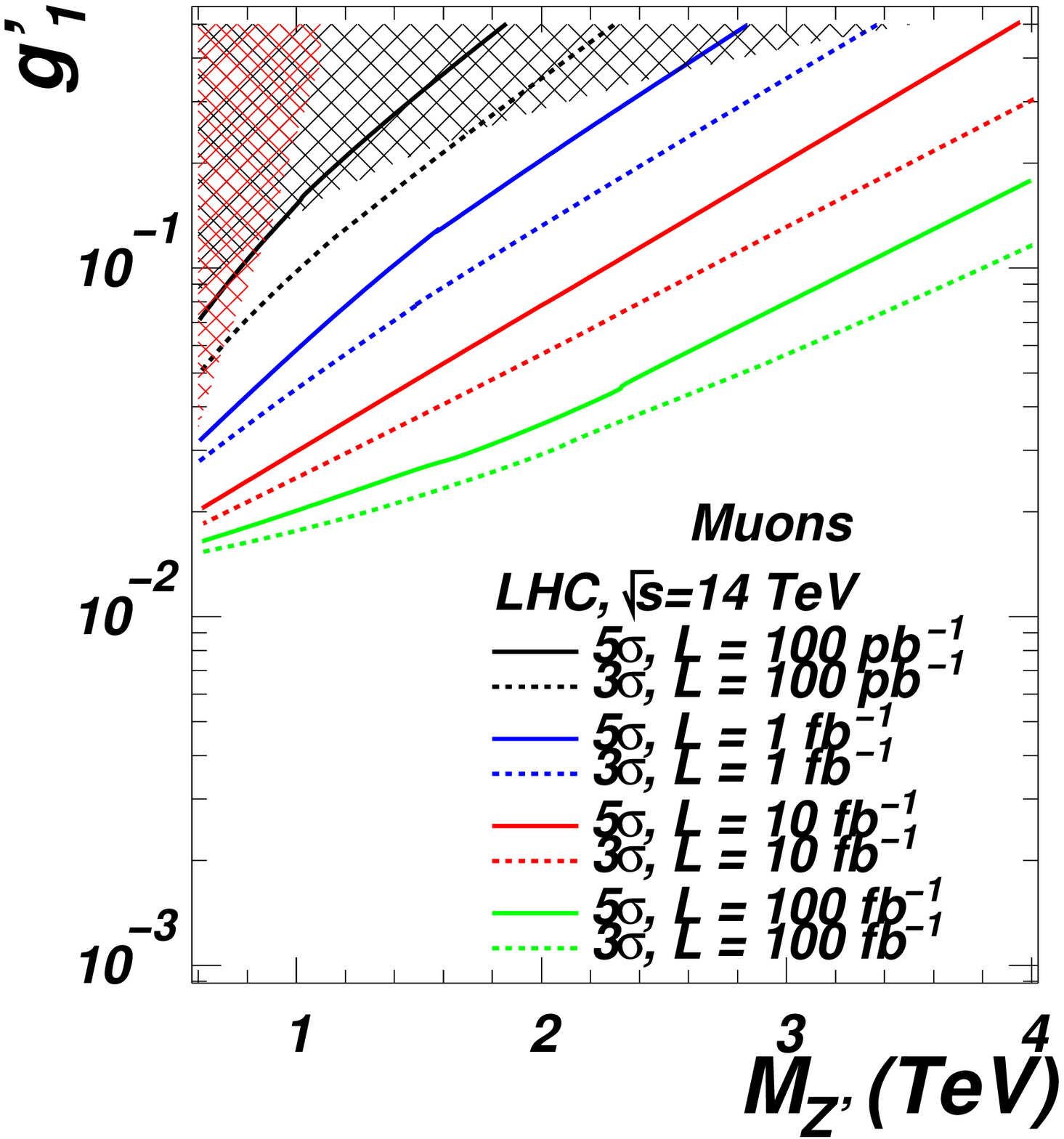}}
  \vspace*{-0.25cm}
  \caption{\it Significance contour levels plotted against $g_1'$
and $M_{Z'}$ at the LHC for $\sqrt{s}~=~14$~TeV for several integrated luminosities for (\ref{contour14_ecal}) electrons and (\ref{contour14_pt}) muons. The shaded areas correspond to the region of parameter space excluded
experimentally, in accordance with eq.~(\ref{LEP_bound}) (LEP bounds, in black) and table~\ref{mzp-low_bound} (Tevatron bounds, in red).}
  \label{contour14}
\end{figure}

\begin{figure}[!h]
  \subfloat[]{ 
  \label{contour14_ecal_Lumi}
  \includegraphics[angle=0,width=0.48\textwidth ]{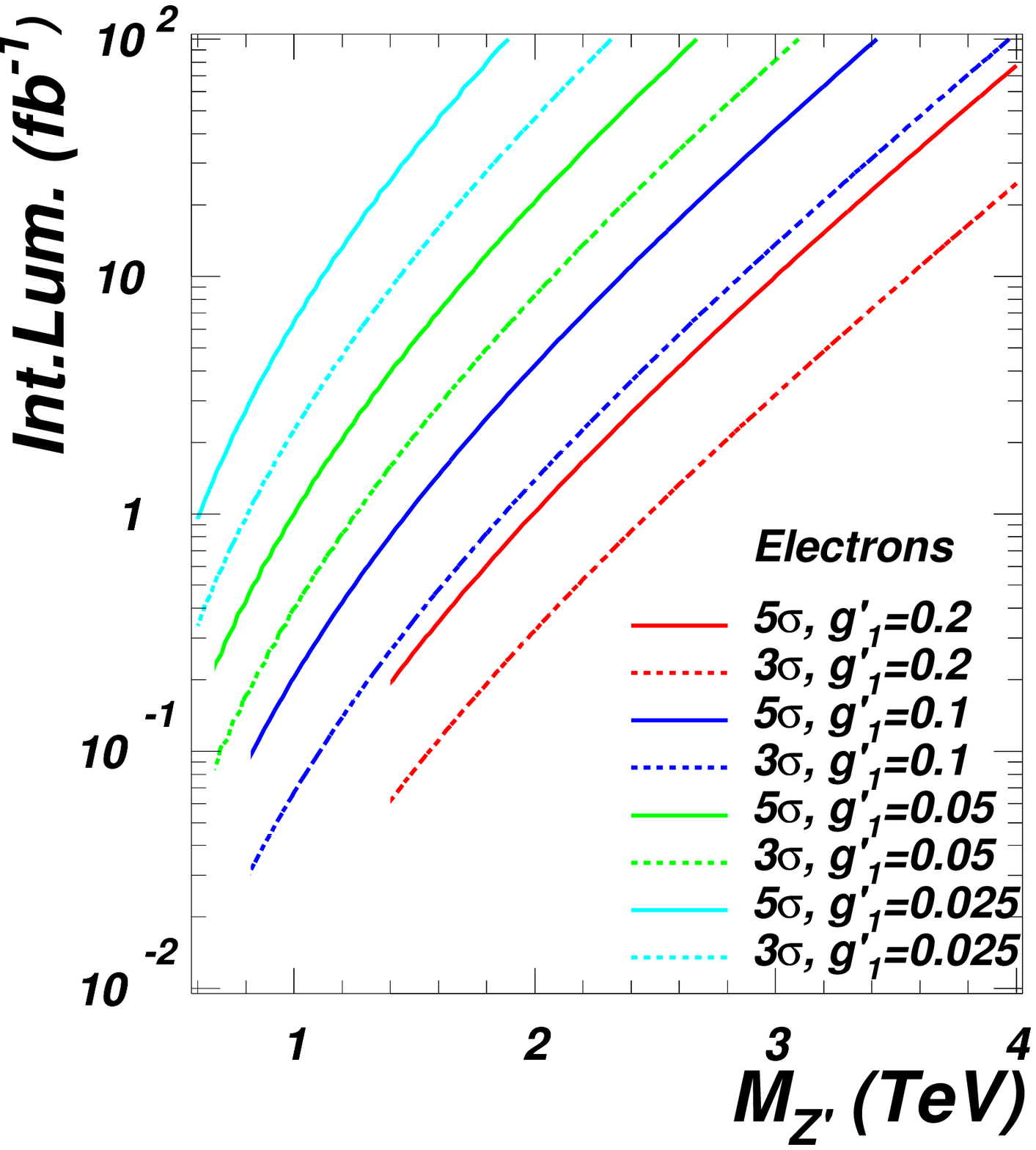}}
  \subfloat[]{
  \label{contour14_pt_Lumi}
  \includegraphics[angle=0,width=0.48\textwidth ]{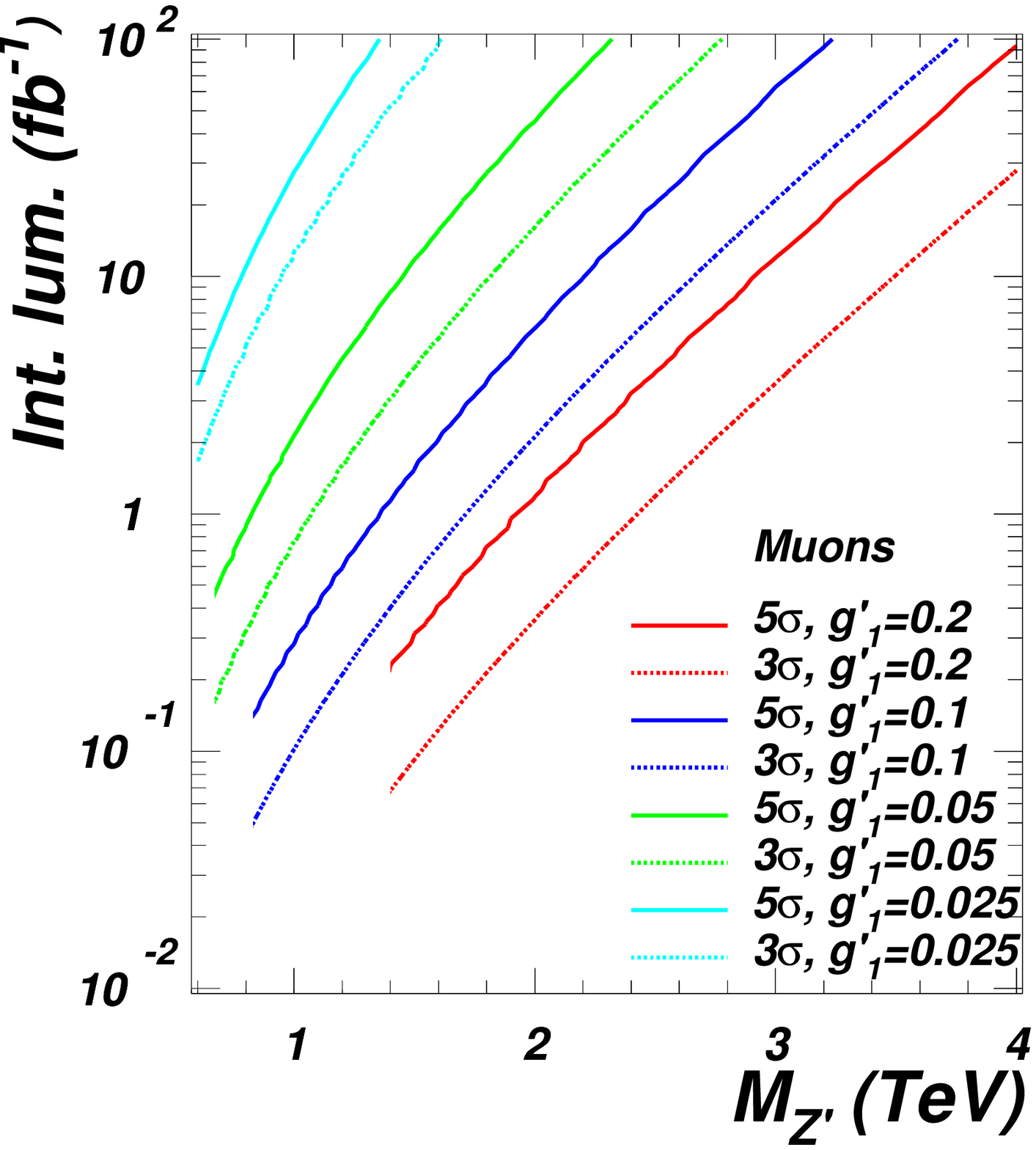}}
  \vspace*{-0.25cm}
  \caption{\it Integrated luminosity required for observation at $3\sigma$ and $5\sigma$ vs. $M_{Z'}$ for selected values of $g_1'$ at the LHC for $\sqrt{s}~=~14$ TeV for (\ref{contour14_ecal_Lumi}) electrons and (\ref{contour14_pt_Lumi}) muons. Only allowed combination of masses and couplings are shown, in accordance with eq.~(\ref{LEP_bound}) and table~\ref{mzp-low_bound}.}
  \label{lumi_vs_mzp_14TeV}
\end{figure}

We consider here the design performance, i.e., $\sqrt{s}=14$ TeV of CM energy with large luminosity, $ \mathcal{L} = 100$ fb$^{-1}$. Figure~\ref{contour14} shows the discovery potential for the $Z'_{B-L}$ boson under these conditions, while figure~\ref{lumi_vs_mzp_14TeV} shows the integrated luminosity required for $3\sigma$ evidence as well as for $5\sigma$ discovery as a function of the $Z'_{B-L}$ boson mass for selected values of the coupling at $\sqrt{s}=14$ TeV. We consider the integrated luminosity in the range between $10$ pb$^{-1}$ up to $100$ fb$^{-1}$. After some years of data analysis, the performances of the detector will be better understood. We therefore use the resolutions for both electrons and muons quoted in eqs.~(\ref{LHC_ris_el_imp}) and (\ref{LHC_ris_mu_imp}), respectively.

From figure~\ref{contour14}, we can see that the LHC at $\sqrt{s}=14$ TeV will start probing a completely new region of the parameter space for $ \mathcal{L} \geq 1$ fb$^{-1}$.
For $ \mathcal{L} \geq 10$ fb$^{-1}$ a $Z'_{B-L}$ gauge boson can be discovered up to masses of $4$ TeV and for couplings as small as $0.01(0.02)$ if we are dealing with electrons(muons). At $ \mathcal{L} = 100$ fb$^{-1}$, the coupling can be probed down to values of $8\,\cdot 10^{-3}$ in the electron channel, while couplings smaller than $1.6\,\cdot 10^{-2}$ cannot be accessed with muons. The mass region that can be covered extends towards $5$ TeV irrespectively of the final state.

As before, figure~\ref{lumi_vs_mzp_14TeV} shows the integrated luminosity required for $3 (5)\sigma$ evidence (discovery) of the $Z'_{B-L}$ boson as a function of
the mass, for selected values of the coupling. We explore the range in luminosities, from $10$ pb$^{-1}$ to $100$ fb$^{-1}$. However, just the configuration with $g'_1=0.1$ can be probed with very low luminosity, requiring $30(100)$ pb$^{-1}$ and $50(150)$ pb$^{-1}$ at $3\sigma(5\sigma)$ considering electrons and muons in the final state, respectively. For values of the coupling such as $0.05$ and $0.2$, $90(220)$ pb$^{-1}$ and $60(200)$ pb$^{-1}$ are the integrated luminosities required to start to be sensitive (at $3(5)\sigma$) if electrons are considered, while $160(500)$ and $70(220)$ pb$^{-1}$ are the least integrated luminosity required, respectively, if instead we look at muons. It is worth to emphasise here that the first couplings that will start to be probed at the LHC are those around $g'_1=0.1$.

The better resolution in the case of electrons reflects in a better sensitivity to smaller $Z'_{B-L}$ masses with respect to muons. For $M_{Z'}=600$ GeV, the LHC with $\sqrt{s}=14$ TeV requires $1.0$ fb$^{-1}$ to be sensitive at $5\sigma$ to a value of the coupling of $0.025$ in the electron channel. If we are considering muons, $3.5$ fb$^{-1}$ is the required luminosity to probe at $5\sigma$ the same value of the coupling.

The $5\sigma$ discovery potential for the LHC at $\sqrt{s}=14$ TeV is summarised in table~\ref{5sigma_at_14TeV}, for selected values of $Z'_{B-L}$ masses and couplings.

\begin{table}[h]
\begin{center}
\scalebox{0.77}{
\begin{tabular}{|c||c|c|c||c|c|c|}
\hline
$\sqrt{s}=14$ TeV & \multicolumn{3}{|c||}{$pp\rightarrow e^+ e^-$} & \multicolumn{3}{|c|}{$pp\rightarrow \mu^+ \mu^-$} \\
\hline
$g'_1$ & $M_{Z'}=1$ TeV & $M_{Z'}=2$ TeV  & $M_{Z'}=3$ TeV & $M_{Z'}=1$ TeV & $M_{Z'}=2$ TeV  & $M_{Z'}=3$ TeV \\
\hline
0.025 & 2.5(7.0)  & 50($>$100)& $>$100($>$300)& 15(30)  & $>$100($>$100)& $>$300($>$300) \\
0.05  & 0.4(1.0)  &  9(20)    & 80($>$100)    & 0.8(2.5)&   20(50)      & $>$100($>$100)\\
0.1   & 0.07(0.2) & 1.5(4.0)  & 15(50)        & 0.1(0.3)&   2.0(6.0)    &  20(65)  \\
0.2   & $-$($-$)  & 0.3(1.0)  & 3(10)         & $-$($-$)&   0.4(1.2)    &  3(12)\\
\hline  
\end{tabular}}
\end{center}
\vskip -0.5cm
\caption{\it Minimum integrated luminosities (in fb$^{-1}$) for a $3\sigma$($5\sigma$) discovery for selected $Z'_{B-L}$ boson masses and $g_1'$ couplings for the $B-L$ model. No numbers are quoted for already excluded configurations.}
\label{5sigma_at_14TeV}
\end{table}

\begin{figure}[!h]
  \subfloat[]{ 
  \label{14TeV_3sigma_el}
  \includegraphics[angle=0,width=0.48\textwidth ]{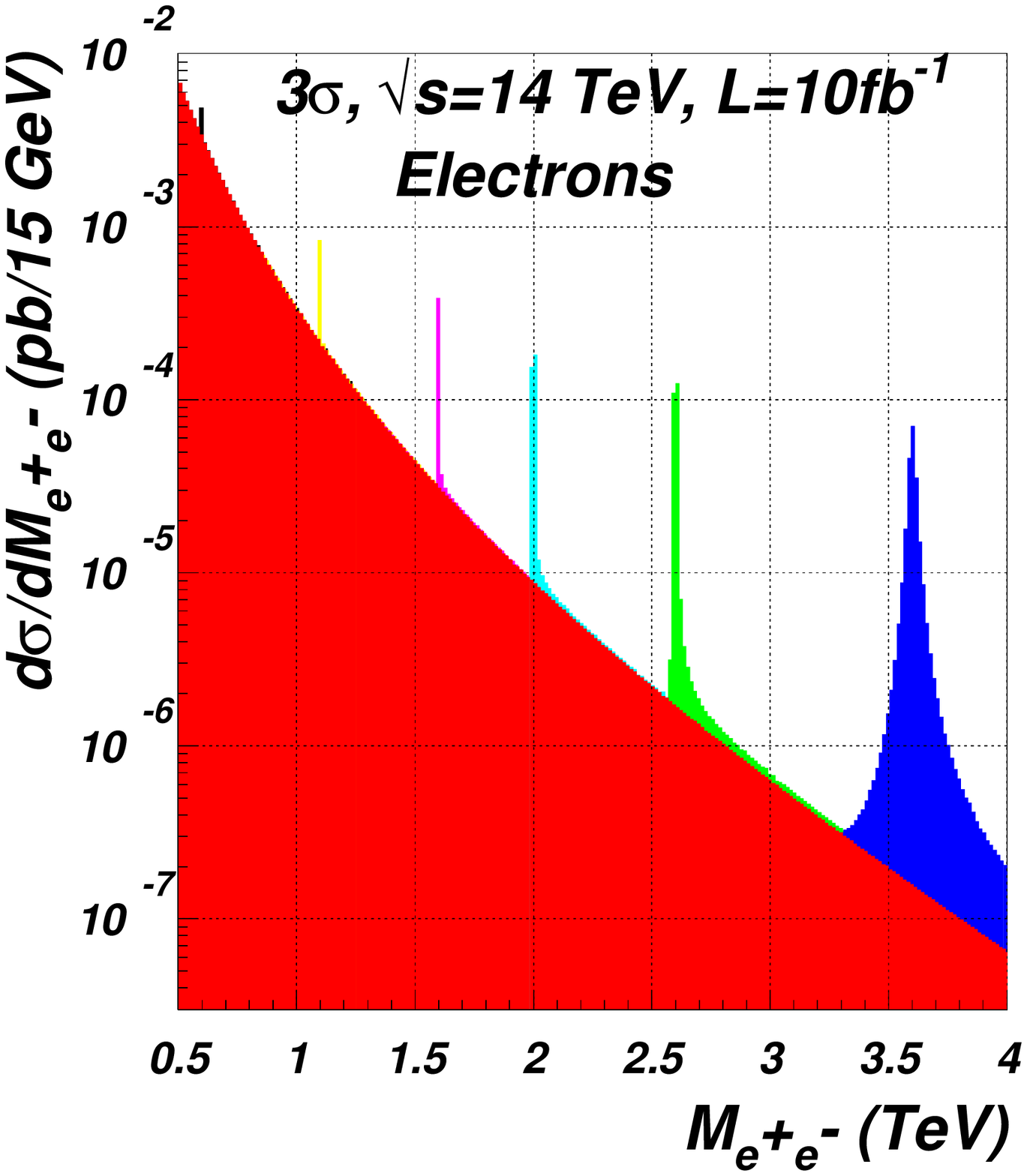}}
  \subfloat[]{
  \label{14TeV_5sigma_el}
  \includegraphics[angle=0,width=0.48\textwidth ]{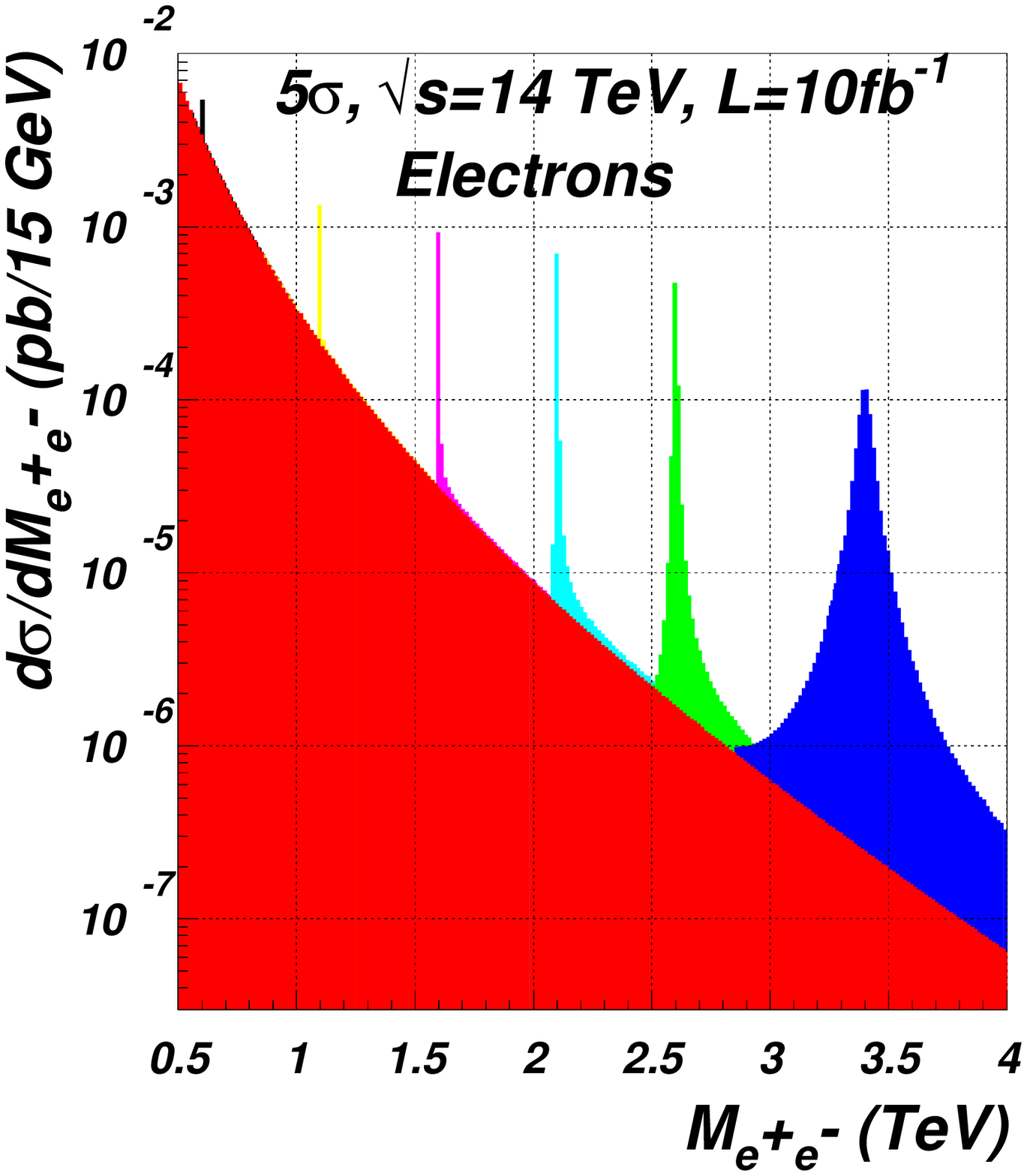}}
  \vspace*{-0.25cm}
  \caption{\it $\frac{d\sigma}{dM_{\ell \ell}}(pp\rightarrow \gamma ,Z,Z'_{B-L} \rightarrow e ^+e ^-)$ for several masses and couplings ($M_{Z'}/$TeV, $g'_1$, $\Gamma _{Z'}/GeV$): (\ref{14TeV_3sigma_el}) ($0.6$, $0.0075$, $0.006$), ($1.1$, $0.015$, $0.05$), ($1.6$, $0.025$, $0.21$), ($2.0$, $0.04$, $0.67$), ($2.6$, $0.07$, $2.7$) and ($3.6$, $0.2$, $31$); (\ref{14TeV_5sigma_el}) ($0.6$, $0.009$, $0.009$), ($1.1$, $0.02$, $0.09$), ($1.6$, $0.04$, $0.53$), ($2.1$, $0.07$, $2.2$), ($2.6$, $0.12$, $7.9$) and ($3.4$, $0.3$, $61$), from the $3\sigma$ and $5\sigma$ lines at $10$ fb$^{-1}$ of figure~\ref{contour14} ($\sqrt{s}=14$ TeV), respectively, using 15 GeV binning. Notice that the asymmetry of the peaks is the result of our choice to consider the full interference structure.}
  \label{14TeV_3-5sigma_el}
\end{figure}

\begin{figure}[!h]
  \subfloat[]{ 
  \label{14TeV_3sigma_mu}
  \includegraphics[angle=0,width=0.48\textwidth ]{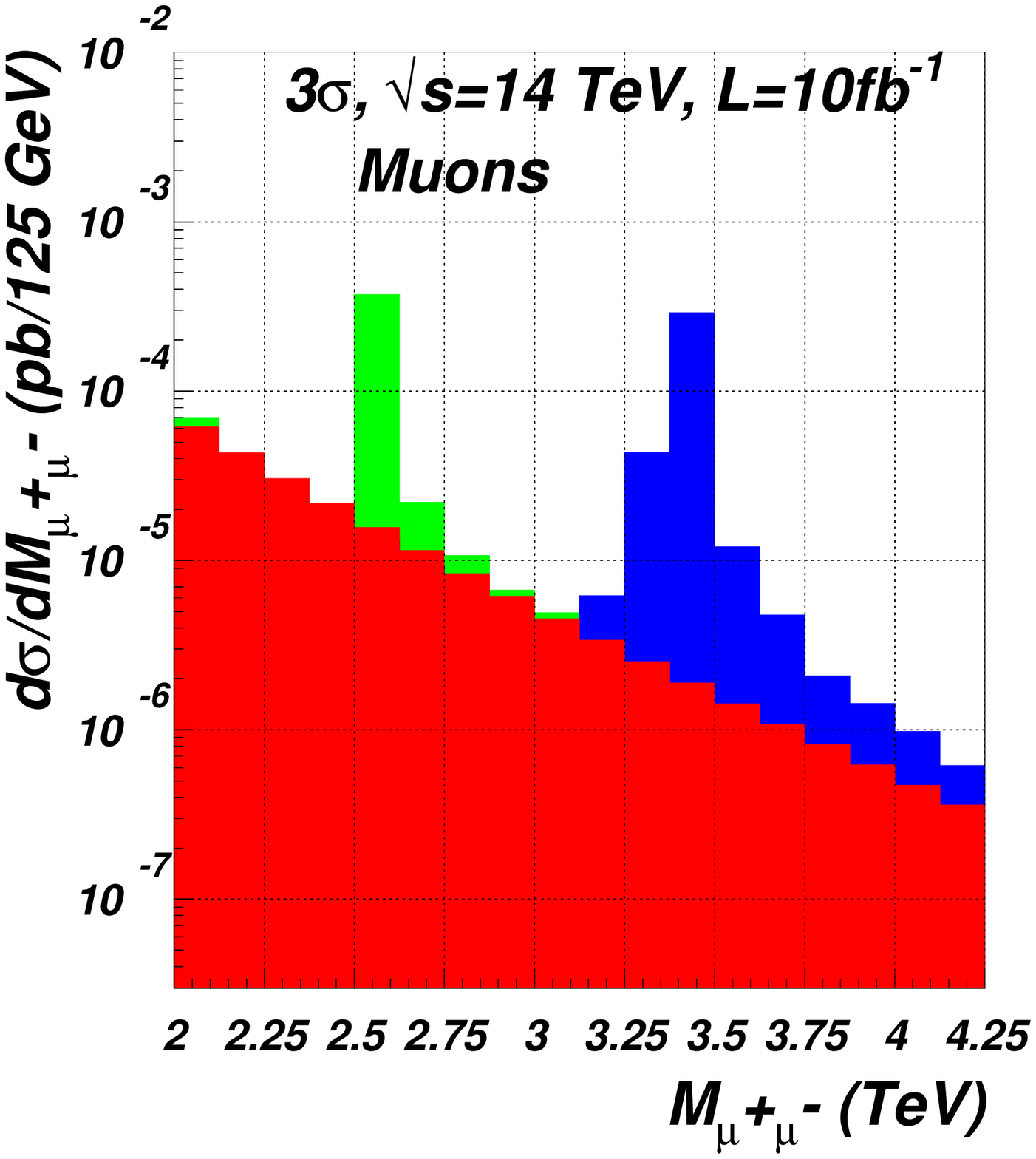}}
  \subfloat[]{
  \label{14TeV_5sigma_mu}
  \includegraphics[angle=0,width=0.48\textwidth ]{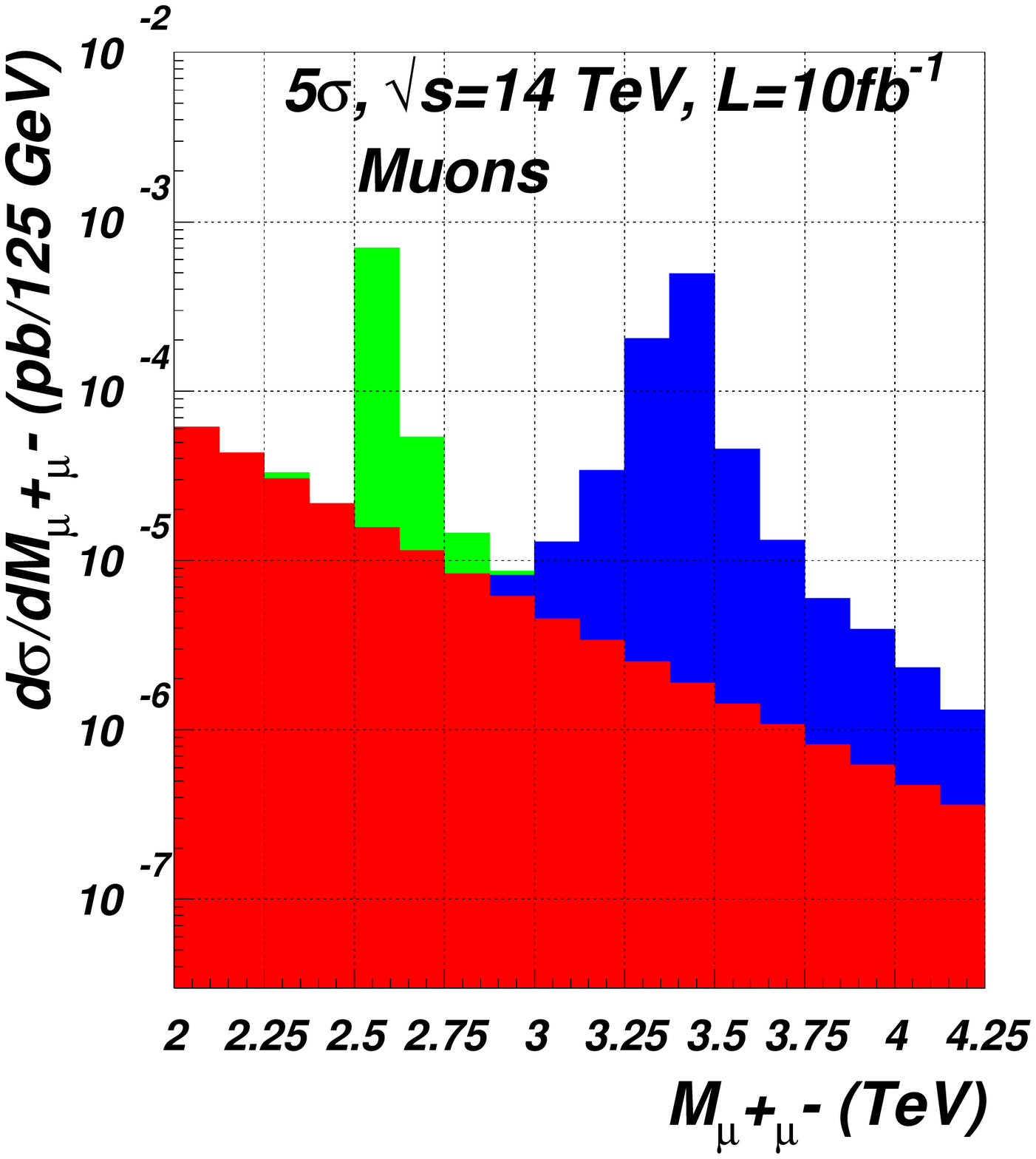}}
  \vspace*{-0.25cm}
  \caption{\it $\frac{d\sigma}{dM_{\ell \ell}}(pp\rightarrow \gamma ,Z,Z' \rightarrow \mu ^+\mu ^-)$ for some masses and couplings ($M_{Z'}/$TeV, $g'_1$, $\Gamma _{Z'}/GeV$): (\ref{14TeV_3sigma_mu}) ($2.6$, $0.07$, $2.7$) and ($3.4$, $0.2$, $29$) and (\ref{14TeV_5sigma_mu}) ($2.6$, $0.12$, $8$) and ($3.4$, $0.3$, $65$), from the $3\sigma$ and $5\sigma$ lines at $10$ fb$^{-1}$ of figure~\ref{contour14} ($\sqrt{s}=14$ TeV), using $125$ GeV binning. Notice that the asymmetry of the peaks is the result of our choice to consider the full interference structure. }
  \label{14TeV_3-5sigma_mu}
\end{figure}

Figures~\ref{14TeV_3-5sigma_el} and \ref{14TeV_3-5sigma_mu} show a pictorial representation of the $Z'$ properties and line-shapes (widths and cross sections) for selected benchmark points on the $3\sigma$ and $5\sigma$ lines for $10$ fb$^{-1}$ of data at $\sqrt{s}=14$ TeV, plotting the di-lepton invariant mass to which just the cuts of eq.~(\ref{LHC_cut}) have been applied (without selecting any mass window). The binning is equal to the typical resolution in each case. \index{Line-shape of the $Z'$ boson}

The improved resolution for electrons allows a measure of the $Z'_{B-L}$ boson width not only at high masses, but also opens the possibility of a measurement even for smaller masses.

\paragraph{Exclusion power}
~

\vspace*{0.3cm}
\noindent If no evidence for a signal is found at the LHC either at its designed energy and luminosity configuration, very strong $95\%$ C.L. exclusion limits can be derived.

\begin{figure}[!h]
  \subfloat[]{ 
  \label{contour14_excl}
  \includegraphics[angle=0,width=0.48\textwidth ]{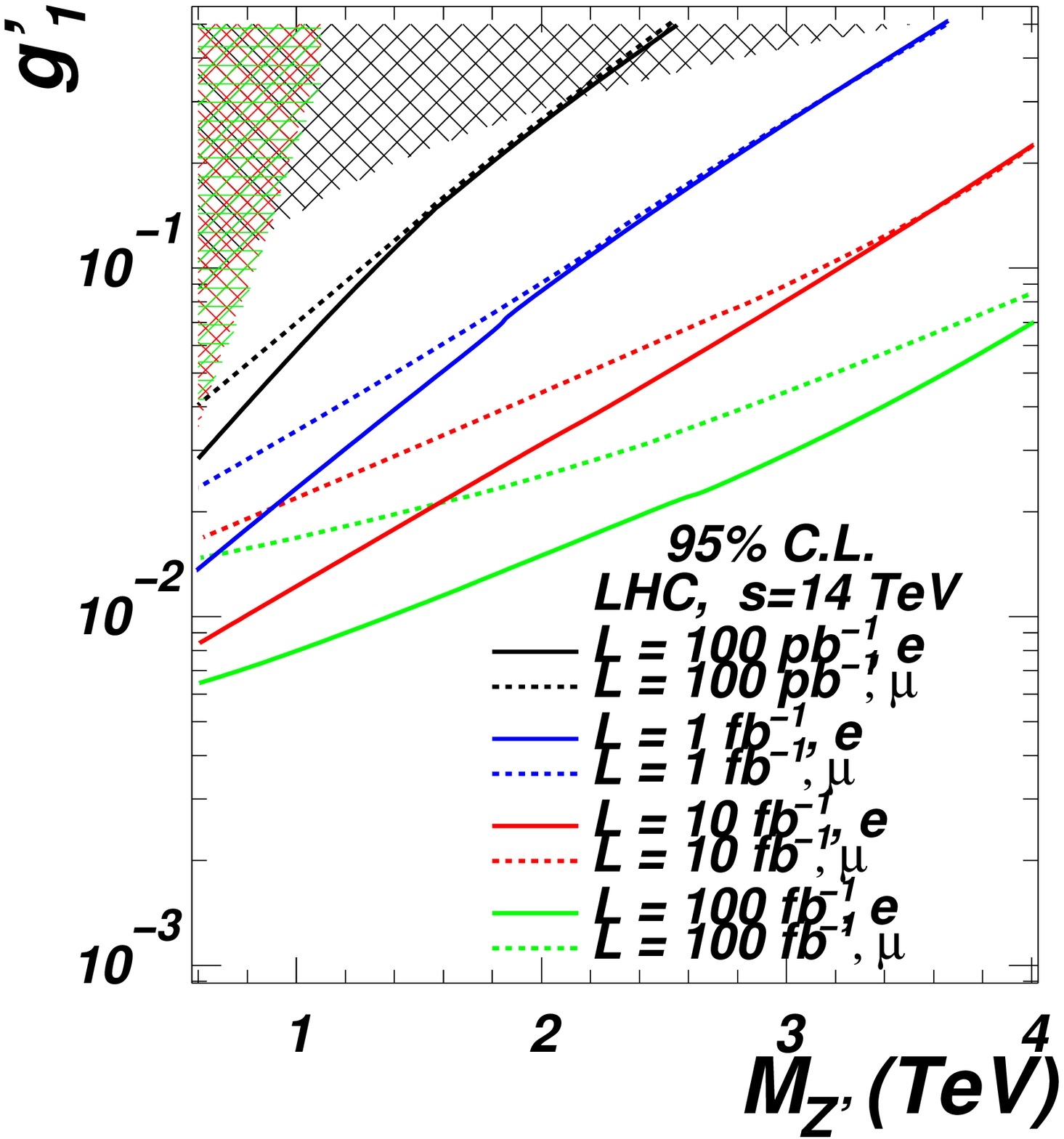}}
  \subfloat[]{
  \label{lumi14_excl}
  \includegraphics[angle=0,width=0.48\textwidth ]{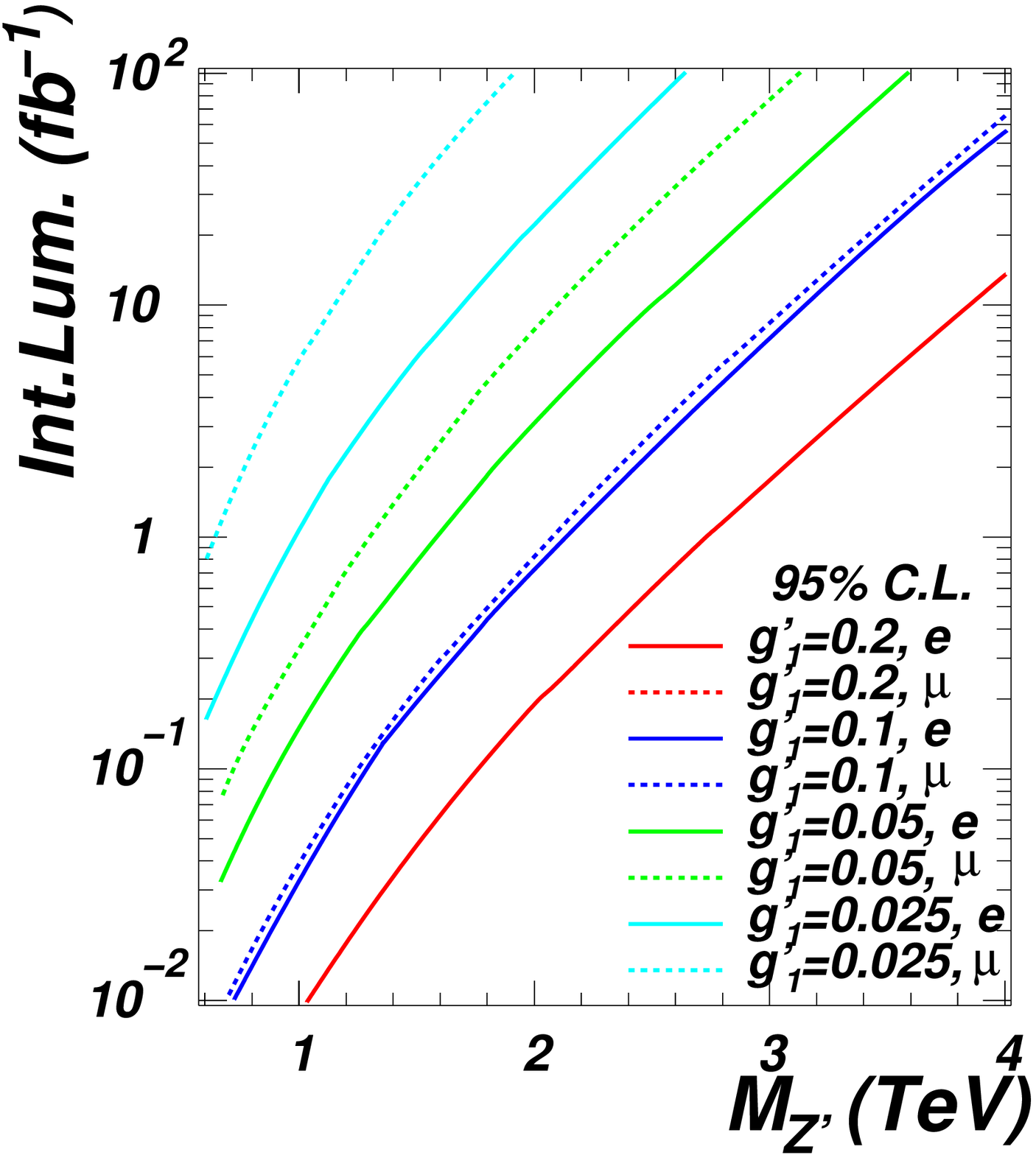}}
  \vspace*{-0.25cm}
  \caption{\it (\ref{contour14_excl}) Contour levels for $95\%$ C.L. plotted against $g'_1$ and $M_{Z'}$ at the LHC for selected integrated luminosities and (\ref{lumi14_excl}) integrated luminosity required for observation at $3\sigma$ and $5\sigma$ vs. $M_{Z'}$ for selected values of $g_1'$ (in which only the allowed combination of masses and couplings are shown), for  $\sqrt{s}=14$ TeV, for both electrons
and muons.
The shaded areas and the allowed $(M_{Z'},g'_1)$ shown are in accordance with eq.~(\ref{LEP_bound}) (LEP bounds, in black) and table~\ref{mzp-low_bound} (Tevatron bounds, in red for electrons and in green for muons).}
  \label{excl_14}
\end{figure}

Due to the improved resolutions for both electrons and muons,
they have very similar exclusion powers for couplings $g'_1\gtrsim 0.1$, therefore setting similar constraints. Depending on the amount of data that will be collected, several maximum bounds can be set (see figure~\ref{contour14_excl}): i.e., with $10$ fb$^{-1}$ of data, the LHC at $14$ TeV can exclude at $95\%$ C.L. up to a mass of roughly $5$ TeV for a value of the coupling $g'_1=0.5$, that we arbitrary take as the biggest allowed value for the consistency of the model
from a RGE analysis of its gauge sector, as in section~\ref{subsubsect:RGE_gauge}. For $100$ fb$^{-1}$ and for the same value of the coupling, the LHC can exclude at $95\%$ C.L. masses up to roughly $6$ 
TeV. For $10$ fb$^{-1}$ it will be possible to exclude a $Z'_{B-L}$ boson for $M_{Z'}=600$ GeV if the coupling is greater than $1.8\,\cdot 10^{-2}(9\,\cdot 10^{-3})$ for muons(electrons), and values of the coupling greater than $1.5\,\cdot 10^{-2}(6.5\,\cdot 10^{-3})$ for an integrated luminosity of $100$ fb$^{-1}$.
Figure \ref{lumi14_excl} shows the integrated luminosity that is required to excluded a certain $Z'_{B-L}$ boson mass for fixed values of the coupling. As previously noticed, electrons and muons set the same limits for $g'_1 \geq 0.1$.
 An integrated luminosity of $10$ fb$^{-1}$ is required to exclude a $Z'_{B-L}$ mass up to $3.8$ TeV for $g'_1=0.2$, instead $40$ fb$^{-1}$ reduces this to
$g'_1=0.1$.
For an integrated luminosity of $10$ fb$^{-1}$ the LHC experiments will be able to exclude the $Z'_{B-L}$ boson for masses up to $3.1(3.1)$ TeV for $g'_1=0.1$, $2.5(2.1)$ TeV for $g'_1=0.05$ and $1.7(1.1)$ TeV for $g'_1=0.025$, when considering decays into electrons(muons). With $100$ fb$^{-1}$ of data, more stringent bounds can be derived: for $g'_1=0.05(0.025)$ the $Z'_{B-L}$ boson can be excluded for masses up to $3.6(2.6)$ TeV in the electron channel, and up to $3.1(1.9)$ TeV in the muon channel. 

The $95\%$ C.L. exclusions for the LHC at $\sqrt{s}=14$ TeV are summarised in  table~\ref{2sigma_at_14TeV}, for selected values of $Z'$ masses and couplings.
\begin{table}[h]
\begin{center}
\scalebox{0.77}{
\begin{tabular}{|c||c|c|c||c|c|c|}
\hline
$\sqrt{s}=14$ TeV & \multicolumn{3}{|c||}{$pp\rightarrow e^+ e^-$} & \multicolumn{3}{|c|}{$pp\rightarrow \mu^+ \mu^-$} \\
\hline
$g'_1$ & $M_{Z'}=1$ TeV & $M_{Z'}=2$ TeV  & $M_{Z'}=3$ TeV & $M_{Z'}=1$ TeV & $M_{Z'}=2$ TeV  & $M_{Z'}=3$ TeV \\
\hline
0.025 & 1.0   & 20   & $>$100	& 7    & $>$100 & $>$300    \\  
0.05  & 0.15  & 3    & 30	& 0.40 &  8     &   80    \\ 
0.1   & 0.04  & 0.7  & 7 	& 0.04 & 0.8    &   9    \\ 
0.2   & $-$   & 0.2  & 2  	& $-$  & 0.2    &   2    \\ 
\hline
\end{tabular}}
\end{center}
\vskip -0.5cm
\caption{\it Minimum integrated luminosities (in fb$^{-1}$) for a $95\%$ C.L. exclusion for selected $Z'_{B-L}$ boson masses and $g_1'$ couplings in the $B-L$ model. No numbers are quoted for already excluded configurations.}
\label{2sigma_at_14TeV}
\end{table}

\section{Conclusion}\label{ch3:concl}
In this chapter, we have presented the results of our investigation of the gauge sector.

First, we have presented the experimental constraints on the free parameter in this sector, the $Z'$ boson mass and gauge coupling $g'_1$. We have shown that the LEP indirect limits and the Tevatron direct ones are complementary, being the former(latter) constraining more for heavy(light) $Z'$ boson. We have also shown that a constraint on $g'_1$ can be derived from a RGE analysis of the running gauge couplings, whose equations decouple from the others. The RGE analysis gives an upper bound on $g'_1$ at the EW scale, to avoid Landau poles up to a certain cut-off energy scale.

We have then moved on, to study the phenomenology of the $Z'$ boson in the pure $B-L$ model. We have presented production cross sections, decay widths and BRs for $2$-body decay modes. On the parameter space that we have considered, the $Z'$ boson is a rather narrow resonance, broadening up for high masses and big couplings. The decay into heavy neutrinos is certainly interesting, with a total BR up to $\sim 20 \%$, at most. We will study it with great details in the next chapter.

We have shown that the $Z'$ boson is dominantly coupled to leptons (their total BR sums up to $3/4$). It only holds vectorial coupling to Dirac fermions, while neutrinos, being Majorana particles, couple to the $Z'$ boson with pure axial couplings.

Finally, we have presented the discovery potential for the $Z'_{B-L}$ gauge boson at the LHC for CM energies of $\sqrt{s}=7$ and $14$ TeV, using the integrated luminosity expected at each stage. This has been done for both the $Z'_{B-L}\rightarrow e^+e^-$ and $Z'_{B-L}\rightarrow \mu ^+\mu ^-$ decay modes, and includes the most up-to-date constraints coming from LEP and Tevatron. We have proposed an alternative analysis that has the potential to improve experimental sensitivities.
 We also haved looked in detail at the different resolutions, showing that electrons and muons present very similar discovery power for values of the coupling bigger than roughly $0.1$, for $\sqrt{s}=14$ TeV.

We are overall confident that the inclusion of further background, as well as a realistic detector simulation, will not have a considerable impact on the results we have presented. In fact, as noted in section~\ref{subsect:disc_power}, all detector effects can be casted in the form of a {\it signal acceptance}, including also the effect of kinematic and angular acceptance cuts. By looking at Refs.~\cite{Abazov:2010ti,Aaltonen:2011gp,Khachatryan:2010fa}, we estimate an overall acceptance factor of $\sim 70\%$, which we found to be approximately constant over the mass regions considered, to be applied to our parton level results [once the cuts of eqs.~(\ref{LHC_cut}) and (\ref{Tev_cut}) are considered]. This acceptance is mainly related to the lepton identification, both at Tevatron and at the LHC. On the significance, as in eq.~(\ref{signif}), the reduction is then of $\sim 84\%$.

A general feature is that greater sensitivity to the $Z'_{B-L}$ resonance is provided by the electron channel. At the LHC this has better energy resolution than the muon channel. A further consequence of the better resolution of electrons is that an estimate of the gauge boson width would eventually be possible for smaller values of the $Z'_{B-L}$ boson mass than in the muon channel. The simultaneous measure of the $Z'$ boson mass and gauge coupling (for instance, through a line-shape fit) will also enable to infer the $B-L$ breaking VEV, given the simple formula that relates them.
Limits from existing data imply that the first couplings that will start to be probed at the LHC are those around $g'_1=0.1$. Increased luminosity will enable both larger and smaller couplings to be probed.

Our comparison has shown that, for an integrated luminosity of $10$ fb$^{-1}$, Tevatron is still competitive with the LHC in the electron channel and in the small mass region, being able to probe the coupling at the level of $5\sigma$ down to a value of $4.2 \cdot 10^{-2}$. The LHC will start to be competitive in such a region only for integrated luminosities close to $1$ fb$^{-1}$ at $\sqrt{s}=7$ TeV. Also, at $\sqrt{s}=7$ TeV, the mass reach will be extended from the Tevatron value of $M_{Z'}=850$ GeV, with electrons, up to $1.25(1.20)$ TeV for electrons(muons).
The muon channel at Tevatron needs more than $10$ fb$^{-1}$ to start probing the $Z'_{B-L}$ at $3\sigma$. Hence, it has not been studied.

When the data from the high energy runs at the LHC becomes available, the discovery reach of $Z'_{B-L}$ boson will be extended towards very high masses and small couplings in regions of parameter space well beyond the reach of  Tevatron and comparable in scope with those accessible at a future LC \cite{Basso:2009hf}.

If no evidence is found at any energies, $95\%$ C.L. limits can be derived, and, given their better resolution, the bounds from electrons will be more stringent than those from muons, especially at smaller masses.
\chapter{Fermion sector}\label{Ch:4}
\ifpdf
    \graphicspath{{Chapter4/Chapter4Figs/PNG/}{Chapter4/Chapter4Figs/PDF/}{Chapter4/Chapter4Figs/}}
\else
    \graphicspath{{Chapter4/Chapter4Figs/EPS/}{Chapter4/Chapter4Figs/}}
\fi

In this chapter we study the phenomenology at the LHC of the new particles in the fermion sector of the pure $B-L$ model, namely, the heavy neutrinos.

We start in section~\ref{sec:expbounds:neutr} by presenting the existent constraints on the fermion sector. Special mention is for the light neutrinos masses, for which no absolute measure exists, with the strongest bounds coming from cosmological observables.

In section~\ref{sect:nu_h} the properties of the heavy neutrinos are presented, i.e., their production cross sections, widths, and BRs. The possibility for these to be long-lived particles is rather interesting, allowing for signatures with distinctive displaced vertices, in which a high energetic and isolated pair of leptons point to a different vertex than the primary one. Bringing pieces together, the process $pp \to Z' \to \nu_h \nu_h$ is studied. In particular, we show that a suitable and peculiar signature to investigate is the tri-lepton decay of the $Z'$ boson via heavy neutrino pairs. This signature provides a powerful insight in the fermion sector of the model (and in its interplay with the gauge sector), and it allows for the measure of the heavy neutrino masses. It is remarkable that with the simultaneous measure of the neutrino lifetime (through the displacement of its decay vertex) and mass, light neutrino masses can be inferred.
For two benchmark points, in section~\ref{subsect:trilep} we present a detailed study of the tri-lepton signature (the main result of this Thesis), where signal and backgrounds are fully simulated and analysed (at the parton level and at the detector level), showing that it yields a good signal over background ratio that makes it observable at the LHC, especially as rather generic cuts for background rejection are considered.

Finally, we draw the conclusions for this chapter in section~\ref{ch4:concl}.

The work in this section is published, in part, in \cite{Basso:2008iv}.

\section{Experimental constraints}\label{sec:expbounds:neutr}

LEP-I data \cite{:2005ema} have established the existence of exactly $3$ neutrinos that couple to the SM $Z$ boson, with masses below $M_{Z}/2$. Therefore, to avoid this bound, we require the heavy neutrinos to be heavier than $M_{Z}/2 \simeq 46$ GeV.

For the SM (or light) neutrinos, no absolute mass measure exists (see \cite{pdg2010}). Nonetheless, oscillation experiments provide a measure for squared mass differences \cite{Fogli:2008jx,Altarelli:2010fk}:
\begin{eqnarray} \label{dm_12}
\delta m_{12}^{\phantom{12}2} &\sim &7.67^{+0.16}_{-0.19} \cdot 10^{-5} \, \mbox{ eV}^2\, , \\ \label{dm_23}
\delta m_{23}^{\phantom{23}2} &\sim &2.39^{+0.11}_{-0.08} \cdot 10^{-3}\, \mbox{ eV}^2\, .
\end{eqnarray}
and mixing angles \cite{Fogli:2008jx,Altarelli:2010fk}:
\begin{eqnarray}\label{mixing_angle_12}
(\sin{\vartheta _{12}})^2 &=& 0.312^{+0.019}_{-0.018}\, ,\\ \label{mixing_angle_23}
(\sin{\vartheta _{23}})^2 &=& 0.466^{+0.073}_{-0.058}\, ,\\ \label{mixing_angle_13}
(\sin{\vartheta _{13}})^2 &=& 0.016 \pm 0.010\, .
\end{eqnarray}
A cosmological upper bound of $\displaystyle \sum_l m_{\nu _l}<0.58$ eV also exists \cite{Jarosik:2010iu}. Ultimately, light neutrinos have been taken to be degenerate and with a mass of $m_{\nu_l}~=~10^{-2}$~eV.


\section{Phenomenology of the heavy neutrinos}\label{sect:nu_h}

After the diagonalisation of the neutrino mass matrix realising the see-saw mechanism, as in section~\ref{sect:neutrino_masses}, we obtain three very light neutrinos ($\nu_l$), which are the SM-like neutrinos, and three heavy neutrinos ($\nu_h$). The latter have an extremely small mixing with the  $\nu_l$'s thereby providing very small but non-vanishing  couplings to  gauge and Higgs bosons.

As already stated, we are interested here in delineating their impact onto the collider phenomenology, especially concerning the interaction with the gauge sector. Therefore, the analysis of the signatures from heavy neutrinos is mainly described in the light of discovery at hadronic colliders, in general, and at the LHC, in particular. In this section, we consider as discovery channel the pair production via the $Z'$ boson.

In this respect, to maximise and highlight the interesting patterns that can be observed, heavy neutrinos will be taken relatively light (with respect to $M_{Z'}$), and, for simplicity, degenerate in mass. As we pointed out in the introduction, the latter is an approximation for illustrative though realistic purposes.

The distinctive features of the $B-L$ model take place because the heavy neutrinos decay predominantly to SM gauge bosons, in association with a lepton (either charged or neutral, depending on the electrical nature of the SM gauge boson). Being these couplings see-saw suppressed, in a large portion of the parameter space heavy neutrinos can be long-lived particles. Also, once heavy neutrinos are pair-produced via the $Z'$ boson, they give rise to novel and spectacular multi-lepton and/or multi-jet decay modes of the intermediate boson. In the following sections are shown the consequences for collider searches.




It is reasonable to conclude that the heavy neutrinos truly carry the hallmarks of the $B-L$ model at colliders.


\subsection{Production cross sections and decay properties}\label{sect:nu_properties}
\begin{figure}[!ht] \begin{center} \scalebox{1.75}{
\begin{picture}(95,79)(0,30)
\unitlength=1.0 pt
\SetScale{1.0}
\SetWidth{0.5}      
\scriptsize    

\Text(-14.0,70.0)[l]{$q$}
\ArrowLine(-8.0,70.0)(12.0,60.0)
\Text(-14.0,50.0)[l]{$\overline{q}$}
\ArrowLine(12.0,60.0)(-8.0,50.0)

\Photon(12.0,60.0)(50.0,60.0){2.5}{5}
\Text(27.0,70.0)[l]{$Z'$}

\Text(72.0,70.0)[l]{$\nu _h$}
\ArrowLine(50.0,60.0)(70.0,70.0) 
\Text(72.0,50.0)[l]{$\nu _h$}
\ArrowLine(70.0,50.0)(50.0,60.0) 
\end{picture}}
\end{center}
\vspace*{-1cm}
\caption{\it Feynman diagram for heavy neutrino pair production. \label{hnu_pic}}
\end{figure}
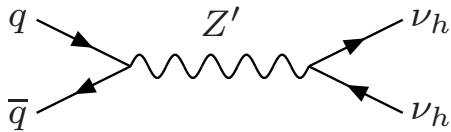

\begin{figure}[!h]
  \subfloat[]{ 
  \label{contour7_nuh}
  \includegraphics[angle=0,width=0.48\textwidth ]{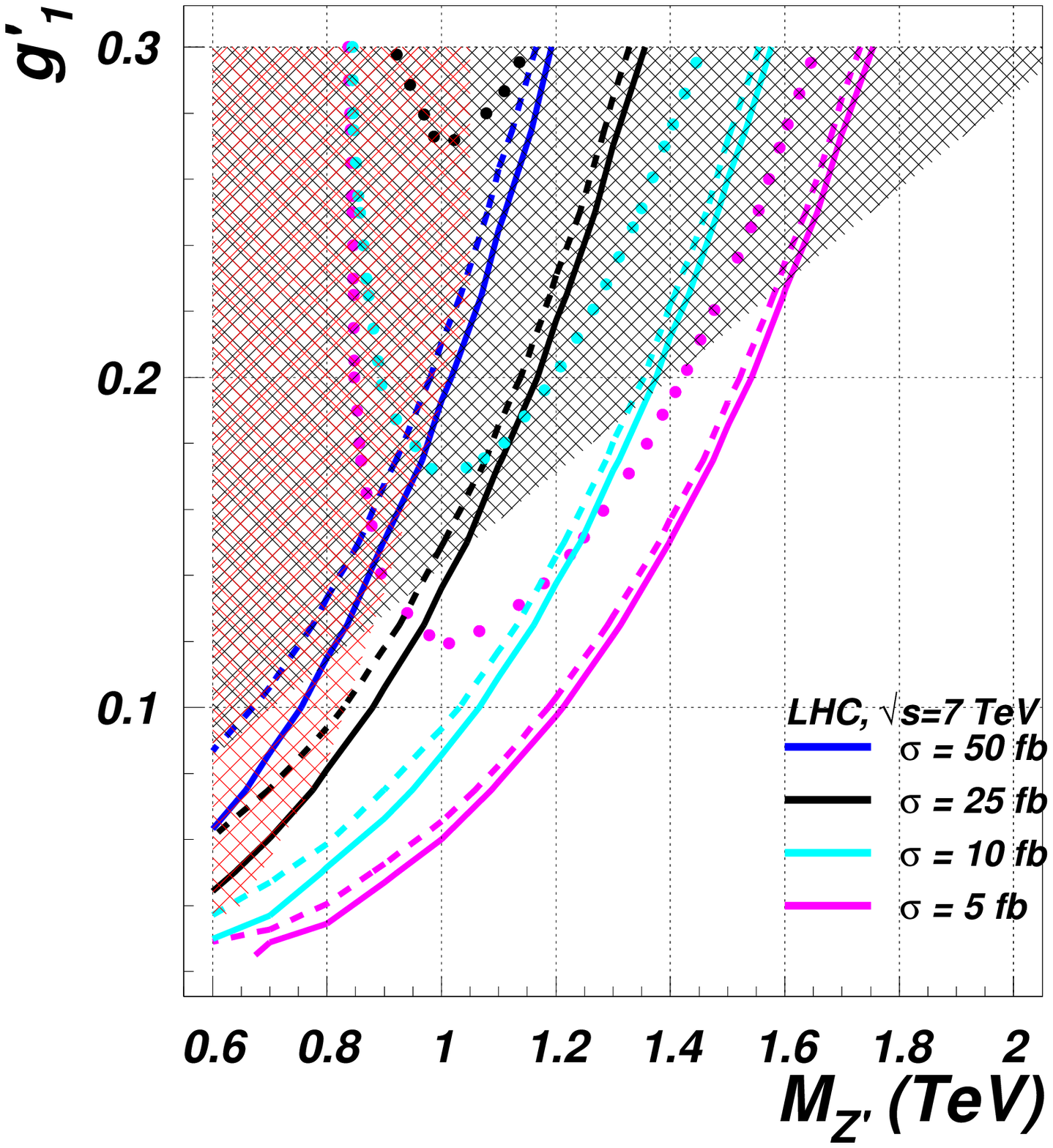}}
  \subfloat[]{
  \label{contour14_nuh}
  \includegraphics[angle=0,width=0.48\textwidth ]{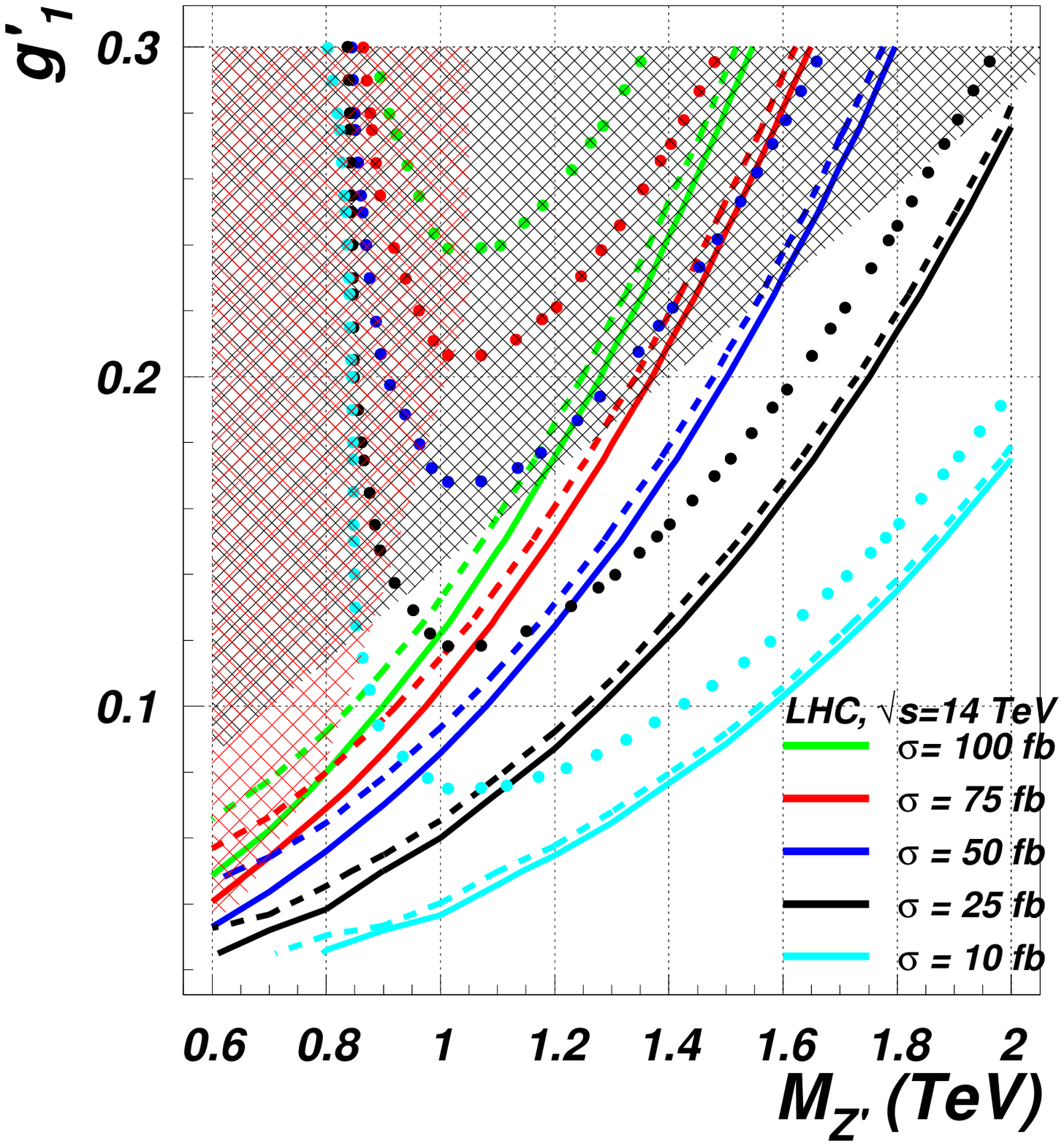}}
  \caption{\it Heavy neutrino pair production cross sections at the LHC for (\ref{contour7_nuh}) $\sqrt{s}=7$ TeV and for (\ref{contour14_nuh}) $\sqrt{s}=14$ TeV, as a function of $m_{\nu _h}$, for several $Z'$ masses and $g'_1$ couplings. The shaded areas shown are experimentally excluded in accordance with eq.~(\ref{LEP_bound}) (LEP bounds, in black) and table~\ref{mzp-low_bound} (Tevatron bounds, in red). Heavy neutrinos have been summed over the generations.}
  \label{nuh_7-14}\index{Production cross sections!Heavy neutrinos}
\end{figure}

We focus here in the pair production via $Z'$ gauge boson exchange in the s-channel, whose Feynman diagram is in figure~\ref{hnu_pic}. Figure~\ref{nuh_7-14} shows the cross sections for the process
\begin{equation}\label{nuh_proc}
pp \rightarrow Z'_{B-L} \rightarrow \sum _i \nu ^i _h \nu ^i _h \,
\end{equation}
at the LHC, for $\sqrt{s}=7$ and $14$ TeV CM energies, in the $M_{Z'}-g'_1$ plane, for several heavy neutrino masses: $m_{\nu _h} = 100$ GeV (straight line), $200$ GeV (dashed line), and $400$ GeV (dotted line).

\vspace*{-2cm}

\begin{figure}[!h]
\centering
  \includegraphics[angle=0,width=0.8\textwidth ]{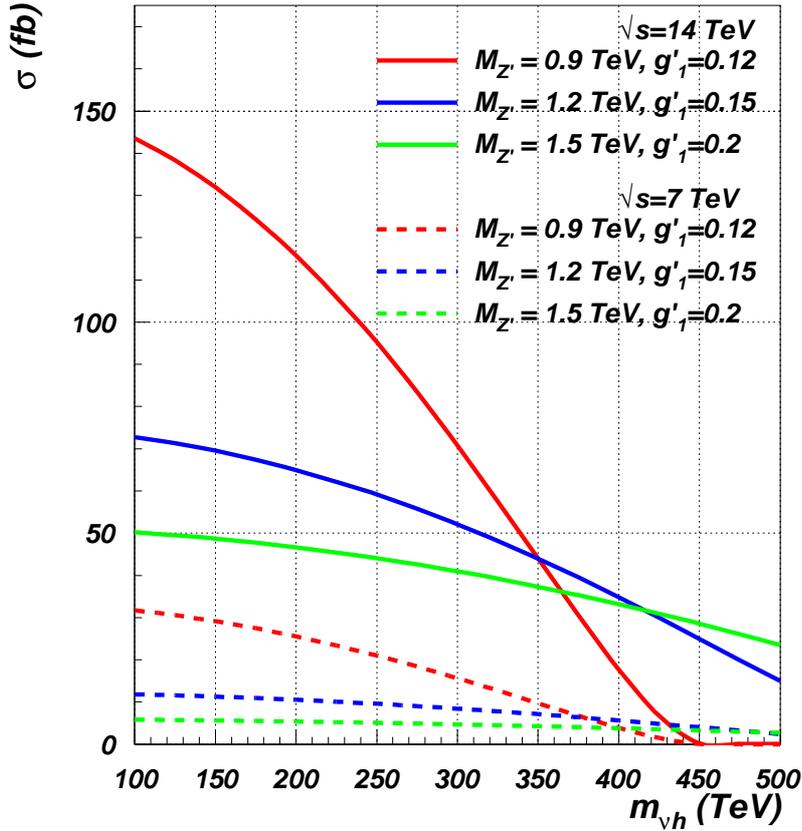}
  \caption{\it Heavy neutrino pair production cross sections at the LHC (for $\sqrt{s}=7$ and $14$ TeV) for several $Z'$ masses and $g'_1$ couplings. Heavy neutrinos have been summed over the generations.}
  \label{nuh_scan}
\end{figure}

Complementary to this, figure~\ref{nuh_scan} shows the production cross sections for the process of eq.~(\ref{nuh_proc}) as a function of the heavy neutrino mass, for a choice of $M_{Z'}$ and $g'_1$, at the LHC, again comparing the two foreseen CM energies. A similar plot, for $\sqrt{s}=14$ TeV only, can be found in Ref.~\cite{Huitu:2008gf}.

It seems clear that the process of eq.~(\ref{nuh_proc}) can hardly be tested at 
the LHC in its early stage (i.e., for $\sqrt{s}=7$ TeV CM energy), where the total integrated luminosity will be not above $1$ fb$^{-1}$. Even supposing the maximum value for the cross section, of $\sim 50$ fb for $M_{Z'}\sim 900$ GeV, $g'_1 \sim 0.13$ and $m_{\nu _h} \sim 100$ GeV, only around $50$ events are produced, and considerably wiped out once detector geometrical and kinematical cuts are included, even before selecting any particular decay mode.

On the contrary, this mechanism is suitable for testing at the LHC for its design performances (i.e., for $\sqrt{s}=14$ TeV CM energy), where, depending on the $Z'$ boson mass and the value of the $g'_1$ coupling, there exist configurations in the parameter space allowing for thousands (or tens of thousands) of heavy neutrino events. In section~\ref{subsect:trilep}, a detailed analysis of a particular decay mode of the neutrino pair will be shown, namely, the tri-lepton mode, for $M_{Z'}=1.5$ TeV and $g'_1=0.2$ (green solid curve of figure~\ref{nuh_scan}).

\subsubsection{Decay properties}
As clear from eqs.~(\ref{nu_mixing}) and (\ref{nu_mix_angle_masses}), heavy neutrinos are predominantly RH states, with a tiny [$\mathcal{O}(10^{-6}$)] component of LH helicity, providing very small but non-vanishing  couplings to  gauge and Higgs bosons. In turn, it is this tiny component that enables the following $\nu_h$ decays, when kinematically allowed: \index{Decay modes!Heavy neutrino}
\begin{eqnarray} \label{nuh_lW}
 \nu_h &\rightarrow& \ell^\pm\, W^\mp\, , \\ \label{nuh_nuZ}
 \nu_h &\rightarrow& \nu_l\,  Z\, , \\
 \nu_h &\rightarrow& \nu_l\, h_1\, , \\
 \nu_h &\rightarrow& \nu_l\, h_2\, , \\
\nu_h &\rightarrow& \nu_l\, Z'\, .
\end{eqnarray}

\begin{figure}[ht]
\begin{center}
\includegraphics[angle=0,height=0.6\textwidth,width=0.80\textwidth]{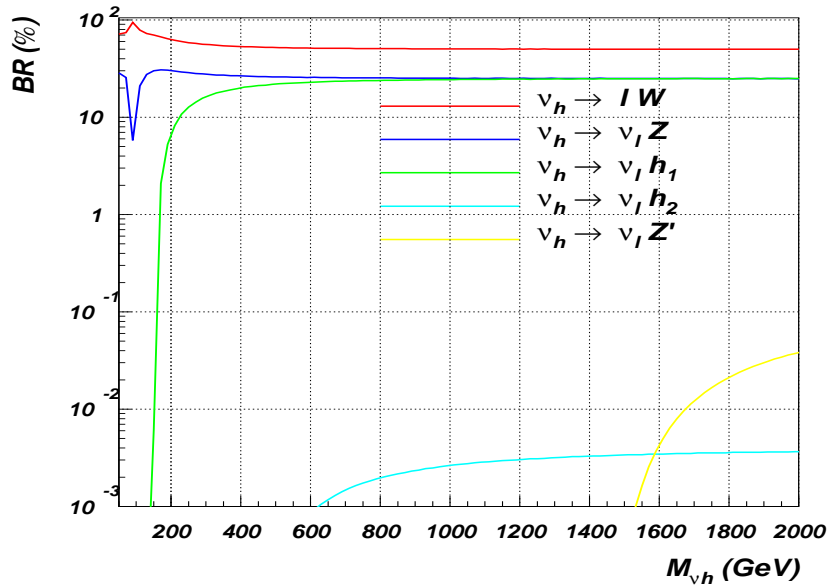}
\caption{\it Heavy neutrino BRs versus its mass for the fixed $M_{Z'}=1.5$~TeV, $m_{h_1}=150$ GeV, $m_{h_2}=450$ GeV, $\alpha\sim 10^{-3}$ rads
and $g'_1=0.2$. Here, $W$ means the sum over $W^+$ and $W^-$.
\label{Hn_BR}
}
\end{center}
\end{figure}\index{BRs!Heavy neutrinos}
Figure~\ref{Hn_BR} presents the corresponding BRs versus 
the heavy neutrino mass for the values of the other relevant $B-L$ parameters
given in the caption. One can see that the BR$\left( \nu_h \rightarrow \ell^\pm W^\mp \right)$ is dominant and reaches the $1/2$ level in the  $m_{\nu _h} \gg M_W, M_Z, m_{h_1}$ limit, while BR$\left( \nu_h\rightarrow \nu_l Z \right)$
and BR$\left( \nu_h\rightarrow \nu_l h_1\right)$ both reach the $1/4$ level in this regime (for $\alpha \to 0$). Schematically, as in Ref.~\cite{Perez:2009mu},
\begin{eqnarray}
\mbox{BR}(\nu_h \rightarrow \ell^+ W^-) &\approx&
\mbox{BR}(\nu_h \rightarrow \ell^- W^+) \approx
\mbox{BR}( \nu_h \rightarrow \nu_l  Z)\, , \\
&\approx& \mbox{BR}(\nu_h \rightarrow \nu_l h_1) + \mbox{BR}(\nu_h \rightarrow \nu_l h_2)\, .
\end{eqnarray}
In contrast, the $\nu_h \rightarrow \nu_l Z'$ decay channel (when kinematically open) is well below the percent level and is negligible for our study. To simplify the discussion in this section, we assume that the heavy neutrino masses are smaller than both Higgs boson masses. Under this assumption $\nu_h\rightarrow \nu_l h_i$ ($i=1,2$) is not kinematically possible and BR$\left( \nu_h\rightarrow \ell^\pm W^\mp \right)$[BR$\left( \nu_h\rightarrow \nu_l Z\right)$] reaches the $2/3$[$1/3$] level in the  $m_{\nu _h} \gg M_W, M_Z$ limit.

Nonetheless, the possibility of producing the light Higgs boson from a heavy neutrino decay is quite interesting and peculiar, as in once it involves new particles from all the sectors of the $B-L$ model. In section~\ref{sect:Higgs}, we will show the cross section for this process, highlighting its importance for a future LC.

We recall that the heavy neutrino couplings to the weak gauge bosons are
proportional to the (squared root of the) ratio of light to heavy neutrino masses (see the Appendix~\ref{App:calchep}), which is extremely small [$\mathcal{O}(10^{-6}$) for $m_{\nu _h} \sim 100$ GeV]. Hence, the decay width of the heavy neutrino is correspondingly small and its lifetime \index{Heavy neutrinos!Lifetime} large. The heavy neutrino can therefore be a long-lived particle and, over a large portion of parameter space, its lifetime can be comparable to or exceed that of the $b$ quark. (In fact, for $m_{\nu_l}=10^{-2}$ eV and $m_{\nu _h}=200$ GeV they are equal).

\begin{figure}[!ht]
\begin{center}
\includegraphics[angle=0,width=0.8\textwidth]{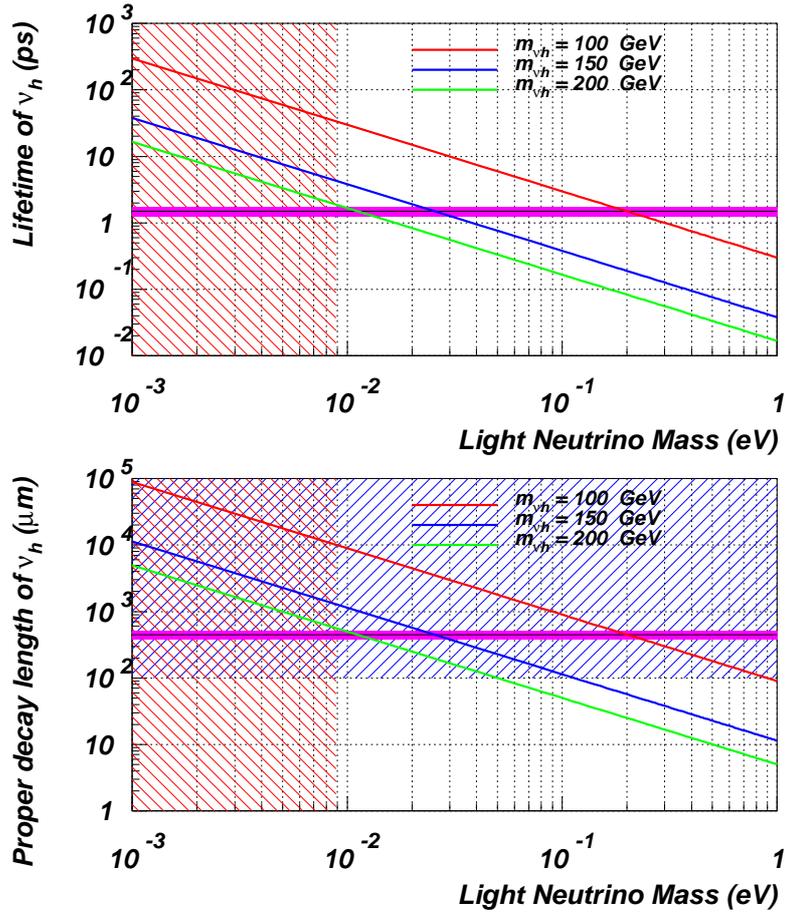}
\caption
{\it Heavy neutrino lifetime (top panel) and proper decay length (or mean path) $c\tau_0$ (bottom panel) as a function of the light neutrino mass for three different choices of heavy neutrino masses. The purple (horizontal) band presents the proper decay length of the $b$ quark while  the blue 
band indicates the range of a typical microvertex detector. The red band shows the region excluded by neutrino oscillation direct measurements.}
\label{life_time}
\end{center}
\end{figure}

In figure~\ref{life_time} we present the heavy neutrino lifetime
(top panel) in picoseconds and the proper decay length (or mean path)
(bottom panel) in micrometers as a function of the light neutrino mass
for three different choices of heavy neutrino masses
($m_{\nu _h}=100$, $150$ and $200$ GeV). The proper decay length is defined as $c\tau_0$, where $\tau_0$ is the lifetime of the heavy neutrino.  
The purple (horizontal) band presents the 
proper decay length of the $b$ quark \cite{pdg2010}, while the blue band indicates the range of a typical microvertex detector \cite{Bayatian:2006zz}.
 The red band shows the region of light neutrino masses excluded by direct measurements of neutrino oscillations [see eq.~(\ref{dm_12})], by taking the lightest neutrino to be massless (thus the other neutrinos
cannot populate this region, only the lightest one if massive). 
One should also note that the lifetime and the proper decay length of the heavy neutrinos in the laboratory frame will actually be equal to those given in
figure~\ref{life_time} times the Lorentz factor equal to
$p_{\nu_h}/m_{\nu _h}$ defined by the ratio $M_{Z'}/m_{\nu _h}$ which
can be as large as about a factor of $10$.  We can then see that there
exists a region where the heavy neutrino lifetime is of the same order
as that of the $b$ quark (shown as a purple band).  The mean path
and the respective lifetime of heavy neutrinos can therefore be
measured from a displaced vertex inside the detector.  
%
%
The heavy neutrino can however be distinguished from a $b$ hadron through the
observation of vertices consisting of only two isolated leptons. (Such a
SM $B$ meson decay, while possible, would have a very small BR $\lesssim 10^{-8}$ \cite{pdg2010}.)

An experimentally resolvable non-zero lifetime along with a mass
determination for the heavy neutrino also enables a determination of
the light neutrino mass. The lifetime measurement allows the small
heavy-light neutrino mixing to be determined and, as one can see from 
eqs.~(\ref{nu_mass_matrix})--(\ref{nu_mixing}), this, along with 
the heavy neutrino mass, gives the light neutrino mass. 
Considering only one generation for simplicity, this is expanded upon below. 
\index{Heavy neutrinos!Light neutrino mass determination}



Neutrino mass eigenstates are related to gauge eigenstates by
eq.~(\ref{nu_mixing}), hence the eigenvalues are given by solving the
equation
\begin{displaymath}
\left( \begin{array}{cc} m_{\nu _l} & 0 \\ 0 & m_{\nu _h} \end{array} \right) = \left( \begin{array}{cc} c_\nu 	& s_\nu \\ -s_\nu &c_\nu \end{array} \right)     \left( \begin{array}{cc} 0 & m_D \\ m_D & M \end{array} \right)     \left( \begin{array}{cc} c_\nu 	& -s_\nu \\ s_\nu &c_\nu \end{array} \right)\, ,
\end{displaymath}
which yields
\begin{eqnarray}\label{nul_mass}
m_{\nu _l} &=& \sin\, 2\alpha _\nu\, m_D + \sin^2 \alpha _\nu\, M\, , \\ \label{nuh_mass}
m_{\nu _h} &=& -\sin\, 2\alpha _\nu\, m_D + \cos^2 \alpha _\nu\, M\, ,
\end{eqnarray}
with $\alpha _\nu$ given by eq.~(\ref{nu_mix_angle}).
We have then three parameters ($m_D$, $M$, and $\alpha _\nu$) and a constraint
[given by eq.~(\ref{nu_mix_angle})],
that can be used to eliminate one parameter from the above equations.

The Feynman rules given in section~\ref{Appsect:nuh_feynman_rules} demonstrate that heavy neutrino interactions are determined entirely by the (sine of the) mixing angle $\alpha_\nu$, as is the total width (and therefore the mean decay
length). Hence, it is convenient to keep $m_{\nu _h}$ and $\alpha_\nu$
as independent model parameters eliminating $m_D$ from
eq.~(\ref{nu_mix_angle}),
\begin{equation}\label{elimin_1}
m_D = m_D(\alpha _\nu , m_{\nu _h})\, .
\end{equation}
By measuring the heavy neutrino mass we can also invert eq.~(\ref{nuh_mass})
\begin{equation}\label{elimin_2}
M = M(\alpha _\nu, m_{\nu _h})\, ,
\end{equation}
to finally get a fully known expression for the SM light neutrino mass
as a function of our input parameters $m_{\nu _h}$ and $\alpha _\nu$
(that can be measured independently), by inserting eqs.~(\ref{elimin_1})--(\ref{elimin_2}) into eq. (\ref{nul_mass}),
\begin{equation}\label{SMnu_mass}
m_{\nu _l} (m_D,M) = m_{\nu_l}( \alpha _\nu, m_{\nu _h})\, .
\end{equation}
This simple picture shows that  within the $B-L$ model
we have an indirect way of accessing the SM light neutrino mass by measuring 
the mass of the heavy neutrino and the kinematic 
features of its displaced  vertex. If the whole structure of
mixing is taken into account, including inter-generational
mixing in the neutrino sector (as in Ref.~\cite{Perez:2009mu}), the task of determining the
light neutrino masses in this way would become more complicated
but the qualitative features and the overall strategy would
remain  the same, thereby
providing one with a unique link between  very large and very small
scale physics. This in turn will also be a direct test of the specific realisation of the see-saw mechanism within the gauged $B-L$ framework, as described in section~\ref{sect:neutrino_masses}.

In the next subsection we will combine the heavy neutrino decay patterns with the pair production mechanism via the $Z'$ boson. The overall process can be seen as non-standard decay patterns of the $Z'$ boson. We will analyse their BRs by distinguishing the number of charged leptons in the final state, individuating the most suitable one for a detailed study, aimed to demonstrating that there exists a signature that enables the heavy neutrino mass to be measured at the LHC. 


\subsection{Phenomenology of the neutrino pair production}\label{sect:Zp_to_nuh}
The possibility of decays of the $Z'$ gauge boson into pairs of heavy neutrinos  is one of the most significant results of this work, since, in addition to the clean SM-like di-lepton signature, it provides multi-lepton signatures where backgrounds can be strongly suppressed.  In order to address this quantitatively, we first determine the relevant BRs.


As in eqs.~(\ref{nuh_lW})--(\ref{nuh_nuZ}), a single heavy
neutrino decay will produce a signature of $0$, $1$, or $2$ charged leptons, 
depending on whether the heavy neutrino decays via a charged or neutral
current and on the subsequent decays of the SM $W^\pm$ and $Z$ gauge bosons. We
can have both chains
\begin{equation}\label{Hn-W}
\nu_h \rightarrow \ell^\pm\, W^\mp \rightarrow \ell^\pm +\left\{ \begin{array}{c} \ell^\mp\,\nu _l\\ 
{\rm hadrons} \end{array}\right.
\end{equation}
{and}
\begin{equation}\label{Hn-Z}
\nu_h \rightarrow \nu _l\, Z \rightarrow \nu _l +\left\{ \begin{array}{c} \ell^+\ell^-\\ \nu_l\,\nu_l 
/{\rm hadrons}   
\end{array} \right.\, .
\end{equation}
The pattern in eq.~(\ref{Hn-W}) provides $1$ or $2$ charged leptons, while
that in eq.~(\ref{Hn-Z}), $0$ or $2$. Hence, multi-lepton signatures may
arise when the $Z'$ gauge boson decays into a pair of heavy neutrinos,
producing up to $4$ charged leptons in the final state. 
There exist in total $5$ different topologies, distinguished by the number of charged leptons in the final state, that are: \index{Decay modes!Z$'$ boson via heavy neutrinos}
\begin{eqnarray} \label{0l}
0 \mbox{ lepton mode:} &Z' \to & \left\{ \begin{array}{ll}
6\nu_l & \mbox{via $2Z$ bosons}\\
4\nu_l\; 2j & \mbox{via $2Z$ bosons} \\
2\nu_l\; 4j & \mbox{via $2Z$ bosons}
\end{array} \right.\\ \label{1l}
1 \mbox{ lepton mode:} &Z' \to & \left\{ \begin{array}{ll}
\ell^\pm \; 3\nu_l\; 2j  & \mbox{via $W+Z$ bosons}\\
\ell^\pm \; \nu_l\; 4j  & \mbox{via $W+Z$ bosons}
\end{array} \right. \\ \label{2l}
2 \mbox{ lepton mode:} &Z' \to & \left\{ \begin{array}{ll}
\ell^\pm \; \ell^\mp\; 4\nu_l & \mbox{via $2Z$ or $W+Z$ bosons}\\
\ell^\pm \; \ell^\mp\; 2\nu_l\; 2j & \mbox{via $2Z$ or $W+Z$ bosons}\\
\ell^\pm \; \ell^\mp\; 4j & \mbox{via $2W$ bosons}
\end{array} \right. \\ \label{3l}
3 \mbox{ lepton mode:} &Z' \to & 2\ell^\pm\; \ell^\mp\; 2j\; \nu_l \qquad \mbox{via $2W$ or $W+Z$ bosons}\\ \label{4l}
4 \mbox{ lepton mode:} &Z' \to & 2\ell^\pm\; 2\ell^\mp\; 2\nu_l   \qquad \mbox{via $2W$, $2Z$ or $W+Z$ bosons}
\end{eqnarray}

\begin{figure}[!t]
\begin{center}
\includegraphics[angle=0,width=\textwidth]{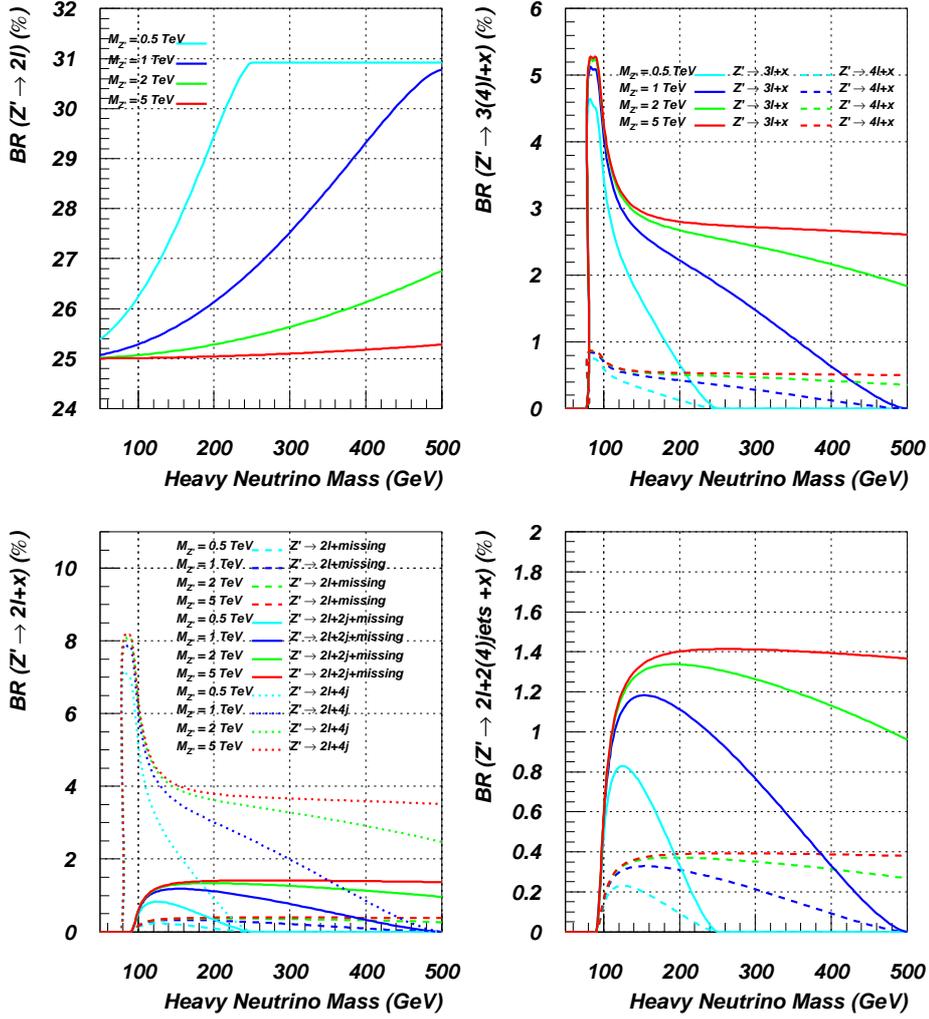}
\vspace*{-2truecm}
\caption{\it $Z'$ boson BRs, as a function of $m_{\nu _h}$, into: $2$ leptons (both $e$ and $\mu$, top-left panel); $3$ and $4$ leptons + $X$
(both $e$ and $\mu$, top-right panel);
$2$ leptons + $X$ jets (both $e$ and $\mu$, bottom-left panel); zoom of
the previous plot with same legend (bottom-right panel).}
\label{3-4lept}
\end{center}
\end{figure}

Eq.~(\ref{0l}) shows that the $Z'$ boson could decay completely invisibly, or in a multi-jet final state with large missing energy. As well, a mono-leptonic decay is possible, as in eq.~(\ref{1l}), with multi-jets and, again, possibly large missing energy. Although rather interesting (especially the completely invisible decay), they are usually overwhelmed by the backgrounds. Therefore, we consider only multi-lepton decay modes, i.e., with $2$ or more charged leptons in the final state. Figure~\ref{3-4lept} shows the BRs of a $Z'$ boson decaying into $2$ (top-left panel) and $3$ or $4$ (top-right panel) leptons (plus possibly missing transverse momentum and/or jets, as appropriate) as a function of $m_{\nu _h}$, where a lepton can be either an electron or a muon and these contributions are summed.  While the former are clearly dominant, the latter are not at all negligible.
For $M_{W^\pm}<m_{\nu _h}<M_Z$, the $\nu_h \rightarrow \ell^\pm\, W^\mp$
decay is the only one kinematically possible, whereas for
$m_{\nu _h}<M_{W^\pm}$ the heavy neutrino can decay only via an off
shell $W/Z$ and can therefore be very long-lived (depending on $m_{\nu _l}$).  For a very massive $Z'$ boson ($2$ TeV $< M_{Z'}< 5$ TeV) the multi-leptonic BRs are roughly $2.5\% $ in the case of $Z' \rightarrow 3\ell$ and $0.5\% $ in the case of $Z' \rightarrow 4\ell$, for a wide range of heavy neutrino masses.

Finally, as in eq.~(\ref{2l}), also non-standard di-lepton decays are possible, with the $Z'$ boson that decays into $2$ leptons plus a large amount of missing transverse momentum and/or highly energetic jets, see figure~\ref{3-4lept} (bottom-left panel). Particularly interesting is the decay into $2$ leptons and $4$ jets (only via $W$ bosons), since here there is no missing transverse momentum at all and its BR is rather large with respect to the other non-SM signatures, as we can see in figure~\ref{3-4lept} (bottom-left panel).


\subsection{Tri-lepton signature analysis (parton level vs. detector level)}\label{subsect:trilep}
The non-standard decay modes of the $Z'$ boson (via the decay into heavy neutrino pairs) have been presented in section~\ref{sect:Zp_to_nuh}. We have already discussed that $Z'$ decay patterns with less than $2$ charged leptons have no scope at the LHC, being overwhelmed by the backgrounds. On the other side, non-standard $2$ leptons decay modes are meagre with respect to the prompt di-lepton decay modes, and their analysis would be a more refined study than the one of section~\ref{subsect:disc_power}.
We think that only the $2\ell+4j$ deserves a closer look at, as backgrounds can be effectively suppressed and neutrino masses reconstructed, although the very high jet multiplicity might complicate the analysis. Ref.~\cite{CuhadarDonszelmann:2008jp} showed that the task could be feasible, although in the slightly different context of a $4^{th}$ family extension of the SM, where the heavy neutrino pair is produced via SM $Z$ and Higgs bosons, therefore in a completely different kinematic region.

The $4$ leptons decay mode of eq.~(\ref{4l}) has the advantage of being rather background free. However, its corresponding BR is small [$\mathcal{O}(0.5\%$)]. In the context of the $B-L$ model it has been studied in Ref.~\cite{Huitu:2008gf}, showing that it can be observable for $ \mathcal{L} = 300$ fb$^{-1}$, and the intermediate heavy neutrino mass poorly measurable.

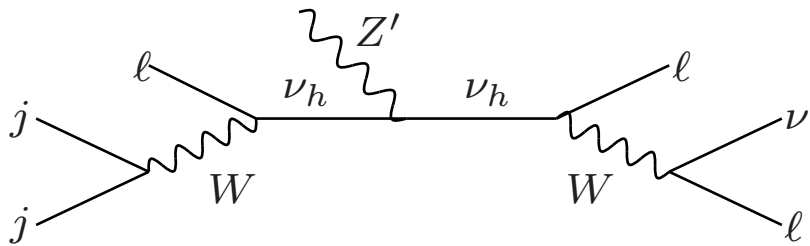
\begin{figure}[!ht] \centering \scalebox{2}{
\begin{picture}(95,79)(0,30)
\unitlength=1.0 pt
\SetScale{1.0}
\SetWidth{0.5}      
\scriptsize    
\Text(60.0,65.0)[r]{$\nu _h$}
\Line(40.0,60.0)(68.0,60.0) 
\Text(90.0,70.0)[l]{$\ell$}
\Line(68.0,60.0)(89.0,70.0) 
\Text(70.0,47.0)[l]{$W$}
\Photon(68.0,60.0)(89.0,50.0){2.0}{4}
\Text(35,80.0)[t] {$Z'$}
\Photon(20.0,80.0)(40.0,60.0){2.0}{4}
\Line(12.0,60.0)(40.0,60.0) 
\Text(17.0,65.0)[l]{$\nu _h$}
\Line(89.0,50.0)(110.0,40.0) 
\Text(111.0,40.0)[l]{$\ell$}
\Line(89.0,50.0)(110.0,60.0) 
\Text(111.0,60.0)[l]{$\nu$}
\Text(-11.0,70.0)[l]{$\ell$}
\Line(12.0,60.0)(-8.0,70.0) 
\Text(3.0,47.0)[l]{$W$}
\Photon(12.0,60.0)(-8.0,50.0){2.0}{4}
\Line(-8.0,50.0)(-29.0,40.0) 
\Text(-34.0,40.0)[l]{$j$}
\Line(-8.0,50.0)(-29.0,60.0) 
\Text(-34.0,60.0)[l]{$j$}
\end{picture} }
\vspace*{-0.5cm}
\caption{\it Feynman diagram for the tri-lepton decay mode of the $Z'$ boson ($\ell=e,\,\mu$). \label{3lep_pic}}
\end{figure}

As this last reference was in progress, we looked in details at the signature with $3$ charged leptons, in eq.~(\ref{3l}), as first reported in Ref.~\cite{Basso:2008iv}. For the price of considering bigger backgrounds, the tri-lepton signature has higher cross section, thereby a lower integrated luminosity is sufficient for its discovery, if compared with the $4$ leptons mode. It is crucial to note that the $2$ jets come always from a $W$ boson, and this is fundamental for reducing the backgrounds, as we will show. A pictorial representation of the tri-lepton decay mode of the $Z'$ boson is in figure~\ref{3lep_pic}. \index{Tri-lepton signature!Definition}

\vspace*{-2cm}

The tri-lepton decay mode also offers a nice framework to study the heavy neutrino properties. It is remarkable that in this signature the heavy neutrino flavours can be univocally identified, as the charged lepton that comes along the jets can be unambiguously determined, and so its flavour \footnote{Notice that the same is true also for the $2\ell +4j$ mode, but in general it is not for the other multi-lepton decay modes.}. Just by counting, the relative BR and the possible mixing in the heavy neutrino sector can also be potentially measured. Nonetheless, we leave these features for future investigation and focus here in delineating the strategy for the discovery, in the approximation of degenerate and non-mixed heavy neutrinos.  For simplicity, we also limit ourselves to the case without leptonically decaying $\tau$'s, leaving also this case for future investigation.

The chosen benchmark points, at the LHC for $\sqrt{s}=14$ TeV and $ \mathcal{L} = 100$ fb$^{-1}$, are for $M_{Z'} = 1.5$ TeV, $g'_1=0.2$ [for which, from figure~\ref{zpxs_cont}, $\sigma (pp\to Z') = 0.272$ pb], and two different heavy neutrino masses: $m_{\nu _h} = 200$ GeV, for which the heavy neutrino pair production cross section via $Z'$ boson is $46.7$ fb [also, from figure~\ref{Zp_BR}, second panel, BR$(Z'\to \sum \nu_h\nu_h)\sim 17.1 \%$], and $m_{\nu _h} = 500$ GeV, for which, again, $\sigma(pp\to Z' \to \sum \nu_h \nu_h) = 23.4$ fb [and BR$(Z'\to \sum \nu_h\nu_h)\sim 8.7 \%$].\index{Tri-lepton signature!Benchmark points}

These two benchmark points provide two very kinematically different examples. In the first case, the heavy neutrinos are much lighter than the $Z'$ boson, producing highly boosted events. In the second case, their mass is comparable to $M_{Z'}/2$, hence close to their production threshold, resulting in minimal boost. From a merely kinematic point of view, all other cases will be somewhere between these two.

When the heavy neutrino decays via the $\ell^\pm W^\mp$ mode, with a
subsequent leptonic decay of the $W^\pm$, the charged pair of leptons
can carry an invariant mass equal to or lower than the heavy neutrino
mass, with the maximum invariant mass configuration occurring when the
light neutrino is produced at rest, so that the edge in this
distribution corresponds to the $\nu_h$ mass. 
A peak in such a distribution corresponding to the SM-like $Z$ boson,
coming from the $\nu Z$ decay mode for the heavy neutrino, will also be
present in this distribution (see, e.g., plots in Ref.~\cite{Basso:2008iv}). 

While the invariant mass distribution can provide some insights into
the mass of the intermediate objects, this is not the best observable
in the case of the tri-lepton signature, because the final state light neutrino
escapes detection. A more suitable distribution to look at
is the transverse mass defined in Ref.~\cite{Barger:1987du}, i.e., \index{Tri-lepton signature!Transverse mass distribution}
\begin{equation}
m^2_T = \left( \sqrt{M^2(vis)+P^2_T(vis)}+\left| \mpt \right| \right) ^2
	- \left( \vec{P}_T(vis) + \mptv\right) ^2\, ,
\end{equation}
where $(vis)$ means the sum over the visible particles.  For the
final state considered here, if we sum over the $3$ leptons and the $2$
jets, this distribution will peak at the $Z'$ mass. We can
also see evidence for the presence of a heavy neutrino by just
considering the $2$ closest (in $\Delta R$, where $\Delta R \equiv \sqrt{\Delta \eta ^2 + \Delta \phi ^2}$) leptons and the missing transverse
momentum, since this is the topology relevant to a $\nu_h$ decay.  The
results show that this transverse mass peak for the heavy neutrino is
likely to be the best way to measure its mass \cite{Basso:2008iv}. The striking signature of this model is that both of the above peaks occur simultaneously.

We stress that by choosing event by event the $2$ closest (in $\Delta R$) leptons,
the peak corresponding to the heavy neutrino is well reconstructed, as clear from figure~\ref{fig:set3}. As we pointed out previously, this prescription also allows for the almost unambiguous identification of the single lepton coming from the heavy neutrino that decays semi-hadronically. Its flavour can therefore be identified.

In the evaluation of the background we considered three sources (including generation cuts, to improve efficiency):\index{Tri-lepton signature!Backgrounds}
\begin{itemize}
\item[-] $WZjj$ associated production ($\sigma _{3\ell} = 246.7$ fb, $\ell=e,\mu ,\tau$; $\Delta R_{jj}>0.5$, 
$P^T_{j_{1,2}}>40\mbox{ GeV}$, $\left| \eta_{j_{1,2}} \right| <3)$;
\item[-] $t\overline{t}$ pair production ($\sigma _{2\ell} = 29.6$ pb, $\ell=e,\mu$ ($b$ quark not decayed); 
QCD scale $=M_t/2$ to emulate the NLO cross section; no cuts applied);
\item[-] $t\overline{t}\ell\nu$ associated production ($\sigma _{3\ell} = 8.6$ fb, $\ell=e,\mu ,\tau$; QCD scale $=\sqrt{\hat s}$, $P^T_\ell>20$ GeV).
\end{itemize}

In the case of $WZjj$ associated production, $3$ leptons come from the subsequent leptonic decays of the two gauge bosons. This is the main source of background.
From $t\overline{t}$ pair production one can obtain $2$ isolated leptons via the decay of the $W^\pm$'s produced from (anti)top decays and produce a further lepton from a semi-leptonic $B$ meson (emerging from one of the $b$'s) decay. This lepton though will not be isolated, because of the large boost of the $b$ quark, and this property is used to suppress this potential background.  (In practice, the requirement of {\it isolated} leptons, made on the angular separation of a lepton from the nearest jet, strongly suppresses $t\bar t$.)
Finally, $t\overline{t}\ell\nu$ will produce $3$ isolated leptons, $2$ of which from the $W^\pm$'s arising from the (anti)top decays (as above), resulting in a significant background despite the small production cross section.

In the following subsections we present the result of the analysis at the parton level and at the detector level, respectively, of the tri-lepton signature and of the backgrounds. We define the cuts and show their effect in enhancing the signal-to-background ratio.


\subsubsection{Parton level analysis}\label{sect:trilep_parton}
This subsection presents the analysis of signal and backgrounds for the tri-lepton signature at the parton level.

The first set of cuts we use is designed to impose generic detector angular acceptances, lepton and jet transverse momentum minimal thresholds and to provide isolation for leptons and jets:

\centerline{\underbar{\large\bf Selection $1$}}
\begin{eqnarray}  \nonumber
\left| \eta_{\ell_{1,2,3}} \right| &<& 2.5,\\ \nonumber
\left| \eta_{j_{1,2}} \right| &<& 3;\\ \nonumber
P^T_{\ell_1} &>& 15 ~{\rm GeV},\\ \nonumber
P^T_{\ell_{2,3}} &>& 10 ~{\rm GeV},\\ \nonumber
P^T_{j_{1,2}} &>& 40~{\rm  GeV};\\ \nonumber
\Delta R _{\ell j}  &>& 0.5\qquad \forall \ell=1\dots 3, ~j=1,2,\\ \nonumber
\Delta R_{\ell,\ell '} &>& 0.2\qquad \forall \ell,\ell '=1\dots 3,\\
\Delta R_{j,j} &>& 0.5.
\label{eq:cuts1}
\end{eqnarray}

\begin{figure}[htb]
\includegraphics[angle=0,width=\textwidth,height=1\textwidth]{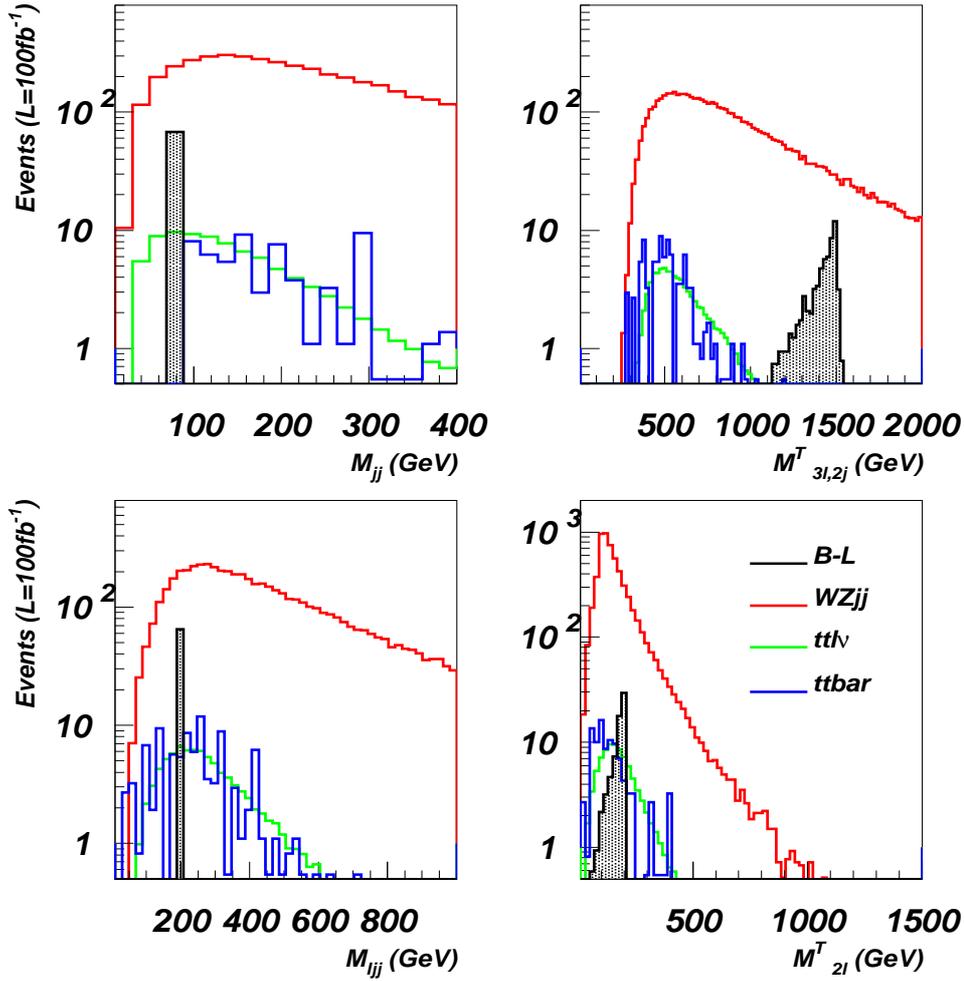}
  \vskip -1cm
  \caption{\it Signal ($m_{\nu _h}=200$ GeV) and background distributions after the 
  Selection $1$ cuts.  (Events per $\mathscr{L}=100$ fb${}^{-1}$.)}\label{fig:set1a}
\end{figure}

\begin{figure}[htb]
\includegraphics[angle=0,width=\textwidth,height=1\textwidth]{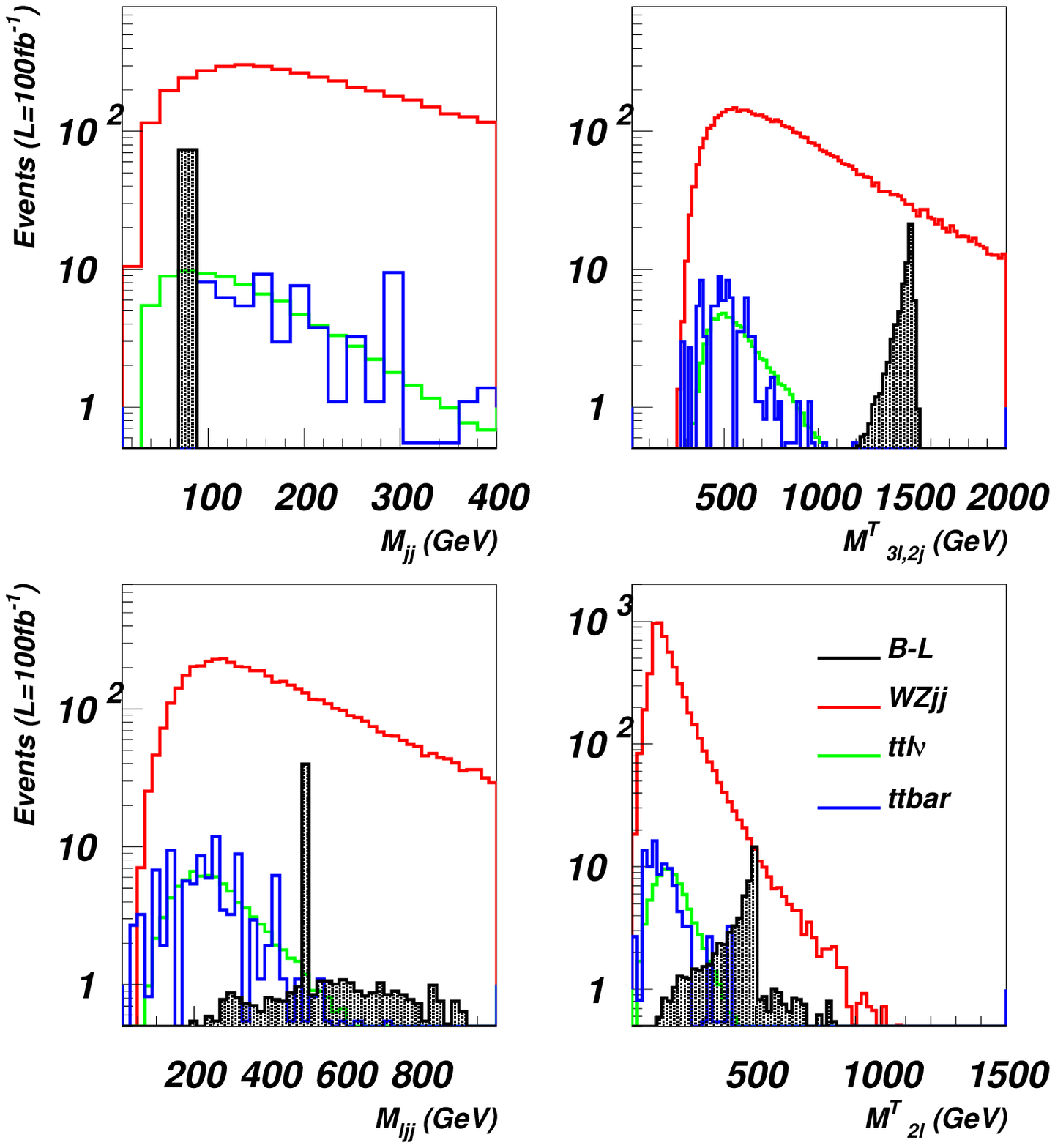}
  \vskip -1cm
  \caption{\it Signal ($m_{\nu _h}=500$ GeV) and background distributions after the  
  Selection $1$ cuts.  (Events per $\mathscr{L}=100$ fb${}^{-1}$.)}\label{fig:set1b}
\end{figure}

Special care should be devoted to the treatment of the $t\bar t$ background, given its large production rate. However, as previously mentioned, we expect that this noise can be eliminated efficiently by enforcing a suitable lepton-jet separation. The impact of the first set of cuts on the signals and $t\overline{t}$ background is illustrated in table~\ref{tab-eff}. The $\Delta R _{\ell j}$ requirement is indeed extremely effective and reduces this background by a factor of $2 \cdot 10^{-3}$. The loss of signal due to this cut is instead minimal. Also note that the signal events with the smaller boost have a higher efficiency for passing the angular isolation cuts.

\begin{center}
\begin{table}[h]
\scalebox{0.85}{
\begin{tabular}{|c|cr|cr|cr|} \hline
&&&&&&\\
${\rm Cut}$ & \multicolumn{2}{c|}{$m_{\nu _h}=200 ~{\rm GeV}$} 
          & \multicolumn{2}{c|}{$m_{\nu _h}=500 ~{\rm GeV}$} &\multicolumn{2}{c|}{$ t\overline{t}$} \\ 
&
$\# {\rm ~of~events} $&$   {\rm Efficiency}~ \%$&
$\# {\rm ~of~events} $&$   {\rm Efficiency}~ \%$&
$\# {\rm ~of~events} $&$   {\rm Efficiency}~ \%$\\
&&&&&&\\\hline
$\mbox{No cuts}		$&$  482.0	$&$  100	$&$ 239.0  $&$ 100	$&$1.28 \cdot 10^{6} 	$&$ 100	$\\
$\mbox{$\eta$ cuts}	 $&$ 346.0	 $&$ 71.8	$&$ 171.0  $&$ 71.4	 $&$5.1 \cdot 10^{5} 	$&$ 43.9$	\\
$\mbox{$\Delta R$+$P_T$ cuts}	$&$ 68.0 	$&$ 14.1	$&$ 73.7 $&$ 30.8	$&$99.7	 	$&$ 0.014$	\\ \hline  
\end{tabular}}
\caption{\it Efficiencies of the Selection $1$ cuts for the two benchmark signals and the $t\overline{t}$ background, for events with three leptons and with two or more jets in the final state for $\mathscr{L}=100$ fb${}^{-1}$. In the case of $\Delta R_{jj}<0.5$ partons were merged into one `jet' at the very beginning of the selection.}
\label{tab-eff}
\end{table}
\end{center}

Figures~\ref{fig:set1a}--\ref{fig:set1b} show the distributions in $M_{jj}$, $M^T_{3\ell 2j}$, $M_{\ell jj}$ and $M^T_{2\ell}$ after  Selection $1$ cuts, for the signal with the two heavy neutrino masses, 200 and 500 GeV, that we are considering, and for the backgrounds. 

In signal events both jets come from the $W^\pm$ therefore we apply the
following constraint:

\vspace*{0.15cm}

\centerline{\underbar{\large\bf  Selection $2$}}
\begin{equation} \label{cut_2}
\left| M_{jj} - M_W \right| < 20~{\rm GeV}.
\end{equation}

\begin{figure}[htb] \includegraphics[angle=0,width=\textwidth,height=1\textwidth]{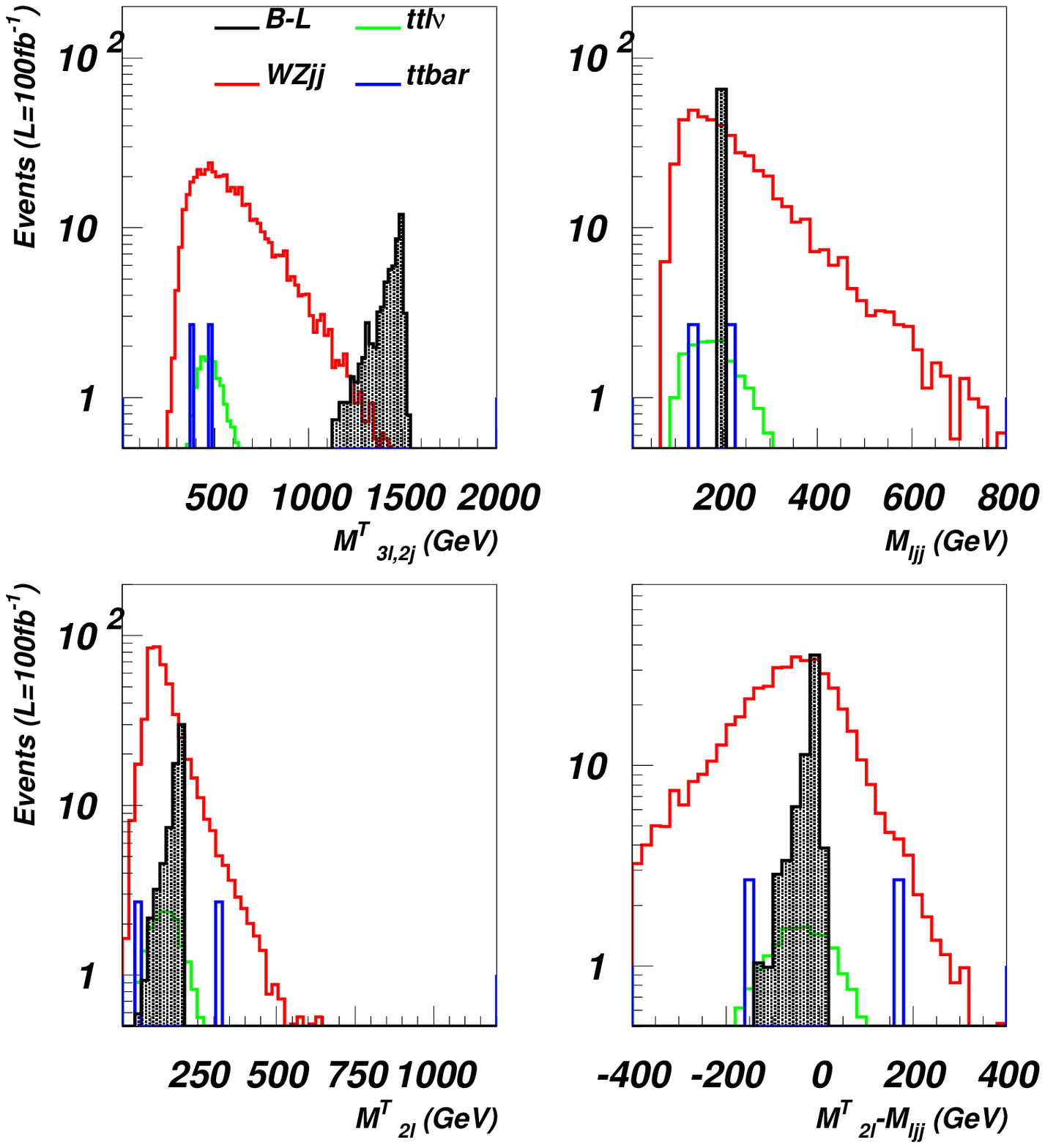}
  \vskip -1cm
  \caption{\it Signal ($m_{\nu _h}=200$ GeV) and background distributions after the Selection $1$ and $2$ cuts. (Events per $\mathscr{L}=100$ fb${}^{-1}$.)}\label{fig:set2a}
\end{figure}

\begin{figure}[htb] \includegraphics[angle=0,width=\textwidth,height=1\textwidth]{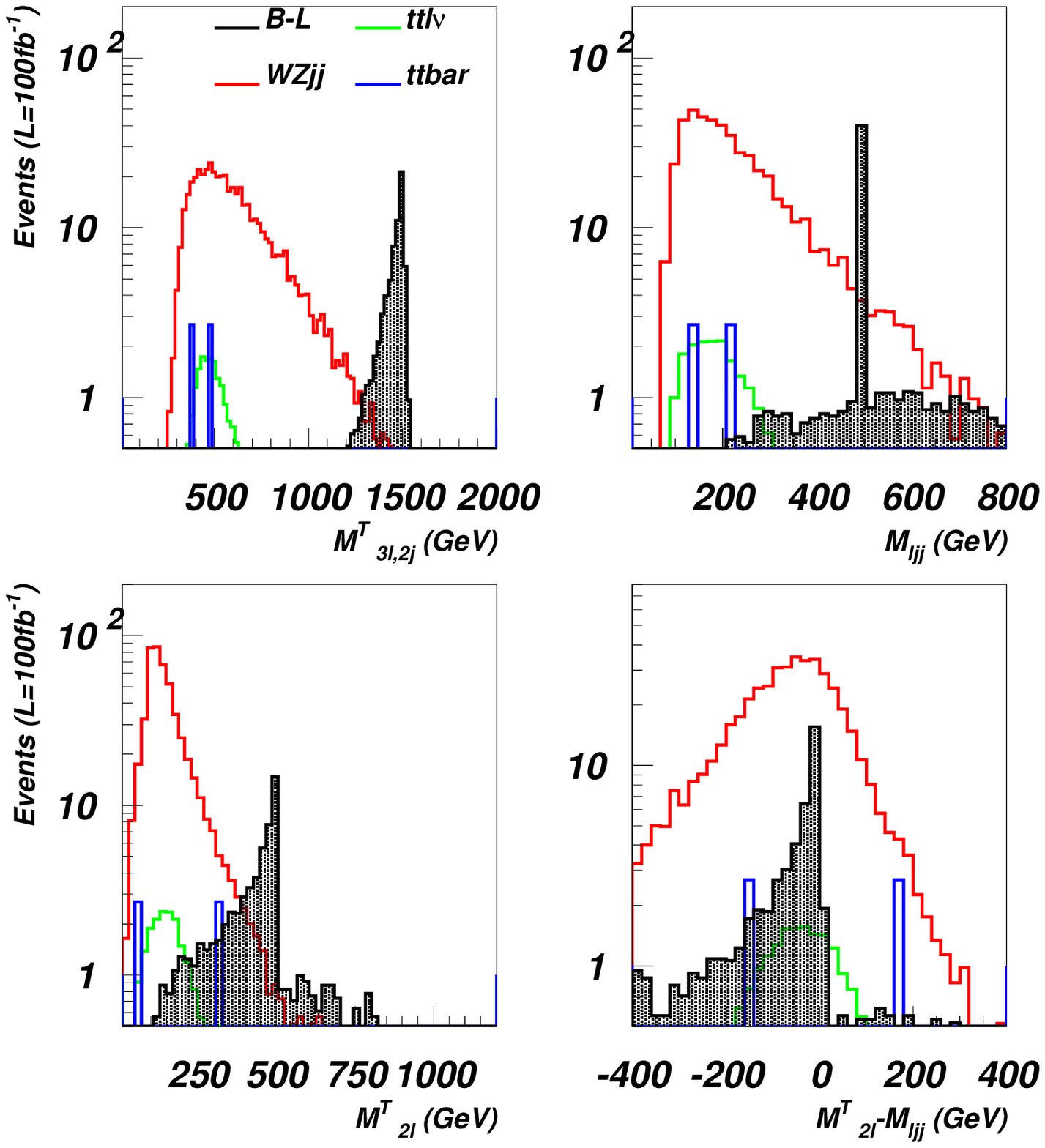}
  \vskip -1cm
  \caption{\it Signal ($m_{\nu _h}=500$ GeV) and background distributions after the
  Selection $1$ and $2$ cuts. (Events per $\mathscr{L}=100$ fb${}^{-1}$.)}\label{fig:set2b}
\end{figure}

After the application of this cut the other distributions considered are shown in figures~\ref{fig:set2a} and \ref{fig:set2b} for the $200$ and $500$ GeV heavy neutrino masses, respectively (here, we now also show the difference between the $M^T_{2\ell}$ and $M_{\ell jj}$ distributions). From these plots it is clear that transverse mass $M^T_{3\ell 2j}$ provides good discrimination between signal and background. The following cut is then used to further suppress the background: 
\vspace*{0.5cm}

\centerline{\underbar{\large\bf  Selection $3$}}
\begin{equation} \label{cut_3}
\left| M^T_{3\ell 2j} - M_{Z'} \right| < 250~{\rm GeV}.
\end{equation}

\begin{figure}[htb]
  \includegraphics[angle=0,width=\textwidth,height=1\textwidth]{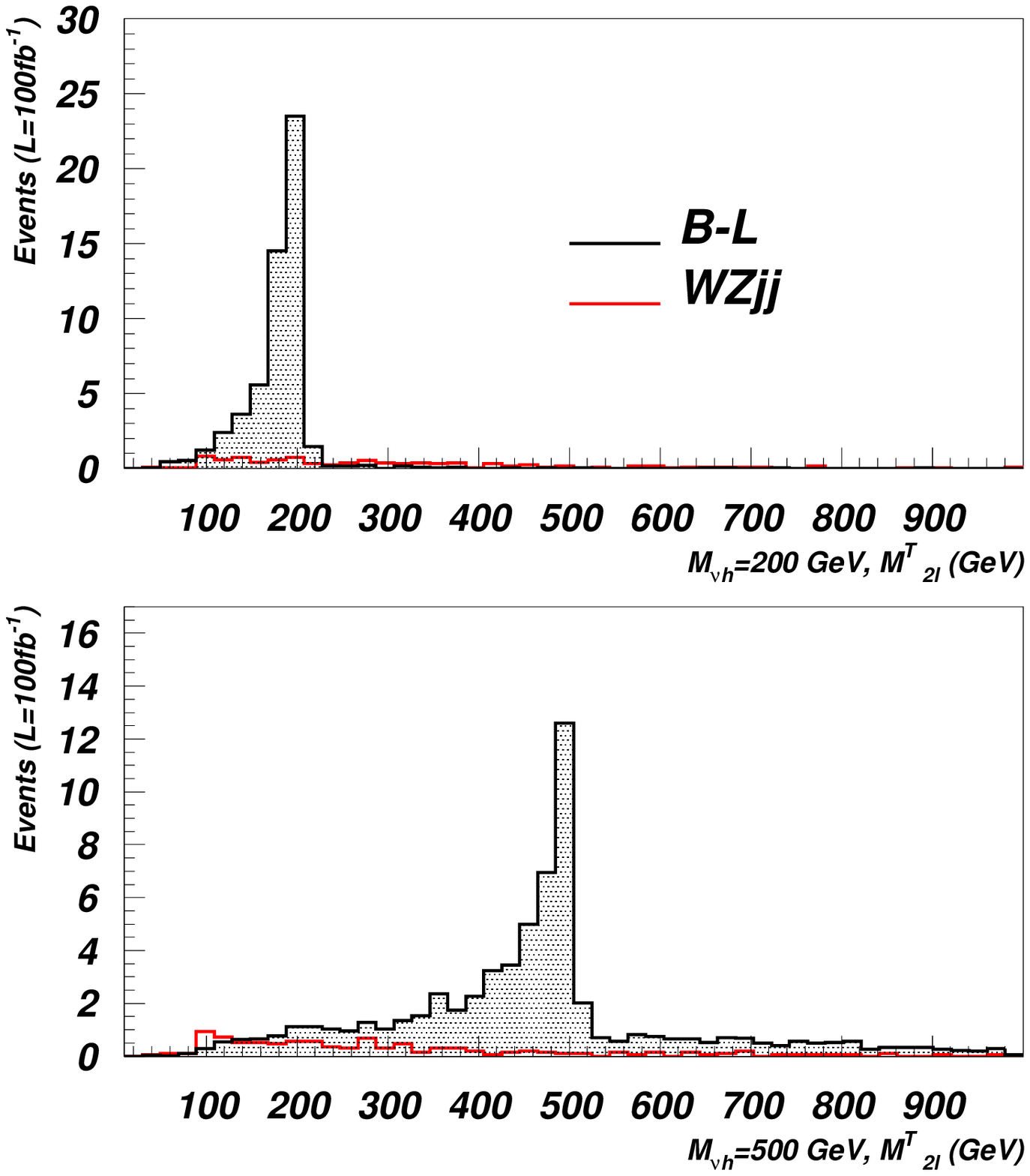}
  \vskip -1cm
  \caption{\it Signal ($m_{\nu _h}=200$ GeV, top panel, and $m_{\nu _h}=500$ GeV, bottom panel)   and background distributions after the Selection $1$, $2$ and $3$ cuts. (Events per $\mathscr{L}=100$ fb${}^{-1}$.) \label{fig:set3}}
\end{figure}

After this set of cuts we end up with a very clean signal for both a $200$ and $500$ GeV $\nu_h$ mass in the di-lepton transverse mass distribution, in fact practically free from background, as shown in figure~\ref{fig:set3}. Notice that this $M_{2\ell}^T$ variable was formed from the two closest (in $\Delta R_{\ell \ell}$) leptons since they are likely to originate from the same boosted $\nu_h$.

In order to establish the signal, we finally select events around the visible $M^T_{2\ell }$ peak, by requiring:\\
\vspace*{0.5cm}
\centerline{\underbar{\large\bf  Selection $4$}}
\vspace*{-0.45cm}
\begin{equation} \label{cut_4}
0< M^T_{2\ell} < 250~{\rm{GeV}} \qquad \mbox{ or } \qquad 400~{\rm{GeV}}< M^T_{2\ell} < 550~{\rm{GeV}},
\end{equation}
depending on the benchmark signal under consideration. A fit of the signal in these regions will finally give the heavy neutrino mass. The efficiencies of the Selection $1$--$4$ cuts, are given in table~\ref{tab:alleff}.

\begin{table}[htb]
\begin{center}
{$m_{\nu _h}=200$ GeV}\\ \vspace{0.2cm}
\scalebox{0.85}{
\begin{tabular}{|c|cr|cr|cr|cr|c|} \hline
${\rm Cuts} $&$ {\rm Signal} $&$ {\rm Efficiency} $&$ WZjj $&$ {\rm Efficiency}$&$ t\overline{t} $&$ {\rm Efficiency}$&$ ~t\overline{t}\ell\nu $&$ {\rm Efficiency}$&$ S/\sqrt{B} $\\
 & $ {\rm events} $&$ (\%) $&$ {\rm events} $&$ (\%) $&$ {\rm events} $&$ (\%) $&$ {\rm events}  $&$ (\%) $& \\ \hline
$1	$&$ 68.0 $&$ 100	  $&$ 5875    $&$ 100  $&$ 99.6    $&$ 100	     $&$ 89.1          $&$ 100   $&$ 0.87 $\\
$2 	$&$ 68.0 $&$ 100	  $&$ 498.8   $&$ 8.5  $&$ 5.38    $&$ 5.4	     $&$ 19.3          $&$ 21.8  $&$ 2.97 $\\
$3	$&$ 58.8 $&$ 86.5	  $&$ 10.58   $&$ 12.7 $&$ 0       $&$ 0.8	     $&$ 0.0667        $&$ 2.2   $&$ 18.0 $\\
$4	$&$ 56.0 $&$ 94.1	  $&$ 4.48    $&$ 67.6 $&$ 0       $&$ 56.4	     $&$ 0.0305        $&$ 64.8  $&$ 26.3 $\\	\hline
\end{tabular}}
\end{center}
\vspace*{0.250cm}
\begin{center}
{$m_{\nu _h}=500$ GeV}\\ \vspace{0.2cm}
\scalebox{0.85}{
\begin{tabular}{|c|cr|cr|cr|cr|c|} \hline
${\rm Cuts} $&$ {\rm Signal} $&$ {\rm Efficiency} $&$ WZjj $&$ {\rm Efficiency}$&$ t\overline{t} $&$ {\rm Efficiency}$&$ ~t\overline{t}\ell\nu $&$ {\rm Efficiency}$&$ S/\sqrt{B} $\\
 & $ {\rm events} $&$ (\%) $&$ {\rm events} $&$ (\%) $&$ {\rm events} $&$ (\%) $&$ {\rm events}  $&$ (\%) $& \\ \hline
$1	$&$ 73.6     $&$ 100  $&$ 5875    $&$ 100  $&$ 99.7          $&$ 100	    $&$ 89.1          $&$ 100   $&$ 0.95 $\\
$2 	$&$ 73.6     $&$ 100  $&$ 498.8    $&$ 8.5  $&$ 5.38          $&$ 5.4	    $&$ 19.3          $&$ 21.8  $&$ 3.22 $\\
$3	$&$ 68.8     $&$ 93.4 $&$ 10.58    $&$ 12.7 $&$ 0 	    $&$ 0.8	    $&$ 0.0667        $&$ 2.2   $&$ 21.1 $\\
$4	$&$ 46.3     $&$ 66.0 $&$ 2.879    $&$ 27.1  $&$ 0 	    $&$ 8.7	    $&$ 0.00952       $&$ 10.1  $&$ 27.6 $\\	\hline
\end{tabular}}
\end{center}
\caption{\it Signal ($m_{\nu _h}=200$ GeV at the top and $m_{\nu _h}=500$ GeV at the bottom)  and background events per ${\mathscr{L}}=100\;{\rm fb}^{-1}$ and efficiencies following the sequential application of  Selection $1$--$4$ cuts.}\label{tab:alleff} \end{table}

We should stress that during all the step of our analysis we have included  detector effects and applied these to the parton level events simulated using \textsc{calchep}. We have used Gaussian smearing  of leptons and quarks energies to simulate a typical electromagnetic energy resolution given by $0.15/\sqrt{E}$
and a typical hadronic energy resolution given by $0.5/\sqrt{E}$, to mimic the ATLAS~\cite{:1999fq} and CMS~\cite{Bayatian:2006zz} detector performances. The resolution in missing transverse momentum was derived in turn from  the above mentioned smearing of leptons and quarks energies. The final figure~\ref{fig:set3} and table~\ref{tab:alleff} present results taking into account these effects. One should also mention that, since we deal with a multi-lepton signature in the final state, detector resolution effects are typically smaller or of the order of the chosen $20$ GeV bin width for the final transverse mass distributions. Finally, we have verified that all results presented here are actually very stable against the implementation of electromagnetic  and hadronic calorimeter energy  resolution effects.

In the next section we will present the study of the tri-lepton signature in a more realistic framework, with a full detector simulation in which also jet hadronisation and showering are included. The main difference with respect to the analysis presented here concerns the jet multiplicity, that at the parton level we cannot emulate. Also, the lepton isolation requirement will be treated in a more comprehensive way than the simple approximation used here. In the analysis, we will adopt the same strategy and we will use cuts emulating those presented here. 

It is interesting to note here that, in the strategy described in this section, we did not impose any requirements on the sign of the leptons in order to reconstruct the heavy neutrinos, nor on the overall balance of their total sign. Also, flavour violating processes, clear indication of new physics, are possible. We decided to not include any of these features in our analysis to be the most model-independent possible, highlighting that simple cuts are enough for heavy neutrino discovery.

\subsubsection{Detector level analysis}\label{sect:trilep_detector}
This subsection presents the analysis of signal and backgrounds for the tri-lepton signature at the detector level. Its aim is to validate the analysis strategy outlined in the previous subsection within a more realistic framework. The parton level events generated for the previous analysis have been here hadronised and showered with \textsc{pythia}, version $6.2.40$~\cite{Sjostrand:2006za}; the CMS detector has been emulated with the fast detector simulator \textsc{delphes}, version $1.9$~\cite{Ovyn:2009tx}; the jets have been reconstructed with the built-in \textsc{SIScone} algorithm, version $1.3.3$~\cite{Salam:2007xv}, with cone size $0.5$.

Most of the Selection $1$ cuts of eq.~(\ref{eq:cuts1}) is built-in in the object definition. The minimum $p_T$ for leptons is $10$ GeV, while the detector angular acceptance is $|\eta _\ell|<2.5$($2.4$) for electrons(muons). Regarding the jets, minimum $p_T$ is $20$ GeV and $|\eta _j|<3$. Our choice here of a cone size of $0.5$ matches the last statement of eq.~(\ref{eq:cuts1}). The tri-lepton signature is finally defined by the presence of exactly $3$ leptons in the final state, both electrons and muons, and we require {\it at least} $2$ jets. The $2$ most energetic jets are requested having $p_T > 40$ GeV. A summary of the Selection $1$ cuts here used is in eq.~(\ref{eq:cuts1_det}).

\newpage

\centerline{\underbar{\large\bf Selection $1$}}
\begin{eqnarray}  \nonumber
\#\ell &\equiv& 3,\\ \nonumber
\#j &\geq& 2;\\ \nonumber
\left| \eta_{e(\mu)} \right| &<& 2.5(2.4),\\ \nonumber
\left| \eta_{j} \right| &<& 3;\\ \nonumber
P^T_{\ell _{1,2,3}} &>& 10 ~{\rm GeV},\\ \nonumber
P^T_{j_{1,2}} &>& 40~{\rm  GeV};\\ \nonumber
\mbox{jet cone radius} &=& 0.5,\\ 
\mbox{isolated leptons} &=& true.
\label{eq:cuts1_det}
\end{eqnarray}

The \textsc{delphes} package allows for the lepton isolation to be more carefully defined. To match the CMS definitions, we define a lepton to be `isolated' if no other tracks in a cone of radius $0.3$ within the tracker has $p_T>2$ GeV. The latter appears more restrictive than the corresponding $\Delta R _{\ell \ell '}$ requirement we used at the parton level.

In the previous section we highlighted that the requirement of isolated leptons was crucial to suppress the $t\bar t$ background, in which a further lepton stems from a $b$ quark decay. Table~\ref{tab-eff_iso} shows the effect of the Selection $1$ cuts for both signal points and the $t\bar t$ background. We see that, in the detector framework, the requirement of isolated leptons reduces the $t\bar t$ background by a further factor $2.4\%$, an order of magnitude less efficient than at the parton level.
Regarding the signal, it still holds that the isolation requirement suppresses more the low $m_{\nu _h}$ point, where final state objects are more boosted.

\begin{center}
\begin{table}[h]
\scalebox{0.85}{
\begin{tabular}{|c|cr|cr|cr|} \hline
&&&&&&\\
${\rm Cut}$ & \multicolumn{2}{c|}{$m_{\nu _h}=200 ~{\rm GeV}$} 
          & \multicolumn{2}{c|}{$m_{\nu _h}=500 ~{\rm GeV}$} &\multicolumn{2}{c|}{$ t\overline{t}$} \\ 
&
$\# {\rm ~of~events} $&$   {\rm Efficiency}~ \%$&
$\# {\rm ~of~events} $&$   {\rm Efficiency}~ \%$&
$\# {\rm ~of~events} $&$   {\rm Efficiency}~ \%$\\
&&&&&&\\\hline
$\mbox{Basic cuts}	$&$  429.4	$&$  100	$&$ 214.4  $&$ 100	$&$4.31 \cdot 10^{5} 	$&$ 100	$\\
$\mbox{Isolation cut}	$&$  178.3 	$&$ 41.5	$&$ 117.9  $&$ 55.0	$&$1.04	\cdot 10^{4} 	$&$ 2.4 $	\\ \hline  
\end{tabular}}
\caption{\it Efficiencies of the isolation cut for the two benchmark signals and the $t\overline{t}$ background, for events with three leptons and with two or more jets in the final state, for $\mathscr{L}=100$ fb${}^{-1}$. `Basic cuts' refer to the Selection $1$ cuts of eq.~(\ref{eq:cuts1_det}) without the isolation requirement.}
\label{tab-eff_iso}
\end{table}
\end{center}

\index{Tri-lepton signature!Jet pair identification}
At the parton level there is no ambiguity concerning the jets, since, for the signal, two and only two are present. At the detector level, instead, the average jet multiplicity is around $4$, thus we need to identify the two jets to include in our distributions. As pointed out, for the signal only, the topology of the tri-lepton signature imposes the jets to come exclusively from a $W$ boson. Hence, on a event by event basis, we can choose the pair of jets that better reconstructs the $W$ boson (meaning, whose invariant mass is closest to  $M_W=80.4$ GeV). Selection $2$ cut is hence defined as a cut in the invariant mass distribution of {\it this} jet pair, jets that will subsequently enter in the definition of our distributions, such as $M^T_{3\ell 2j}$. To avoid to spoil considerably the signal, while still rejecting most of the background, the cut of eq.~(\ref{cut_2}) has to be relaxed to
\vspace*{0.5cm}

\centerline{\underbar{\large\bf  Selection $2$}}
\begin{equation} \label{cut_2_det}
\left| M_{jj} - M_W \right| < 30~{\rm GeV}.
\end{equation}

It shall be noted that this is not the most efficient way to identify the pair of jets to be used. We comment on this at the end of this section.

\begin{figure}[!h]
  \subfloat[]{ 
  \label{mt3l2j_200_det}
  \includegraphics[angle=0,width=0.48\textwidth ]{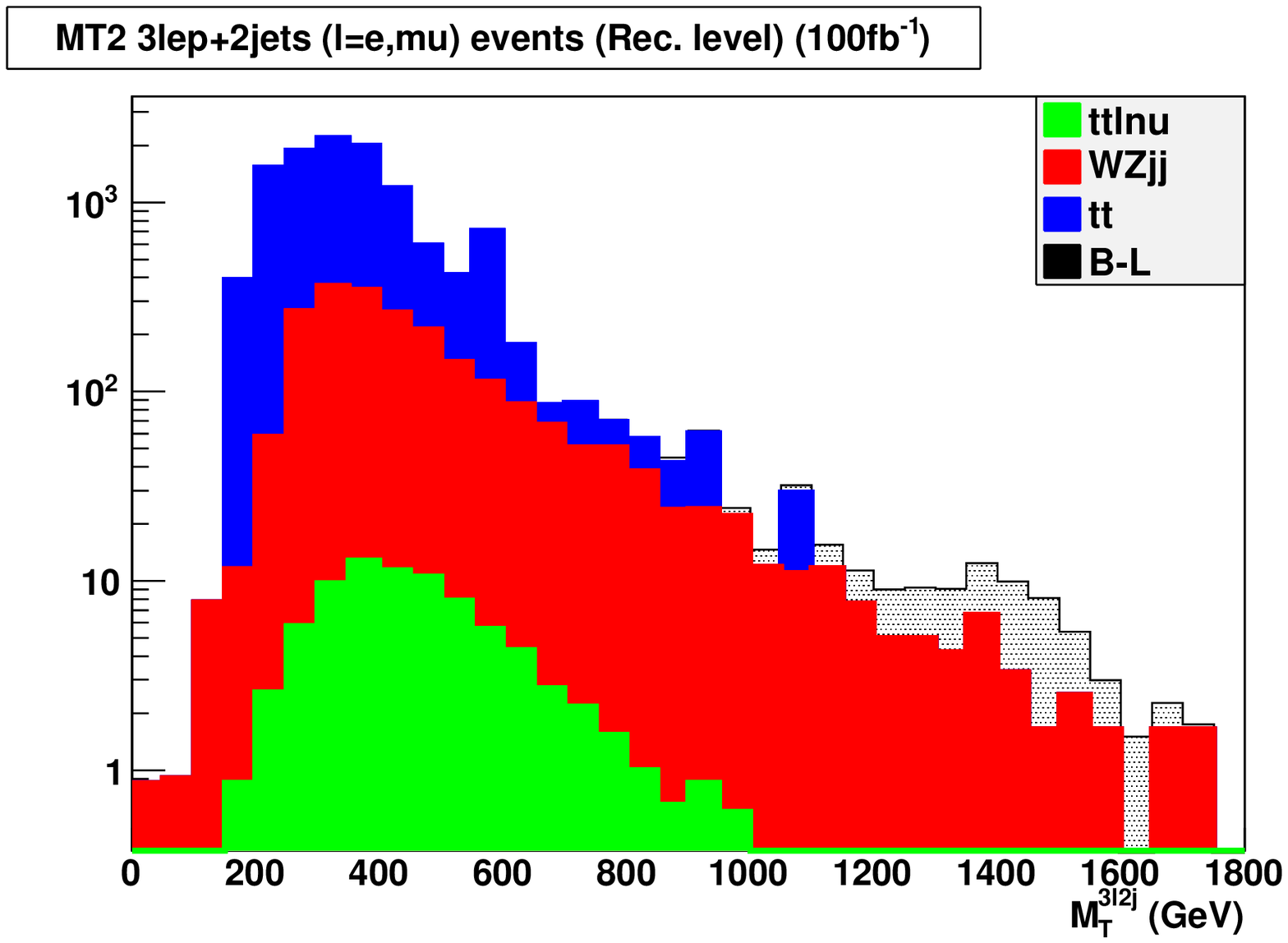}}
  \subfloat[]{
  \label{mt3l2j_500_det}
  \includegraphics[angle=0,width=0.48\textwidth ]{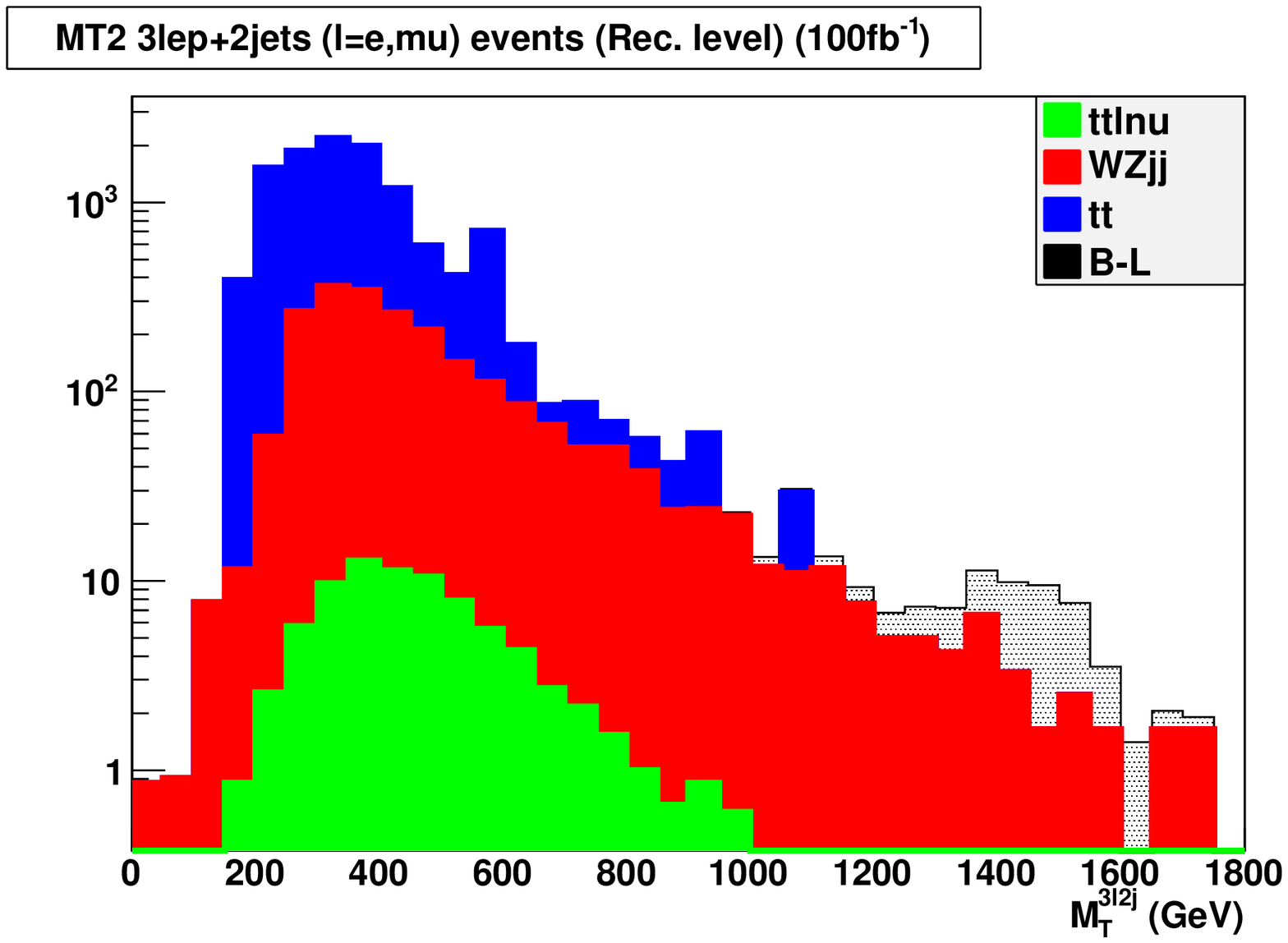}}\\
    \subfloat[]{ 
  \label{mt3l2j_200_det_signal}
  \includegraphics[angle=0,width=0.48\textwidth ]{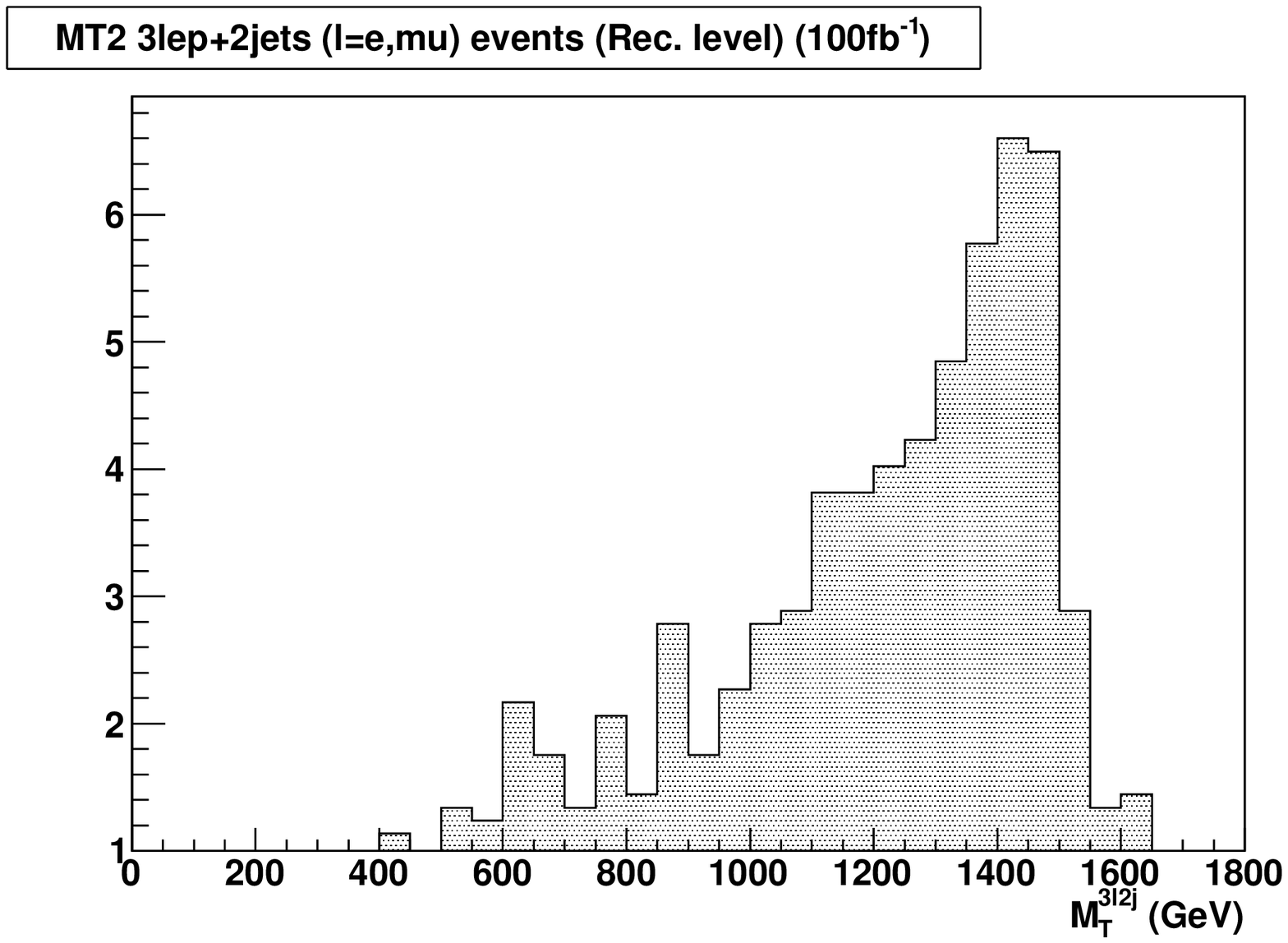}}
  \subfloat[]{
  \label{mt3l2j_500_det_signal}
  \includegraphics[angle=0,width=0.48\textwidth ]{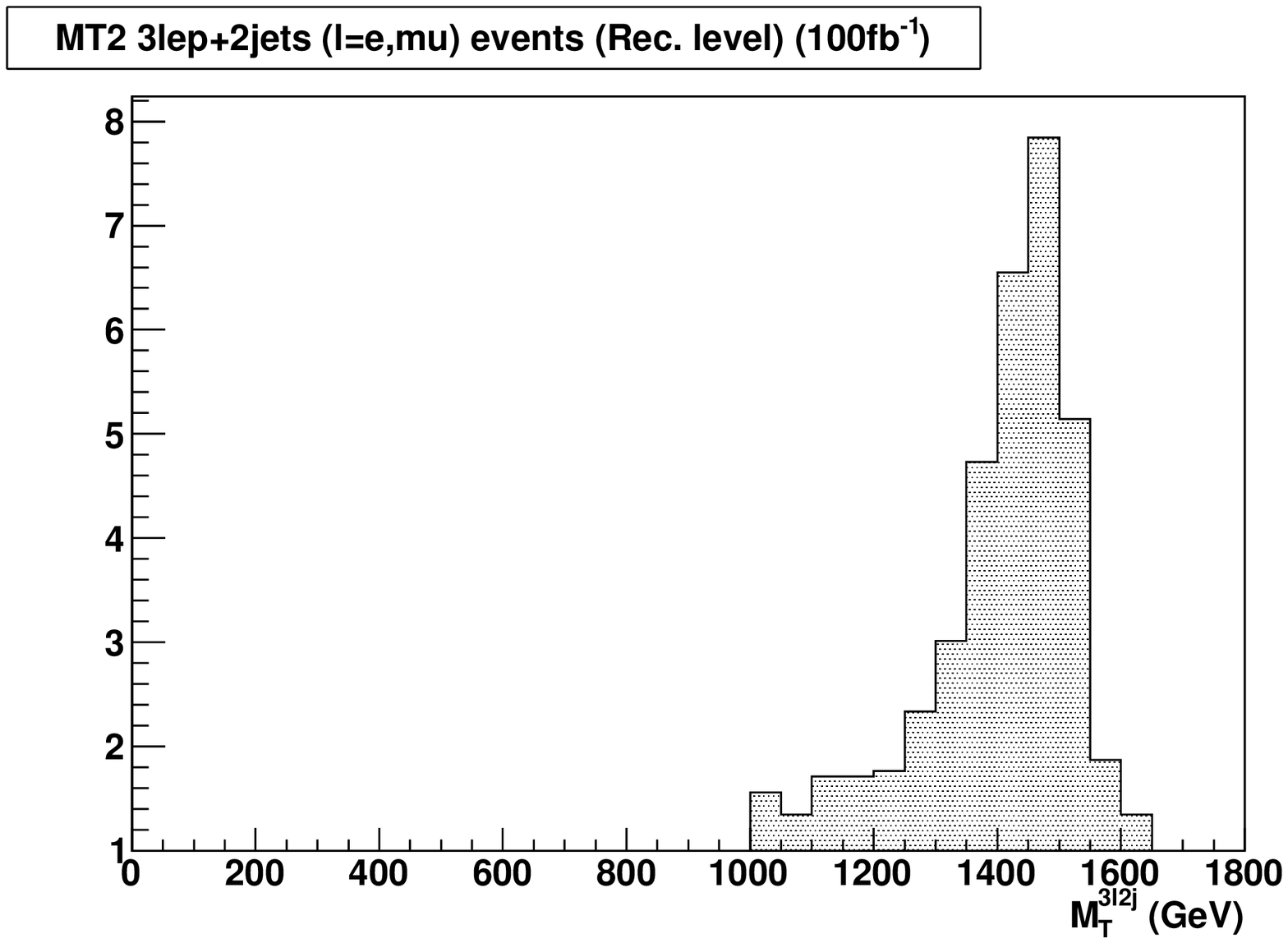}}
  \caption{\it Distribution of $M^T_{3\ell 2j}$ for signal (\ref{mt3l2j_200_det}) $m_{\nu _h}=200$ GeV and (\ref{mt3l2j_500_det}) $m_{\nu _h}=500$ GeV, summed with the backgrounds, and for signal only, in (\ref{mt3l2j_200_det_signal}) and (\ref{mt3l2j_500_det_signal}), respectively, after the Selection $2$ cuts. (Events per $\mathscr{L}=100$ fb${}^{-1}$).}
  \label{mt3l2j_det_lev}
\end{figure}

Once the pair of jets has been identified, we can use it to plot the $M^T_{3\ell 2j}$ distribution, shown in figure~\ref{mt3l2j_det_lev} (summing up all backgrounds and signal, and for the signal only, in the top panels and in the bottom panels, respectively). The events for the signal are visible as an excess over the background, peaking at the $Z'$ boson mass, as clear from figures~\ref{mt3l2j_200_det_signal} and \ref{mt3l2j_500_det_signal} (in figures~\ref{mt3l2j_200_det} and \ref{mt3l2j_500_det} the peak is not evident only due to the log scale).

\begin{figure}[!h]
  \subfloat[]{
  \label{mt2l_200_det}
  \includegraphics[angle=0,width=0.48\textwidth ]{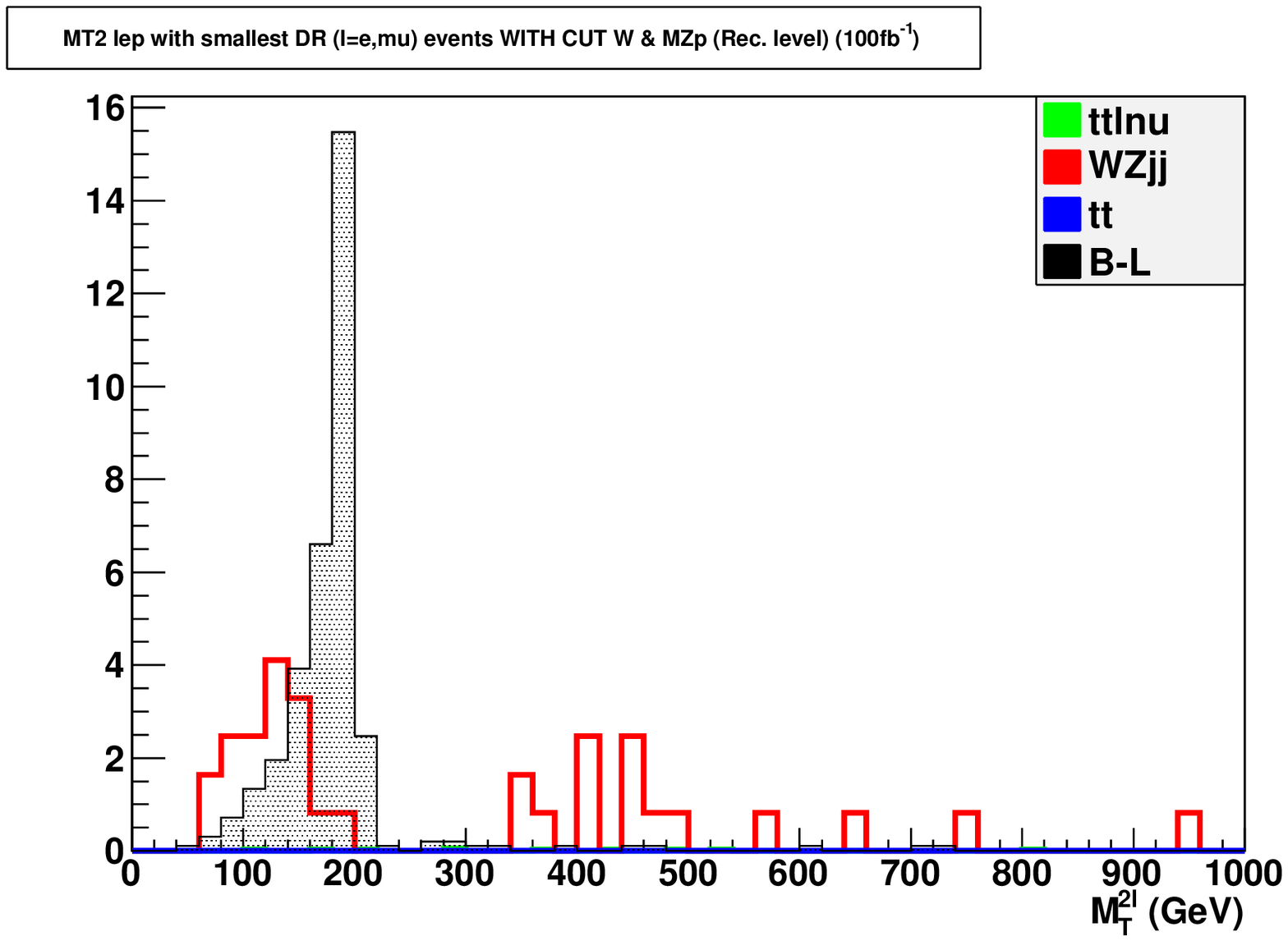}}  
  \subfloat[]{
  \label{mt2l_500_det}
  \includegraphics[angle=0,width=0.48\textwidth ]{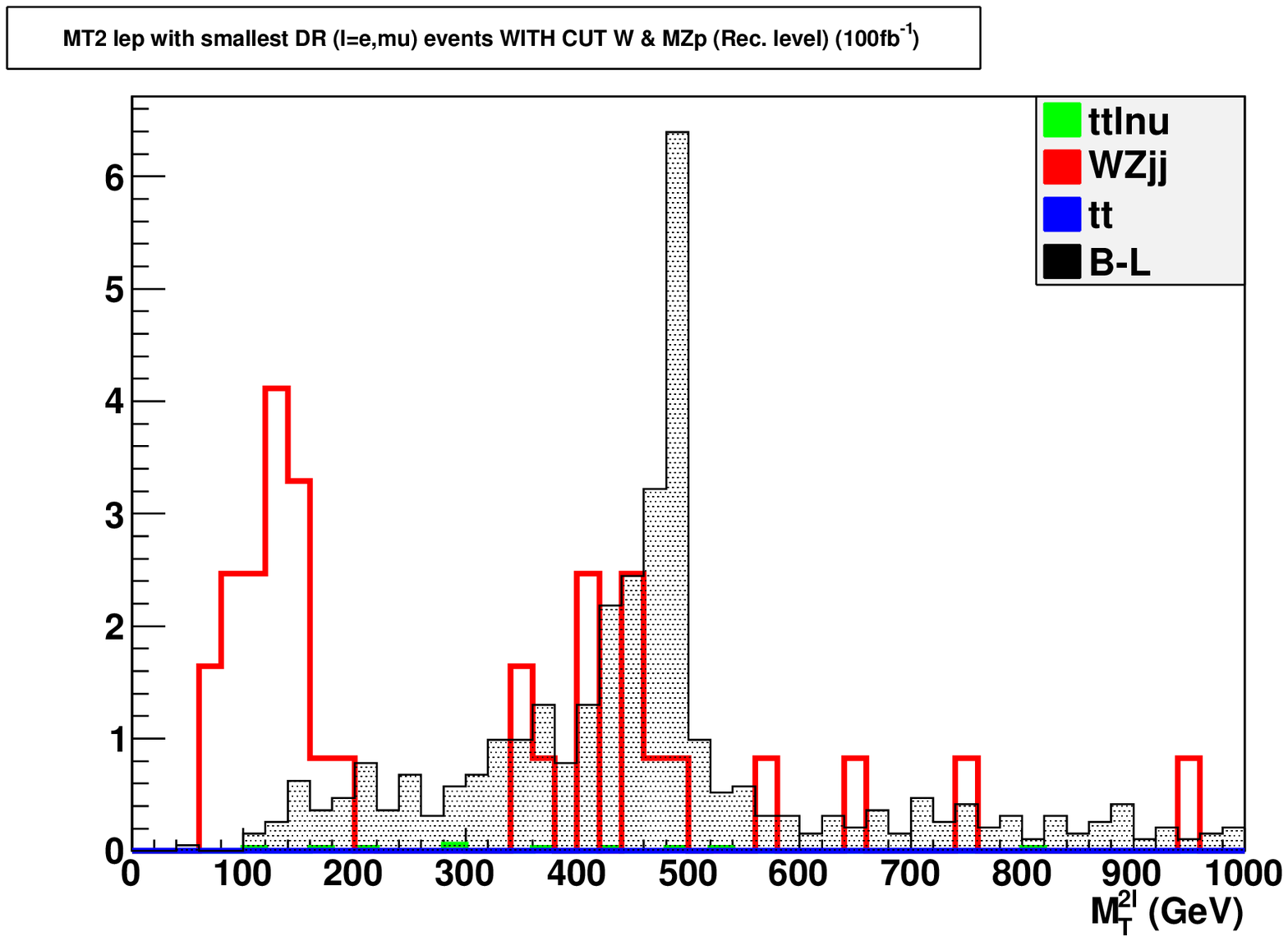}}  
  \caption{\it Distribution of $M^T_{2\ell}$ for signal (\ref{mt2l_200_det}) $m_{\nu _h}=200$ GeV and (\ref{mt2l_500_det}) $m_{\nu _h}=500$ GeV, superimposed to the backgrounds after the Selection $3$ cuts. (Events per $\mathscr{L}=100$ fb${}^{-1}$).}
  \label{mt2l_det_lev}
\end{figure}

As in the previous section, we define the following cut to further suppress the background:
\vspace*{0.5cm}

\centerline{\underbar{\large\bf  Selection $3$}}
\begin{equation} \label{cut_3_det}
\left| M^T_{3\ell 2j} - M_{Z'} \right| < 250~{\rm GeV},
\end{equation}
in which we consider the $Z'$ mass as known, measured from Drell-Yan processes (see section~\ref{subsect:disc_power}). The $M^T_{2\ell}$ distribution after this cut is shown in figure~\ref{mt2l_det_lev}.

In order to establish the signal, we finally select events around the visible $M^T_{2l}$ peak, by requiring:\\
\vspace*{0.5cm}
\centerline{\underbar{\large\bf  Selection $4$}}
\vspace*{-0.45cm}
\begin{equation} \label{cut_4_det}
100< M^T_{2\ell} < 250~{\rm{GeV}} \qquad \mbox{ or } \qquad 300~{\rm{GeV}}< M^T_{2\ell} < 550~{\rm{GeV}},
\end{equation}
depending on the signal benchmark under consideration. Again, a fit of the signal in these regions will finally give the heavy neutrino mass. The efficiencies of the Selection $1$--$4$ cuts are given in table~\ref{tab:alleff_det}. Notice that the number of events for the signal, especially at the beginning, is bigger than at the parton level. This is because all the other multi-lepton signatures can impinge in the tri-lepton one, but they do not survive the whole chain of cuts. This summary clearly confirms the feasibility of the extraction of the signal even with less than $100$ fb$^{-1}$ of integrated luminosity. \index{Heavy neutrinos!Mass measurement}

\begin{table}[htb]
\begin{center}
{$m_{\nu _h}=200$ GeV}\\ \vspace{0.2cm}
\scalebox{0.85}{
\begin{tabular}{|c|cr|cr|cr|cr|c|} \hline
${\rm Cuts} $&$ {\rm Signal} $&$ {\rm Efficiency} $&$ WZjj $&$ {\rm Efficiency}$&$ t\overline{t} $&$ {\rm Efficiency}$&$ ~t\overline{t}\ell\nu $&$ {\rm Efficiency}$&$ S/\sqrt{B} $\\
 & $ {\rm events} $&$ (\%) $&$ {\rm events} $&$ (\%) $&$ {\rm events} $&$ (\%) $&$ {\rm events}  $&$ (\%) $& \\ \hline
$1	$&$ 178.3 $&$ 100	  $&$ 3154   $&$ 100   $&$ 1.04	\cdot 10^{4}$&$ 100  $&$ 112.5       $&$ 100   $&$ 1.52 $\\
$2 	$&$ 74.2  $&$ 41.6	  $&$ 2189   $&$ 69.4  $&$ 9.36 \cdot 10^{3}$&$ 89.8 $&$ 84.4        $&$ 75.0  $&$ 0.69 $\\
$3	$&$ 34.3  $&$ 46.3	  $&$ 27.96  $&$ 1.28  $&$ <1       $&$ <0.1	     $&$ 0.286       $&$ 0.34  $&$ 6.46 $\\
$4	$&$ 31.9  $&$ 92.8	  $&$ 11.51  $&$ 41.18 $&$ 0       $&$ -	     $&$ 0.086       $&$ 30.0  $&$ 9.36 $\\	\hline
\end{tabular}}
\end{center}
\vspace*{0.250cm}
\begin{center}
{$m_{\nu _h}=500$ GeV}\\ \vspace{0.2cm}
\scalebox{0.85}{
\begin{tabular}{|c|cr|cr|cr|cr|c|} \hline
${\rm Cuts} $&$ {\rm Signal} $&$ {\rm Efficiency} $&$ WZjj $&$ {\rm Efficiency}$&$ t\overline{t} $&$ {\rm Efficiency}$&$ ~t\overline{t}\ell\nu $&$ {\rm Efficiency}$&$ S/\sqrt{B} $\\
 & $ {\rm events} $&$ (\%) $&$ {\rm events} $&$ (\%) $&$ {\rm events} $&$ (\%) $&$ {\rm events}  $&$ (\%) $& \\ \hline
$1	$&$ 117.9 $&$ 100   $&$ 3154   $&$ 100   $&$ 1.04	\cdot 10^{4}$&$ 100  $&$ 112.5       $&$ 100   $&$ 1.01 $\\
$2 	$&$ 48.9  $&$ 41.5  $&$ 2189   $&$ 69.4  $&$ 9.36 \cdot 10^{3}$&$ 89.8 $&$ 84.4        $&$ 75.0  $&$ 0.45 $\\
$3	$&$ 33.5  $&$ 68.5  $&$ 27.96  $&$ 1.28  $&$ <1       $&$ <0.1	     $&$ 0.286       $&$ 0.34  $&$ 6.31 $\\
$4	$&$ 22.1  $&$ 66.0  $&$ 9.05   $&$ 32.4  $&$ 0 	    $&$ -	    $&$ 0.114       $&$ 40.0  $&$ 6.50 $\\	\hline
\end{tabular}}
\end{center}
\caption{\it Signal ($m_{\nu _h}=200$ GeV at the top and $m_{\nu _h}=500$ GeV at the bottom)  and background events per ${\mathscr{L}}=100\;{\rm fb}^{-1}$ and efficiencies following the sequential application of  Selection $1$--$4$ cuts.}\label{tab:alleff_det} \end{table}

As we specified earlier, table~\ref{tab:alleff_det} shows that the procedure we used to identify the pair of jets to include in our distributions is not the most efficient. In fact, the Selection $2$ cut has a lower efficiency for the signal ($\sim 40\%$, irrespectively of the benchmark point) than for the background ($\sim 70\% \div 90\%$). The reason for this is that we are shaping the background as the signal, forcing to select those jets that mimic it. Nonetheless, mimicking the $W$ boson mass is not sufficient to imitate all the signal properties. In fact, the subsequent Selection $3$ cut strongly suppresses only the background.

\section{Conclusion}\label{ch4:concl}
In this chapter, we have presented the results of our investigation of the fermion sector.

First, we have summarised the existing experimental constraints on the neutrino masses. We have also discussed the approximation used in the phenomenological study, i.e., to take heavy neutrinos relatively light (with respect to the $Z'$ boson) and degenerate.

Then, we have presented a detailed study of the heavy neutrino properties (cross sections, decay widths and  BRs). We have shown that the production cross sections via the $Z'$ boson, the channel of interest, are up to $50$ fb for the LHC in its early stage ($\sqrt{s}=7$ TeV), value that induced us to consider this channel as viable only for the LHC in its designed performances ($\sqrt{s}=14$ TeV and full luminosity).

Regarding the decay properties of the heavy neutrinos, we have shown that the decay into a SM gauge boson (together with a lepton) is preferred. This partial width is very small, and, hence, the heavy neutrino has a very small intrinsic width.
The decay patterns of the heavy neutrino pair production via $Z'$ boson have been analysed in details. They can be thought of as multi-lepton (and multi-jet) decays of the $Z'$ boson. Up to $4$ charged leptons in the final state can be produced.

We have concluded that for a large portion of the parameter space, the heavy neutrinos are rather long-lived particles, so that they produce displaced vertices in the LHC detectors, that can be distinguished from those induced by $b$ quarks. In addition, from the simultaneous measurement of both the heavy
neutrino mass and decay length one can estimate the absolute mass of the parent light neutrino, for which at present, only limits exist.

To address the neutrino mass measurement, we have chosen to study the so-called tri-lepton decay mode, in which the $Z'$ boson decays into $3$ charged leptons (either electrons or muons) and $2$ jets, always stemming from a $W$ boson.

For benchmark scenarios of the $B-L$ model, we have chosen two that should be accessible at the LHC, having a $Z'$ boson mass and fermion couplings not far beyond the ultimate reach of Tevatron and LEP and displaying two extreme relative conditions between the $Z'$ boson and heavy neutrinos, that is, one
with the latter produced at rest and the other highly boosted in the direction of the $Z'$ boson.

A detailed parton level simulation of the aforementioned signal benchmarks has been performed, including all the relevant backgrounds. We have shown that rather generic cuts are already suitable for an efficient background rejection. This analysis strategy has, then, been validated with a full detector simulation of both signal and background. To do so, a strategy for the jet identification has been proposed, based on the signal topology at the parton level (in which, as mentioned, the jets come exclusively from the $W$ boson).

The transverse mass has been found to be the suitable distribution for the tri-lepton signature analysis. By carefully choosing what particles to consider for its definition, peaks for both the $Z'$ boson and for the heavy neutrino emerge. Their simultaneous observation is a striking signature of the $B-L$ model. In particular, the sharpness of the peak corresponding to the heavy neutrino, as well as its definition well above the background, as shown, makes the heavy neutrino mass well measurable.

As a result, the tri-lepton signature holds a very good signal over background ratio that makes it observable at the LHC, for $\sqrt{s}=14$ TeV and $\mathcal{L} \leq 100$ fb$^{-1}$.

\chapter{Scalar sector}\label{Ch:5}
\ifpdf
    \graphicspath{{Chapter5/Chapter5Figs/PNG/}{Chapter5/Chapter5Figs/PDF/}{Chapter5/Chapter5Figs/}}
\else
    \graphicspath{{Chapter5/Chapter5Figs/EPS/}{Chapter5/Chapter5Figs/}}
\fi

In this chapter we study the scalar sector of the $B-L$ model. The new states here are two {\it CP}-even Higgs bosons, whose masses are free parameters. Another free parameter is the angle $\alpha$ that controls their mixing.

Section~\ref{sect:Higgs_constraints} presents the constraints on the free parameters in this sector. We will first review in section~\ref{sec:expbounds:Higgs} the existent experimental limits, coming from both direct and indirect searches at LEP. In section~\ref{sect:theo_constraints} is then presented the analysis of the theoretical bounds on the scalar sector, i.e., the so-called unitarity bound and the triviality and vacuum stability bounds, the latter two coming from the study of the RGEs.

In section~\ref{sect:Higgs}, the properties of the scalar sector are delineated (i.e., production cross sections, intrinsic widths, BRs), and the capabilities for Higgs discovery at the LHC are summarised. The main focus is on the impact of the gauge and fermion sectors. In fact, although no new consistent production mechanisms arise, the decay patterns in this model are rather peculiar, such as the decay into pairs of $Z'$ bosons and into heavy neutrinos (besides the decay of the heavy Higgs boson into pairs of the light scalar bosons). Cross sections for some full processes are presented in section~\ref{subsect:event_rates}.

Finally, we  draw the conclusions for this chapter in section~\ref{ch5:concl}.

The study of the unitarity bound is published in \cite{Basso:2010jt}. The RGE study (triviality and vacuum stability) is published in \cite{Basso:2010jm}. The analysis of the phenomenology of the Higgs bosons is published in \cite{Basso:2010yz}.

\section{Constraints}\label{sect:Higgs_constraints}
In this section is delineated the viable parameter space for the new independent parameters in the scalar sector, i.e., $m_{h_1}$, $m_{h_2}$, and the mixing angle $\alpha$. 

First, we summarise the experimental constraints from LEP, from direct searches and from the analysis of the precision tests. Regarding the former, as discussed in the introduction, we did not pursue a complete beyond the tree-level analysis and we will refer to the literature.

Finally, we present our analysis of the theoretical constraints, i.e., the unitarity bound and the constraints coming from the study of the RGEs.

\subsection{Experimental constraints}\label{sec:expbounds:Higgs}

\begin{figure}[!h]
  \centering
  \includegraphics[angle=0,width=0.68\textwidth ]{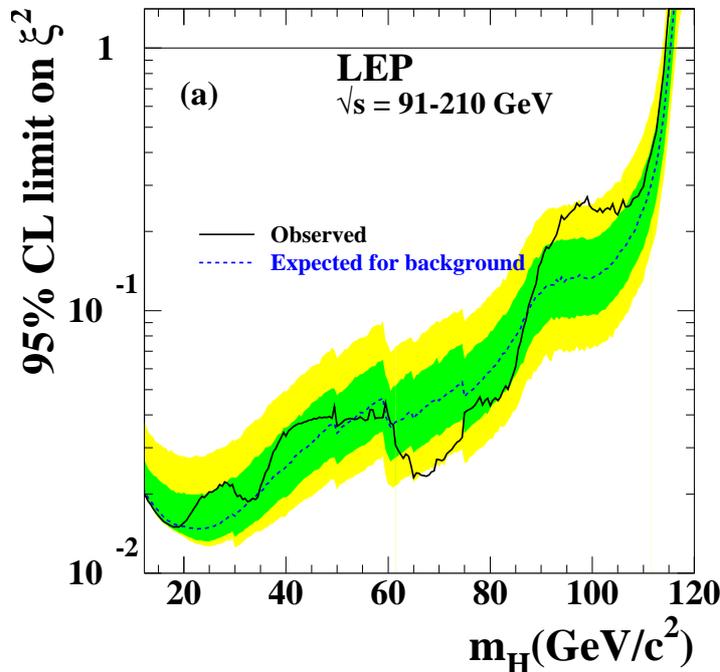}
  \vspace*{-0.5cm}
  \caption{\it The $95\%$ C.L. upper bound on $\xi=g_{HZZ}/g^{SM}_{HZZ}$ \cite{Barate:2003sz}. In the $B-L$ model, $\xi=\cos{\alpha}(\sin{\alpha})$ for $H=h_1(h_2)$. The solid line represents observed values; the dotted line represents values expected for the background. \label{LEP_lim}}
\end{figure}

Past and current experiments have set limits on the parameters of the scalar sector in the SM as well as in various extensions of it: see, for example Ref.~\cite{Barate:2003sz} for LEP and 
Ref.~\cite{Aaltonen:2011gs} for Tevatron. 
For the model discussed here, the relevant analysis is summarised in figure~\ref{LEP_lim} (figure~10a in the LEP combined analysis of Ref.~\cite{Barate:2003sz}), in which a generic overall factor $\xi$ has been introduced. 
Such parameter is defined as the coupling(s) to the $Z$ boson of the Higgs particle(s) in the considered extension 
normalised to the SM:
\begin{equation}
\xi\equiv \frac{g_{HZZ}}{g^{SM}_{HZZ}}\, ,
\end{equation}
hence it parametrises the deviations of the new model with respect to the SM.

 \index{LEP constraints!Higgs bosons}

In the minimal $U(1)$ extension of the SM, the argument of this Thesis, two real scalar degrees of freedom exist: 
the one coming from the Higgs singlet, required to break the extra $U(1)_{B-L}$ gauge factor 
(and therefore giving the $Z'$ gauge boson a mass), and the one coming from the Higgs doublet, 
required to break the EW gauge symmetry to give masses to the $W$ and $Z$ bosons. In all generality, these two scalars will mix (as, for example, in any other extension of the SM with one scalar singlet in addition to the Higgs doublet).
With reference to eq.~(\ref{scalari_autostati_massa}), $h_1$ is the lightest eigenstate, 
that couples to the $Z$ boson proportionally to $\cos{\alpha}$, $h_2$ is the heaviest one, that couples to the $Z$ boson proportionally to $\sin{\alpha}$. Hence, the LEP lower bounds on 
the scalar masses of the $U(1)_{B-L}$ extension here considered are read straightforwardly from figure~\ref{LEP_lim} 
by considering:
\begin{equation}
\left\{ \begin{array}{cc} 
	\xi=\cos{\alpha} &\qquad \mbox{for } H = h_1\, ,\\
	\xi=\sin{\alpha} &\qquad \mbox{for } H = h_2\, ,
	\end{array} \right. 
\end{equation}
i.e., the limit for $h_1$($h_2$) is extracted by considering $\xi$ as the cosine(sine) of the mixing angle 
in the scalar sector [see eq.~(\ref{scalari_autostati_massa})].

Figure~\ref{LEP_lim} shows the lower bound on the Higgs boson mass as a function of $\xi$. The SM Higgs boson is recovered by the condition $\xi =1$. We see that we can have significant deviations from the SM Higgs boson mass limit, $m_h > 114.4$ GeV, only for values of the angle $\alpha > \pi /4$, for the lightest state $h_1$. For example, for $\alpha = \pi /3$, the LEP limit on the lightest Higgs state reads as $m_{h_1} > 100$ GeV. That is, in this model, a light Higgs boson with mass smaller than the SM limit can exist only if it is highly mixed, i.e., the light Higgs boson is mostly the singlet state. For the same value of the angle, the limit for $m_{h_2}$ is more stringent than the condition $m_{h_2} > m_{h_1}$, in fact for $\alpha = \pi /3$, $m_{h_2} \gtrsim 114$ GeV must be fulfilled.

\begin{figure}[!h]
  \centering
  \includegraphics[angle=0,width=0.68\textwidth ]{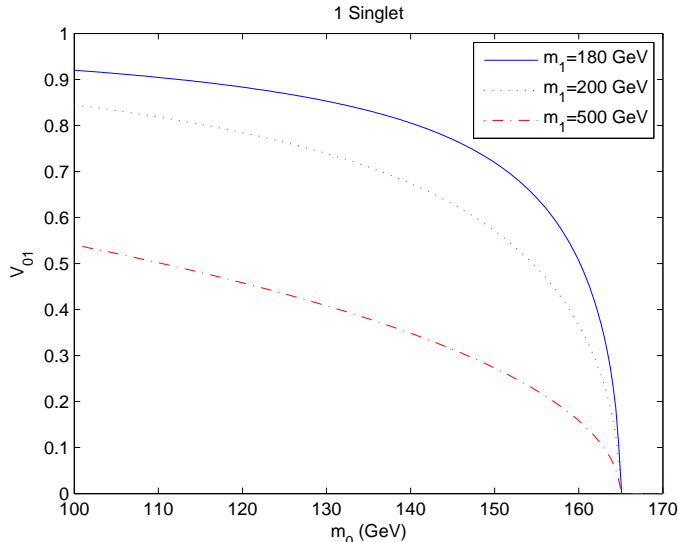}
  \vspace*{-0.5cm}
  \caption{\it Allowed region (at $95\%$ confidence level) in a model with one additional singlet in addition to the usual $SU(2)_L$ doublet. The lightest (heavier) scalar mass is $m_0$ ($m_1$) and the mixing matrix is defined in eq.~($4$) of Ref.~\cite{Dawson:2009yx}. In our notation, $\left|V_{01}\right| = \sin{\alpha}$. The region below the curves is allowed by fits to $S$, $T$ and $U$. \label{Prec_meas_constraint_Higgs}}
\end{figure}

Direct searches have put limits on scalar masses in the $B-L$ model below the LEP limit of $114.4$ GeV (for the SM). The existence of scalar bosons heavier than this limit can also be restricted. One way is by analysing  their impact on the EW precision measurements \index{LEP constraints!Precision measurements} \cite{:2005ema}, encoded in the $S$, $T$ and $U$ parameters \cite{Peskin:1991sw}. In the case of a SM scalar sector augmented by just one scalar singlet, the constraints on the scalar masses are analysed in Ref.~\cite{Dawson:2009yx}, and summarised in figure~\ref{Prec_meas_constraint_Higgs}. 

Generally speaking, the scalar sector of the $B-L$ model is equivalent to the one of generic scalar singlet extensions of the SM, so that we can straightforwardly interpret the results of Ref.~\cite{Dawson:2009yx} as valid also here. It ought to be noticed, though, that this statement relies on the simplification that the impact of the extra matter content of the $B-L$ model, i.e., $Z'$ boson and heavy neutrinos, on the precision parameters is negligible.
This is indeed the case for the direct contribution of the $Z'$ boson, as it has vanishing mixing with the SM $Z$ boson at the tree level \cite{Cacciapaglia:2006pk} (and the running on $\widetilde{g}$, turning this mixing on at the NLO, is a two-loops effect).
Nonetheless, the $Z'$ boson and the heavy neutrinos can still alter the scalar boson total widths entering in the precision parameters' calculation. An increment in the total width will relax the results of Ref.~\cite{Dawson:2009yx}, although marginally as the total width changes are mild for values of the mixing angle bounded therein.

With reference to~\cite{Dawson:2009yx}, the mixing matrix $V$ corresponds to the matrix in eq.~(\ref{scalari_autostati_massa}), so that $\left|V_{01}\right| = \sin{\alpha}$. Important consequences of this analysis are that there is an upper bound on the light scalar, $m_{h_1} < m^{MAX}_{h_1} \equiv 165$ GeV, and that the complete inversion in the scalar sector, i.e.,  for $\alpha = \pi /2$, is forbidden by precision data for $m_{h_2} > m^{MAX}_{h_1}$.

\subsection{Theoretical constraints}\label{sect:theo_constraints}
In this section we present the analysis of the theoretical constraints on the new parameters in the scalar sector, namely the unitarity bound and those coming from the study of the RGEs. The unitarity assumption \cite{Basso:2010jt} relies on demanding that the cross sections for the scattering of all the particles in the model are unitary, i.e., with an occurring probability less than one. The RGE constraints \cite{Basso:2010jm} come from the model to be well-defined at one loop, assuming that the evolved parameters do not hit any Landau pole or destabilise the vacuum of the model. The latter request is known as the vacuum stability condition, while the former is the triviality condition.

\subsubsection{Unitarity bound}\label{subsect:Unitarity}\index{Unitarity (perturbative)}
It is generally not possible within the SM
framework (or any of its non-supersymmetric extensions encompassing
the Higgs mechanism) to predict the mass of the Higgs boson. Hence, several theoretical methods have been developed to constrain its value
(see \cite{Dicus:1992vj}, \cite{Lee:1977eg}, \cite{Dashen:1983ts}).
For example, to stay with the SM,
the pioneeristic work of Ref.~\cite{Lee:1977eg} showed 
that, when the Higgs boson mass is greater than a critical value of around $1$ TeV (known as unitarity bound), the spherical partial wave describing the
elastic scattering of the longitudinally polarised vector bosons at very high energy ($\sqrt{s} \rightarrow \infty$) violates unitarity at the tree level and
the theory ceases to be valid from a perturbative point of view.

In the high energy limit ($s \gg m^2_{W^{\pm},Z,Z'}$), the
amplitude involving the (physical) longitudinal polarisation of
gauge bosons approaches the one involving the (unphysical) scalar one
(equivalence theorem, see \cite{Chanowitz:1985hj}), and, following the {\textsc{BRS}} invariance \cite{Becchi:1975nq}, the amplitude for emission or absorption of a `scalarly' polarised gauge boson becomes equal to the amplitude for emission or absorption of the related Goldstone boson. 
Since it is the
elastic scattering of longitudinally polarised vector bosons that gives rise to unitarity violation, the analysis of the perturbative unitarity of two-to-two
particle scatterings in the gauge sector can be performed, in the high energy limit, by exploiting the Goldstone sector.

Hence, in the high energy limit, we can substitute the vector boson and Higgs boson sectors with the related (would-be) Goldstone and Higgs boson sectors. We will therefore focus on the scalar
interacting Lagrangian of the Higgs and would-be Goldstone sectors (in the
Feynman gauge), i.e., the scalar Lagrangian of eqs.~(\ref{new-scalar_L}), (\ref{BL-potential}) and (\ref{Higgs_goldstones}), neglecting here the gauge couplings in the covariant derivative of eq.~(\ref{cov_der}), and we will
calculate tree-level amplitudes for all two-to-two processes 
involving the full set of possible (pseudo)scalar fields \footnote{Moreover, while evaluating scalar bosons' scattering amplitudes, it has
been explicitly verified that, in the search for the Higgs boson mass limits, the contribution that arises from intermediate vector boson exchange
is not relevant.}.
The relevant Feynman rules are in section~\ref{sect:feym_rules}, where $V_F$ is the Goldstone counterpart of the vector $V$.

\paragraph{Evaluation of the unitarity bound}\label{subsubsect:Uniev}
~


\vspace*{0.3cm}
\noindent As already intimated, the equivalence theorem allows
one to compute the amplitude of any process with longitudinal vector bosons
$V_L$ ($V = W^\pm,Z,Z' $), in the limit $m^2_V\ll s$,
by substituting each one of them with the related Goldstone bosons $v
= w^\pm,z,z'$, and its general validity is proven
\cite{Chanowitz:1985hj}; schematically, if we consider a
process with four longitudinal vector bosons: $\mathcal{M}(V_L V_L \rightarrow
V_L V_L) = \mathcal{M}(v v \rightarrow v v)+ O(m_V^2/s)$.

Given a tree-level scattering amplitude between two spin-$0$ particles,
$\mathcal{M}(s,\theta)$, where $\theta$ is the scattering (polar) angle, 
the partial wave amplitude with angular
momentum $J$ is given by
\begin{eqnarray}\label{integral}
a_J = \frac{1}{32\pi} \int_{-1}^{1} d(\cos{\theta}) P_J(\cos{\theta})
\mathcal{M}(s,\theta),
\end{eqnarray}
where $P_J$ are Legendre polynomials. It has been proven
in Ref.~\cite{Luscher:1988gk} that, in order to preserve unitarity, each
partial wave must be bounded by the condition
\begin{eqnarray}\label{condition}
\Big|\textrm{Re}\big( a_J(s) \big)\Big|\leq \frac{1}{2}.
\end{eqnarray}
It turns out that only $J=0$ (corresponding to the spherical partial
wave contribution) leads to some bound, so we will not discuss the
higher partial waves any further. 

We have verified that, in the high energy limit, only the four-point vertices
(related to the four-point functions of the interacting potential)
contribute to the $J=0$ partial wave amplitudes, and this
is consistent with many other works that exploit the same methodology
(for example, see \cite{Maalampi:1991fb,Huffel:1980sk,Casalbuoni:1987cz}). 
The main contributions come from the $zz \to zz$ and $z'z' \to z'z'$ channels. To a lesser extent, also $h_1h_1\to h_1h_1$ and $h_2h_2\to h_2h_2$ play a relevant role, being equal to the most constraining one among the main channels in some regions of the parameter space (for $\alpha \to 0$ and $\alpha \to \pi/2$, respectively). Due to this, we will not discuss them any further (details of this can be found in Ref.~\cite{Basso:2010jt}).

Moving to the results, the unitarity bound is, in general, a function of all the parameters that take part to the four-point vertices, i.e., the scalar masses, the mixing angle $\alpha$ and the singlet VEV $x$.
It is convenient to define a `high-mixing' and a `low-mixing' domains in the mixing angle. The former is defined by noticing that for a value of the angle in the range
\begin{equation}
\mbox{arctan}\bigg(\frac{v}{x}\bigg) \leq \alpha \leq \frac{\pi}{2}\, ,
\end{equation}
the allowed parameter space [in the ($m_{h_1}$--$m_{h_2}$) plane] is completely defined by the $zz \to zz$ channel. The `low-mixing' domain is the complementary region. For instance, since $x \geq 3.5$ TeV
as shown in section \ref{sec:expbounds:Zp}, due to the LEP bound \cite{Cacciapaglia:2006pk}, if we choose exactly $x = 3.5$ TeV, the high-mixing
domain, in this case, is the one for $0.07 \leq \alpha \leq \pi/2$
(and, conversely, the low-mixing one is the interval $0 \leq \alpha < 0.07$).

\begin{figure}[!ht]
  \subfloat[]{
  \label{a001} \includegraphics[angle=0,width=0.49\textwidth]{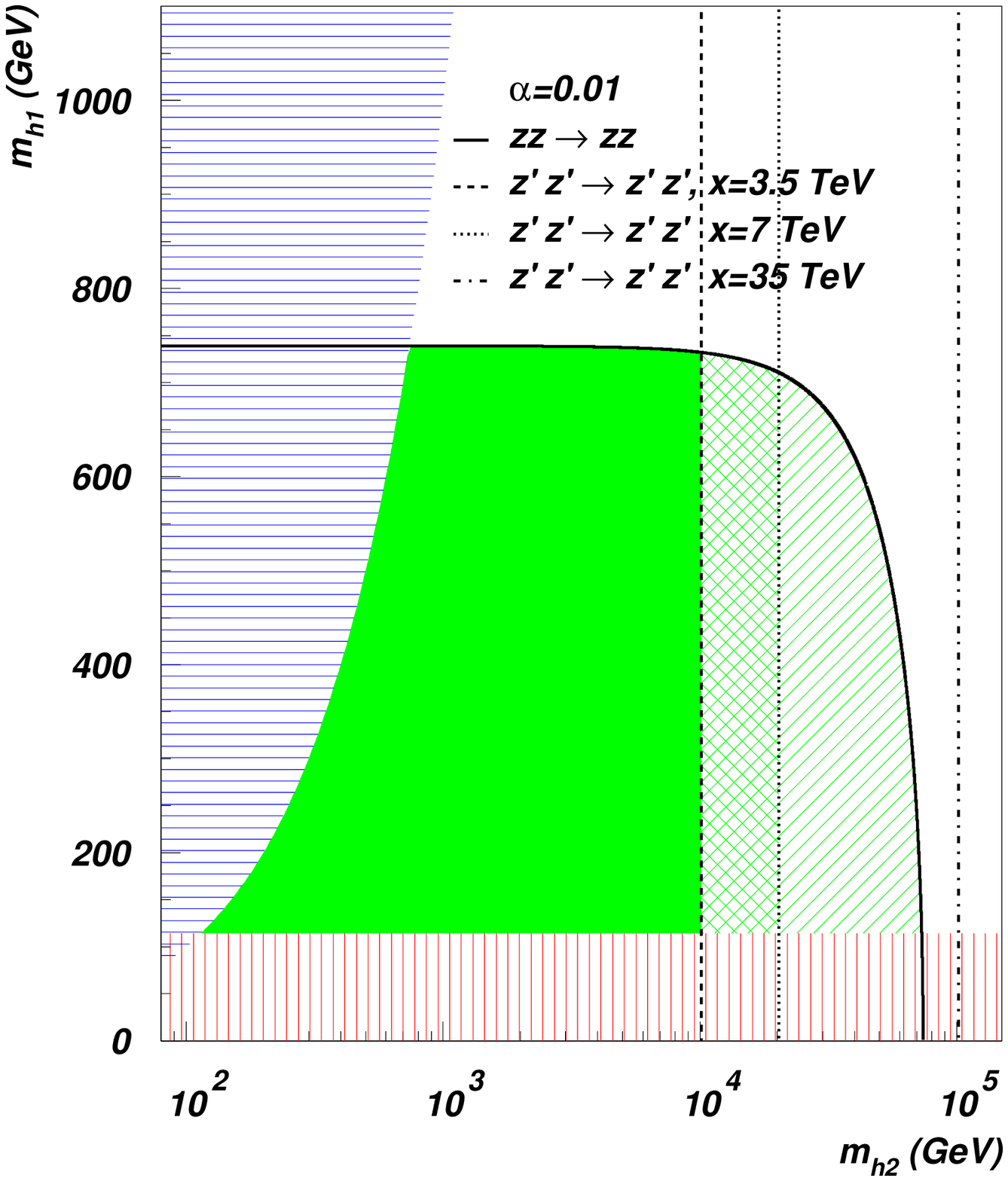}}
  \subfloat[]{
  \label{a01}
  \includegraphics[angle=0,width=0.49\textwidth ]{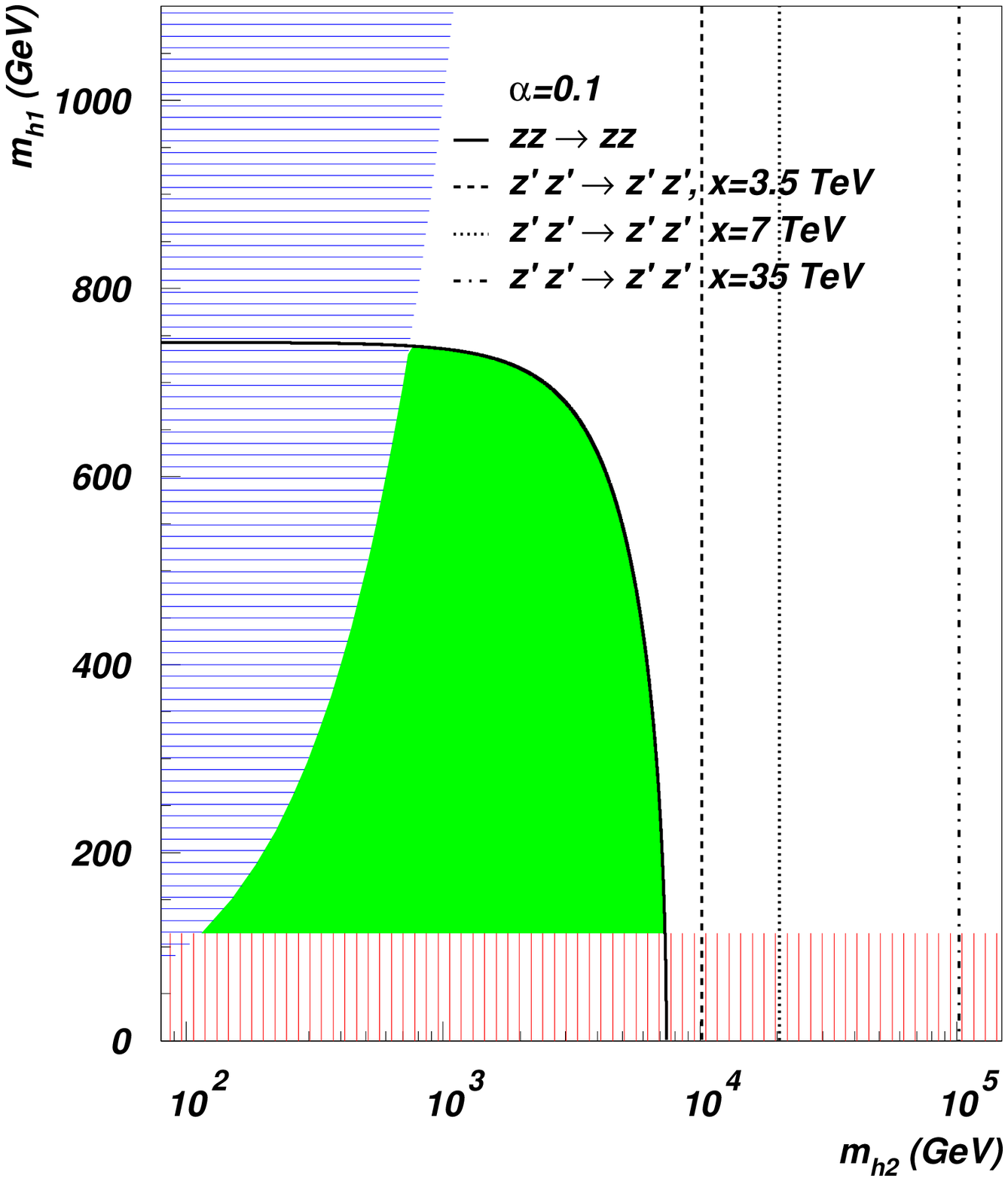}}
\\
  \subfloat[]{
  \label{api4} \includegraphics[angle=0,width=0.49\textwidth]{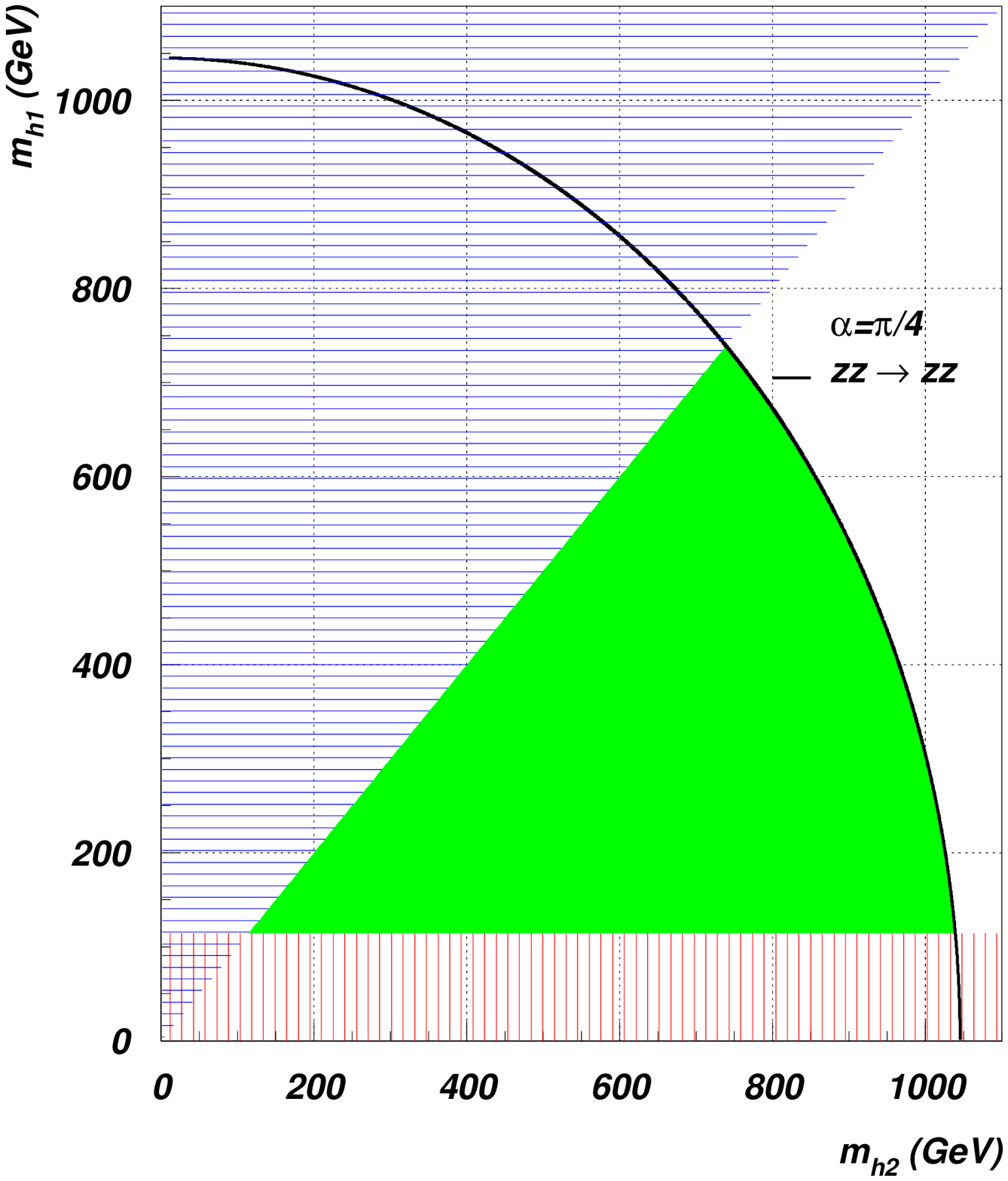}}
  \subfloat[]{
  \label{api2} \includegraphics[angle=0,width=0.49\textwidth]{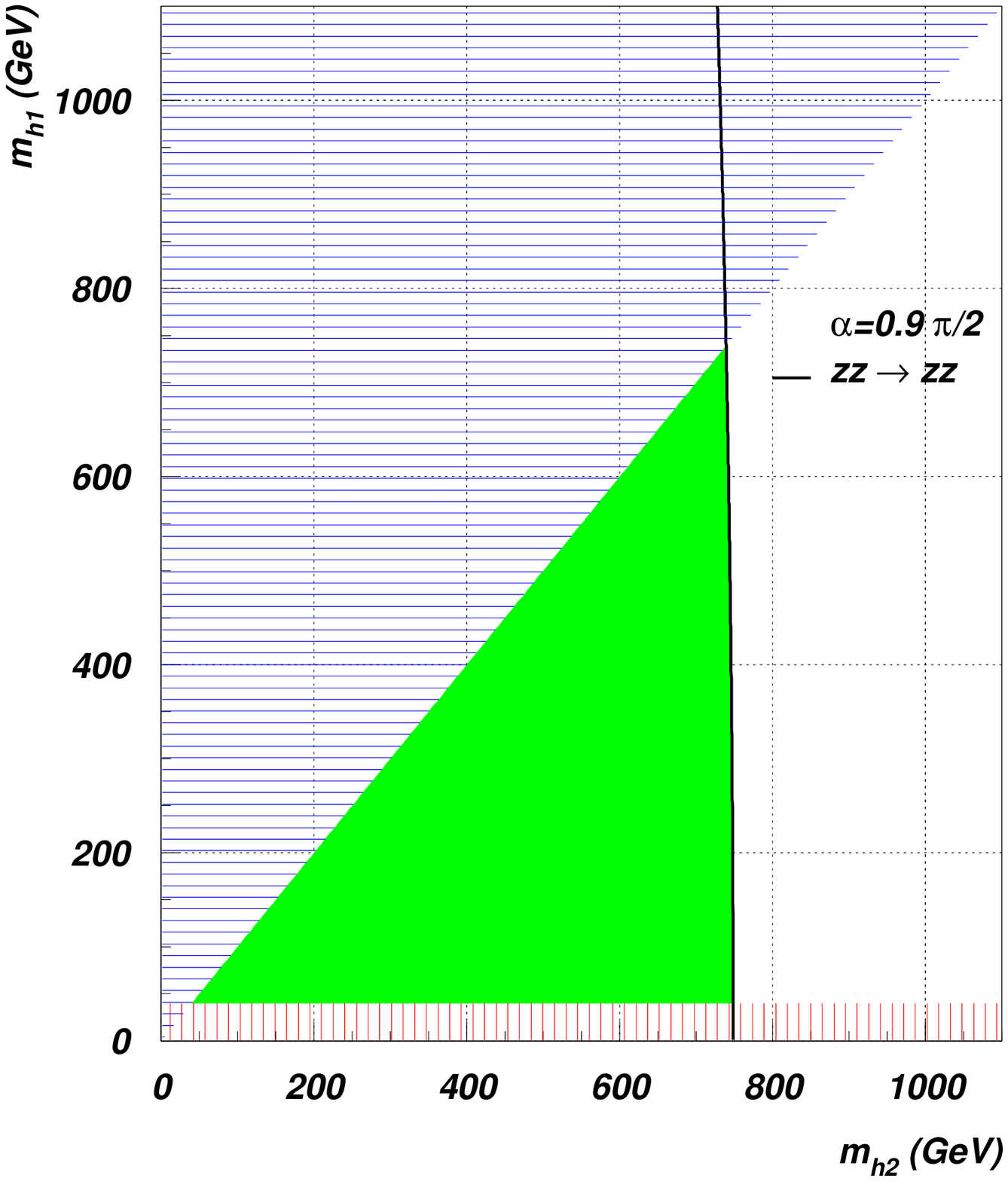}}
  \vspace*{-0.25cm}
  \caption{\it Higgs bosons mass limits in the $B-L$ model coming from
  the unitarity condition $|\textrm{Re}(a_0)|\le \frac{1}{2}$ applied
  to the $zz~\rightarrow~zz$ and $z'z'~\rightarrow~z'z'$ scatterings for
  several values of $x$: (\ref{a001}) for $\alpha=0.01$, (\ref{a01}) for $\alpha=0.1$, (\ref{api4}) for $\alpha=\pi/4$, and (\ref{api2}) for $\alpha=0.9\ \pi/2$. The (blue) horizontal shadowed region is forbidden by our convention $m_{h_2}~>~m_{h_1}$. The (red) vertical shadowed region is excluded by the LEP experiments.}
  \label{Unit:mh1vsmh2}
\end{figure}

Figure~\ref{Unit:mh1vsmh2} shows the allowed regions in the ($m_{h_1}$--$m_{h_2}$) plane, for a value of the angle in the low-mixing
domain (figure~\ref{a001}), and for values in the high-mixing one (figures~\ref{a01}, \ref{api4} and \ref{api2}), for several values of the singlet VEV $x$ (affecting only the limit coming from the $z'z'~\to~z'z'$ channel).

We see that in both cases, as expected, the upper bound on the light Higgs boson mass does not exceed the SM one (which is around $700$~GeV~\cite{Luscher:1988gk}), 
and it approaches the experimental lower limit from LEP (according
to~\cite{Barate:2003sz}) as the heavy Higgs boson mass increases. This is because the two Higgs bosons `cooperate' in the unitarisation of the scattering channels, so that, if one mass tends to grow, the other one must become lighter and lighter to keep the scattering matrix elements unitarised.

In the high-mixing domain, as shown in
figures \ref{a01}, \ref{api4} and \ref{api2} (for $\alpha=0.1$, $\alpha=\pi/4$, $\alpha=0.9 \pi/2$, respectively \footnote{For the last of these values of the mixing angle, the lower limit from LEP experiments on the light Higgs boson mass is $m_{h_1}>40$ GeV \cite{Barate:2003sz}, while for the other ones it is almost equal to the SM lower limit ($m_{h_1}>114.4$~GeV), as illustrated in figure~\ref{Unit:mh1vsmh2}.}), the allowed region is completely determined by the $zz\to~zz$ scattering, irrespectively of the value of the singlet VEV $x$.
The maximum allowed value for the heavy Higgs boson mass only depends on the mixing angle
\begin{eqnarray}\label{high-maxmh2}
{\rm{Max}}(m_{h_2})=2\sqrt{\frac{2}{3}}
\ \frac{m_W}{\sqrt{\alpha_W}\sin{\alpha}}.
\end{eqnarray}

Only in the low-mixing domain, as in figure~\ref{a001} (for $\alpha=0.01$), we are able to appreciate some interplay between the two scattering processes in setting the unitarity bound. In this case, while the $zz~\to~zz$
scattering channel would allow the existence of a heavy Higgs boson with mass of more than $10$ TeV, the $z'z'~\to~z'z'$ channel sharply limits the allowed mass region, with a `cut-off' on the maximum $m_{h_2}$ almost insensible
to the light Higgs boson mass (and to the value of the mixing angle), that scales linearly with the singlet VEV $x$
\begin{eqnarray}\label{low-maxmh2}
{\rm{Max}}(m_{h_2})\simeq 2\sqrt{\frac{2\pi}{3}}x.
\end{eqnarray} 
This is pictorially shown in figure~\ref{a001}: the solid green area
represents the allowed portion of the $m_{h_1}$-$m_{h_2}$ space at $x=3.5$~GeV,  that, at $x=7$~TeV, increases until the (green) crossed shadowed region, to relax even further to the (green) single line shadowed region for $x=35$~TeV.

To summarise, given a value of the scalar mixing angle $\alpha$, the upper bound on the light Higgs boson mass varies between the SM limit and the experimental lower limit from LEP as the upper bound for the heavy Higgs boson mass increases. Moreover, when $\alpha$ assumes values in the high-mixing domain, the strongest bound comes exclusively from the $z$-boson scattering, independently  from the chosen singlet VEV $x$, while, in the low-mixing domain, the $z'$ boson scattering can also be important, imposing a cut-off, linear in $x$, on the heavy Higgs boson mass (when more constraining than the $zz\rightarrow zz$ scattering).

In the following subsection we investigate the constraints coming from the analysis of the RGEs, collected in appendix~\ref{App:RGE}.


\subsubsection{Renormalisation group equations}\label{subsubsect:RGE_scalar}\index{RGE!Scalar sector}

The RGE evolution can constrain the parameter space of the scalar sector in two complementary ways. From one side, the couplings must stay perturbative. This condition reads:
\begin{equation}\label{cond_1}
0 < \lambda _{1,2,|3|}(Q') < 1 \qquad \forall \; Q' \leq Q\, ,
\end{equation}
and it is usually referred to as the `triviality' \index{Triviality} condition. Notice that $\lambda _{|3|}\equiv |\lambda _3|$.
On the other side, the vacuum of the theory must be well-defined at any scale, that is, to guarantee the validity of eqs.~(\ref{inf_limitated}) and (\ref{positivity}) at any scale $Q'\leq Q$:
\begin{equation}\label{cond_2}
0 < \lambda _{1,2}(Q') \qquad \mbox{and} \qquad
4\lambda _1(Q')\lambda _2(Q')-\lambda _{3}^2(Q') > 0 \qquad \forall \; Q' \leq Q\, .
\end{equation}
Eq.~(\ref{cond_2}) is usually referred to as the `vacuum stability'  \index{Vacuum stability} condition. In contrast to the SM, in which it is sufficient that the Higgs self-coupling $\lambda$ be positive, in the case of this model the vacuum stability condition [and especially the second part of eq.~(\ref{cond_2})] can be violated even for positive $\lambda _{1,2,3}$.

One should notice that our conventional choice $m_{h_1} < m_{h_2}$, as noted previously, allows us to consider $\alpha$ and $-\alpha$ as two independent solutions, although the theory is manifestly invariant under the symmetry $\alpha\rightarrow -\alpha$. These two solutions are complementary, meaning that the region excluded by the choice $m_{h_1} < m_{h_2}$ at a certain value of the angle $\alpha$ is precisely the allowed one for the complementary angle $\pi /2 - \alpha$. The special case $\alpha=\pi /4$ is symmetric, and corresponds to maximal mixing between the scalars. $\alpha=0$ corresponds to a SM scalar sector totally decoupled from the extended one, and $h_1$ is the usual SM Higgs boson. $\alpha=\pi /2$ is the specular case, in which $h_2$ plays the role of the SM Higgs boson.

Notice also that, again in contrast to the SM in which the gauge couplings have a marginal effect, in our case the RH neutrinos play for the extra scalar singlet the role of the top quark for the SM Higgs in the vacuum stability condition \footnote{Also notice that we have three RH neutrinos, as we have three colours for the top quark. However, they are Majorana particles rather than Dirac ones, so they carry half the (independent) degrees of freedom of the top quark.}. Their RGEs are then controlled by the Yukawa couplings with a negative contribution coming from $g'_1$ [see eq.~(\ref{RGE_nu_r_maj})]. Therefore, in some regions of the parameter space, the impact of the gauge sector is not marginal and can effectively stabilise the otherwise divergent evolution of the Majorana Yukawa couplings for the RH neutrinos.

A final remark is in order about eq.~(\ref{RGE_lamda3}), the evolution of $\lambda _3$,  the mixing parameter of the scalar potential [see eq.~(\ref{BL-potential})]. This RGE is almost proportional to $\lambda _3$ itself, so a vanishing boundary condition is almost stable \footnote{From the last line of eq.~(\ref{inversion}), setting $\lambda _3 =0$ corresponds to $\alpha =0$, but not vice versa.}. Non-proportional terms arise from the new gauge couplings ($\widetilde{g}$ and $g'_1$), i.e., deviations from the vanishing boundary conditions are of the order of the gauge couplings, hence quite small. They are particularly negligible in the pure $B-L$ model, as also $\widetilde{g}$ has a vanishing boundary condition, with a weak departure from it due to the mixing in the gauge coupling sector \cite{BL_master_thesis}. Nonetheless, other benchmark models in our general parameterisation could show different behaviours.

The results we are going to present are obtained by analysing the RGEs in appendix~\ref{App:RGE}.
For their numerical study, we put boundary conditions at the EW scale on the physical observables: $m_{h_1},\, m_{h_2},\, \alpha,\, v, M_{Z'}, g'_1, \widetilde{g}, m^{1,2,3}_{\nu_h}$, that we trade for $m,\, \mu$, $\lambda _1,\, \lambda _2,\, \lambda _3,\, x, y^M_{1,2,3}$ using, for the relevant parameters therein, eq.~(\ref{inversion}). Notice that $\alpha$$\backslash$$\widetilde{g}$ denotes the mixing angle$\backslash$coupling in the scalar$\backslash$gauge sector. Where stated in the text, we impose boundary conditions on some parameters of the Lagrangian rather than on the physical observables. This is done for consistency of those studies.

For the pure $B-L$ model, object of the numerical analysis in this Thesis, the definition $\widetilde{g}=0$ holds, and as a consequence, we also have that the $B-L$ breaking VEV $x$ can be easily related to the new $Z'$ boson mass by $\displaystyle x=\frac{M_{Z'}}{2g'_1}$. Here we fix $g'_1=0.1$. Regarding the heavy neutrinos, for simplicity we consider them degenerate and we fix their masses to $m^{1,2,3}_{\nu_h} \equiv m_{\nu_h} = 200$ GeV (whenever not specified otherwise), a value that can lead to some interesting phenomenology (see section~\ref{sect:nu_h}). The free parameters in this study are then $m_{h_1}$, $m_{h_2}$, $\alpha$ and $x$. The general philosophy is to fix in turn some of the free parameters and scan over the other ones, individuating the allowed regions fulfilling the conditions of eqs.~(\ref{cond_1}) and (\ref{cond_2}).

\paragraph{Triviality and vacuum stability bounds}
~

\vspace*{0.3cm}
\noindent Given the simplicity of the scalar sector in the SM, the triviality and vacuum stability conditions can be studied independently and they both constrain the Higgs boson mass, providing an upper bound and a lower bound, respectively. In more complicated models as the one considered here, it might be more convenient to study the overall effect of eqs.~(\ref{cond_1})--(\ref{cond_2}), since there are regions of the parameter space in which the constraints are evaded simultaneously. This is the strategy we decided to follow.

\begin{figure}[!h]
  \subfloat[]{ 
  \label{mh1_mh2_a0}
  \includegraphics[angle=0,width=0.48\textwidth ]{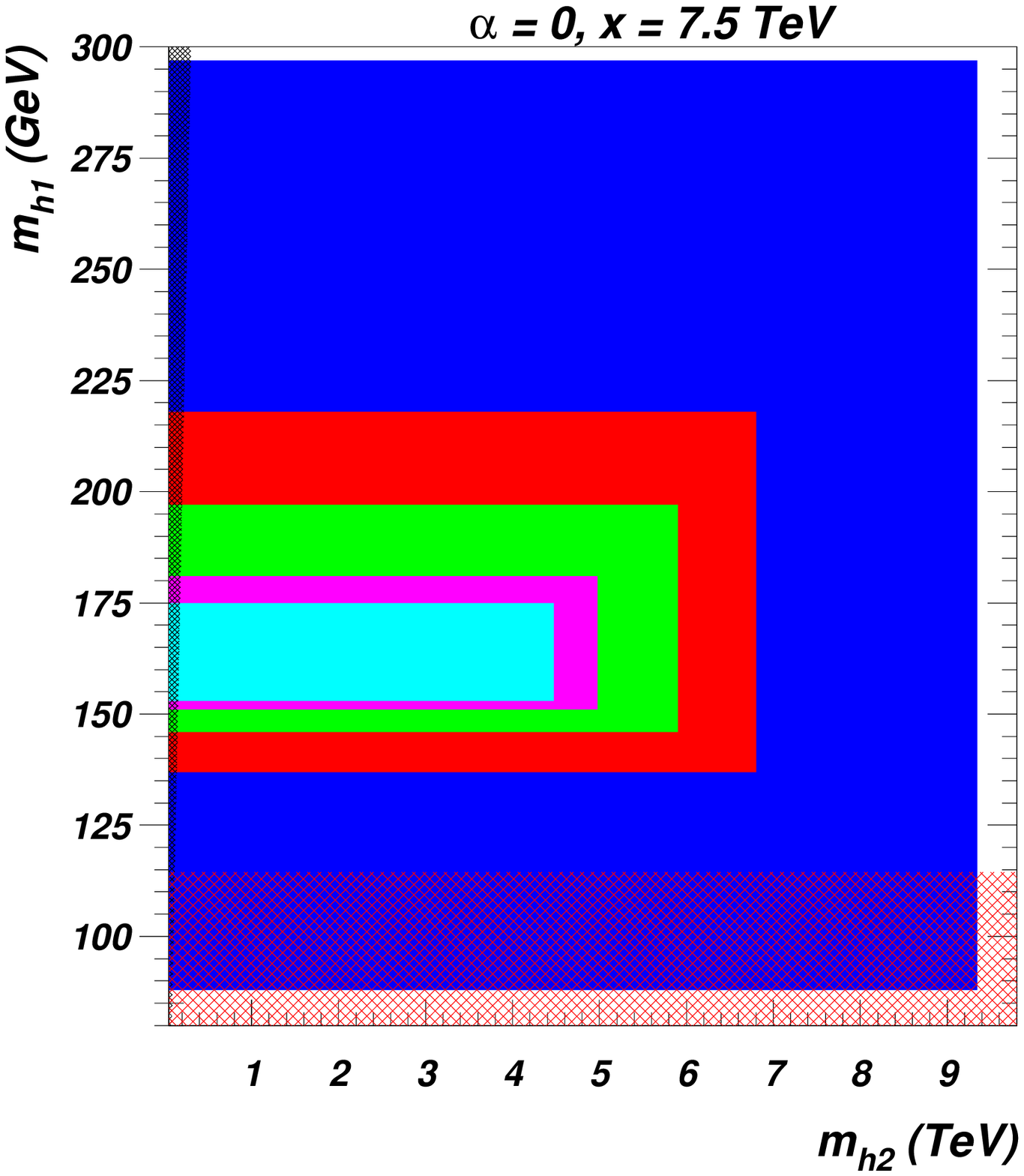}}
  \subfloat[]{
  \label{mh1_mh2_a01}
  \includegraphics[angle=0,width=0.48\textwidth ]{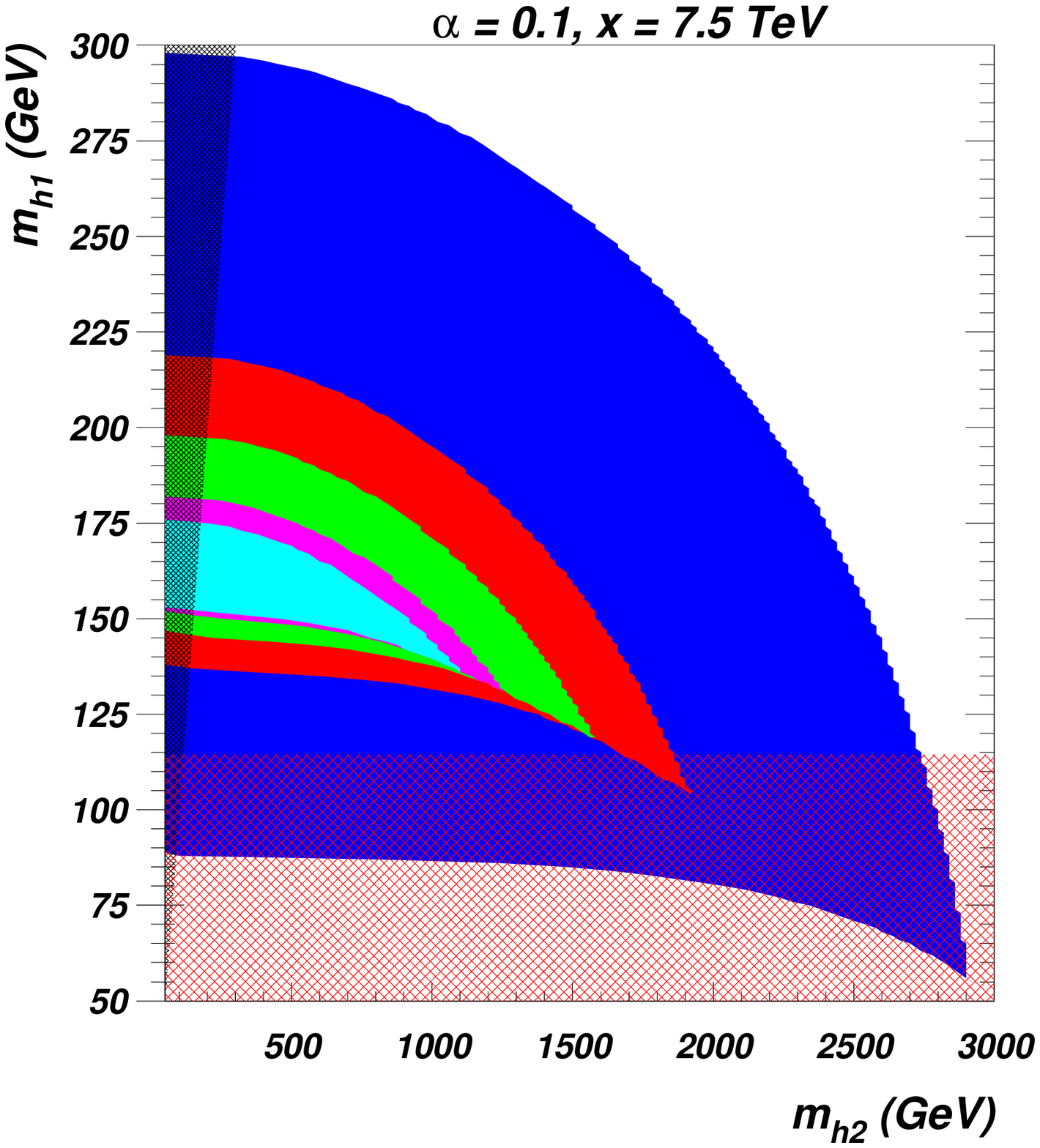}}
  \\
  \subfloat[]{
  \label{mh1_mh2_api4}
  \includegraphics[angle=0,width=0.48\textwidth ]{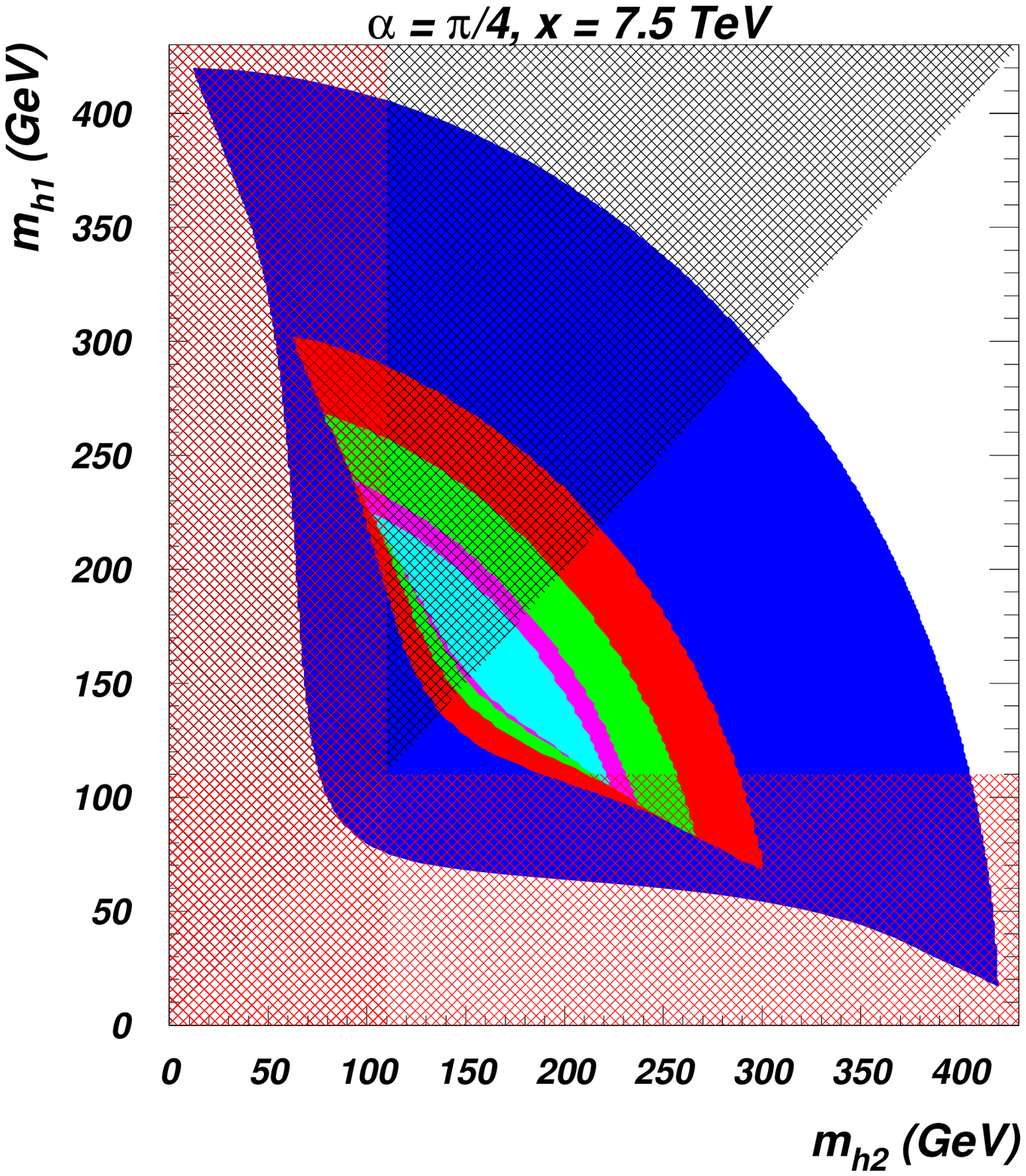}}
  \subfloat[]{
  \label{mh1_mh2_api3}
  \includegraphics[angle=0,width=0.48\textwidth ]{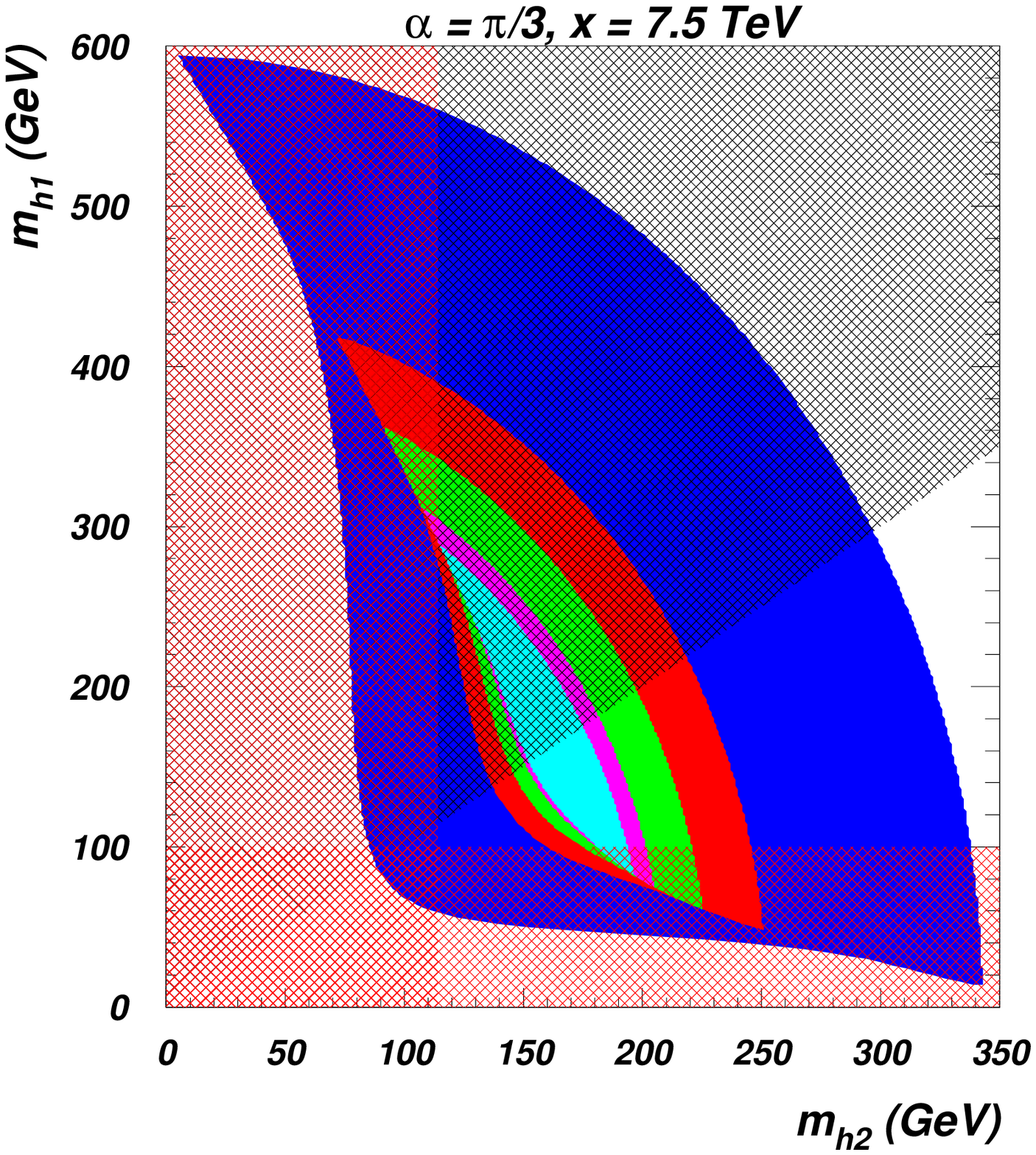}}
  \vspace*{-0.5cm}
  \caption{\it Allowed values in the $m_{h_1}$ vs. $m_{h_2}$ space in the $B-L$ model by eqs.~(\ref{cond_1}) and (\ref{cond_2}), for (\ref{mh1_mh2_a0}) $\alpha =0$, (\ref{mh1_mh2_a01}) $\alpha =0.1$, (\ref{mh1_mh2_api4}) $\alpha =\pi /4$ and (\ref{mh1_mh2_api3}) $\alpha =\pi /3$. Colours refer to different values of $Q/$GeV: blue ($10^{3}$), red ($10^{7}$), green ($10^{10}$), purple ($10^{15}$) and cyan ($10^{19}$). The shaded black region is forbidden by our convention $m_{h_2} > m_{h_1}$, while the shaded red region refers to the values of the scalar masses forbidden by LEP. Here: $x=7.5$ TeV,
$m_{\nu_h}=200$ GeV.  \label{mh1_mh2}}
\end{figure}

Figure~\ref{mh1_mh2} shows the allowed region in the ($m_{h_1}$--$m_{h_2}$) parameter space, for increasing values of the mixing angle $\alpha$, for fixed VEV $x=7.5$ TeV and heavy neutrino masses $m_{\nu_h}=200$ GeV, corresponding to Yukawa couplings whose effect on the RGE running can be considered negligible.
For $\alpha=0$, the allowed values for $m_{h_1}$ are the SM ones and the extended scalar sector is completely decoupled. The allowed space is therefore the simple direct product of the two, as we can see in figure~\ref{mh1_mh2_a0}. When there is no mixing, the bounds we get for the new heavy scalar are quite loose, allowing a several TeV range for $m_{h_2}$, depending on the scale of validity of the theory. We observe no significant lower bounds (i.e., $m_{h_2}>0.5$ GeV), as the impact of the RH Majorana neutrino Yukawa couplings is negligible.

As we increase the value for the angle, the allowed space deforms towards smaller values of $m_{h_1}$. If for very small scales $Q$ of validity of the theory such masses have already been excluded by LEP, for big enough values of $Q$, at a small angle as $\alpha=0.1$, the presence of a heavier boson allows the model to survive up to higher scales for smaller $h_1$ masses if compared to the SM (in which just $h_1$ would exist). Correspondingly, the constraints on $m_{h_2}$ become tighter. Moving to bigger values of the angle, the mixing between $h_1$ and $h_2$ grows up to its maximum, at $\alpha = \pi /4$, where $h_1$ and $h_2$ both contain an equal amount of doublet and singlet scalars. The situation is therefore perfectly symmetric, as one can see from figure~\ref{mh1_mh2_api4}. Finally, in figure~\ref{mh1_mh2_api3}, we see that the bounds on $m_{h_2}$ are getting tighter, approaching the SM ones, and those for $m_{h_1}$ are relaxing. That is, for values of the angle $\pi/4 < \alpha < \pi /2$, the situation is qualitatively not changed, but now $h_2$ is the SM-like Higgs boson. Visually, one can get the allowed regions at a given angle $\pi/2 - \alpha$ by simply taking the transposed about the $m_{h_1}=m_{h_2}$ line of the plot for the given angle $\alpha$.

\begin{figure}[!h]
  \subfloat[]{ 
  \label{mh2_vs_Q_a0}
  \includegraphics[angle=0,width=0.48\textwidth ]{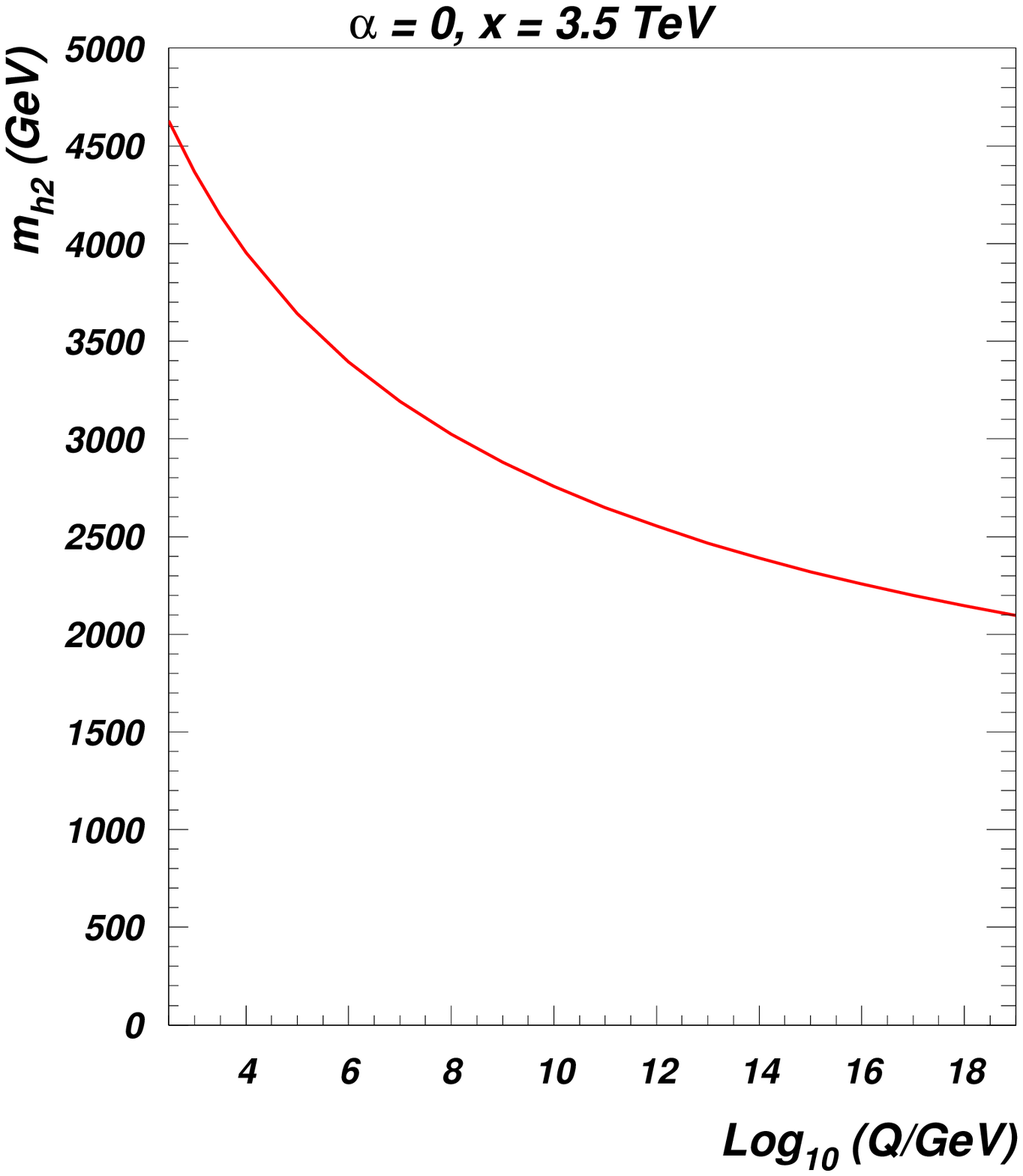}}
  \subfloat[]{
  \label{mh2_vs_Q_a01}
  \includegraphics[angle=0,width=0.48\textwidth ]{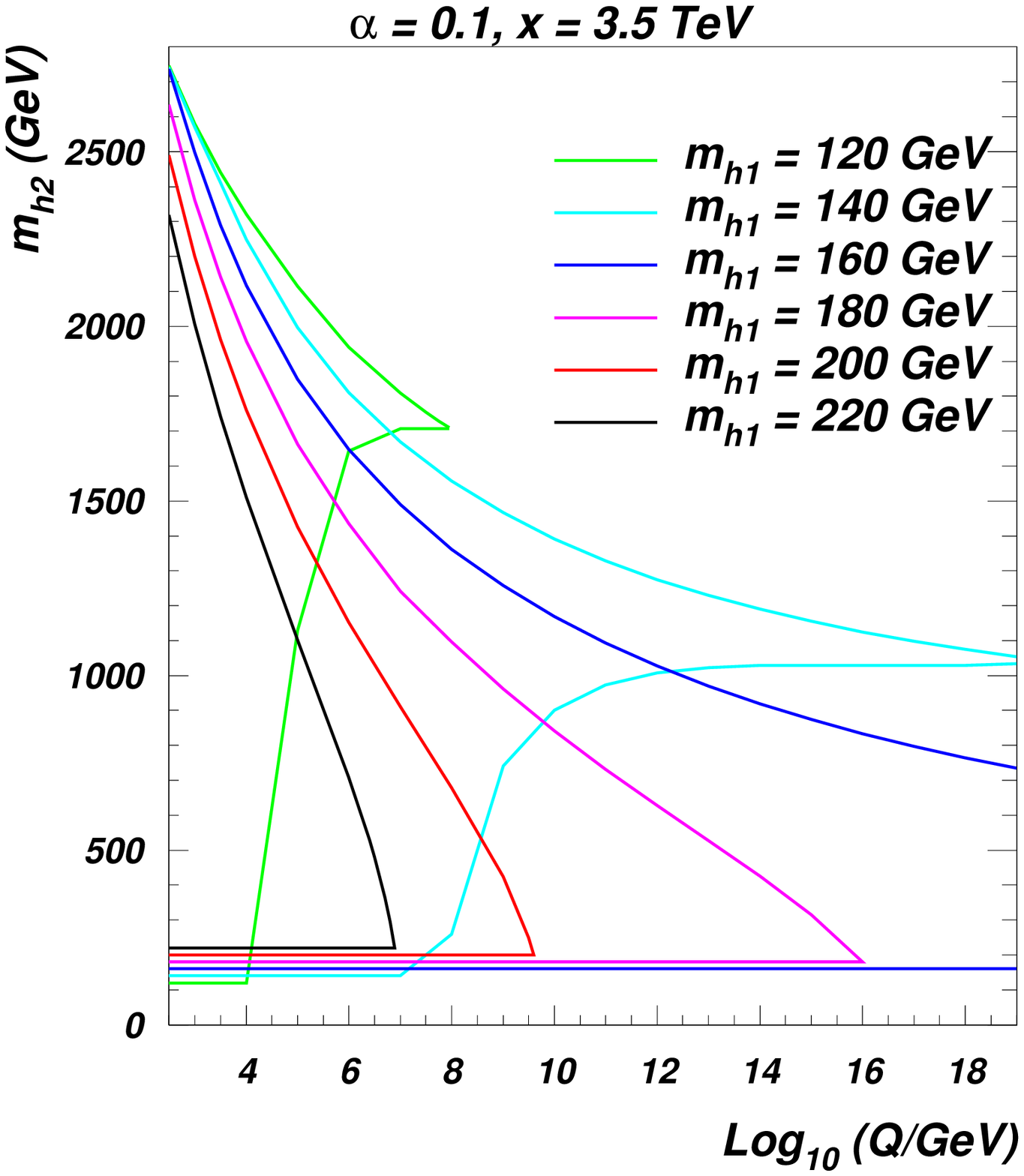}}
  \\
  \subfloat[]{ 
  \label{mh2_vs_Q_pi8}
  \includegraphics[angle=0,width=0.48\textwidth ]{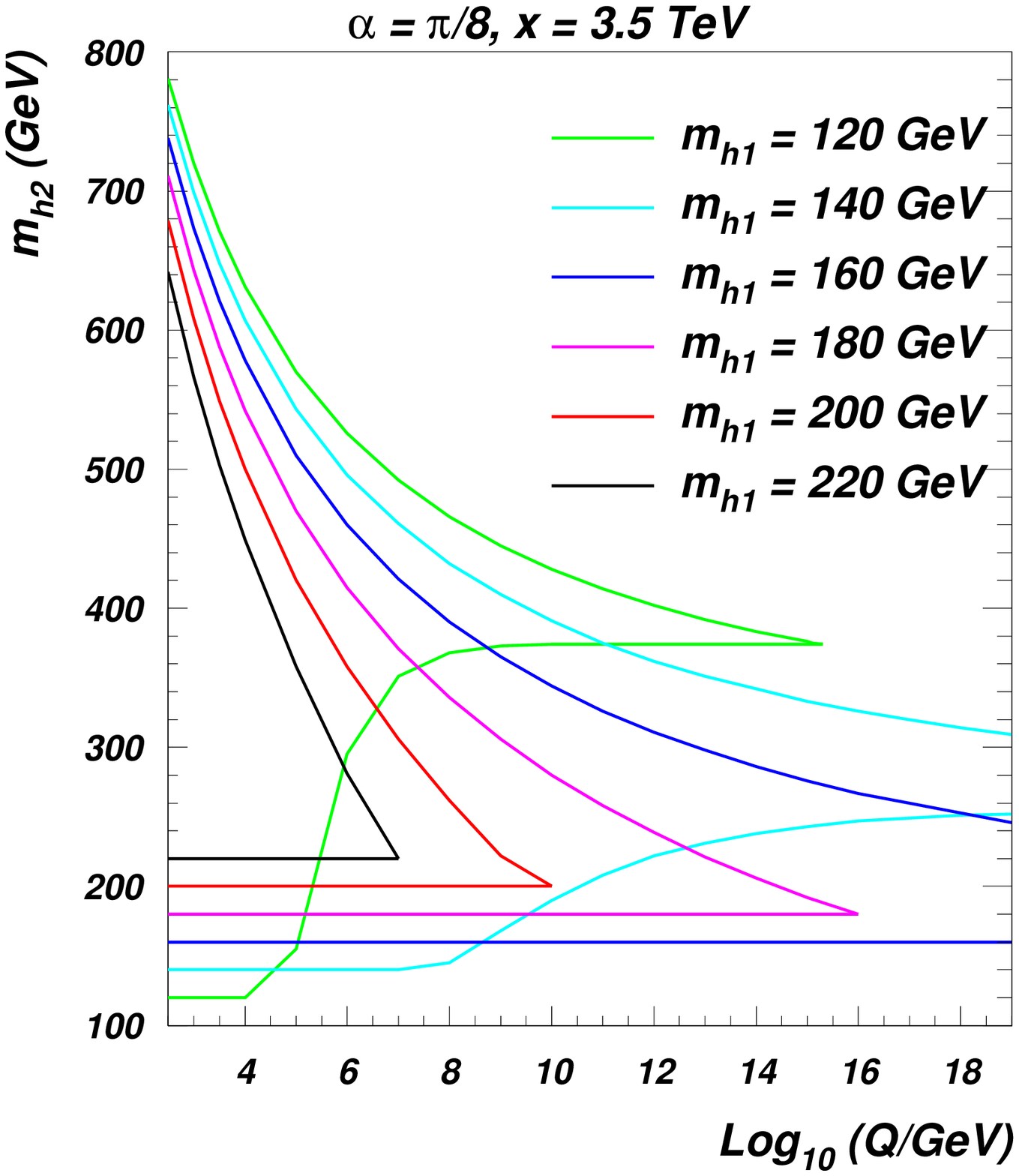}}
  \subfloat[]{
  \label{mh2_vs_Q_api4}
  \includegraphics[angle=0,width=0.48\textwidth ]{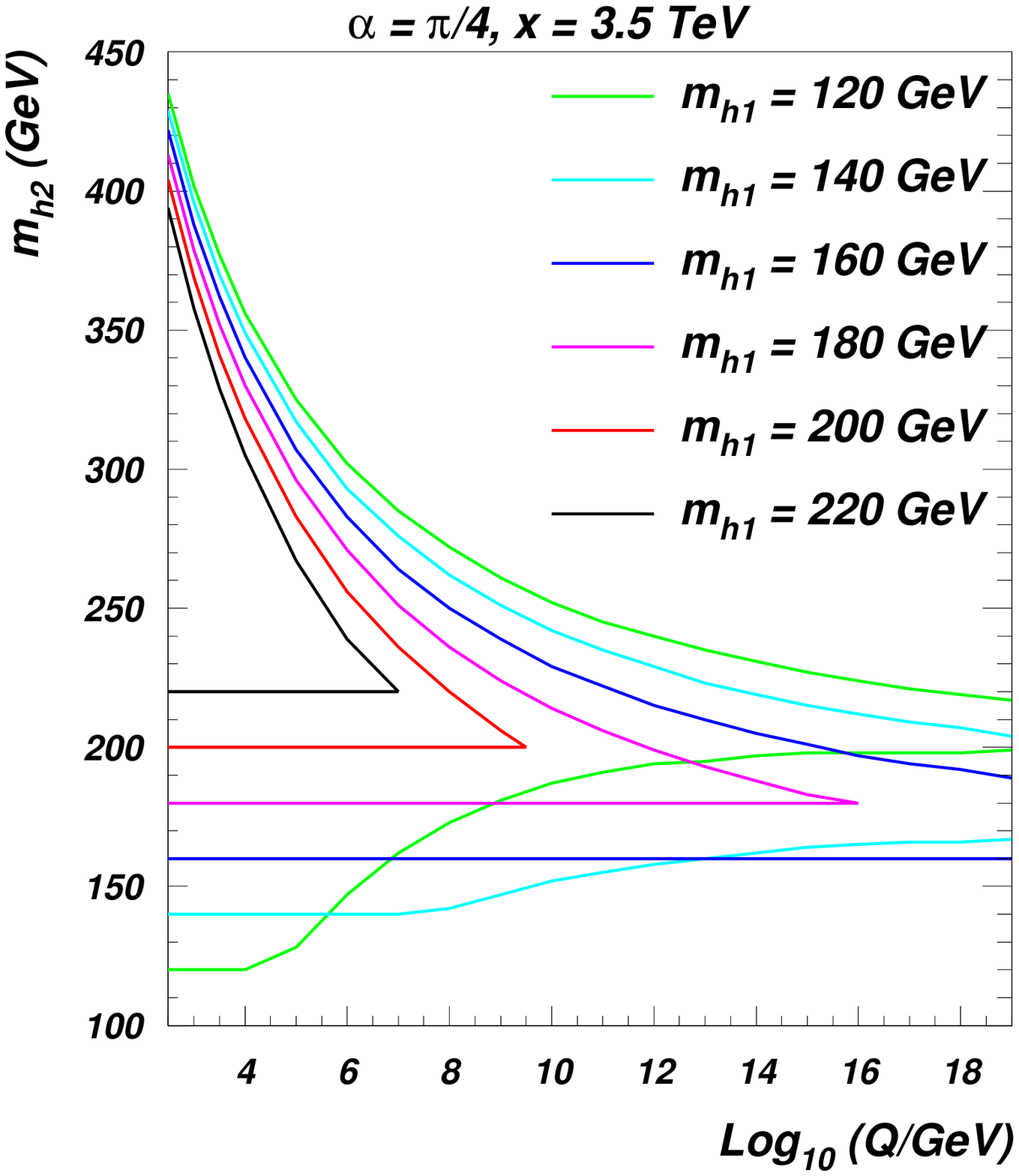}}
    \vspace*{-0.5cm}
  \caption{\it Allowed values (that are those between the same colour lines) for $m_{h_2}$ as a function of the scale $Q$ in the $B-L$ model by eqs.~(\ref{cond_1}) and (\ref{cond_2}), for several values of $m_{h_1}$ and (\ref{mh2_vs_Q_a0}) $\alpha =0$, (\ref{mh2_vs_Q_a01}) $\alpha =0.1$, (\ref{mh2_vs_Q_pi8}) $\alpha =\pi /8$ and (\ref{mh2_vs_Q_api4}) $\alpha = \pi /4$. Also, $x=3.5$ TeV and $m_{\nu_h}=200$ GeV. Only the allowed values by our convention $m_{h_2} > m_{h_1}$ are shown.  \label{mh2_Q}}
\end{figure}

Per each value of the angle, we can then fix the lighter Higgs mass $m_{h_1}$ to some benchmark values (allowed by LEP for the SM Higgs) and plot the allowed mass for the heavier Higgs as a function of the scale $Q$. This is done in figure~\ref{mh2_Q}, where the allowed masses are those contained between the same colour lines. Notice that here the VEV $x$ is fixed to a different value, $x=3.5$ TeV. The effects of changing the VEV $x$ will be described in section~\ref{sect:VEV_eff}.

As previously noticed, the allowed range in $m_{h_2}$ gets smaller as we increase the angle. Apart from the case $\alpha =0$ where there is no dependency at all from $m_{h_1}$, there is a strong effect of $m_{h_1}$ on the bounds on $m_{h_2}$. Not all the allowed regions (for $m_{h_2}$) at a fixed $h_1$ mass are inside the regions that are allowed for a smaller $m_{h_1}$. This is true only for $m_{h_1} > 160$ GeV. For smaller $m_{h_1}$'s, the distortion in the allowed region constraints tightly $m_{h_2}$ for the survival of the model to big scales $Q$. This is because such distortion is just towards smaller $h_1$ masses, see figure~\ref{mh1_mh2}.

\begin{figure}[!h]
  \subfloat[]{ 
  \label{mh2_a_mh1-100}
  \includegraphics[angle=0,width=0.48\textwidth ]{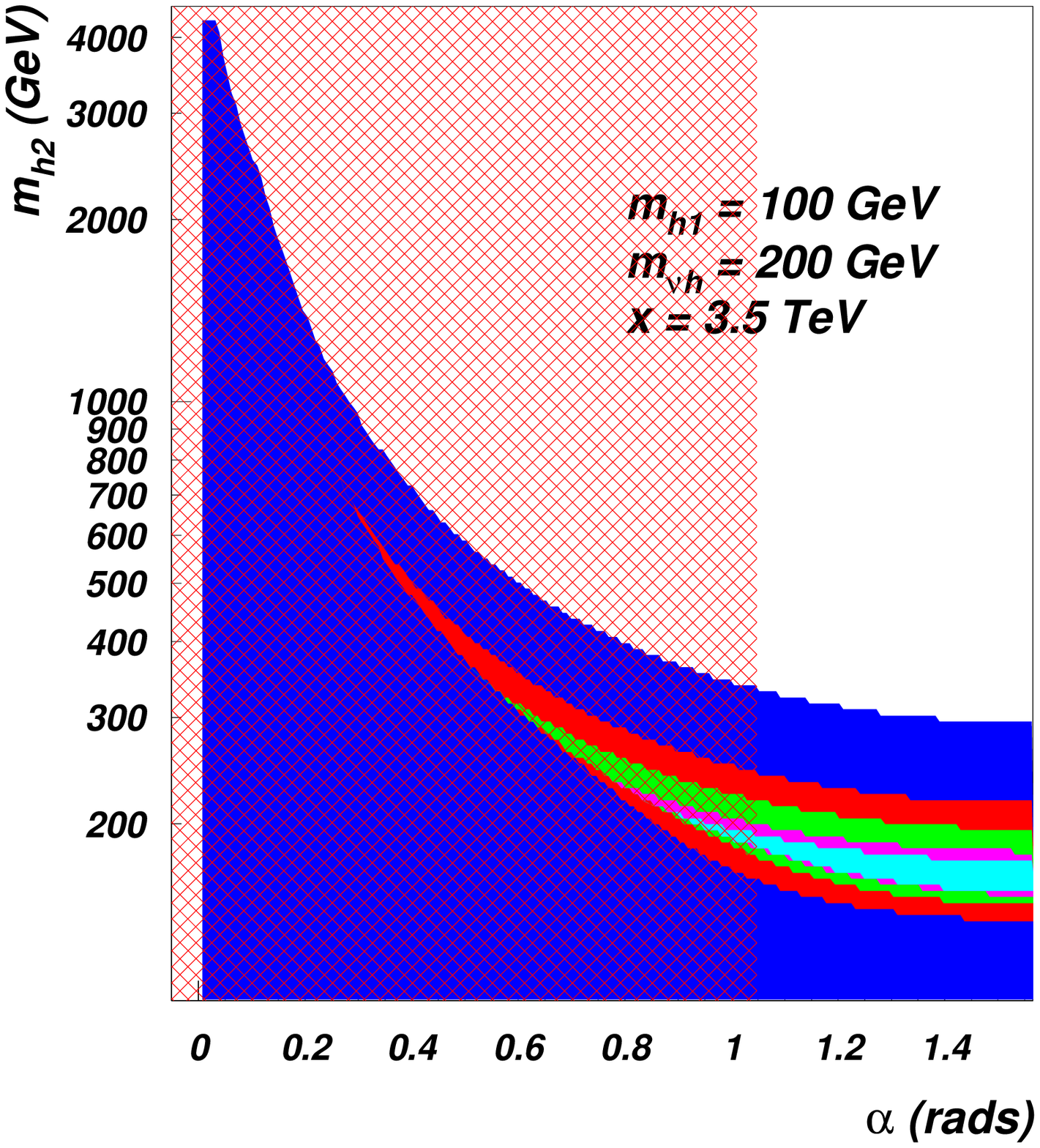}}
  \subfloat[]{
  \label{mh2_a_mh1-120}
  \includegraphics[angle=0,width=0.48\textwidth ]{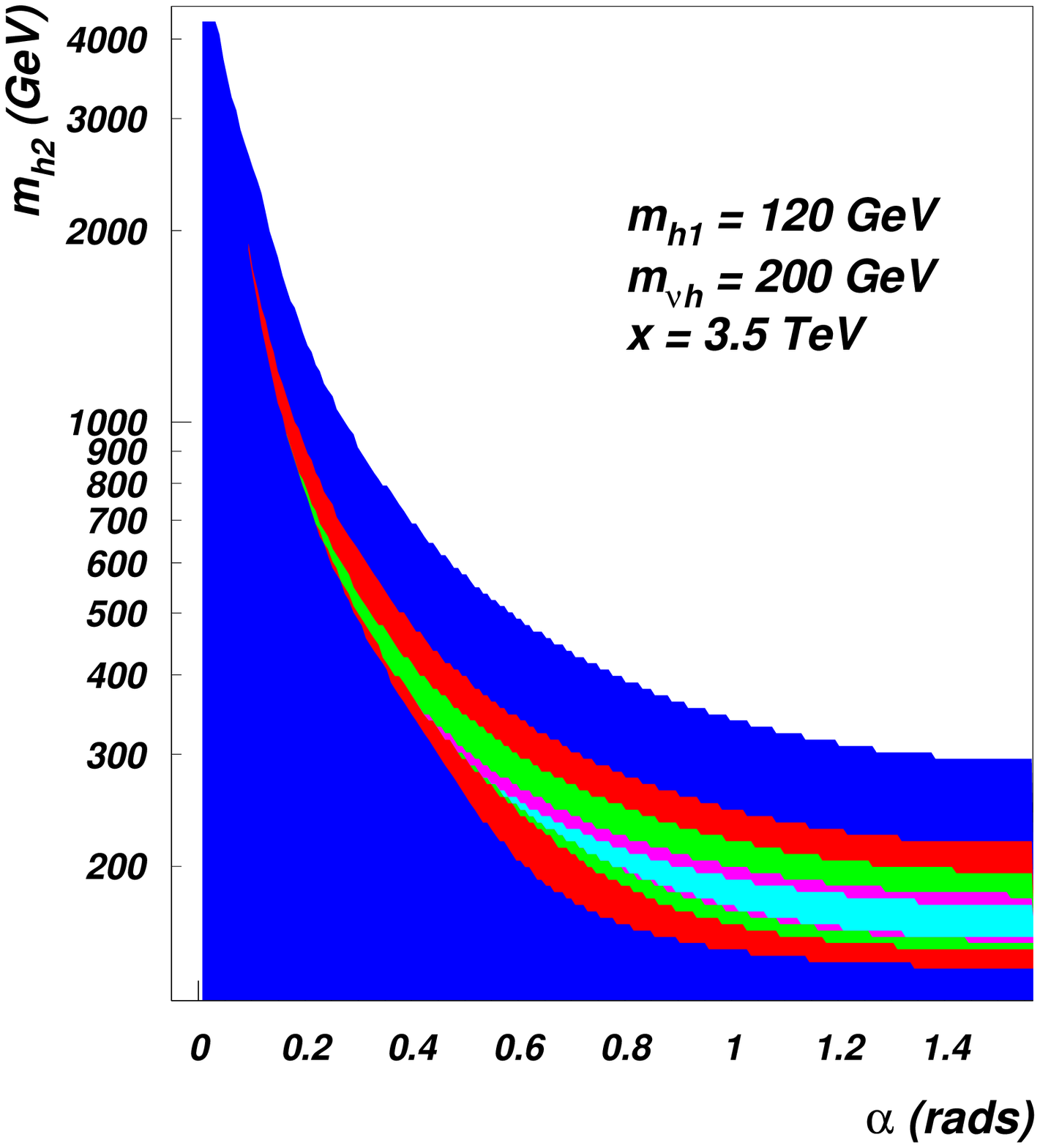}}
\\
  \subfloat[]{
  \label{mh2_a_mh1-160}
  \includegraphics[angle=0,width=0.48\textwidth ]{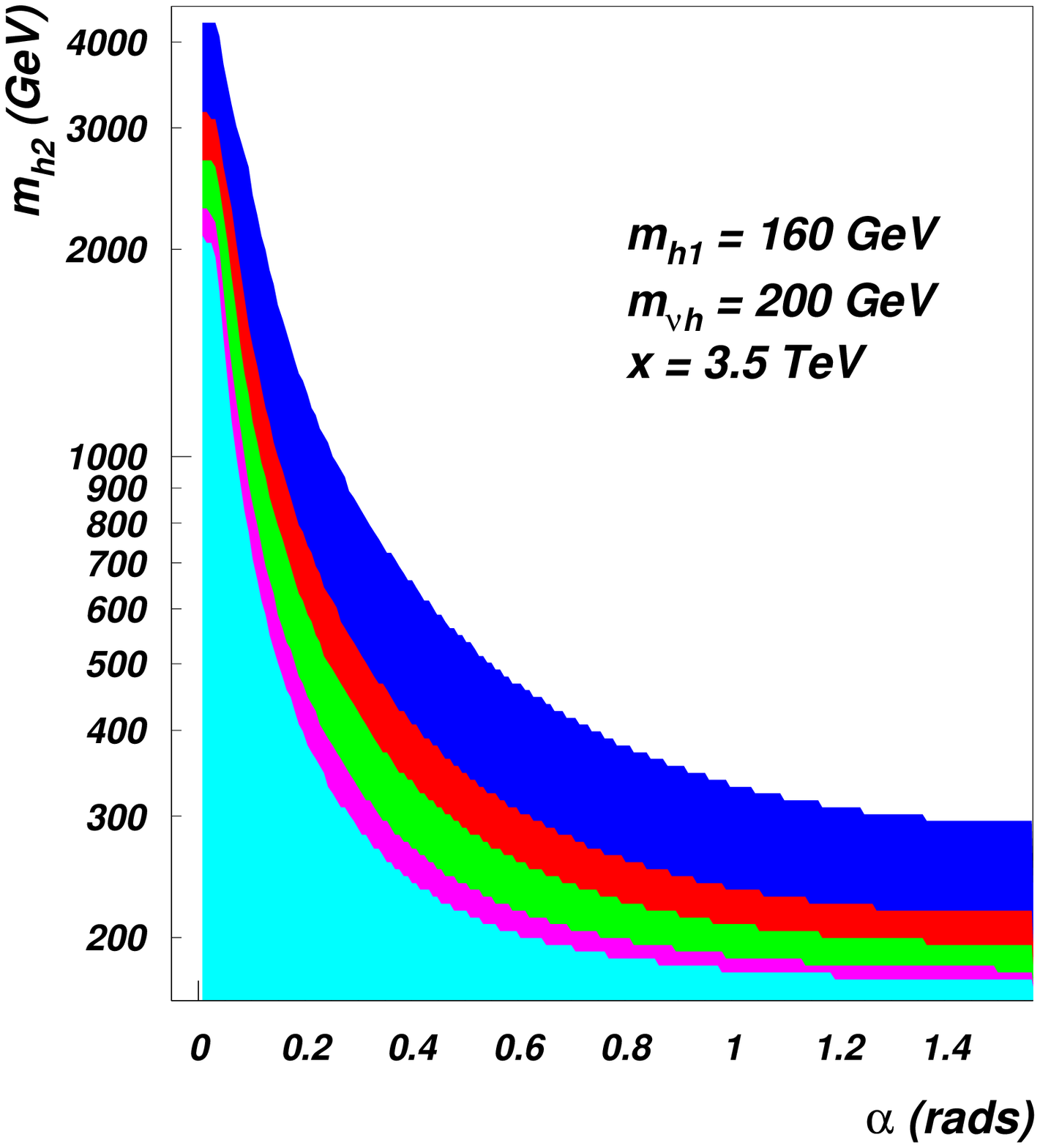}}
  \subfloat[]{
  \label{mh2_a_mh1-180}
  \includegraphics[angle=0,width=0.48\textwidth ]{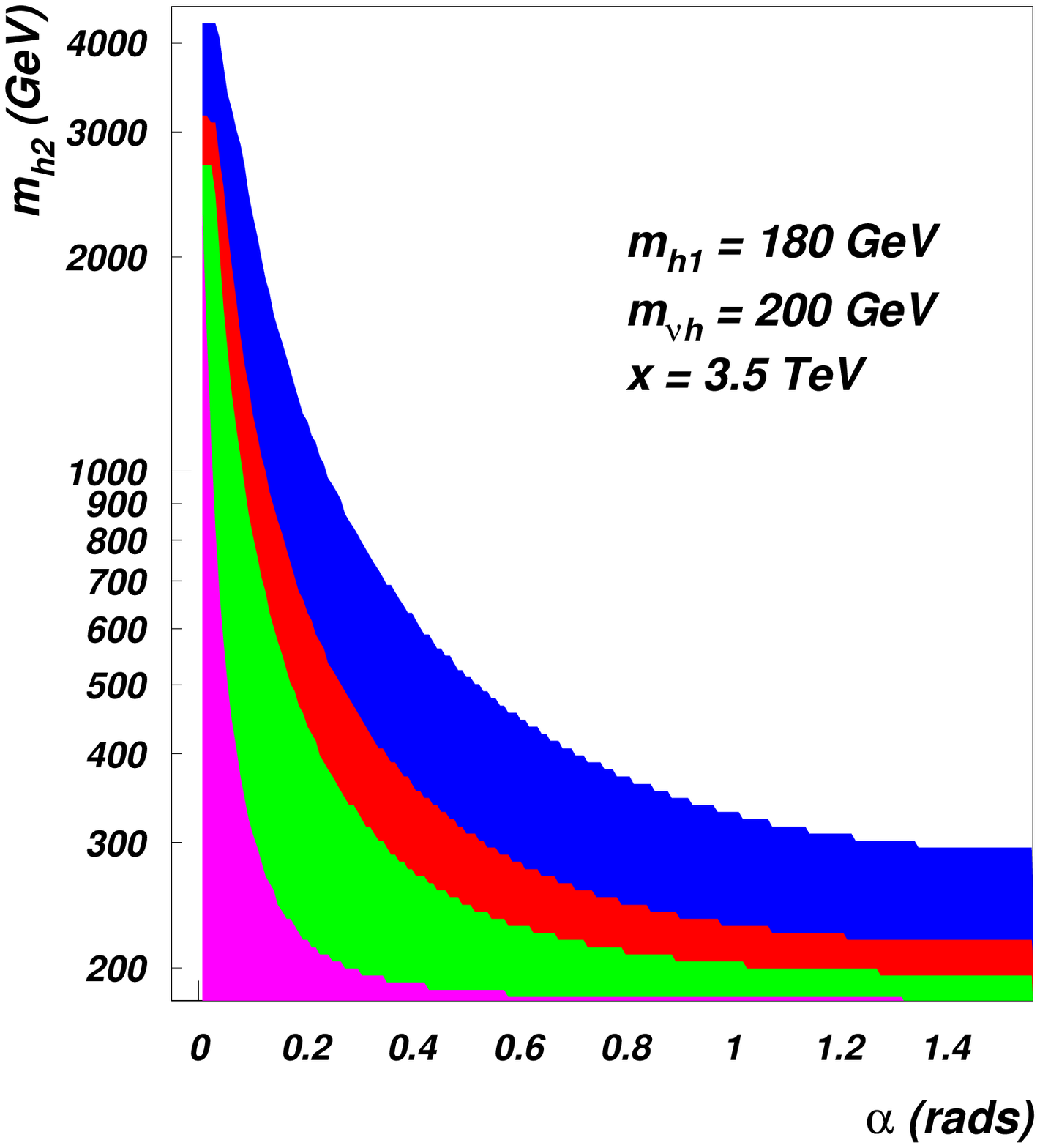}}
  \vspace*{-0.5cm}
  \caption{\it Allowed values in the $m_{h_2}$ vs. $\alpha$ space in the $B-L$ model by eqs.~(\ref{cond_1}) and (\ref{cond_2}), for (\ref{mh2_a_mh1-100}) $m_{h_1}=100$ GeV, (\ref{mh2_a_mh1-120}) $m_{h_1}=120$ GeV, (\ref{mh2_a_mh1-160}) $m_{h_1}=160$ GeV and (\ref{mh2_a_mh1-180}) $m_{h_1}=180$ GeV. Colours refer to different values of $Q/$GeV: blue ($10^{3}$), red ($10^{7}$), green ($10^{10}$), purple ($10^{15}$) and cyan ($10^{19}$). The plots already encode our convention $m_{h_2} > m_{h_1}$ and the shaded red region refers to the values of $\alpha$ forbidden by LEP. Here: $x=3.5$ TeV, $m_{\nu_h}=200$ GeV.  \label{mh2_alpha}}
\end{figure}

Complementary to the previous study, we can now fix the light Higgs mass at specific, experimentally interesting \footnote{The chosen values maximise the probability for the decays $h_1\rightarrow b\overline{b}$, $h_1\rightarrow \gamma \gamma$, $h_1\rightarrow W^+W^-$ and $h_1\rightarrow ZZ$, respectively (see figure~\ref{BR_h1}).}, values, i.e., $m_{h_1} = 100$, $120$, $160$ and $180$ GeV, and show the allowed region in the $m_{h_2}$ vs. $\alpha$ plane. This is done in figure~\ref{mh2_alpha}.

From this figures it is clear the transition of $h_2$ from the new extra scalar to the SM-like Higgs boson as we scan on the angle. As we increase $m_{h_1}$ (up to $m_{h_1} = 160$ GeV), a bigger region in $m_{h_2}$ is allowed for the model to be valid up to the Plank scale (i.e., $Q=10^{19}$ GeV, the most inner regions, in cyan). Nonetheless, such a region exists also for a value of the light Higgs mass excluded by LEP for the SM, $m_{h_1} = 100$ GeV, but only for big values of the mixing angle. No new regions (with respect to the SM) in which the model can survive up to the Plank scale open for $m_{h_1} > 160$ GeV, as the allowed space deforms towards smaller values of $m_{h_1}$.

\paragraph{Heavy neutrino mass influence}\label{sect:neutrino_eff}
~

\begin{figure}[!h]
  \subfloat[]{ 
  \label{mh1_mh2_a0_mhn-1000}
  \includegraphics[angle=0,width=0.48\textwidth ]{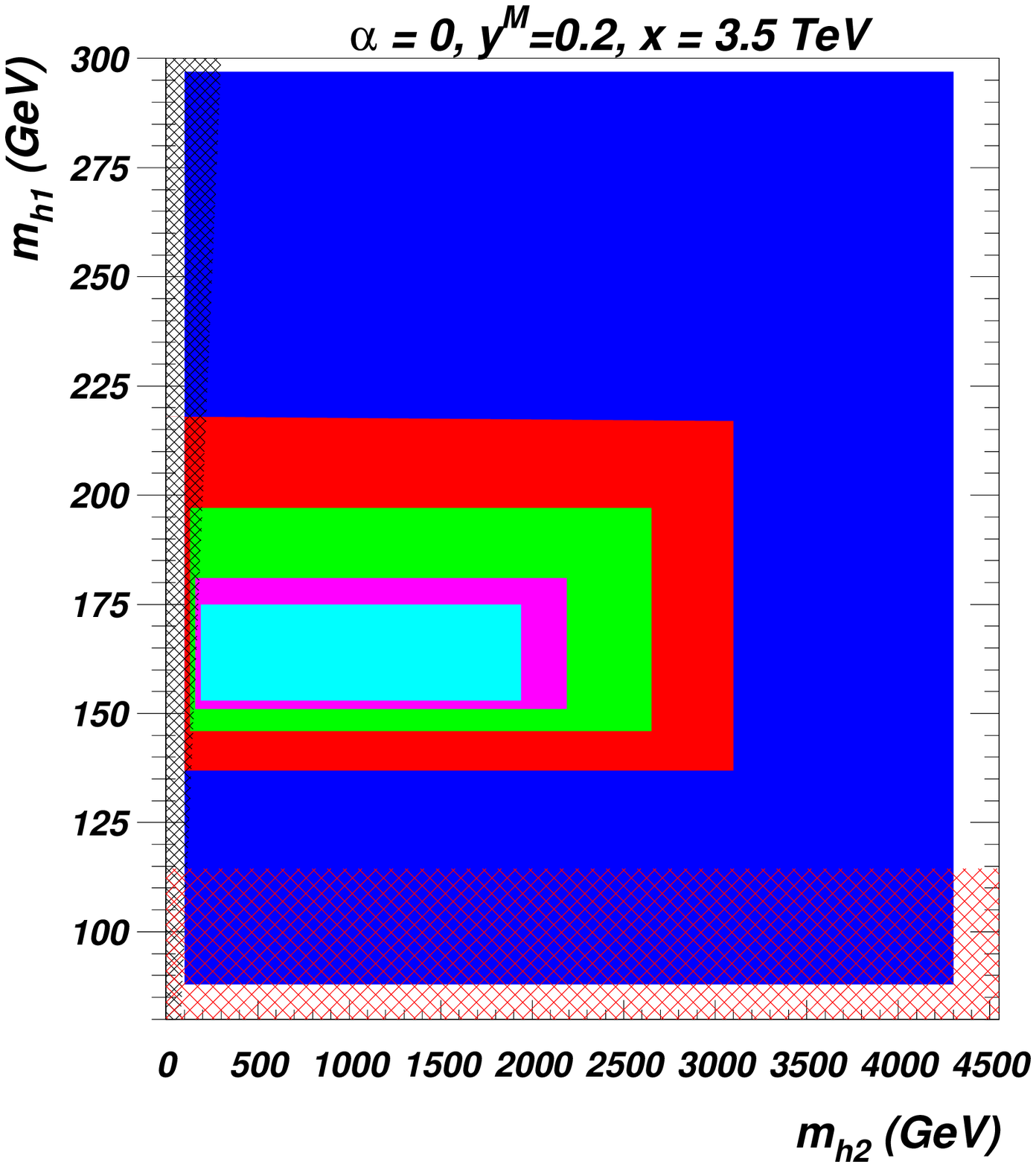}}
  \subfloat[]{
  \label{mh1_mh2_a01_mhn-1000}
  \includegraphics[angle=0,width=0.48\textwidth ]{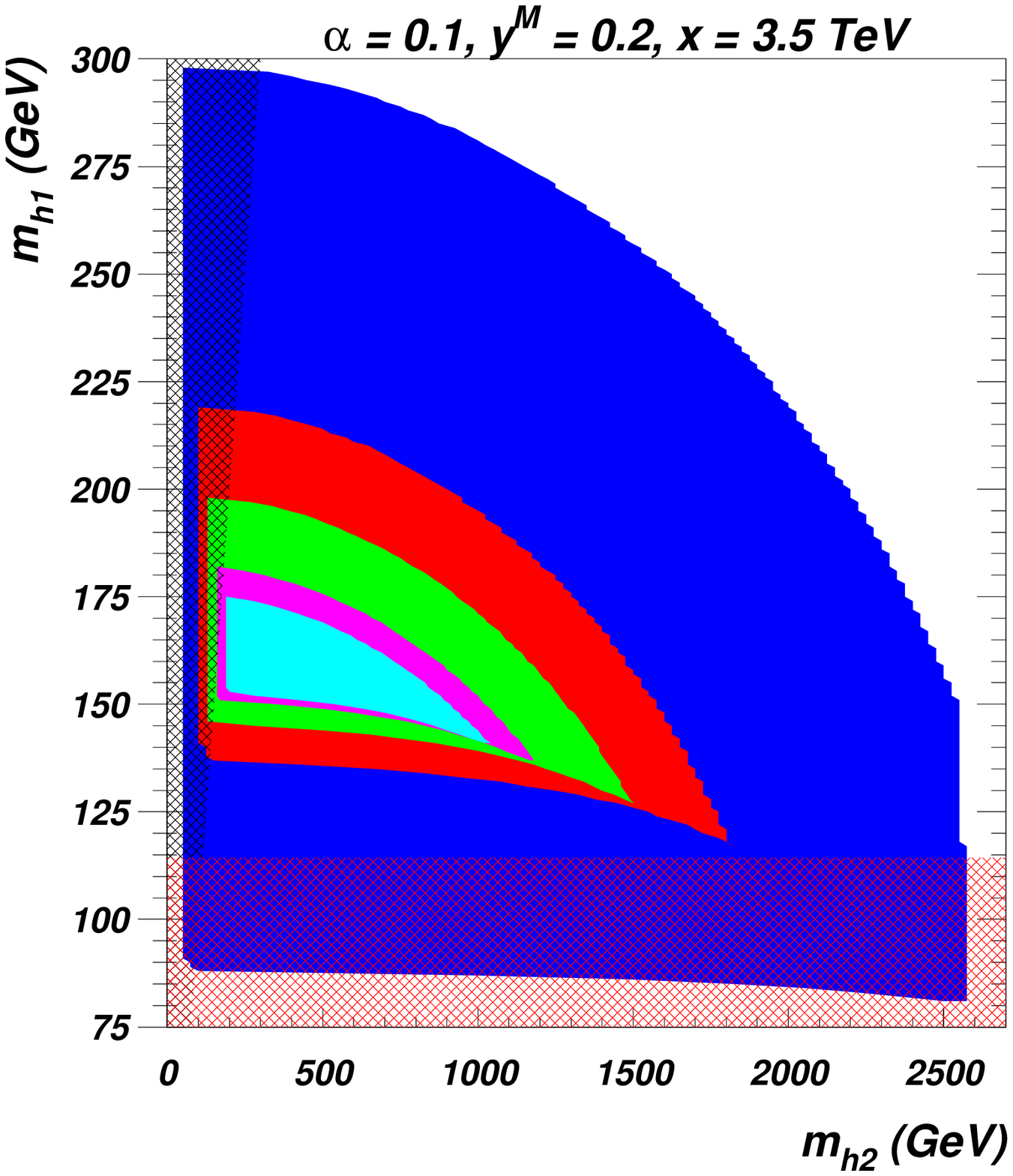}}
\\
  \subfloat[]{
  \label{mh1_mh2_api4_mhn-1000}
  \includegraphics[angle=0,width=0.48\textwidth ]{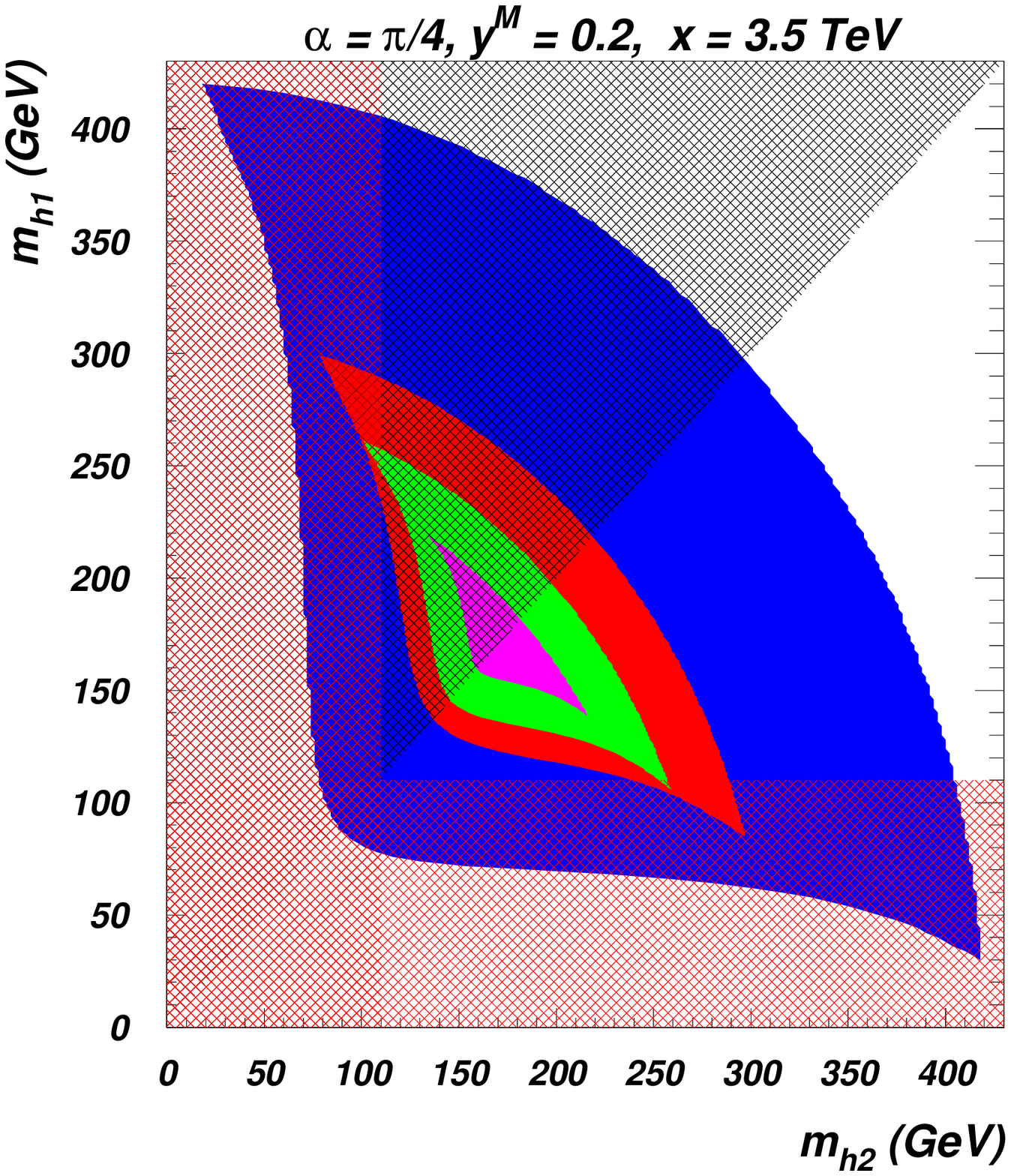}}
  \subfloat[]{
  \label{mh1_mh2_api3_mhn-1000}
  \includegraphics[angle=0,width=0.48\textwidth ]{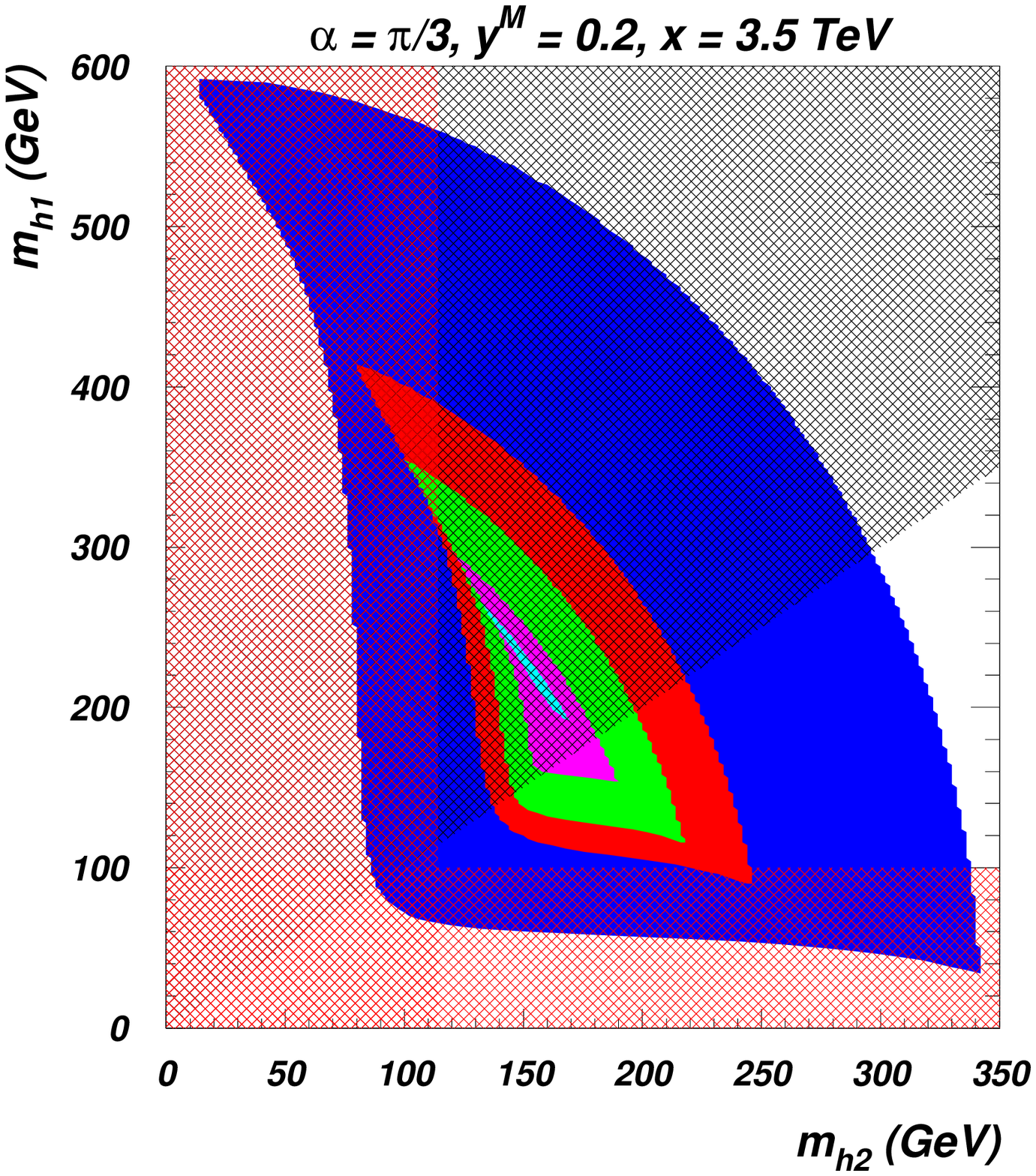}}
  \vspace*{-0.5cm}
  \caption{\it Allowed values in the $m_{h_1}$ vs. $m_{h_2}$ space  by eqs.~(\ref{cond_1}) and (\ref{cond_2}), for (\ref{mh1_mh2_a0_mhn-1000}) $\alpha =0$ and (\ref{mh1_mh2_a01_mhn-1000}) $\alpha =0.1$,  (\ref{mh1_mh2_api4_mhn-1000}) $\alpha = \pi /4$ and (\ref{mh1_mh2_api3_mhn-1000}) $\alpha = \pi /3$, for $m_{\nu_h}=1$ TeV and $x=3.5$ TeV. Colours refer to different values of $Q/$GeV: blue ($10^{3}$), red ($10^{7}$), green ($10^{10}$), purple ($10^{15}$) and cyan ($10^{19}$). The shaded black region is forbidden by our convention $m_{h_2} > m_{h_1}$, while the shaded red region refers to the values of the scalar masses forbidden by LEP.  \label{mh1_mh2_mhn}}
\end{figure}

\vspace*{0.3cm}
\noindent As already intimated, the RH neutrinos play for the extra scalar singlet the role of the top quark for the SM Higgs. This is particularly true for the vacuum stability condition, as the fermions in general provide the negative term that can drive the scalar couplings towards negative values. Figure~\ref{mh1_mh2_mhn} shows how the allowed regions in the ($m_{h_1}$--$m_{h_2}$) plane change for a RH Majorana neutrino Yukawa coupling $y^M=0.2$ (that for $x=3.5$ TeV correspond to $m_{\nu_h} = 1$ TeV), not negligible anymore. For $y^M=0.4$, the changes are even more drastic, shrinking the allowed region even further.

The effect of having non negligible $y^M$ couplings is evident if we compare figure~\ref{mh1_mh2_mhn} to figure~\ref{mh1_mh2}. Notice that also the VEV $x$ is changed (from $7.5$ TeV to $3.5$ TeV), but this is only responsible for the smaller upper bounds of $m_{h_2}$ in figures~\ref{mh1_mh2_a0_mhn-1000} and \ref{mh1_mh2_a01_mhn-1000}. For small values of $\alpha$ it is evident our analogy between the top quark and the RH neutrinos, as now $m_{h_2}$ has a sensible lower bound too. The analogy holds also for bigger values of the angle, as the allowed region of masses is shrunk from below as we increase the RH Majorana neutrino Yukawa coupling, while the upper bound stays unaffected. The effect is even more evident for big values of the scale $Q$, with the Plank scale (i.e., $Q=10^{19}$ GeV) precluded now for whatever Higgs boson masses at $\alpha = \pi /4$ and tightly constraining the allowed ones at $\alpha = \pi /3$.

\begin{figure}[!h]
  \subfloat[]{ 
  \label{mh2_a_mh1-100_mhn1000}
  \includegraphics[angle=0,width=0.48\textwidth ]{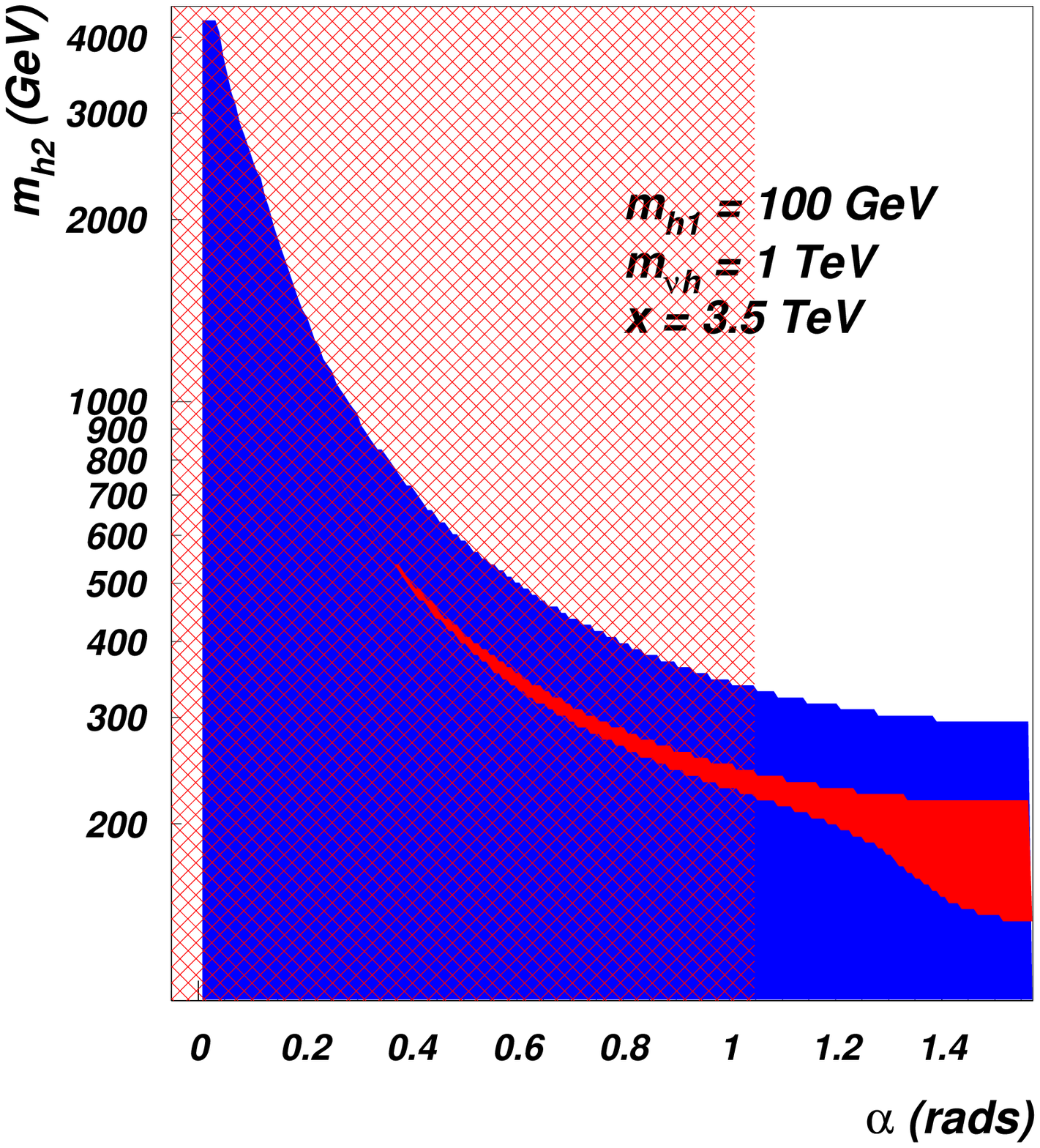}}
  \subfloat[]{
  \label{mh2_a_mh1-120_mhn1000}
  \includegraphics[angle=0,width=0.48\textwidth ]{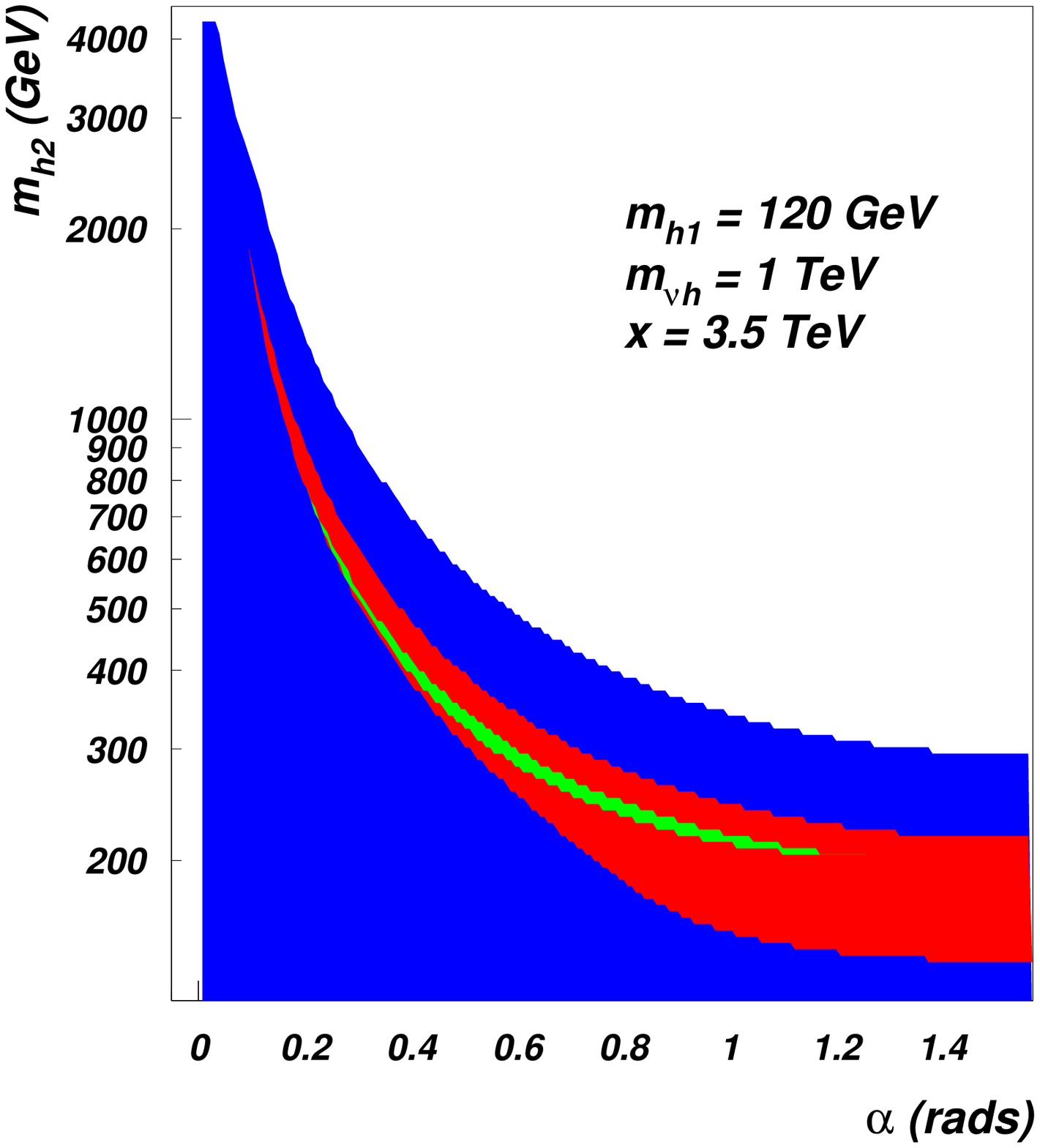}}
\\
  \subfloat[]{
  \label{mh2_a_mh1-160_mhn1000}
  \includegraphics[angle=0,width=0.48\textwidth ]{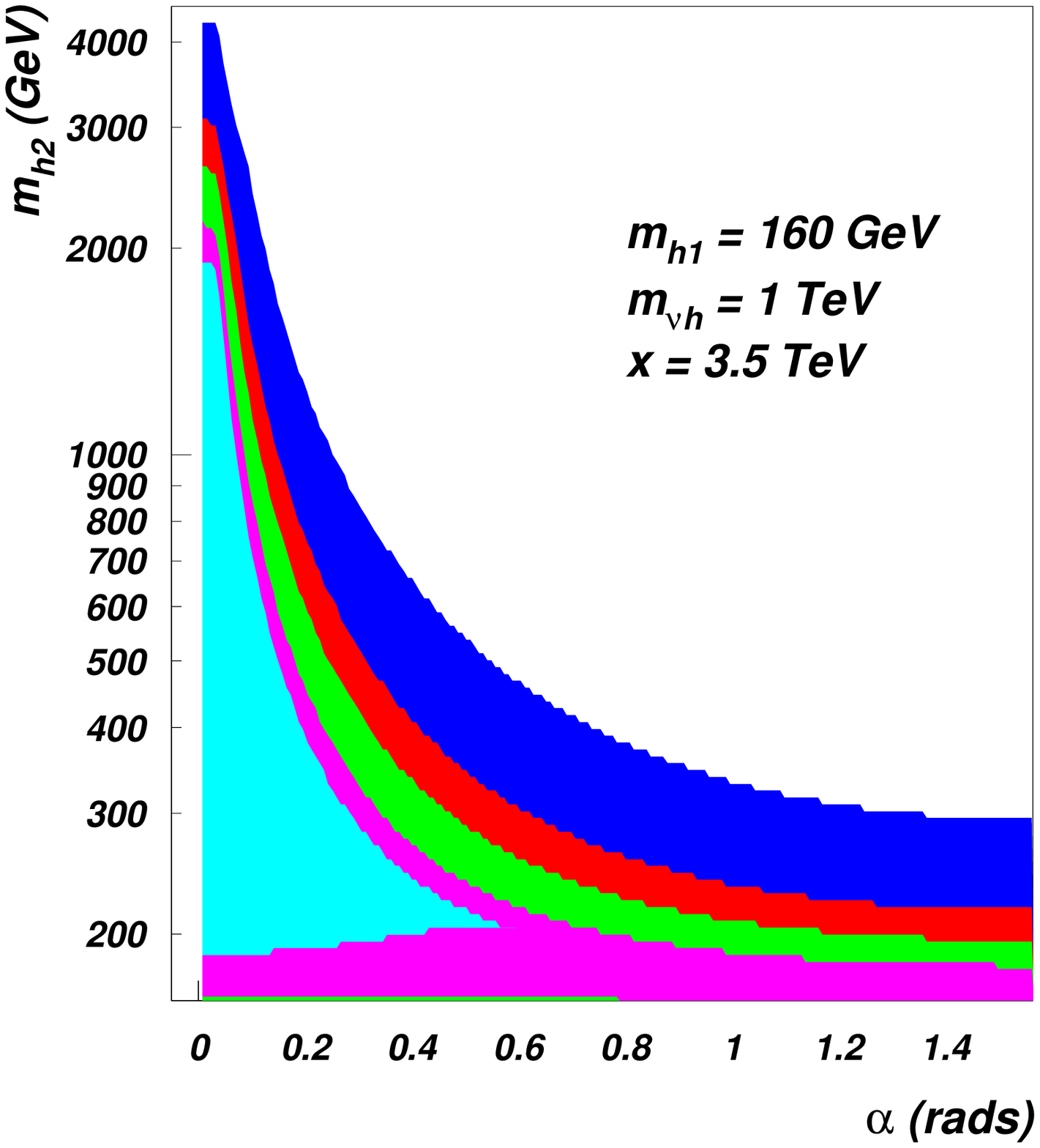}}
  \subfloat[]{
  \label{mh2_a_mh1-180_mhn1000}
  \includegraphics[angle=0,width=0.48\textwidth ]{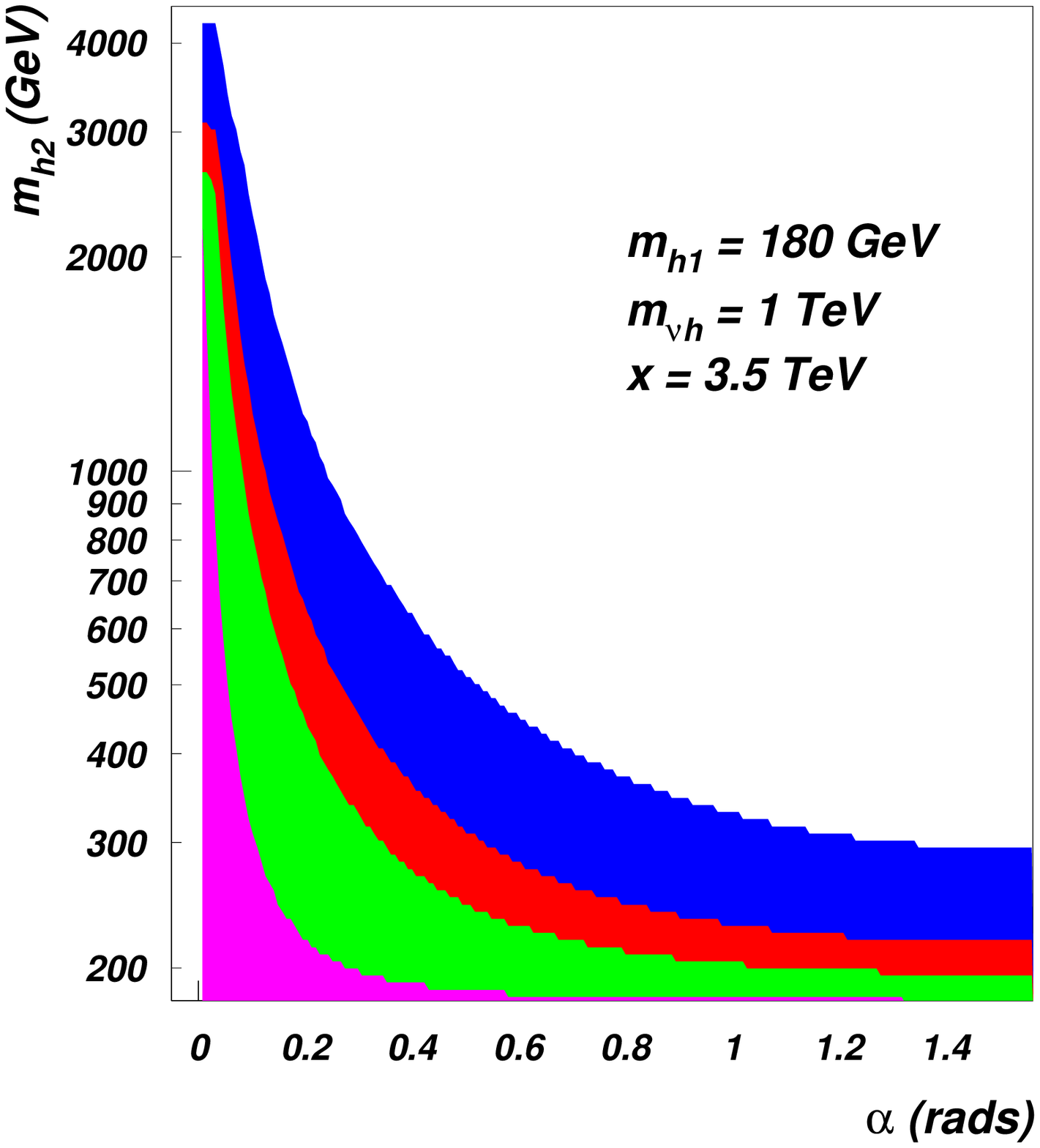}}
  \vspace*{-0.5cm}
  \caption{\it Allowed values in the $m_{h_2}$ vs. $\alpha$ space in the $B-L$ model by eqs.~(\ref{cond_1}) and (\ref{cond_2}), for (\ref{mh2_a_mh1-100_mhn1000}) $m_{h_1}=100$ GeV, (\ref{mh2_a_mh1-120_mhn1000}) $m_{h_1}=120$ GeV, (\ref{mh2_a_mh1-160_mhn1000}) $m_{h_1}=160$ GeV and (\ref{mh2_a_mh1-180_mhn1000}) $m_{h_1}=180$ GeV. Colours refer to different values of $Q/$GeV: blue ($10^{3}$), red ($10^{7}$), green ($10^{10}$), purple ($10^{15}$) and cyan ($10^{19}$). The plots already encode our convention $m_{h_2} > m_{h_1}$ and the shaded red region refers to the values of $\alpha$ forbidden by LEP. Here: $x=3.5$ TeV, $m_{\nu_h}=1$ TeV.  \label{mh2_alpha_mhn}}
\end{figure}

Moving to the ($m_{h_2}$--$\alpha$) scan at fixed $m_{h_1}$ values, figure~\ref{mh2_alpha_mhn} shows the effect of the heavy neutrinos in this case, to be compared to figure~\ref{mh2_alpha}. It is evident that this model can survive until very large scales $Q$ with massive heavy neutrinos (for which, $y^M > 0.2$) only for the light Higgs boson masses allowed in the case of the SM, that is, $m_{h_1} \sim 160$ GeV. The mixing angle must also be small, $\alpha < \pi /5$, providing a tight constraint on $m_{h_2}$. For smaller $h_1$ masses, the effect of a large $y^M$ is to preclude scales $\displaystyle Q \gtrsim 10^7$ GeV almost completely, with for example just a tiny strip for $m_{h_1}=120$ GeV for which there exists a combination of $m_{h_2}$ and $\alpha$ such that the model is consistent up to $\displaystyle Q=10^{10}$ GeV. Finally, figure~\ref{mh2_a_mh1-180_mhn1000} is not visibly different from figure~\ref{mh2_a_mh1-180} just because we are showing only the $m_{h_2} > m_{h_1}$ region, the shrunk region being below.

\paragraph{VEV effect}\label{sect:VEV_eff}
~

\begin{figure}[!h]
  \subfloat[]{
  \label{VEV_effect}
  \includegraphics[angle=0,width=0.48\textwidth ]{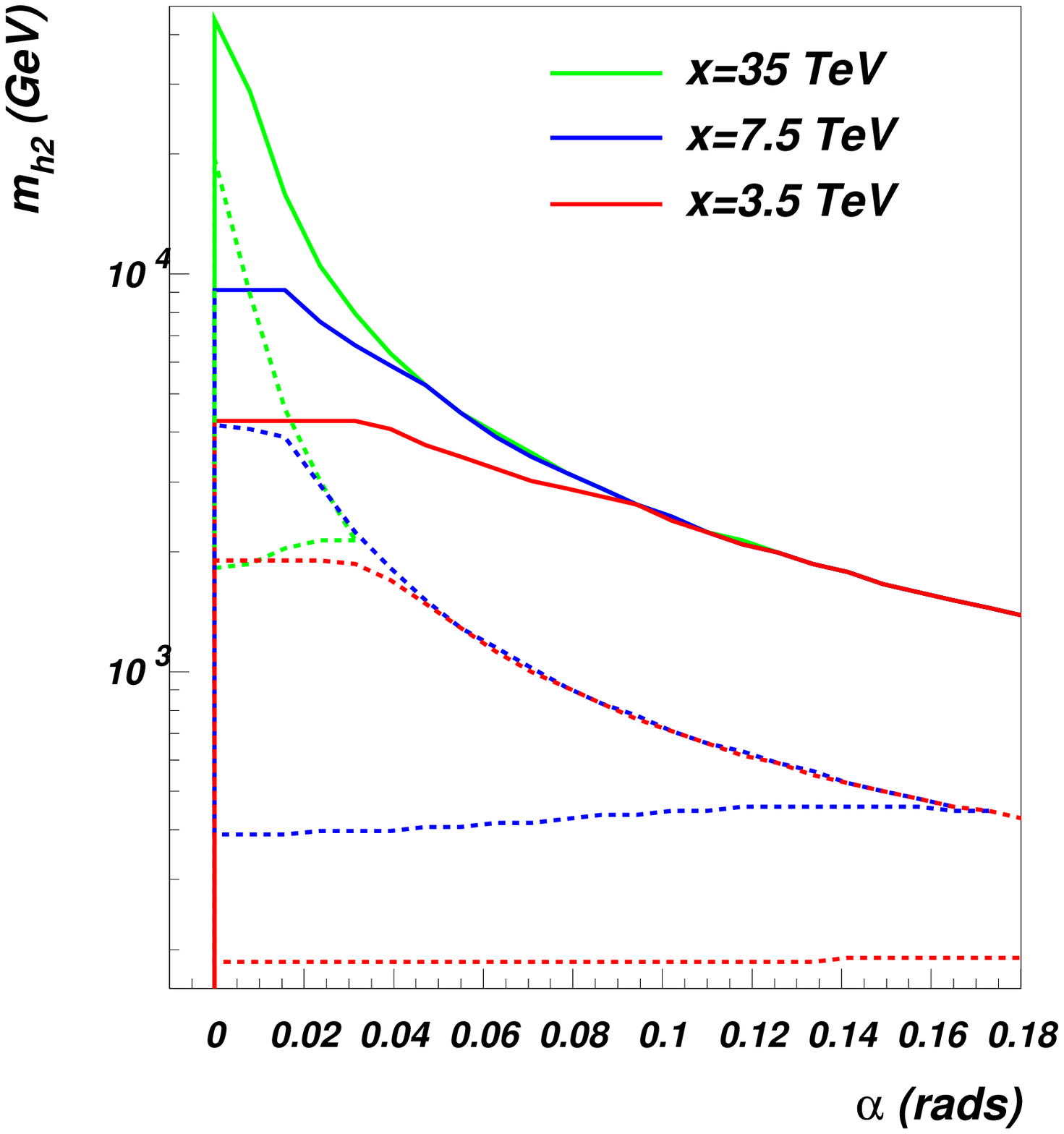}}
  \subfloat[]{
  \label{mh2_mh1_mhn1000_x75}
  \includegraphics[angle=0,width=0.48\textwidth ]{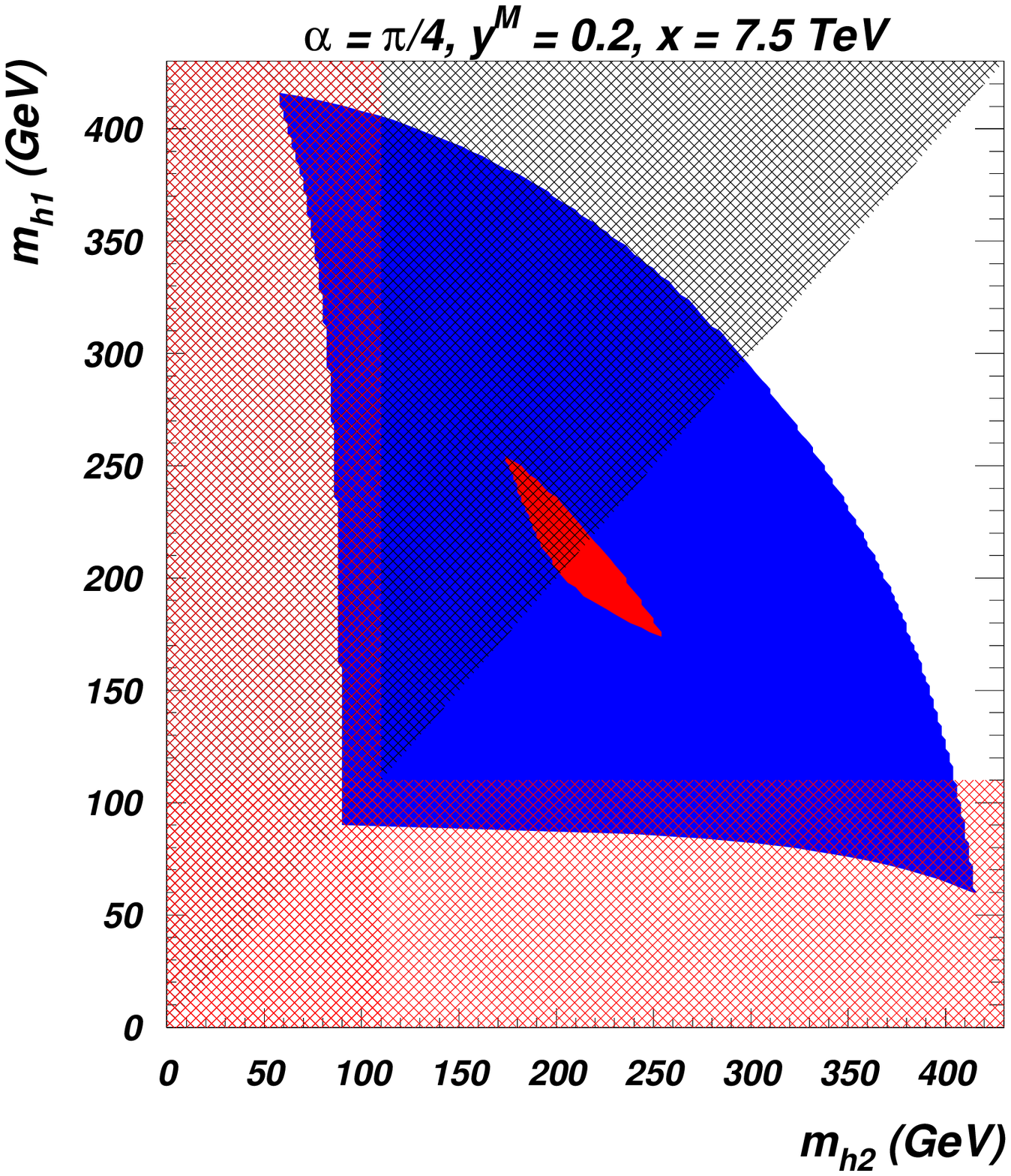}}
  \vspace*{-0.5cm}
  \caption{\it  Allowed values by eqs.~(\ref{cond_1}) and (\ref{cond_2}) (\ref{VEV_effect}) in the $m_{h_2}$ vs. $\alpha$ space for $m_{h_1}=160$ GeV and $y^M=0.2$, for $Q=10^3$ GeV (straight line) and $Q=10^{19}$ GeV (dashed line) for several $B-L$ breaking VEV values ($x=3.5$, $7.5$ and $35$ TeV, giving $m_{\nu _h}=1$, $2$ and $10$ TeV, respectively), and (\ref{mh2_mh1_mhn1000_x75}) in the $m_{h_1}$ vs. $m_{h_2}$ space, for $\alpha = \pi /4$, $x=7.5$ TeV and $y^M=0.2$, where colours refer to different values of $Q/$GeV: blue ($10^{3}$), red ($10^{7}$). The plots already encode our convention $m_{h_2} > m_{h_1}$ and the shaded red region refers to the values of $\alpha$ forbidden by LEP.  \label{mh2_mh1_mhn_api4}}
\end{figure}

\vspace*{0.3cm}
\noindent The last effect to evaluate comes from changing the values for the $B-L$ breaking VEV $x$.
Figure~\ref{VEV_effect} shows the allowed regions in the $m_{h_2}$ vs. $\alpha$ plane for fixed $m_{h_1}=160$ GeV and $y^M = 0.2$ (that is, a particular case that shows all the interesting features at once). As expected, since $\lambda _2$ is a function of $m_{h_2}/x$ [see for instance eq.~(\ref{inversion_lam2})], at $\alpha=0$ the bound on $m_{h_2}$ simply scales linearly with the VEV. Regarding the upper bound, increasing the VEV $x$ naively increases the allowed region for the heavy Higgs mass, but it is remarkable that the effects are present only for small angles, $\alpha < 0.1$ radians, being the bigger angles unaffected. Concerning the lower bound, or the vacuum stability of the model, at fixed $y^M$, increasing the VEV $x$ requires to increase $m_{h_2}$ to keep $\lambda _2$ constant at the EW scale. This explains why, with non negligible $y^M$, the allowed heavy Higgs masses are shrinking from below when we increase the VEV $x$, as one can see in figure~\ref{VEV_effect} and comparing figure~\ref{mh2_mh1_mhn1000_x75} with figure~\ref{mh1_mh2_api4_mhn-1000}, both for $\alpha=\pi /4$ and $y^M=0.2$, but for $x=3.5$ and $x=7.5$ TeV, respectively.

In general, for the model to survive up to very large scales $Q\sim M_{\rm{Planck}}$, it is preferred the heavy neutrinos to be light with respect to the VEV $x$, in such a way that their Yukawa couplings are negligible in the RGE evolution of the scalar sector.

\section{Phenomenology of the Higgs sector}\label{sect:Higgs}
In the previous section we have defined the regions of the scalar sector parameter space that are compatible with experimental and theoretical constraints.
In this section we summarise some of the phenomenology of the Higgs bosons in the $B-L$ model at the LHC, first presented in Ref.~\cite{Basso:2010yz}. 

The scalar Lagrangian of the $B-L$ model is part of a rather general family of extensions of the SM, in which the scalar content is augmented with a scalar singlet (see, e.g.,  Refs.~\cite{BahatTreidel:2006kx,O'Connell:2006wi,Barger:2006sk,Barger:2007im,Bhattacharyya:2007pb,Profumo:2007wc,Barger:2008jx}). The interactions of the Higgs bosons with SM particles are, therefore, the same as in the traditional literature. However, a richer phenomenology arises in the $B-L$ model because of the interplay with the gauge and the fermion sectors, where the new particles, the $Z'$ boson and the heavy neutrinos, interact with the scalar bosons. The interplay of the sectors can be strong, and, in some region of the parameter space, both the $Z'$ boson and the heavy neutrinos can be lighter than the Higgs bosons, allowing for new decays, beside those into SM particles.

Concerning the strength of Higgs interactions, some of the salient 
phenomenological features can be summarised as follows:
\begin{itemize}
\item[--] SM-like interactions scale with $\cos{\alpha}$($\sin{\alpha}$)
for $h_1$($h_2$);
\item[--] those involving the other new $B-L$ fields, like $Z'$ boson and heavy
neutrinos, scale with  
the complementary angle, i.e., with $\sin{\alpha}$($\cos{\alpha}$) for
$h_1$($h_2$); 
\item[--] triple (and quadruple) Higgs couplings are possible and can
induce resonant behaviours, so that, e.g.,  
the $h_2\rightarrow h_1\,h_1$ decay can become dominant if $m_{h_2}>2m_{h_1}$.
\end{itemize}


We first present production cross sections for the two Higgs bosons at the LHC, for CM energies of $\sqrt{s}=7$ and $14$ TeV, as well as their BRs, for some fixed values of the scalar mixing angle $\alpha$. Its values have been chosen in each plot to highlight some relevant phenomenological aspects, such as the decay into the new $B-L$ particles.

It turns out that the most efficient production mechanisms at the LHC are still the SM ones. However, new signatures arise in the decay processes.
Section~\ref{subsect:event_rates} will bring pieces together and present event rates for some phenomenologically viable signatures, as, among those relevant here, four lepton decays of a heavy Higgs boson via pairs of $Z'$
gauge bosons (which, e.g., in the SM also occurs via $ZZ$
but in very different kinematic regions) and heavy neutrino pair production via a light Higgs boson (yielding, e.g., the same multi-lepton signatures discussed in section~\ref{sect:Zp_to_nuh} for the $Z'$ boson).

\subsection{Production cross sections and decay properties}

In figure~\ref{Xs} we present the cross sections for the most relevant
production mechanisms, i.e., the usual SM processes such as
gluon-gluon fusion, vector-boson fusion, $t\overline{t}$ associated
production, and Higgs-strahlung (whose Feynman diagrams are in figure~\ref{SM_Higgs_pic}). For reference, we show in dashed lines the SM case (only for $h_1$), that corresponds to $\alpha =0$.

Comparing figure~\ref{xs_h1_14} to figure~\ref{xs_h1_7}, there is a
factor $3\,\div4$ enhancement passing from $\sqrt{s}=7$ TeV to $\sqrt{s}=14$
TeV CM energy at the LHC.

The cross sections are a smooth function of the mixing angle $\alpha$,
so as expected every subchannel has a cross section that scales with
$\cos{\alpha}$($\sin{\alpha}$), respectively, for $h_1$($h_2$). As a
general rule, the cross section for $h_1$ at an angle $\alpha$ is
equal to that one of $h_2$ for $\pi /2 -\alpha$. In particular, the
maximum cross section for $h_2$ (i.e., when $\alpha =\pi/2$) 
coincides with the one of $h_1$ for $\alpha =0$.

We notice that these results are in agreement with the ones that have
been discussed in
\cite{BahatTreidel:2006kx,Barger:2007im,Bhattacharyya:2007pb} in the
context of a scalar singlet extension of the SM, having
the latter the same Higgs production phenomenology.
Moreover, as already shown in Ref.~\cite{BahatTreidel:2006kx}, also in the
minimal $B-L$ context a high value of the mixing angle could lead
to important consequences for Higgs boson discovery at the LHC: a sort
of rudimentary see-saw mechanism could suppress $h_1$ production below
an observable rate at $\sqrt{s}=7$ TeV and favour just heavy Higgs
boson production, with peculiar final states clearly beyond the SM, or
even hide the production of both (if no more than $1$ fb$^{-1}$ of
data is accumulated). Instead, at $\sqrt{s}=14$ TeV we expect that at
least one Higgs boson will be observed, either the light one or the
heavy one, or indeed both, thus shedding light on the scalar sector of
the $B-L$ extension of the SM. The region of
the parameter space that would allow the scalar sector to be
completely hidden, for example for $\alpha \simeq \pi/2$ and $m_{h_2}$
heavy enough to not be produced, whatever the value of $m_{h_1}$, is
experimentally excluded by precision analyses at LEP
\cite{Dawson:2009yx}.

\begin{figure}[!ht] \centering \scalebox{1}{
\SetWidth{1.1}
\vspace*{2mm}
\begin{picture}(300,100)(0,0)
\ArrowLine(0,25)(40,50)
\ArrowLine(40,50)(0,75)
\Photon(40,50)(90,50){4}{5.5}
\DashLine(90,50)(130,25){4}
\Photon(90,50)(130,75){3.5}{5.5}
\Text(-10,20)[]{$q$}
\Text(-10,80)[]{$\bar{q}$}
\Text(70,65)[]{$V^*$}
\Text(90,50)[]{\blue{\Large\bf $\bullet$}}
\Text(139,20)[]{\bH}
\Text(139,80)[]{$V$}
\ArrowLine(200,25)(240,25)
\ArrowLine(200,75)(240,75)
\ArrowLine(240,25)(290,0)
\ArrowLine(240,75)(290,100)
\Photon(240,25)(280,50){3.5}{5}
\Photon(240,75)(280,50){-3.5}{5}
\DashLine(280,50)(320,50){4}
\Text(280,50)[]{\blue{\Large\bf $\bullet$}}
\Text(190,20)[]{$q$}
\Text(190,80)[]{$q$}
\Text(275,72)[]{$V^*$}
\Text(275,27)[]{$V^*$}
\Text(300,60)[]{\bH}
\Text(300,10)[]{$q$}
\Text(300,90)[]{$q$}
\end{picture} }
\vspace*{-3.mm}
\end{figure}
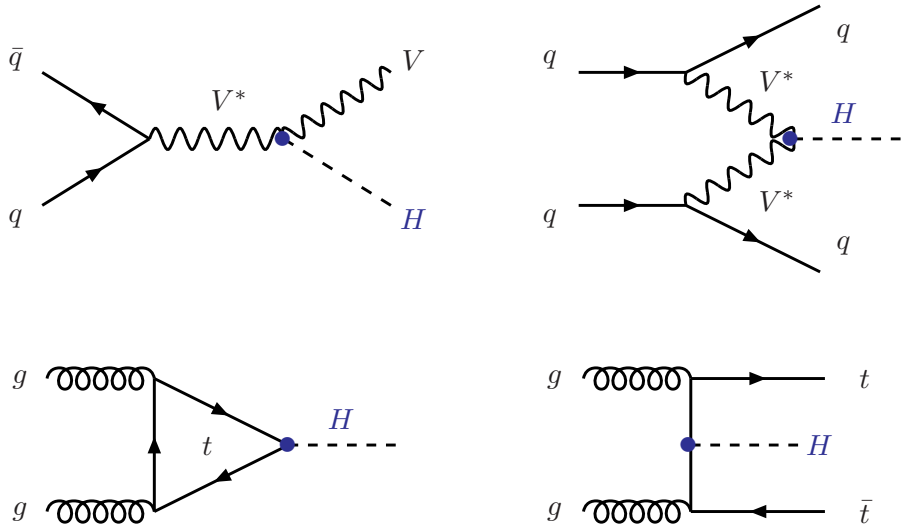
\begin{figure}[!ht] \centering \scalebox{1}{
\SetWidth{1.1}
\begin{picture}(300,100)(0,0)
\Gluon(0,25)(40,25){4}{5.5}
\Gluon(0,75)(40,75){4}{5.5}
\ArrowLine(40,75)(90,50)
\ArrowLine(90,50)(40,25)
\ArrowLine(40,25)(40,75)
\DashLine(90,50)(130,50){4}
\Text(90,50)[]{\blue{\Large\bf $\bullet$}}
\Text(-10,25)[]{$g$}
\Text(-10,75)[]{$g$}
\Text(110,60)[]{\bH}
\Text(60,50)[]{$t$}
\Gluon(200,25)(240,25){4}{5.5}
\Gluon(200,75)(240,75){4}{5.5}
\ArrowLine(290,25)(240,25)
\ArrowLine(240,75)(290,75)
\Line(240,75)(240,25)
\DashLine(240,50)(280,50){4}
\Text(240,50)[]{\blue{\Large\bf $\bullet$}}
\Text(190,25)[]{$g$}
\Text(190,75)[]{$g$}
\Text(289,50)[]{\bH}
\Text(306,75)[]{$t$}
\Text(306,25)[]{$\bar{t}$}
\end{picture}}
\vspace*{-0.3cm}
\caption{\it Feynman diagrams for the dominant SM Higgs boson production mechanisms in hadronic collisions. From top-left, clockwise: Higgs-strahlung, vector boson fusion, $t\overline{t}$ associated production, and gluon-gluon fusion ($V=W,Z$; $H=h_1,h_2$). \label{SM_Higgs_pic}} 
\end{figure}

\begin{figure}[!h]
  \subfloat[]{ 
  \label{xs_h1_7}
  \includegraphics[angle=0,width=0.48\textwidth ]{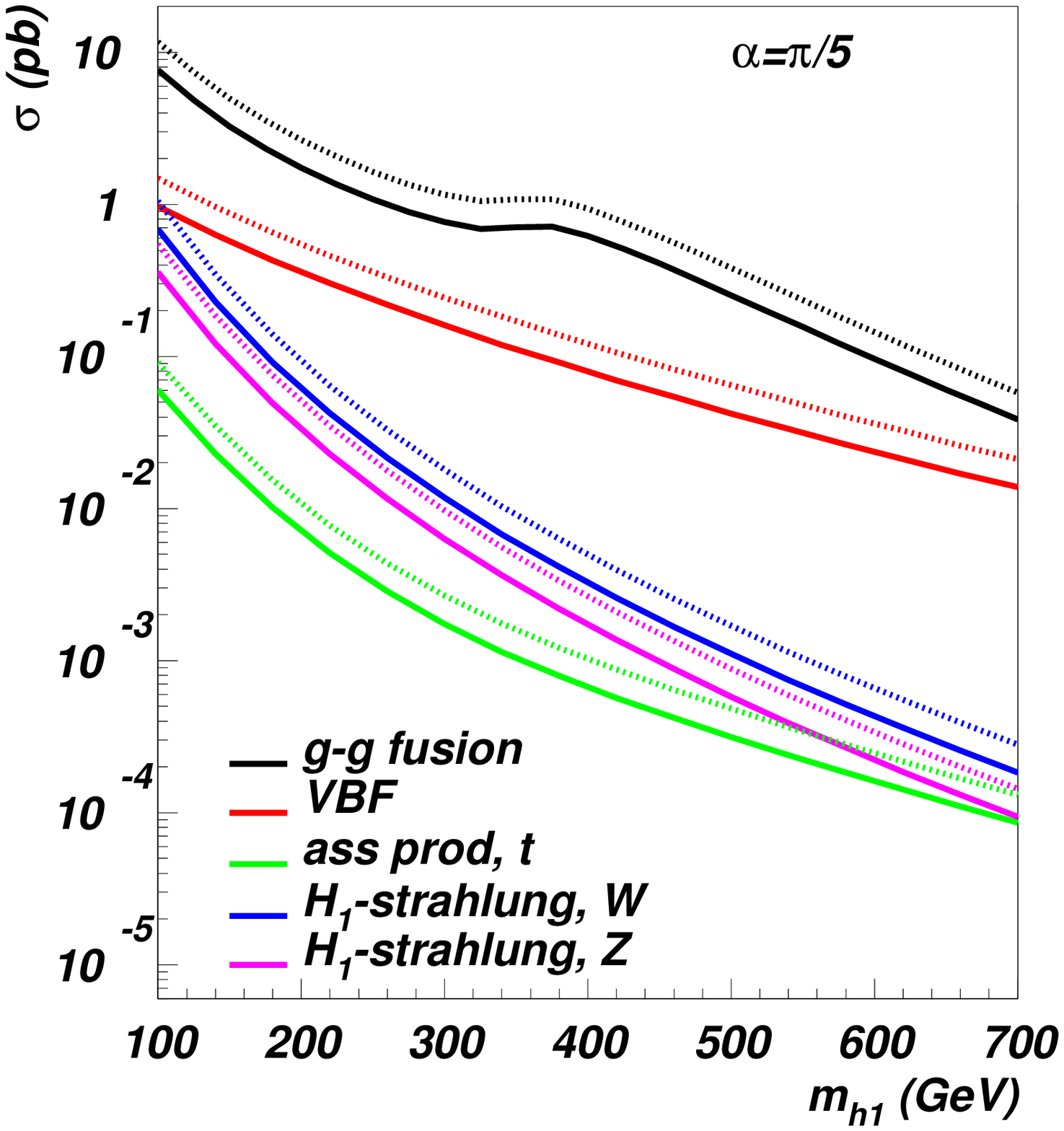}}
  \subfloat[]{
  \label{xs_h2_7}
  \includegraphics[angle=0,width=0.48\textwidth ]{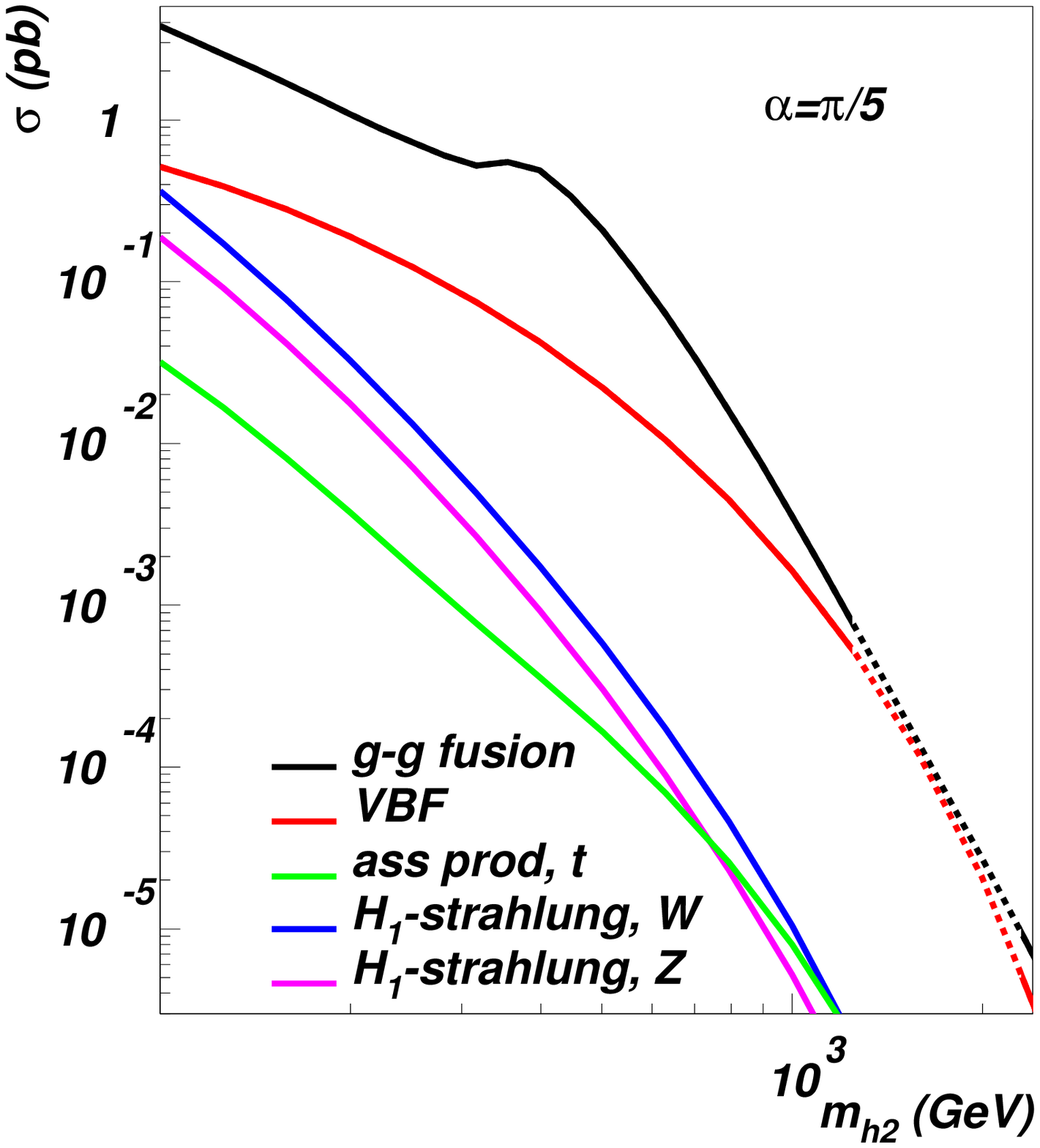}}\\
  \subfloat[]{
  \label{xs_h1_14}
  \includegraphics[angle=0,width=0.48\textwidth ]{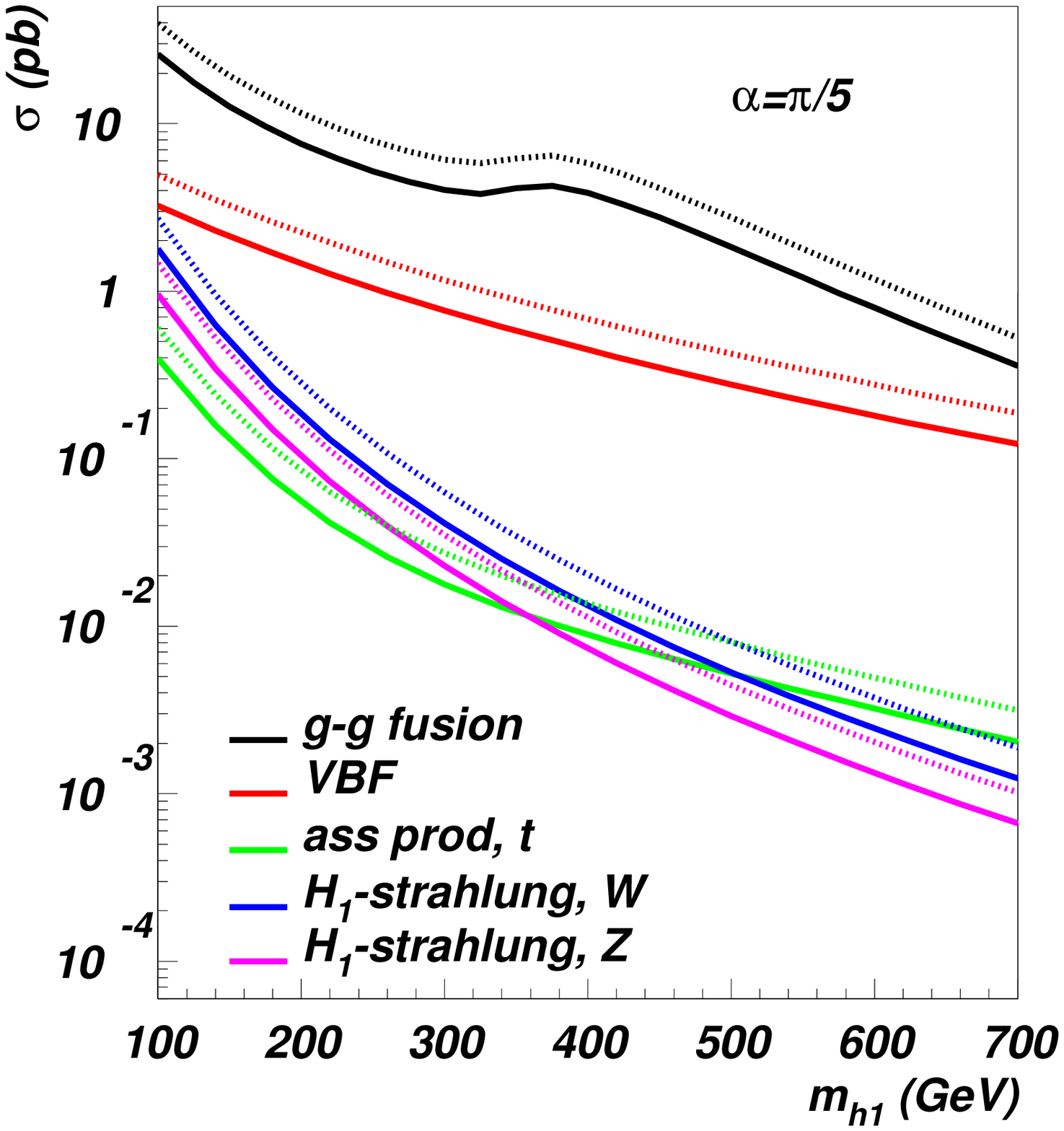}}  
  \subfloat[]{
  \label{xs_h2_14}
  \includegraphics[angle=0,width=0.48\textwidth ]{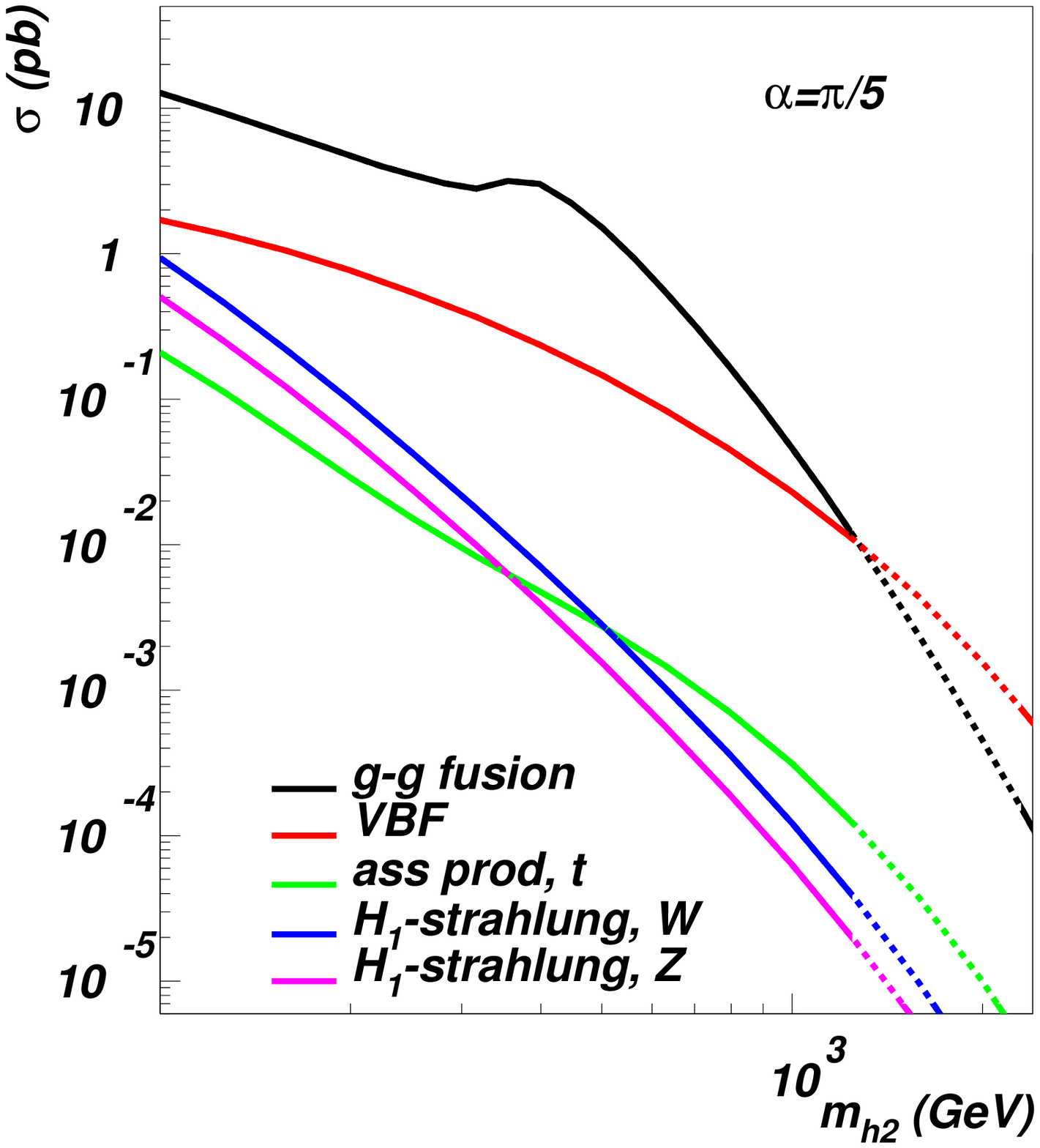}}  
  \vspace*{-0.25cm}
  \caption{\it Cross sections in the $B-L$ model for $h_1$ at the LHC
    (\ref{xs_h1_7}) at $\sqrt{s}=7$ TeV and (\ref{xs_h1_14}) at
    $\sqrt{s}=14$ TeV, and for $h_2$ (\ref{xs_h2_7}) at $\sqrt{s}=7$
    TeV and (\ref{xs_h2_14}) at $\sqrt{s}=14$ TeV. The dashed lines in
    figs.~(\ref{xs_h1_7}) and (\ref{xs_h1_14}) refer to $\alpha
    =0$. The dotted part of the lines in figs.~(\ref{xs_h2_7}) and (\ref{xs_h2_14}) refer to $h_2$ masses excluded by unitarity (see
    Ref.~\cite{Basso:2010jt}).}
  \label{Xs}
\end{figure}\index{Production cross sections!Higgs bosons}

\subsubsection{New production mechanisms}
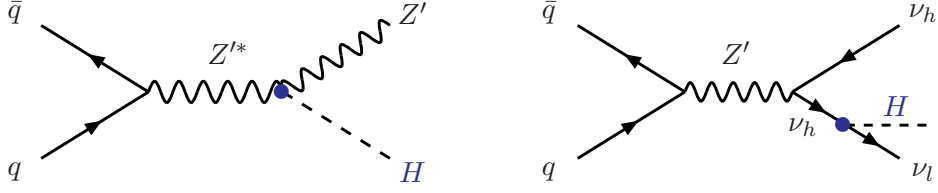
\begin{figure}[!ht] \centering \scalebox{1}{
\SetWidth{1.1}
\vspace*{2mm}
\begin{picture}(300,100)(0,0)
\ArrowLine(0,25)(40,50)
\ArrowLine(40,50)(0,75)
\Photon(40,50)(90,50){4}{5.5}
\DashLine(90,50)(130,25){4}
\Photon(90,50)(130,75){3.5}{5.5}
\Text(-10,20)[]{$q$}
\Text(-10,80)[]{$\bar{q}$}
\Text(70,65)[]{$Z'^*$}
\Text(90,50)[]{\blue{\Large\bf $\bullet$}}
\Text(139,20)[]{\bH}
\Text(139,80)[]{$Z'$}
\ArrowLine(200,25)(240,50)
\ArrowLine(240,50)(200,75)
\ArrowLine(320,75)(280,50)
\ArrowLine(280,50)(300,37.5)
\ArrowLine(300,37.5)(320,25)
\Photon(240,50)(280,50){3.5}{5}
\DashLine(300,37.5)(330,37.5){3.5}
\Text(300,37.5)[]{\blue{\Large\bf $\bullet$}}
\Text(190,20)[]{$q$}
\Text(190,80)[]{$\bar{q}$}
\Text(260,65)[]{$Z'$}
\Text(320,45)[]{\bH}
\Text(330,80)[]{$\nu _h$}
\Text(285,37)[]{$\nu _h$}
\Text(330,20)[]{$\nu _l$}
\end{picture} }
\vspace*{-0.3cm}
\caption{\it Feynman diagrams for the new Higgs boson production mechanisms in hadronic collisions in the $B-L$ model. From left to right: the associated production of a Higgs boson and the $Z'$ boson and Higgs production via heavy neutrino ($H=h_1,h_2$). \label{new_Higgs_pic}} 
\end{figure}

\begin{figure}[!h]
  \subfloat[]{ 
  \label{xs_Zp-h1}
  \includegraphics[angle=0,width=0.48\textwidth ]{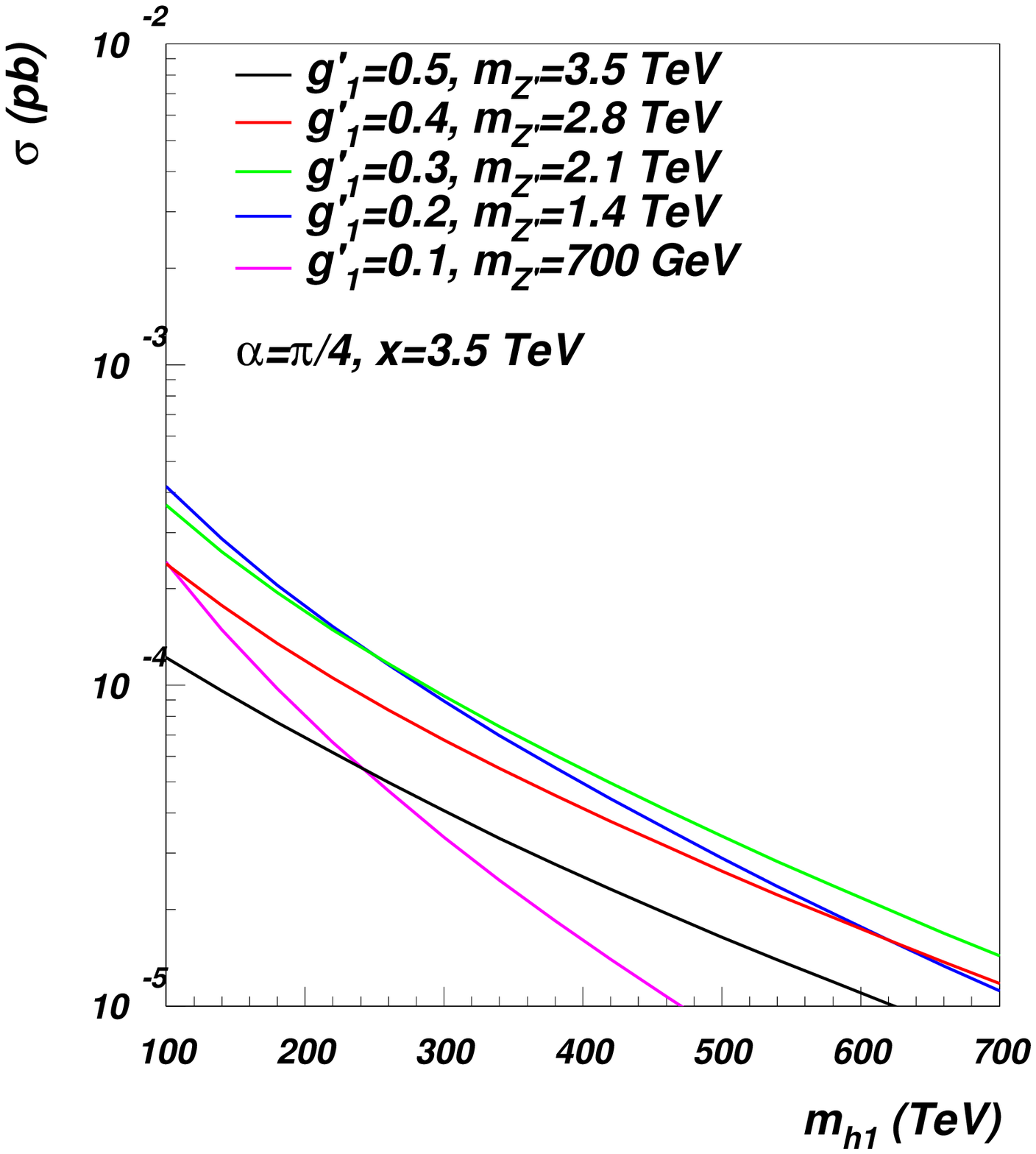}
}
  \subfloat[]{
  \label{xs_Zp-h2}
  \includegraphics[angle=0,width=0.48\textwidth ]{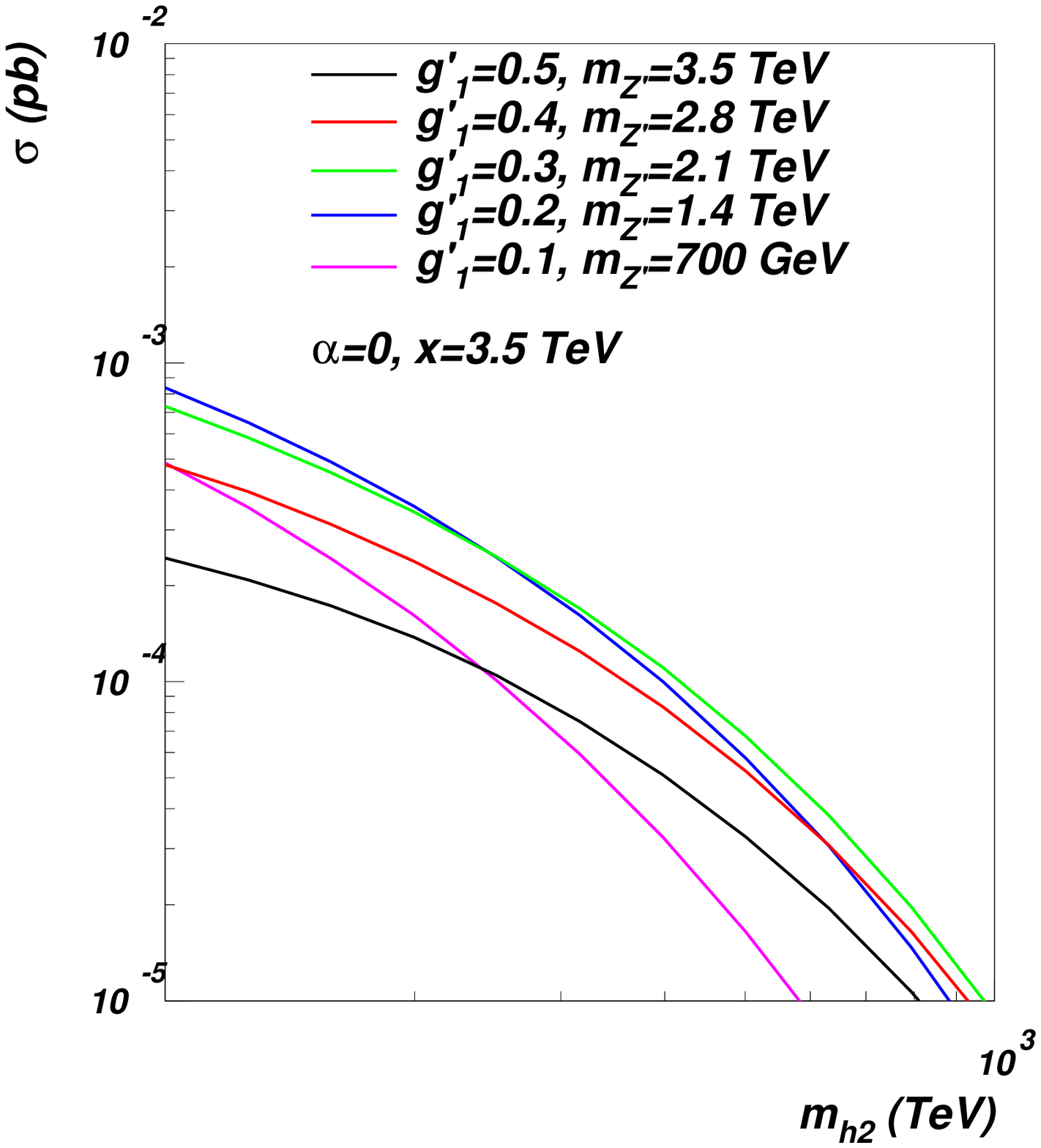}
}
  \vspace*{-0.25cm}
  \caption{\it Cross sections in the $B-L$ model for the associated
    production with the $Z'_{B-L}$ boson (\ref{xs_Zp-h1}) of $h_1$ at
    $\alpha = \pi /4$ and (\ref{xs_Zp-h2}) of $h_2$ at $\alpha = 0$, at $\sqrt{s}=14$ TeV.}
  \label{strah-Xs}
\end{figure}

\begin{figure}[!h]
\centering
  \includegraphics[angle=0,width=0.6\textwidth ]{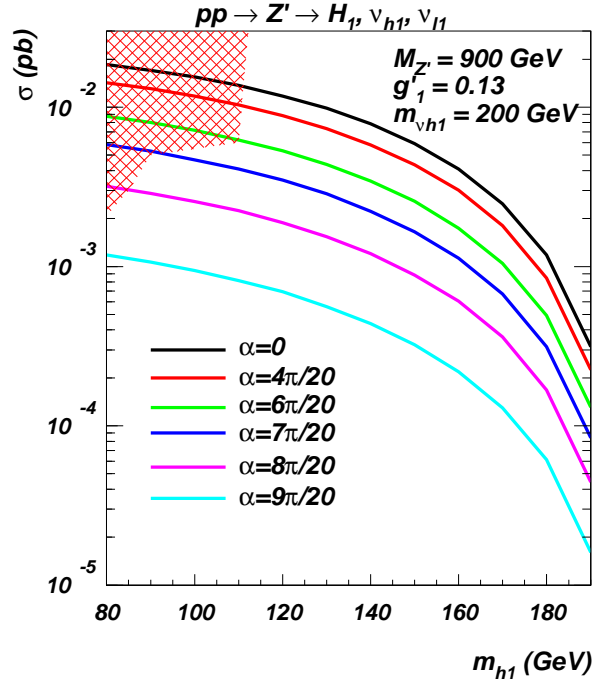}
  \vspace*{-0.5cm}
  \caption{\it Cross sections in the $B-L$ model for the associated
    production of $h_1$ with one heavy and one light
    neutrino at $\sqrt{s}=14$ TeV. The red shading is the region excluded by LEP    constraints (see section~\ref{sec:expbounds:Higgs}).}
  \label{non-std-Xs}
\end{figure}

All the new particles in the $B-L$ model interact with the scalar
sector, so novel production mechanisms can arise considering the
exchange of new intermediate states. Among the new production
mechanisms, the associated production of the scalar boson with the
$Z'$ boson and the decay of a heavy neutrino into a Higgs boson are
certainly the most promising, depending on the specific masses. The Feynman diagrams for these processes are in figure~\ref{new_Higgs_pic}. Figures~\ref{strah-Xs} and \ref{non-std-Xs} show the cross sections for these new production mechanisms, for $\sqrt{s}=14$ TeV and several values of $\alpha$.

Figures~\ref{xs_Zp-h1} and \ref{xs_Zp-h2} show the cross sections for the
associated production with the $Z'$ boson of $h_1$ and of $h_2$,
respectively, for several combinations of $Z'$ boson masses and $g'_1$
couplings. The process is
\begin{equation}
q\,\overline{q}\rightarrow Z'^{\ast} \rightarrow Z'\, h_{1(2)}\, ,
\end{equation}
and it is dominated by the $Z'$ boson's production cross sections (see
section~\ref{sect:Zp_properties}). 
Although never dominant (always below $1$ fb), this channel is the
only viable mechanism to produce $h_2$ in the decoupling scenario,
i.e., $\alpha =0$.

In figure~\ref{non-std-Xs} we plot the cross sections for the other
new production mechanism (the Higgs production via heavy neutrino) against the light Higgs mass, for
several choices of parameters (as explicitly indicated in the labels).
We superimposed the red-shadowed region in order to avoid any mass-angle combination that has been already excluded by LEP constraints, as discussed in section~\ref{sec:expbounds:Higgs}.
The whole process chain is
\begin{equation}
q\,\overline{q}\rightarrow Z' \rightarrow \nu _h\, \nu _h \rightarrow
\nu _h\, \nu _l\, h_{1(2)}\, ,
\end{equation}
and it requires to pair produce heavy neutrinos, again via the $Z'$
boson (see section~\ref{sect:nu_properties}). Although rather
involved, this mechanism has the advantage that
the whole decay chain can be of on-shell particles, beside the
peculiar final state of a Higgs boson and a heavy neutrino. For a
choice of the parameters that roughly maximises this mechanism
($M_{Z'}=900$ GeV, $g'_1=0.13$ and $m_{\nu _h}=200$ GeV, from figure~\ref{contour14_nuh}),
figure~\ref{non-std-Xs} shows that the cross sections for the production
of the light Higgs boson (when only one generation of heavy neutrinos
is considered) are above $10$ fb for $m_{h_1} < 130$ GeV (and small
values of $\alpha$), dropping steeply when the light Higgs boson mass
approaches the kinematical limit for the heavy neutrino to decay into
it. Assuming the transformation $\alpha \rightarrow \pi/2 -
\alpha$, the production of the heavy Higgs boson via this mechanism
shows analogous features.

These new production mechanisms have rather small cross sections at the LHC. Nonetheless, a future LC could be the suitable framework to probe them, as shown in Ref.~\cite{Basso:2010si}. We showed therein that these mechanisms can also be the leading ones for accessing the scalar sector of the $B-L$ model.

\subsection{Branching ratios and total widths}
\begin{figure}[!h]
  \subfloat[]{ 
  \label{BR_h1}
  \includegraphics[angle=0,width=0.48\textwidth ]{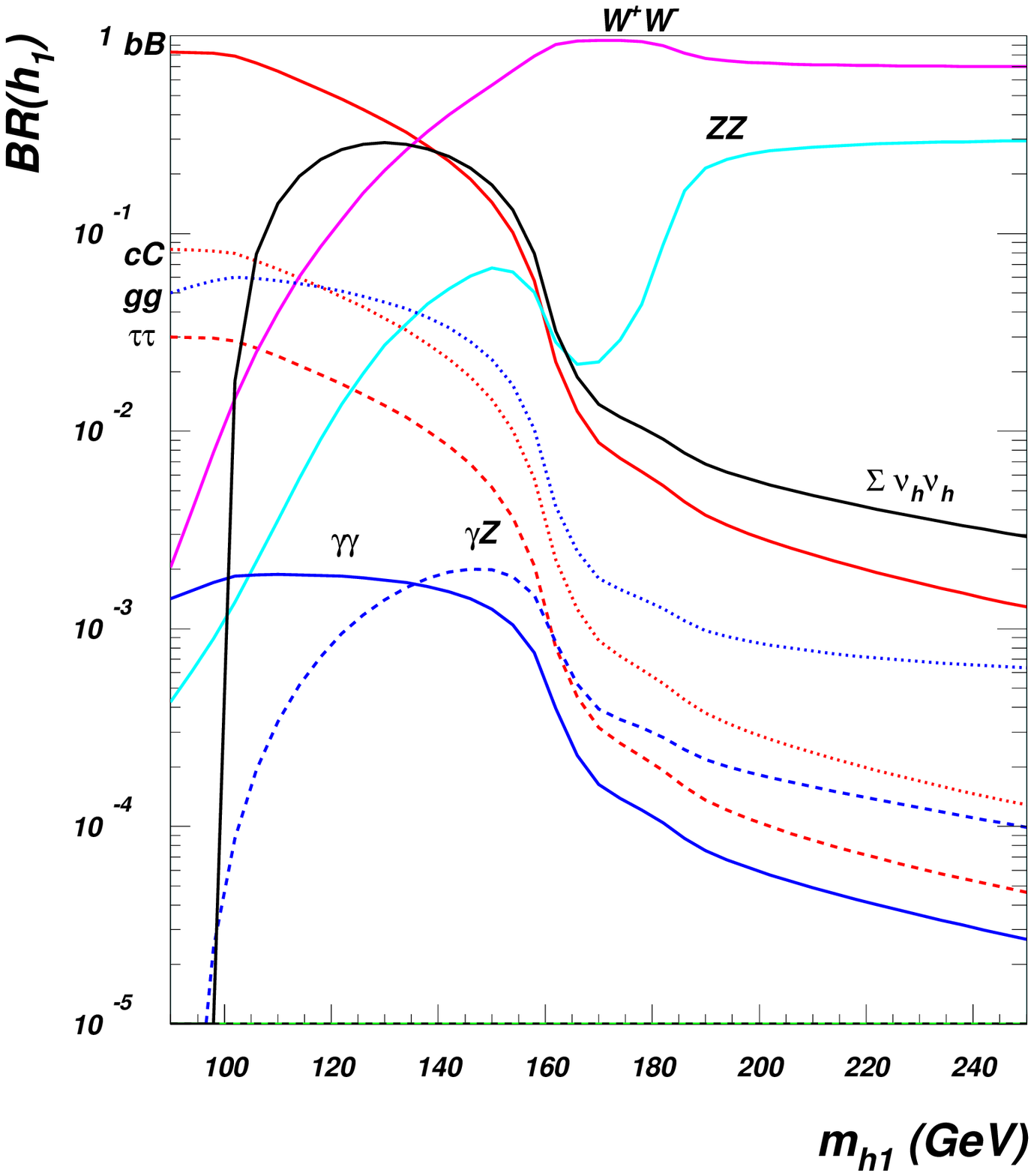}}
  \subfloat[]{
  \label{BR_h2}
  \includegraphics[angle=0,width=0.48\textwidth ]{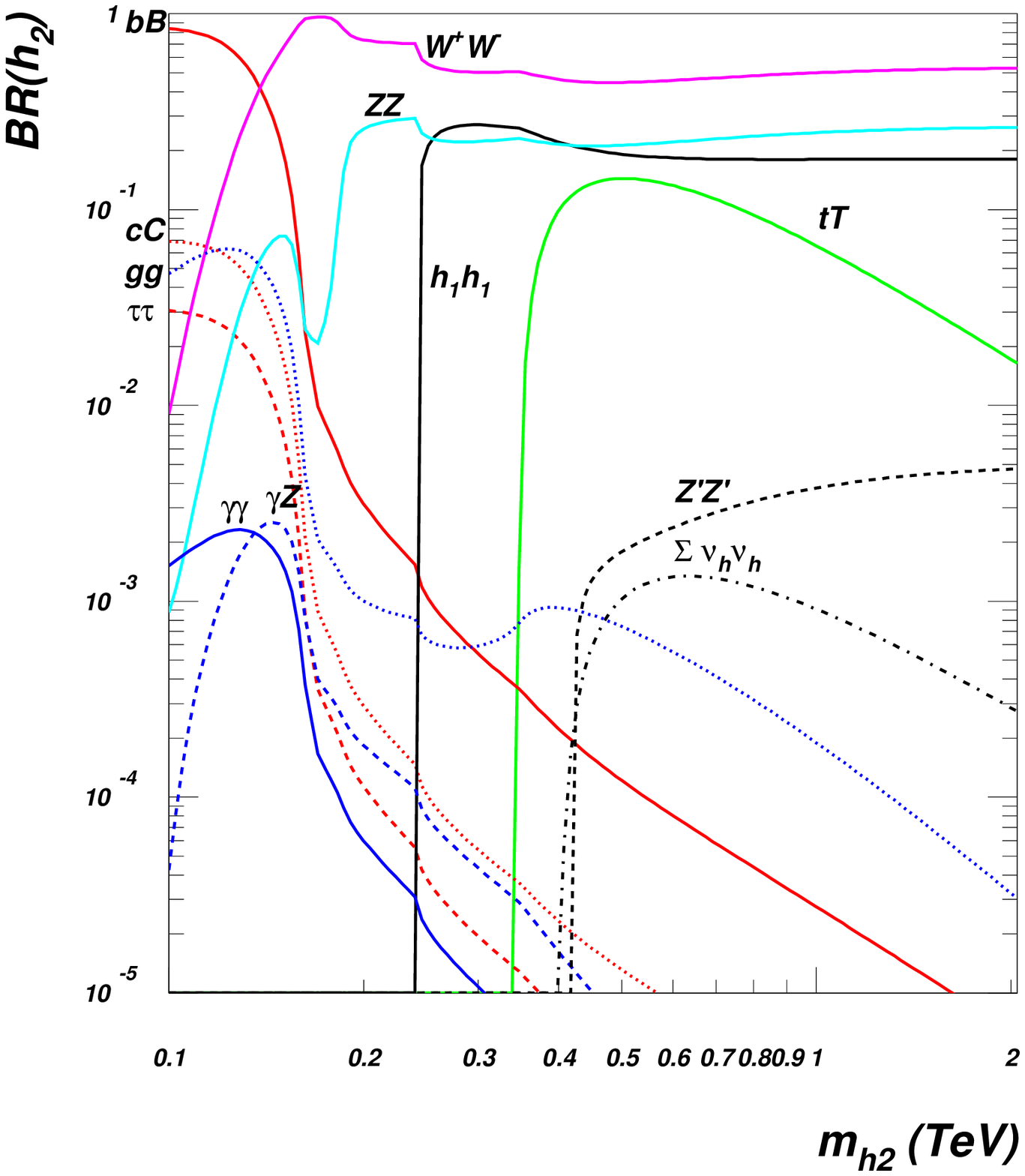}} \\
  \subfloat[]{ 
  \label{H1_TW}
  \includegraphics[angle=0,width=0.48\textwidth ]{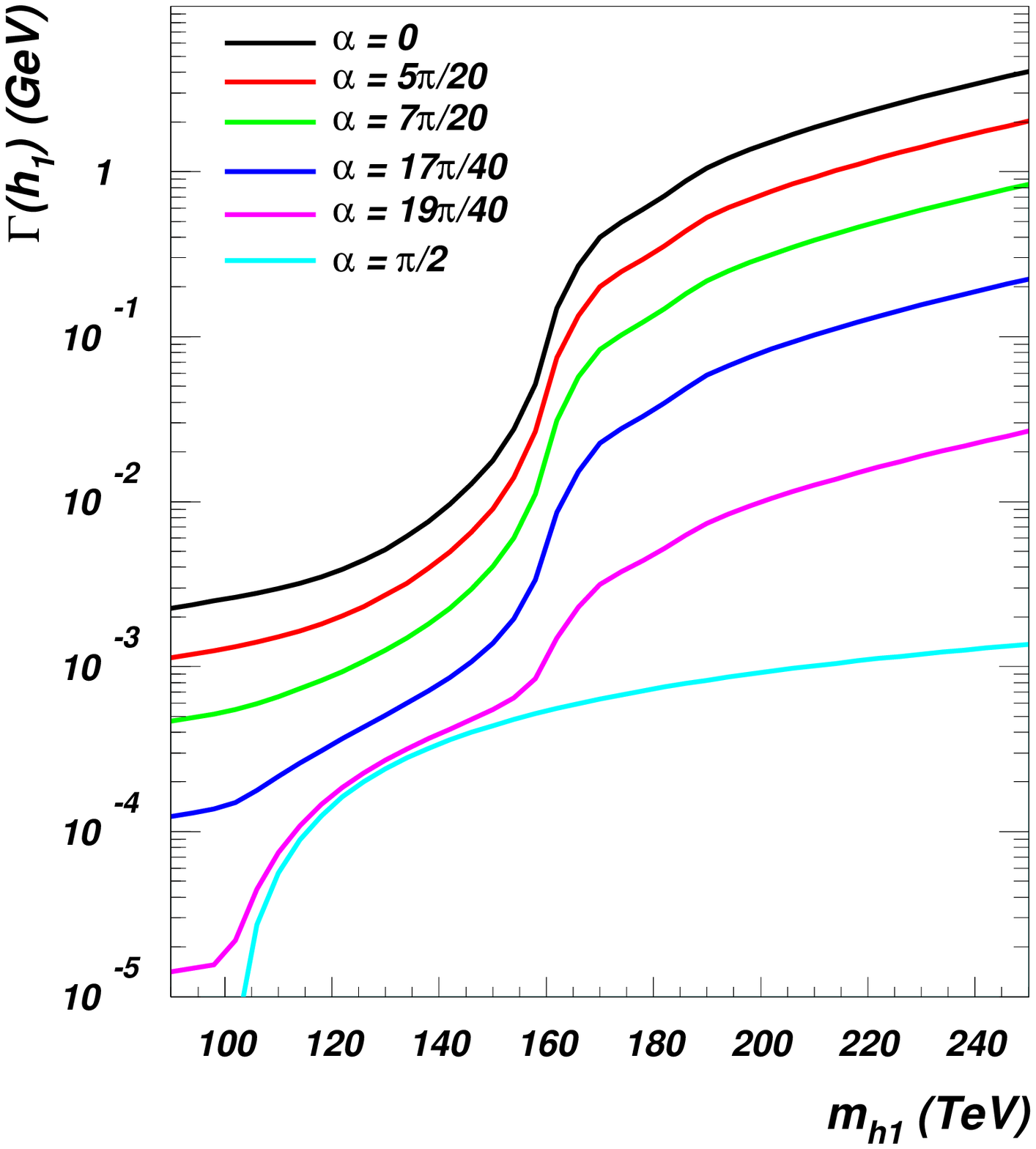}}
  \subfloat[]{
  \label{H2_TW}
  \includegraphics[angle=0,width=0.48\textwidth ]{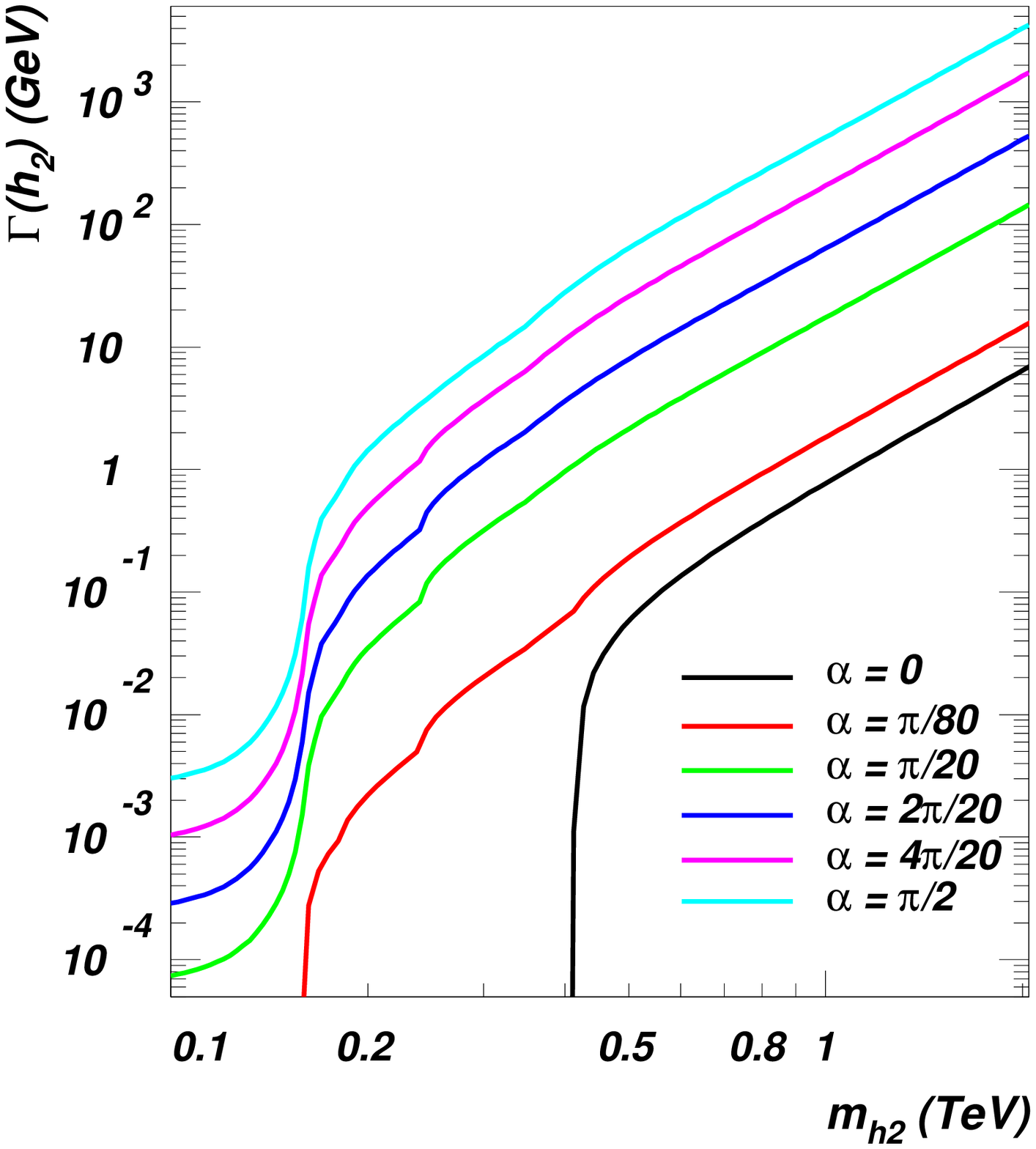}} 
  \vspace*{-0.25cm}
  \caption{\it (\ref{BR_h1}) BRs for $h_1$ for $\alpha =2\pi /5$ and $m_{\nu_h}=50$ GeV and (\ref{H1_TW}) $h_1$ total width for a choice of mixing angles and (\ref{BR_h2}) BRs for $h_2$ for $\alpha =3\pi /20$ and $m_{h_1}=120$ GeV, $M_{Z'}=210$ GeV and $m_{\nu_h}=200$ GeV and (\ref{H2_TW}) $h_2$ total width for a choice of mixing angles.}
  \label{Brs}
\end{figure}

Moving to the Higgs boson decays, figure~\ref{Brs} shows the BRs for both the Higgs bosons, $h_1$ and $h_2$, respectively. Only the $2$-body decay channels are shown here. For a description of the $3$-body decays, see Ref.~\cite{Basso:2010yz}. \index{BRs!Higgs bosons}

Regarding the light Higgs boson, the only new particle it can decay into is the heavy neutrino (we consider a very light $Z'$ boson unlikely and unnatural), if the channel is kinematically open. In figure~\ref{BR_h1} we show this case, for a small heavy neutrino mass, i.e., $m_{\nu_h}=50$ GeV, and we see that the relative BR of this channel can be rather important, as the decay into 
$b$ quark pairs or into $W$ boson pairs, in the range of masses $110$ GeV $\leq m_{h_1} \leq 150$ GeV. Such a range happens to be critical in the SM since here the SM Higgs boson passes from decaying dominantly into 
$b$ quark pairs to a region in masses in which the decay into $W$ boson pairs is the prevailing one. These two decay channels have completely different signatures and discovery methods/powers. The fact that the signal of the Higgs boson decaying into 
$b$ quark pairs is many orders of magnitude below the natural QCD background, spoils its sensitivity. In the case of the $B-L$ model, the decay into heavy neutrino pairs is therefore phenomenologically very important, besides being an interesting feature of the $B-L$ model if $m_{\nu_h} < M_W$, as it allows for multi-lepton signatures of the light Higgs boson. Among them, there is the decay of the Higgs boson into $3\ell$, $2j$ and $\met$ (studied for the $Z'$ case in section~\ref{subsect:trilep}), into $4\ell$ and $\met$ (as, again, already studied for 
the $Z'$ case in Ref.~\cite{Huitu:2008gf}) or into $4\ell$ and $2j$ (as already studied, when $\ell = \mu$, in the $4^{th}$ family extension of the SM \cite{CuhadarDonszelmann:2008jp}). All these peculiar signatures allow the Higgs boson signal to be studied in channels much cleaner than the decay 
into $b$ quark pairs.

In the case of the heavy Higgs boson, further decay channels are possible in the $B-L$ model, if kinematically open. The heavy Higgs boson can decay in pairs of the light Higgs boson ($h_2 \rightarrow h_1\, h_1$) or even in triplets ($h_2 \rightarrow h_1\, h_1\, h_1$), in pairs of heavy neutrinos and $Z'$ bosons. Even for a small value of the angle, figure~\ref{BR_h2} shows that the decay of a heavy Higgs boson into pairs of the light one can be quite sizeable, at the level of the decay into SM $Z$ bosons for $m_{h_1} = 120$ GeV. 

\begin{figure}[!h]
  \subfloat[]{
  \label{H2_BR-a_Hnu}
  \includegraphics[angle=0,width=0.48\textwidth ]{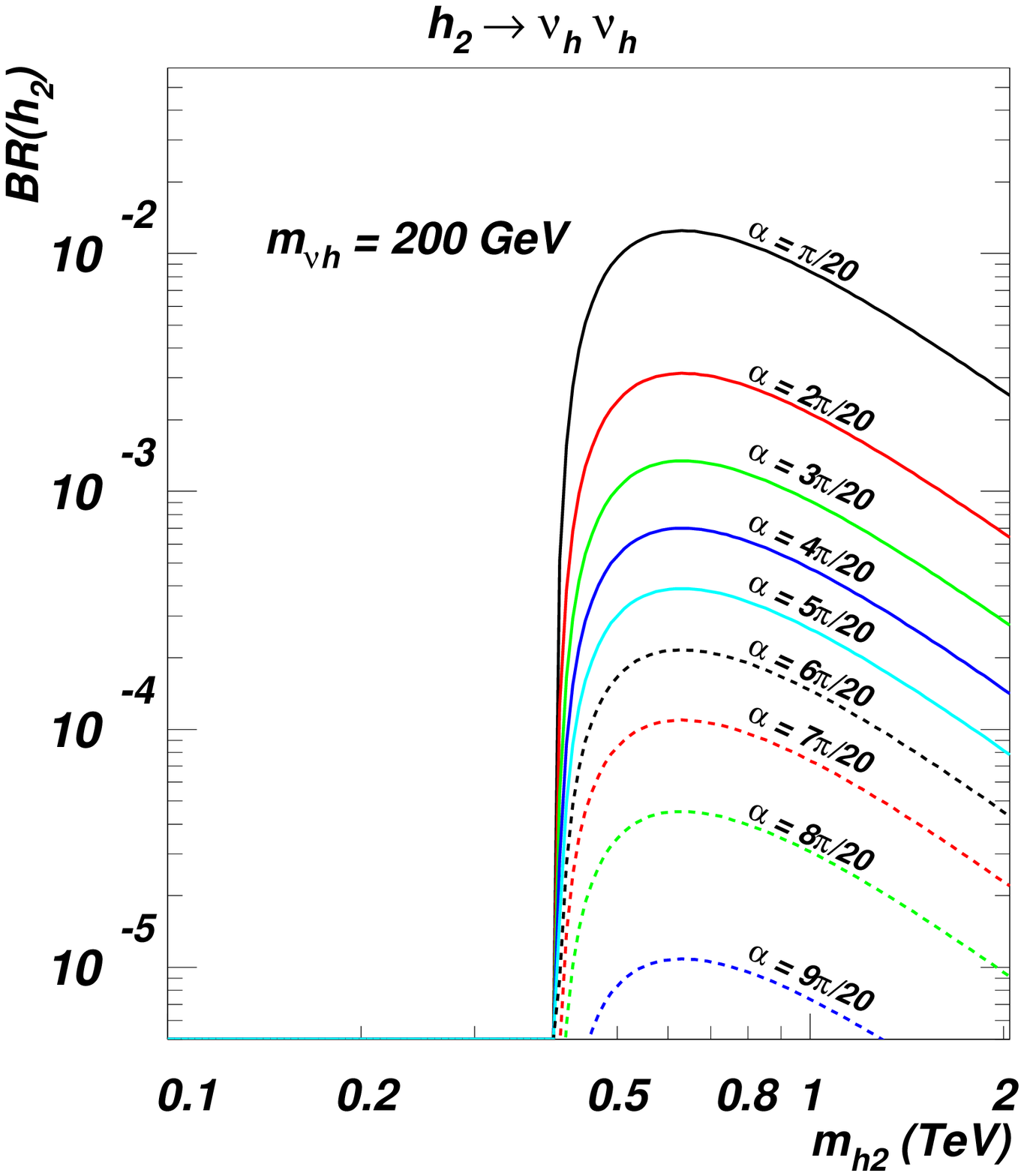}}  
  \subfloat[]{
  \label{H2_BR-a_Zp}
  \includegraphics[angle=0,width=0.48\textwidth ]{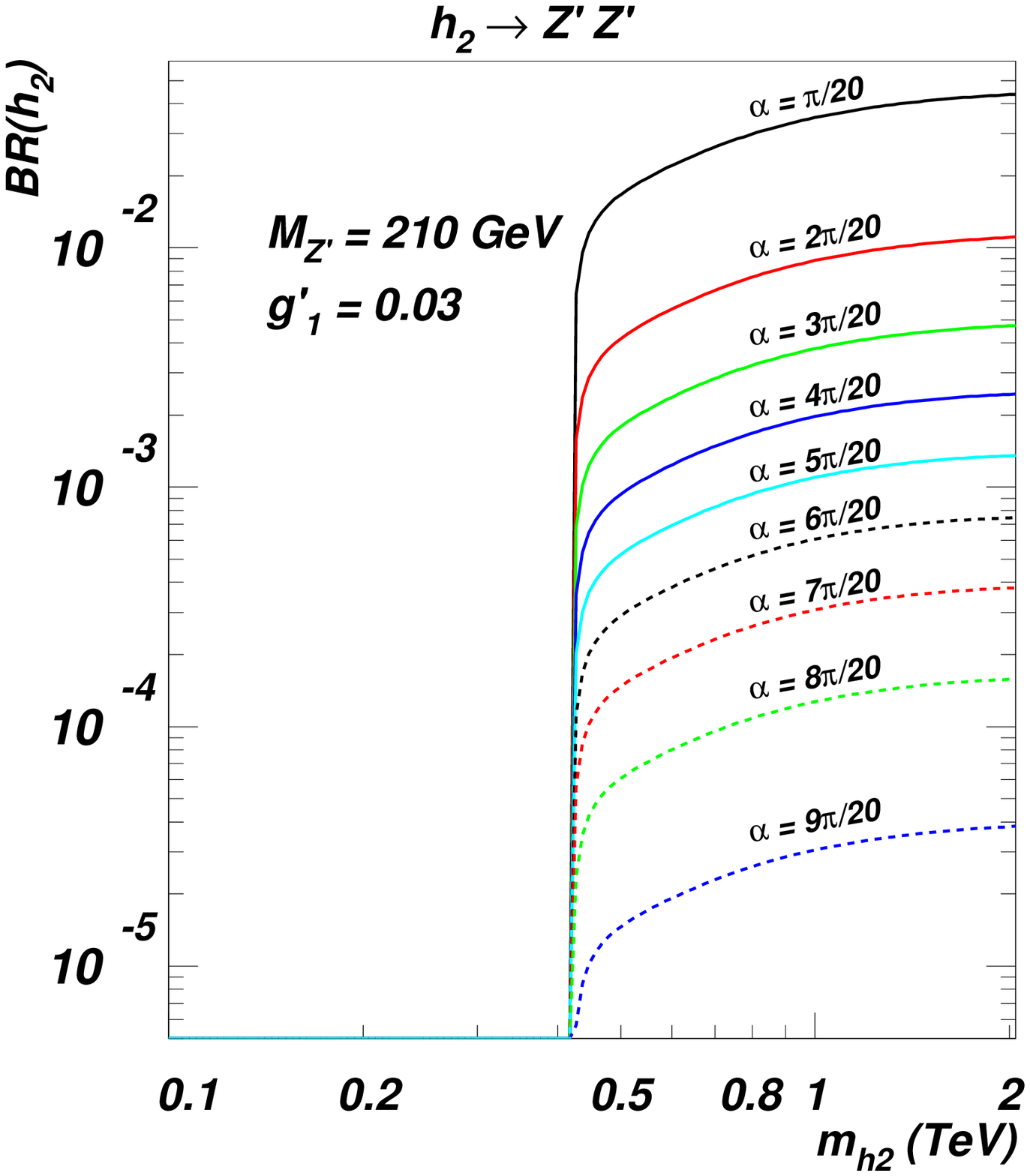}}  
  \vspace*{-0.25cm}
  \caption{\it Dependence on the mixing angle $\alpha$ of (\ref{H2_BR-a_Hnu}) $BR(h_2\rightarrow \nu _h\, \nu _h)$ and of (\ref{H2_BR-a_Zp}) $BR(h_2\rightarrow Z'\, Z')$.}
  \label{Br-alpha}
\end{figure}

The BRs of the heavy Higgs boson decaying into $Z'$ boson pairs and heavy neutrino pairs decrease as the mixing angle 
increases, getting to their maxima (comparable to the $W$ and $Z$ ones) for a vanishing $\alpha$, for which the production 
cross section is however negligible. As usual, and also clear from figure~\ref{BR_h2}, the decay of the heavy Higgs boson into gauge bosons (the $Z'$ boson) is always bigger than the decay into pairs of fermions (the heavy neutrinos, even when summed over the generations, as plotted), when they have comparable masses (here, $M_{Z'}=210$ GeV and $m_{\nu _h} = 200$ GeV).
It is important to note that all these new channels do not have a simple dependence on the mixing angle $\alpha$ (see figure~\ref{Br-alpha} for $Z'$ bosons and neutrinos final states and Ref.~\cite{Basso:2010yz} for the light Higgs boson case).

The other standard decays of both the light and the heavy Higgs bosons are not modified substantially in the $B-L$ model (i.e., the Higgs boson to $W$ boson pairs is always dominant when kinematically open, otherwise 
the decay into $b$ quarks is the prevailing one; further, radiative decays, such as Higgs boson decays into 
pairs of photons, peak at around $120$ GeV, etc.). Only when other new channels open, the standard decay channels 
alter accordingly. This rather common picture could be altered when the mixing angle $\alpha$ 
approaches $\pi/2$, but such situation is phenomenologically not viable \cite{Dawson:2009yx}.

Figures~\ref{H1_TW} and \ref{H2_TW} show the total widths for $h_1$ and $h_2$, respectively. In the first case, few thresholds are clearly recognisable as the heavy neutrino one at $100$ GeV (for angles very close to $\pi /2$ only), the $W$ and the $Z$ ones. Over the mass range considered ($90$ GeV $< m_{h_1} < 250$ GeV), the particle's width is very small until the $W$ threshold, less than $1-10$ MeV, rising steeply to few GeV for higher $h_1$ masses and small angles (i.e., for a SM-like light Higgs boson). As we increase the mixing angle, the couplings of the light Higgs boson to SM particles is reduced, like its total width.

On the contrary, as we increase $\alpha$, the $h_2$ total width increases, as clear from figure~\ref{H2_TW}. Also in this case, few thresholds are recognisable, as the usual $W$ and $Z$ gauge boson ones, the light Higgs boson one (at $240$ GeV) and the 
$t$ quark one (only for big angles, i.e., when $h_2$ is the SM-like Higgs boson). When the mixing angle is small, the $h_2$ total width stays below $1$ GeV all the way up to $m_{h_2} \sim 300\div 500$ GeV, rising as the mass increases towards values for which $\Gamma_{h_2} \sim m_{h_2}\sim 1$ TeV and $h_2$ loses the meaning of 
bound state, only for angles very close to $\pi /2$. Instead, if the angle is small, i.e., less than $\pi /10$, the ratio of width over mass is less than $10 \%$ and the heavy Higgs boson is a well-defined particle. In the decoupling regime, i.e., when $\alpha =0$, the only particles $h_2$ couples to are the $Z'$ boson and the heavy neutrinos. The width is therefore dominated by the decay into pairs of them and it is tiny, as is clear from figure~\ref{H2_TW}.

As already mentioned, figure~\ref{Br-alpha} shows the dependence on the mixing angle $\alpha$ of the BRs of $h_2$ into pairs of $Z'$ bosns and heavy neutrinos, 
$h_2 \rightarrow \nu _h\, \nu _h$ and $h_2 \rightarrow Z'\, Z'$, respectively (not influenced by $m_{h_1}$). As discussed, the interaction of the heavy Higgs boson with SM (or non-SM) particles has an overall $\sin{\alpha}$ (or $\cos{\alpha}$, respectively)
dependence. Nonetheless, the BRs in figure~\ref{Br-alpha} depend also on the total width, that for  $\alpha > \pi/4$ is dominated by the $h_2 \rightarrow W^+ W^-$ decay. Hence, when the angle assumes big values, the angle dependence of the $h_2$ BRs into heavy neutrino pairs and into $Z'$ boson pairs follows a  $\cot{\alpha}$ behaviour. For a study of BR($h_2 \rightarrow h_1\, h_1$), see Ref.~\cite{Basso:2010yz}.

\subsection{Cross sections for processes involving $Z'$ bosons and neutrinos}\label{subsect:event_rates}

In this subsection we combine the results from the Higgs boson cross sections and those from the BR analysis in order to perform a detailed study of typical event rates for some Higgs signatures that are specific to the $B-L$ model.

As in section~\ref{sect:pheno_gauge}, we identify two different experimental scenarios related to the LHC: the `early discovery scenario' (i.e., with CM energy of $\sqrt{s}=7$ TeV and an integrated
luminosity of $L=1$~fb$^{-1}$), and the `full luminosity scenario' (i.e., with CM energy of $\sqrt{s}=14$ TeV and an integrated luminosity of $L~=300$~fb$^{-1}$).

Combining the production cross sections and the decay BRs presented in the previous subsections, the two different scenarios open different possibilities for the detection of peculiar signatures of the model, involving heavy neutrinos and the $Z'$ boson. In the early discovery scenario there is a clear possibility to detect a light Higgs state yielding heavy neutrino pairs, while the full luminosity scenario affords the possibility of looking for the heavy Higgs state decaying into $Z'$ boson pairs, as well as into heavy neutrino pairs. In addition to them, the decays of the heavy Higgs state into light Higgs boson pairs are possible, as discussed in Ref~\cite{Basso:2010yz}.

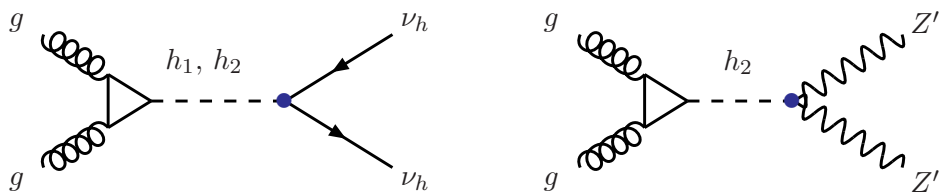
\begin{figure}[!ht] \centering \scalebox{1}{
\SetWidth{1.1}
\vspace*{2mm}
\begin{picture}(300,100)(0,0)
\Gluon(0,25)(24,40){4}{4.5}
\Gluon(24,60)(0,75){4}{4.5}
\DashLine(40,50)(90,50){5}
\ArrowLine(90,50)(130,25)
\ArrowLine(130,75)(90,50)
\Line(24,40)(24,60)
\Line(40,50)(24,60)
\Line(40,50)(24,40)
\Text(-10,20)[]{$g$}
\Text(-10,80)[]{$g$}
\Text(60,65)[]{$h_1,\, h_2$}
\Text(90,50)[]{\blue{\Large\bf $\bullet$}}
\Text(139,20)[]{$\nu _h$}
\Text(139,80)[]{$\nu _h$}
\Gluon(200,25)(224,40){4}{4.5}
\Gluon(224,60)(200,75){4}{4.5}
\DashLine(240,50)(280,50){5}
\Photon(320,75)(280,50){4}{5.5}
\Photon(280,50)(320,25){4}{5.5}
\Text(280,50)[]{\blue{\Large\bf $\bullet$}}
\Line(224,40)(224,60)
\Line(240,50)(224,60)
\Line(240,50)(224,40)
\Text(190,20)[]{$g$}
\Text(190,80)[]{$g$}
\Text(260,65)[]{$h_2$}
\Text(330,80)[]{$Z'$}
\Text(330,20)[]{$Z'$}
\end{picture} }
\vspace*{-0.3cm}
\caption{\it Feynman diagrams for the new Higgs boson discovery mechanisms in hadronic collisions in the $B-L$ model. From left to right: the decay of a Higgs boson into heavy neutrino pairs and into $Z'$ boson pairs. \label{Higgs_sign_pic}} 
\end{figure}

The Feynman diagrams for these processes are shown in figure~\ref{Higgs_sign_pic}, with the gluon-gluon fusion being the dominant production mechanism.

\begin{figure}[!h]
  \subfloat[]{ 
  \label{gg-h1-nunu50}
  \includegraphics[angle=0,width=0.48\textwidth ]{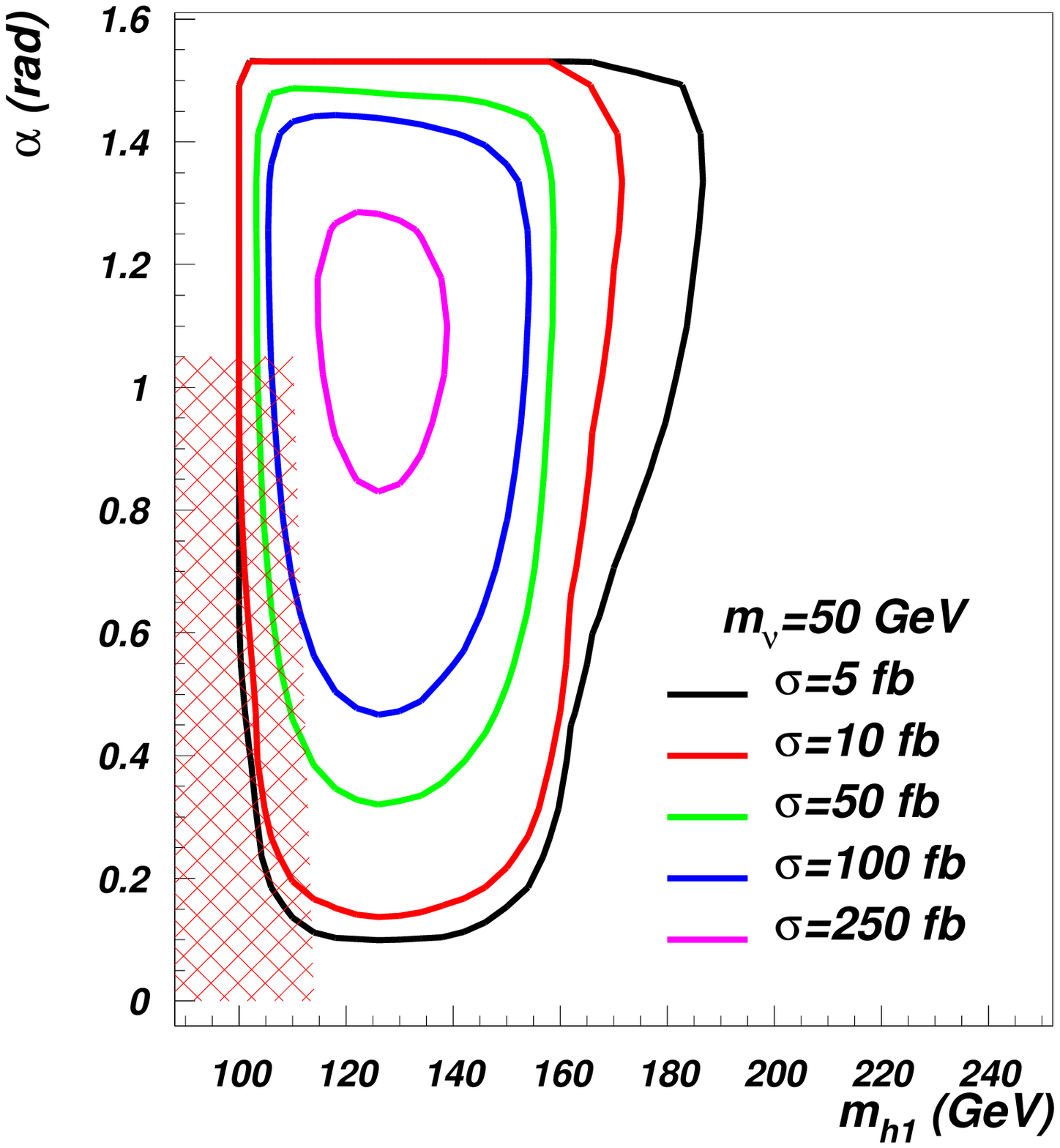}
}
  \subfloat[]{
  \label{gg-h1-nunu60}
  \includegraphics[angle=0,width=0.48\textwidth ]{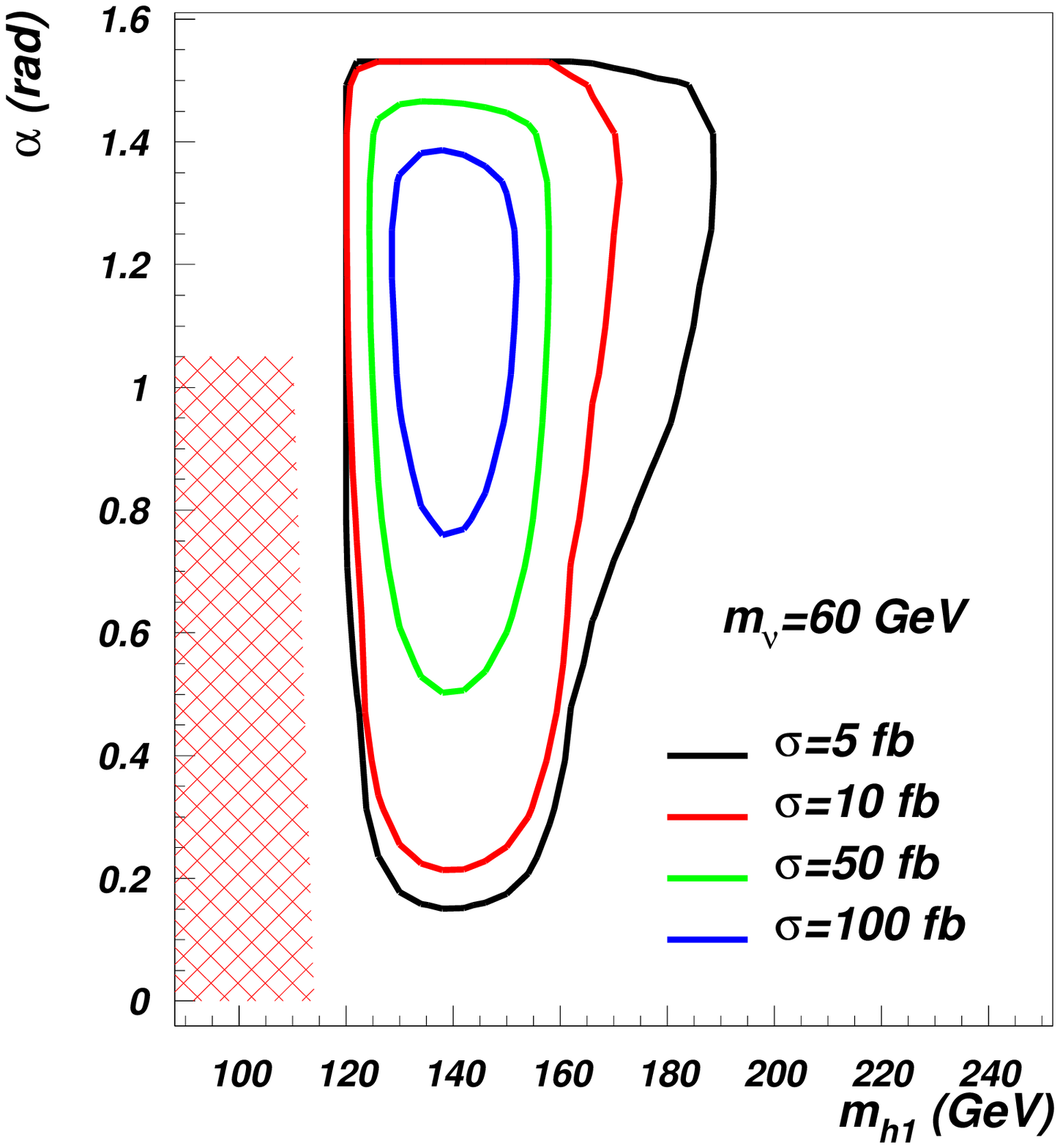}
}
  \vspace*{-0.25cm}
  \caption{\it Cross section times BR contour 
    plot for the process $pp\rightarrow 
    h_1\rightarrow \nu_h \nu_h$ at the LHC for $\sqrt{s}=7$ TeV, plotted
    in the $m_{h_1}$--$\alpha$ plane, with (\ref{gg-h1-nunu50}) $m_{\nu_h}=50$ GeV and  (\ref{gg-h1-nunu60}) $m_{\nu_h}=60$ GeV. The
  red-shadowed region is excluded by the LEP experiments \cite{Barate:2003sz}.}
  \label{ggnunu50-60}
\end{figure}

First, we focus on the early discovery scenario: in this experimental configuration, the most important $B-L$ distinctive process is the heavy neutrino pair production via the light Higgs boson:
\begin{equation}\label{h1-nuhnuh}
pp\rightarrow h_1 \rightarrow \nu_h\nu_h.
\end{equation}
In figure~\ref{ggnunu50-60} we show the cross sections, in the ($m_{h_1}$--$\alpha$) plane, for this process at the LHC for $\sqrt{s}=7$ TeV, for two different heavy neutrino masses ($m_{\nu_h}=50$, $60$ GeV). These plots are the result of the combination of the light Higgs boson production cross sections via gluon-gluon fusion only (that represents the main contribution) and of the BR of the light Higgs boson to heavy neutrino pairs.

At this energy configuration, the expected integrated luminosity is not above 
 $1$ fb$^{-1}$, yielding the production of up to around $200$ (or more) heavy neutrino pairs via the light Higgs boson, for $m_{\nu_h}=50$ GeV, $110$ GeV$<m_{h_1}<150$ GeV and $\pi/4<\alpha<0.42 \pi$. For $m_{\nu_h}=60$ GeV, up to $100$ heavy neutrino pairs can be produced, for $130$ GeV$<m_{h_1}<160$ GeV, and $\pi/4<\alpha<0.45\pi$. Notice that the maximum cross section [of about $250$($100$) fb, for $m_{\nu_h}=50$($60$) GeV] occurs for a mixing angle of around $\pi/3$, due to the interplay between the production mechanism (maximised for $\alpha \to 0$) and the BR into heavy neutrino pairs (maximised for $\alpha \to \pi/2$), with the latter being the dominant part.

In the full luminosity scenario, several important distinctive signatures appear. We focus here on the $Z'$ boson and heavy neutrino pair production via $h_2$,
\begin{eqnarray}\label{h2_hnuhnu}
pp \rightarrow &h_2& \rightarrow \nu_h\nu_h \, , \\ \label{h2_ZpZp}
pp \rightarrow &h_2& \rightarrow Z'Z'\, .
\end{eqnarray}

\begin{figure}[!h]
  \subfloat[]{ 
  \label{gg-h2-nunu150}
  \includegraphics[angle=0,width=0.48\textwidth ]{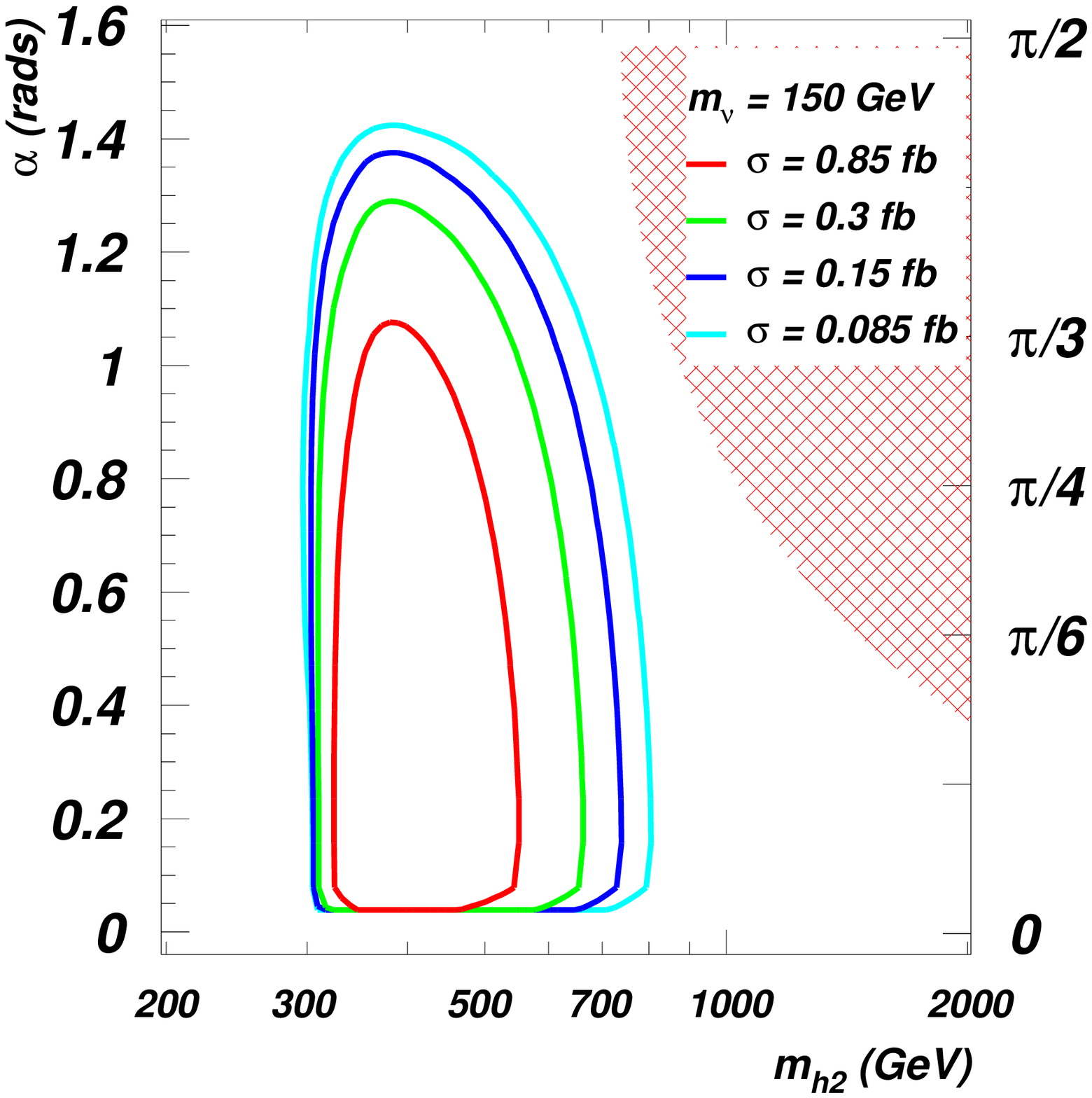}
}
  \subfloat[]{
  \label{gg-h2-nunu200}
  \includegraphics[angle=0,width=0.48\textwidth ]{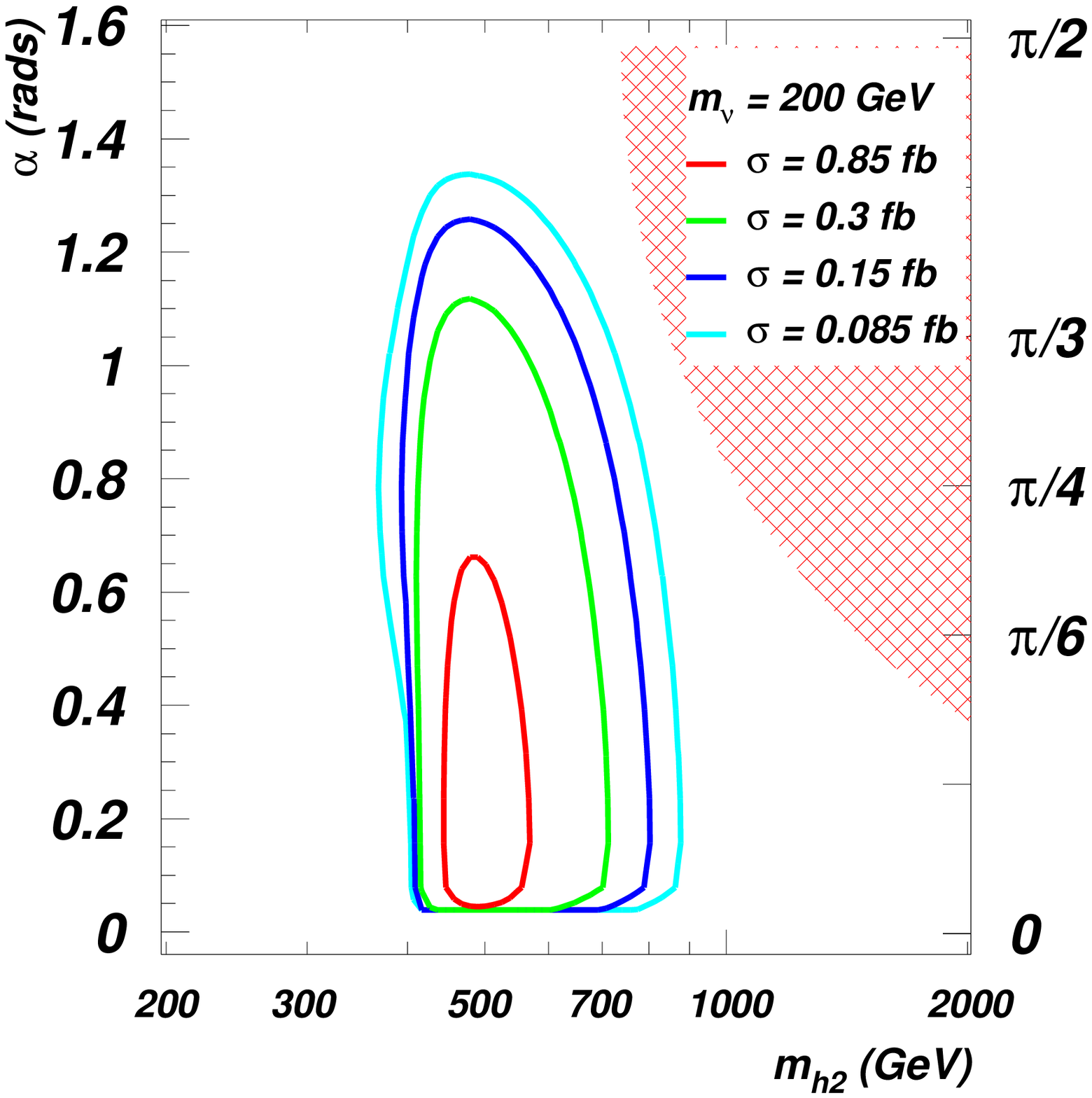}
}
  \vspace*{-0.25cm}
  \caption{\it Cross section times BR contour 
    plot for the process $pp\rightarrow 
    h_2\rightarrow \nu_{h}\nu_{h}$ at the LHC for $\sqrt{s}=14$ TeV, plotted
    in the ($m_{h_2}$--$\alpha$) plane, with (\ref{gg-h2-nunu150}) $m_{\nu_{h}}=150$ GeV and (\ref{gg-h2-nunu200}) $m_{\nu_{h}}=200$ GeV. The
  red-shadowed region is excluded by unitarity constraints (see section~\ref{subsect:Unitarity}).}
  \label{ggnunu150-200}
\end{figure}

In analogy with the previous case, figure~\ref{ggnunu150-200}
shows the cross sections for the process of eq.~(\ref{h2_hnuhnu}) (via the heavy Higgs boson), at the LHC for $\sqrt{s}=14$ TeV, for $m_{\nu_h}=150$ GeV and
$m_{\nu_h}=200$ GeV. In this case, as expected, the maximum cross section is for the complementary angle, i.e., for about $\pi /6$. The expected high integrated luminosity for this energy stage of the LHC is such that a good amount of events can be produced even for cross sections of around $1$ fb,
so that for both choices of the heavy neutrino mass, there exist values of $\alpha$ and $m_{h_2}$ for which the event rate could lead to the observation of this signature. In particular, for $m_{\nu_h}=150$ GeV, we find a cross section times BR of $0.85$ fb (that corresponds to around $250$ events, at most) for $320$ GeV$<m_{h_2}<520$ GeV and $0.03\pi<\alpha<0.33\pi$. For $m_{\nu_h}=200$ GeV, the same value of the cross section is for $450$ GeV$<m_{h_2}<550$ GeV and $0.03\pi<\alpha<0.21\pi$.

\begin{figure}[!h]
  \subfloat[]{ 
  \label{gg-h2-zpzp210}
  \includegraphics[angle=0,width=0.48\textwidth ]{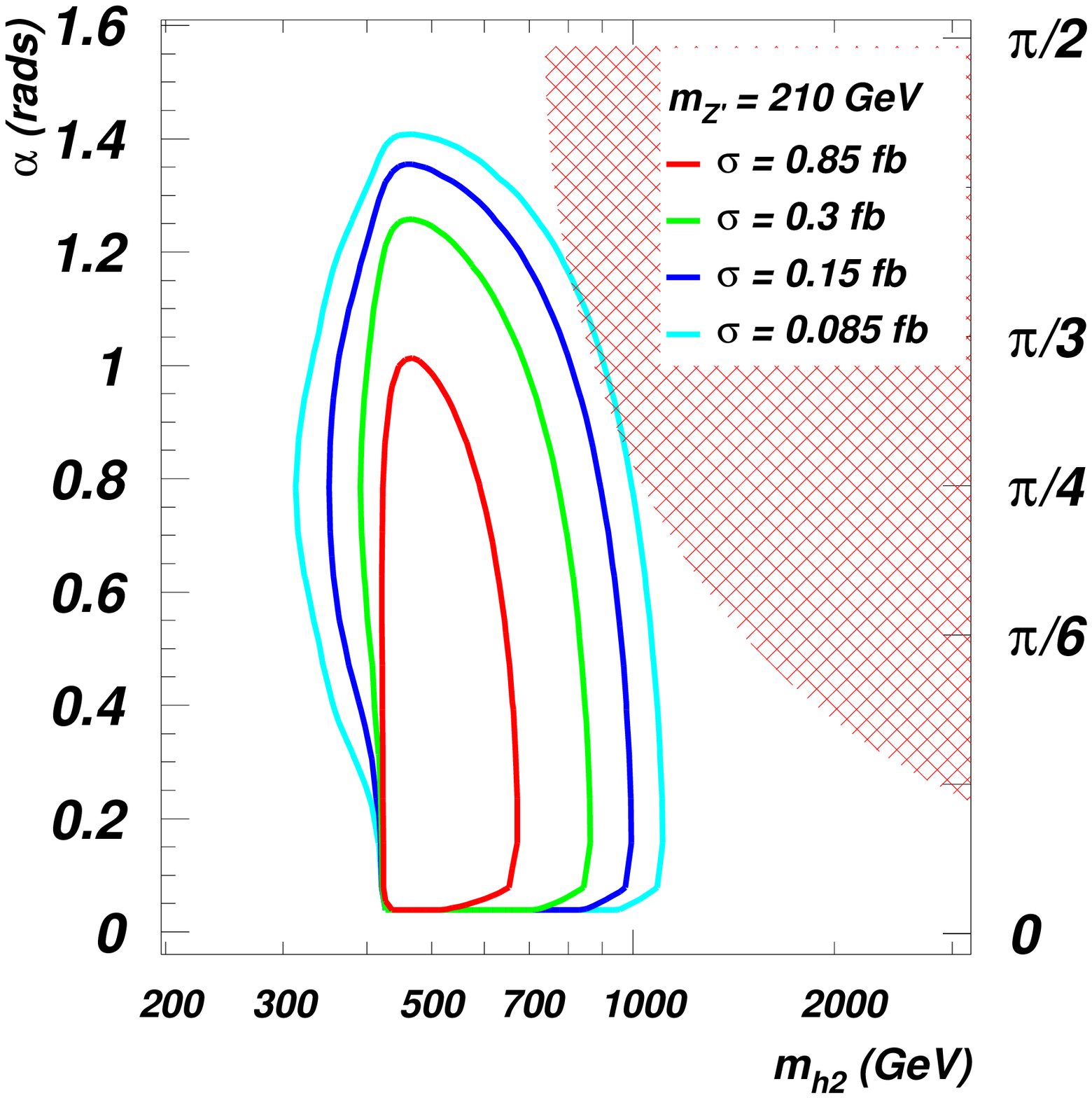}
}
  \subfloat[]{
  \label{gg-h2-zpzp280}
  \includegraphics[angle=0,width=0.48\textwidth ]{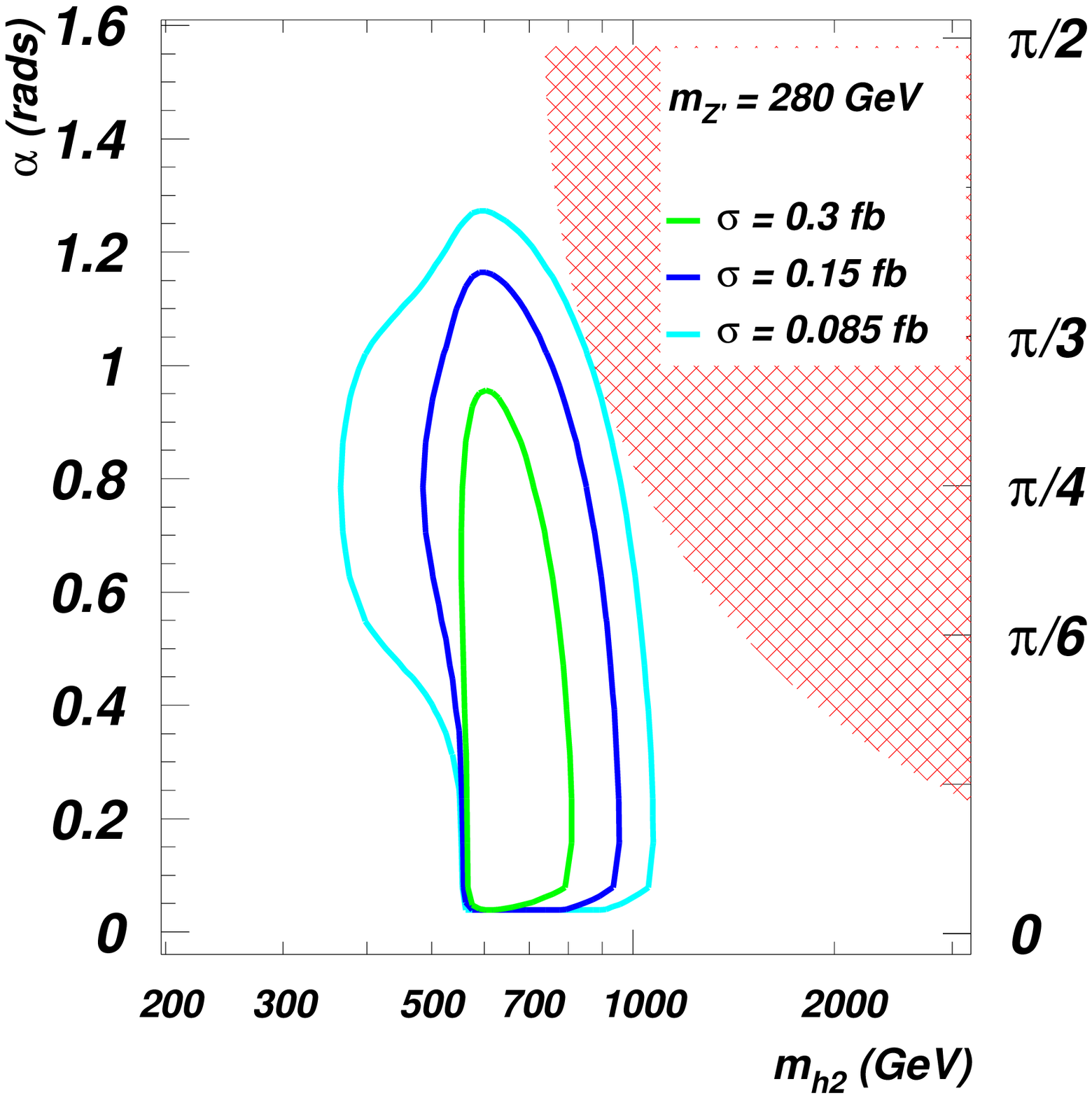}
}
  \vspace*{-0.25cm}
  \caption{\it Cross section times BR contour 
    plot for the process $pp\rightarrow 
    h_2\rightarrow Z' Z'$ at the LHC for $\sqrt{s}=14$ TeV, plotted in the ($m_{h_2}$--$\alpha$) plane, (\ref{gg-h2-zpzp210}) with $m_{Z'}=210$ GeV, and (\ref{gg-h2-zpzp280}) $m_{Z'}=280$ GeV. The red-shadowed region is excluded by unitarity constraints (see section~\ref{subsect:Unitarity}).}
  \label{ggzpzp210-280}
\end{figure}

In figure~\ref{ggzpzp210-280} we show the cross sections, in the ($m_{h_2}$--$\alpha$) plane, for the process of eq.~(\ref{h2_ZpZp}) (i.e., the $Z'$ boson pair production via the heavy Higgs boson) at the LHC for $\sqrt{s}=14$ TeV, for $m_{Z'}=210$ GeV and
$m_{Z'}=280$ GeV. This process is also favoured for small mixing angles, with a maximum cross section of $0.85$ fb for  $420$ GeV$<m_{h_2}<650$ GeV and $0.03\pi<\alpha<0.25\pi$, for the former $Z'$ mass. For a bigger value of the $Z'$ boson mass, such as $M_{Z'}=280$ GeV, a noticeable parameter space is still potentially accessible, with a maximum cross section of $0.3$ fb (corresponding to around $100$ events, at most) for $560$ GeV$<m_{h_2}<800$ GeV and
$0.03\pi<\alpha<0.19\pi$.
The $Z'$ bosons can subsequently decay into pairs of leptons, both electrons and muons, with $\displaystyle \sum _{\ell = e,\mu}$ BR($Z' \to \ell^+ \ell^-$) $\sim 30\%$ (see figure~\ref{Zp_BR}). Thus, in total, up to around $25$ events of heavy Higgs bosons decaying into $4$ leptons could be observed, in a completely different kinematical region from the decay into SM $Z$ bosons.


\section{Conclusion}\label{ch5:concl}
In this chapter, we have presented the results of our investigation of the scalar sector, with focus on the interaction with the gauge and fermion sectors.

First, we have presented a summary of the existing experimental constraints and the full analysis of the high energy theoretical constraints on the scalar sector.

We have briefly summarised the unitarity bound, whose scope was to clarify the role of the two Higgs bosons in the unitarisation of vector and scalar bosons scattering amplitudes, that we know must hold at any energy scale.

Using the equivalence theorem, we have evaluated the spherical partial wave amplitude of all possible two-to-two scatterings in the scalar Lagrangian at an infinite energy, identifying the $zz \rightarrow zz$ and $z'z'\rightarrow z'z'$ processes as the most relevant scattering channels for this analysis ($z^{(')}$ is the would-be Goldstone boson of the $Z^{(')}$ vector boson).

Hence, we have shown that these two channels impose an upper bound on the two Higgs masses: the light one cannot exceed the SM bound while the limit on the heavy one is established by the singlet Higgs VEV, whose value is presently constrained by LEP and may shortly be extracted by experiments following 
the discovery of the $Z'_{B-L}$ boson.


We have then investigated the triviality and vacuum stability conditions of the pure $B-L$ model with a particular view to define the phenomenologically viable regions of the parameter space of the scalar sector, by computing the complete set of RGEs (gauge, scalar and fermion) at the one-loop level in presence of all available experimental constraints. The RGE dependence on the Higgs masses and couplings (including mixings) has been studied in detail for selected heavy neutrino masses and couplings as well as discrete choices of the singlet Higgs  VEV.

Altogether, we have found that there exist configurations of the model for which its validity is guaranteed up to energy scales well beyond those reachable at the LHC while at the same time enabling the CERN hadron collider to probe its scalar sector in Higgs mass and coupling regions completely different from those accessible to the SM. Furthermore, we have shown that investigations of the Higgs sector of this extended scenario may also lead to constraints on other areas, such as the (heavy) neutrino and $Z'$ sectors (the latter indirectly, through the VEV of the singlet Higgs state directly intervening in the scalar RGEs).


Finally, we have investigated the production and decay phenomenology of both Higgs states of the minimal $B-L$ model at the LHC, at both the foreseen energy stages of the LHC (i.e., $\sqrt{s}=7$ and $14$ TeV) and with the corresponding integrated  luminosities. While virtually all relevant production and decay processes of the two Higgs states of the model have been investigated, we have eventually paid particular attention to those that are peculiar to the described $B-L$ scenario. The phenomenological analysis has been carried out in the presence of all available theoretical and experimental constraints and by exploiting numerical programs at the parton level. While many Higgs signatures already existing in the SM could be replicated in the case of its $B-L$ version, in either of the two Higgs states of the latter (depending on their mixing), it is more important to notice that several novel Higgs processes could act as hallmarks of the minimal $B-L$ model. These include Higgs production via gluon-gluon fusion, of either the light or heavy Higgs state, the former produced at the lower energy stage of the CERN collider and decaying in two heavy neutrinos, and the latter produced at the higher energy stage of such a machine and decaying not only in heavy neutrino pairs but also in $Z'$ boson pairs (and eventually in light Higgs boson pairs). For each of these signatures we have in fact found parameter space regions where the event rates are sizable and potentially amenable to discovery.

While, clearly, detailed signal-to-background analyses will have to either confirm or disprove the possibility of the latter, our results have laid the basis for the phenomenological exploitation of the Higgs sector of the minimal $B-L$ model at the LHC.  
\def\baselinestretch{1}
\chapter{Discussion}
\ifpdf
    \graphicspath{{Conclusions/ConclusionsFigs/PNG/}{Conclusions/ConclusionsFigs/PDF/}{Conclusions/ConclusionsFigs/}}
\else
    \graphicspath{{Conclusions/ConclusionsFigs/EPS/}{Conclusions/ConclusionsFigs/}}
\fi

This Thesis studies the phenomenology of a minimal extension of the SM, namely, the $B-L$ model. This well-motivated framework naturally implements the see-saw mechanism for neutrino mass generation, being the RH neutrinos required for anomaly cancellation. It also explains the accidental and non-anomalous $U(1)_{B-L}$ global symmetry in the SM, being this, here, the remnant after spontaneous symmetry breaking of its local counterpart. It finally introduces scope in the Higgs boson sector, that lacks any experimental observation.

We think that the goal of a proficient interaction with experimentalists has been achieved. A joint effort has been fundamental in the studies concerning the signatures of the model and the strategy for their observation. Also, some of the gaps in the preparation towards real data have been filled in this work. The model under discussion has been studied in many of its aspects, identifying
 novel and exciting signatures.

The phenomenology of the new particles of the model has been analysed, possibly, in the most comprehensive way. The peculiarities with respect to the SM, and to some of its extensions, have been described and highlighted, focusing on the most interesting signatures one may observe.

In the attempt of a complete study of the model, our work has defined the allowed regions of its parameter space and presented a detailed study of some of its signatures, in the gauge, fermion and scalar sectors, at the LHC, waiting for the data to confirm or disprove this TeV scale realisation of the minimal $B-L$ model.

\section{Summary of the results} 

We summarise here the most important results we achieved in this Thesis. For a more detailed description, see the partial conclusions for each chapter, in section~\ref{ch3:concl}, section~\ref{ch4:concl}, and section~\ref{ch5:concl}, for the gauge, fermion, and scalar sector, respectively.

The non-observation of any $Z'$ boson nor of Higgs boson at past and recent colliders results in lower bounds on the parameter space of the model. We have presented the most up-to-date constraints for the former at Tevatron in section~\ref{sec:expbounds:Zp_Tev}. Complementary to it, theoretical arguments can mainly give upper bounds on the free parameters. The topics we studied in this Thesis are the unitarity bound of the scalar sector (that we summarised in  section~\ref{subsect:Unitarity}) and the constraints coming from the analysis of the RGEs of the model at one loop (that we calculated and collected in appendix~\ref{App:RGE}). The requirement that the evolution of the parameters does not hit any Landau pole, up to a specific cut-off scale, is the so-called triviality bound, that we studied for the gauge couplings in section~\ref{subsubsect:RGE_gauge}, and for the scalar parameters in section~\ref{subsubsect:RGE_scalar}. In the latter, this analysis has been combined with the study of the vacuum stability condition, leading also to lower bounds on the Higgs boson masses. It is remarkable that there exist configurations of the model for which its validity is guaranteed up to very high energy scales, well beyond those reachable at the LHC, that are completely different form the SM ones. In particular, a $h_1$ boson lighter than the SM limit of $114.4$ GeV is allowed for $\alpha > \pi/3$ and a suitable choice of the other parameters.

We have then investigated the collider signatures of this model at the LHC, focusing on the gauge and fermion sectors. In chapter~\ref{Ch:3} we have shown that the $Z'_{B-L}$ boson is dominantly coupled to leptons, whose total BR sums up to $75\%$. Also, it can decay into heavy neutrino pairs, with a BR up to around $20\%$. We finally presented (in section~\ref{subsect:disc_power}) the parton level discovery potential of the $Z'_{B-L}$ boson at Tevatron and the LHC in Drell-Yan processes. The LHC at $\sqrt{s}=7$ TeV will be able to give similar results to Tevatron, in the electron channel and in the low mass region, and is able to extend its reach for heavier $Z'$ boson, up to masses of $1.25(1.20)$ TeV for electrons(muons). When the data from the high energy runs at the LHC become available, the discovery reach of the $Z'_{B-L}$ boson will be extended towards very high masses and small couplings in regions of parameter space well beyond the reach of Tevatron.
We also have shown that greater sensitivity to the $Z'_{B-L}$ resonance is provided by the electron channel, that at the LHC has better energy resolution than the muon channel. This in turn could enable the estimate of the gauge boson width. Finally, we have commented that the inclusion of further background, as well as a realistic detector simulation, will not have a considerable impact on the results that we have presented.

The phenomenology of heavy neutrinos has been presented in chapter~\ref{Ch:4}. Due to the see-saw mechanism, that allows for a tiny LH component, the main decay modes for the heavy neutrinos (i.e., the decay into SM gauge bosons) have very small partial widths. Hence, in a large portion of the parameter space, the heavy neutrino is a long-lived particle, with a proper lifetime that can be comparable to (or exceed) the $b$ quark's one (for example, they are equal for $m_{\nu _h}=200$ GeV and $m_{\nu _l}=10^{-2}$ eV). We have emphasised that an independent measurement of the heavy neutrino lifetime (through the displacement of its decay vertex) and its mass may lead to infer the absolute mass of the SM-like neutrinos. Therefore, we have proposed a way to measure the heavy neutrino mass. We have described that heavy neutrino pairs, produced by the $Z'$ boson, give rise to multi-lepton decays of the latter and we have identified the tri-lepton one as the most interesting one for a detailed study. We have proposed cuts and a search strategy, identifying the suitable distributions one should look at. A parton level study, in section~\ref{sect:trilep_parton}, has shown that simple cuts can effectively reduce the backgrounds, that the peak corresponding to the heavy neutrino enables a rather precise measure of its mass and that a striking signature for this model is the simultaneous presence of peaks for the intermediate $Z'$ boson and for the heavy neutrinos. Finally, a full detector simulation, in section~\ref{sect:trilep_detector}, has validated the results of the parton level study in a more realistic framework, showing that a discovery could be possible with less than $100$ fb$^{-1}$. 

Finally, in chapter~\ref{Ch:5}, we have laid the basis for the investigation of the Higgs sector. The most appealing result is that multi-lepton decays of a Higgs boson, via heavy neutrino pairs, are at the same time rather peculiar and they might allow for the discovery of the intermediate scalar in the early stage of the LHC.

We can reasonably conclude that the heavy neutrinos truly carry the hallmarks of the $B-L$ model at colliders.

\section{Outlook on future work}

In this Thesis, we have mentioned several topics that were left for future investigations, primarily for time limitations. We further discuss here some of them as a possible continuation of the work undertaken. Also, further developments and new studies are addressed.

When discussing the phenomenology of the $Z'$ boson, in chapter~\ref{Ch:3}, we have discussed that the asymmetries of the decay products stemming from the $Z'_{B-L}$ are trivial at the peak. Hence, we have decided to not study them. However, as already pointed out, asymmetries become important in the interference region, especially just before the $Z'$ boson peak, where the $Z-Z'$ interference will effectively provide an asymmetric distribution somewhat milder than the case in which there is no $Z'$ boson. A first attempt to study them is in Ref.~\cite{Coutinho:2011xb}, though further improvements may be possible, for example by studying the asymmetries of the heavy neutrinos, which, as we have shown, have just axial couplings to the gauge boson.

Another important aspect that needs to be quantitatively addressed is the impact on the model of the electroweak precision observables. Although the impact on the gauge sector has been found to be negligible in the pure $B-L$ model, at one loop (see, e.g., \cite{Cacciapaglia:2006pk}), one is left with the scalar and fermion sectors. The former has been studied in Ref.~\cite{Dawson:2009yx}, though without considering heavy neutrinos, that, as we highlighted, might contribute.

In chapter~\ref{Ch:4}, we have highlighted that the tri-lepton signature via heavy neutrino pairs is one of the most distinctive signatures of the pure $B-L$ model at colliders. We have successfully studied it with a full detector simulation, concluding that a rather precise measurement of the heavy neutrino mass is a realistic task. It might be appealing to
improve the undertaken
 analysis: the implementation of the full light neutrino masses and mixing angles, as well as considering the heavy neutrinos as not degenerate, would be a further step, whose aim is to show the feasibility of extracting the mass of each light neutrino in a realistic case. Further, the precision of the heavy neutrino mass measurement and the impact of fully decayed tau leptons ought to be quantitatively addressed.

Likewise, in chapter~\ref{Ch:5}, it has been pointed out that the neutrino pair production could be important for the light Higgs boson discovery in the pure $B-L$ model, in a range of masses that happens to be critical in the SM, since here the SM Higgs boson passes from decaying dominantly into $b$ quark pairs to a region in masses in which the decay into $W$ boson pairs is the prevailing one. It is also important to note that the former signature provides poor sensibility, being it many orders of magnitude below the natural QCD background.
Therefore, the decay into heavy neutrino pairs is phenomenologically very important, especially at the LHC in its early stage (i.e., for $\sqrt{s}=7$ TeV). Beside being an interesting feature of the $B-L$ model if $m_{\nu_h} < M_W$, as it allows for multi-lepton signatures of the light Higgs boson (some of them being much cleaner than the decay into $b$ quark pairs), these decays could provide the first hints of a fundamental scalar particle. The analysis of the tri-lepton signature for the Higgs boson is certainly a very interesting task to pursue at the LHC at both foreseen CM energies. Furthermore, a full analysis of the kinematic distribution of the decay products is also crucial for the identification of the spin of the intermediate resonance. Although in a very different kinematic region, the $Z'$ boson gives rise to the same decay products, hence the intermediate resonance has to be distinguished, also for model discrimination purposes.

Finally, the embedding of the model in GUTs could be pursued. In this respect, we note that the supersymmetric version of the pure $B-L$ model has already been considered: see, for example, \cite{Khalil:2007dr,Kikuchi:2008xu,Khalil:2008ps,Allahverdi:2008jm,Khalil:2009tm,Khalil:2011tb} and references therein. This implementation combines the virtues of the pure $B-L$ model (namely, the natural and dynamic implementation of neutrino masses) with the elegant solution of the hierarchy problem of the SM, as well as naturally providing a dark matter candidate, that are typical for a supersymmetric framework. Because of the enlarged spectrum, it might be interesting to apply the collider studies of this Thesis to the supersymmetric $B-L$ model, analysing the discovery potential at the LHC of the new particles and the new signatures arising in this enlarged framework.



\label{Ch:concl}

\appendix
\chapter{Model implementation}\label{App:calchep}
\ifpdf
    \graphicspath{{Chapter3/Chapter3Figs/PNG/}{Chapter3/Chapter3Figs/PDF/}{Chapter3/Chapter3Figs/}}
\else
    \graphicspath{{Chapter3/Chapter3Figs/EPS/}{Chapter3/Chapter3Figs/}}
\fi

In this appendix is presented the implementation of the pure $B-L$ model into the numerical tools that have been used. Also, the Feynman rules are collected in section~\ref{sect:feym_rules}.

\section{Implementation of the pure $B-L$ model}

Most of the numerical analysis of this Thesis \footnote{A part from the full detector simulation of the tri-lepton signature (see section~\ref{sect:trilep_detector}), the study of the RGEs (see section~\ref{subsubsect:RGE_scalar}), done with \textsc{mathematica} \cite{mathematica5}, and the study of the unitarity bound (see section~\ref{subsect:Unitarity}).} has been carried out using the \textsc{calchep} package~\cite{Pukhov:2004ca}, an automated tool for tree level calculation of physical observables.

We implemented the pure $B-L$ model, for which the covariant derivative given by eq.~(\ref{cov_der}) reads
\begin{equation}\label{cov_der_pure}
D_{\mu}\equiv \partial _{\mu} + ig_S T^{\alpha}G_{\mu}^{\phantom{o}\alpha} 
+ igT^aW_{\mu}^{\phantom{o}a} +ig_1YB_{\mu} +ig_1'Y_{B-L}B'_{\mu}\, .
\end{equation}
For its straightforward implementation in \textsc{calchep}, we have used the \textsc{lanhep} module \cite{Semenov:1996es}. This package, beside providing the suitable output for \textsc{calchep}, also derives the Feynman rules of the model, that we collected in section~\ref{sect:feym_rules}.

The availability of the model implementation into \textsc{calchep} in both the unitary and t'Hooft-Feynman gauges allowed us to perform powerful cross-checks to test the consistency of the model itself.

The implementation of the gauge sector is quite straightforward. Since there is no mixing between the SM $Z$ and  the $Z'_{B-L}$ bosons, one just needs to define a new heavy neutral gauge boson together with the simplified covariant derivative given by eq.~(\ref{cov_der_pure}), and the charge assignments in table~\ref{tab:quantum_number_assignation}. For the scalar sector, the mixing between mass and gauge eigenstates of the two Higgs bosons, as well as the reformulation of the Lagrangian parameters in terms of the physical quantities, has been done accordingly to eqs.~(\ref{scalari_autostati_massa})--(\ref{inversion}).

Finally, regarding the fermion sector, the neutrinos ought to be carefully handled, as explicitly described in the following subsection.

\subsection{Neutrino eigenstates}
The implementation of the neutrino sector is somewhat more complicated. Majorana-like Yukawa terms are present in eq.~(\ref{L_Yukawa}) for the RH neutrinos, therefore one must implement this sector such that the gauge invariance of the model is explicitly preserved.  This can be done as follows. As a first step we rewrite Dirac neutrino fields in terms of Majorana ones using the following general substitution:
\begin{equation}\label{D-M}
\nu^D = \frac{1-\gamma _5}{2}\nu_L +  \frac{1+\gamma _5}{2}\nu_R\, ,
\end{equation}
where $\nu^D$ is a Dirac field and $\nu_{L(R)}$ are its left(right) Majorana components. If we perform the substitution of eq.~(\ref{D-M}) in the neutrino sector of the SM, we will have an equivalent theory formulated in terms of Majorana neutrinos consistent with all experimental constraints.

The derivation of the mass eigenstates is the next step, and it has been explained in detail in section~\ref{sect:neutrino_masses}.

The last subtle point is the way the Lagrangian has to be written to meet the requirements of \textsc{lanhep}. In particular, this regards the Majorana-like Yukawa terms for the RH neutrinos [the last term in eq.~(\ref{L_Yukawa})]. In order to explicitly preserve gauge invariance, this term has to be written, in two-component notation, as:
\begin{equation}
- y^M \nu ^c \frac{1+\gamma _5}{2} \nu \chi + \rm{h.c.}\, ,
\end{equation}
where $\nu$ is the Dirac field of eq.~(\ref{D-M}), whose Majorana components $\nu_{L,R}$  mix as in eq.~(\ref{nu_mixing}).

Altogether, this implementation is an explicit gauge-invariant formulation of the neutrino sector suitable for the \textsc{lanhep} tool. The specific interactions pertaining to the heavy neutrinos are collected in the following subsection.

\subsection{Feynman rules involving heavy neutrinos}\label{Appsect:nuh_feynman_rules}
Given their importance, and to help the reader, we list here the Feynman rules involving the heavy neutrinos in the pure $B-L$ model. The intervening quantities are defined in the main text. 

\scalebox{1.33}{
\begin{picture}(170,79)(0,30)
\unitlength=1.0 pt
\SetScale{1.0}
\SetWidth{0.1}      
\scriptsize    
\Text(20.0,65.0)[r]{$\nu _h$}
\Line(0.0,60.0)(28.0,60.0) 
\Text(50.0,70.0)[l]{$\ell$}
\Line(28.0,60.0)(49.0,70.0) 
\Text(50.0,50.0)[l]{$W$}
\Photon(28.0,60.0)(49.0,50.0){2.0}{4}
\Text(30,20)[b] {$\displaystyle \frac{\sqrt{2}e}{4\sin{\vartheta _W}}\sin{\alpha _\nu}$}
\end{picture} 
}
\scalebox{1.33}{
\begin{picture}(170,79)(0,30)
\unitlength=1.0 pt
\SetScale{1.0}
\SetWidth{0.1}      
\scriptsize    
\Text(20.0,65.0)[r]{$\nu _h$}
\Line(0.0,60.0)(28.0,60.0) 
\Text(50.0,70.0)[l]{$\nu _l$}
\Line(28.0,60.0)(49.0,70.0) 
\Text(50.0,50.0)[l]{$Z$}
\Photon(28.0,60.0)(49.0,50.0){2.0}{4}
\Text(30,20)[b] {$\displaystyle -\frac{e}{4\sin{\vartheta _W}\cos{\vartheta _W}}\sin{2\alpha _\nu}$}
\end{picture} 
}
\\\\
\scalebox{1.33}{
\begin{picture}(170,79)(0,30)
\unitlength=1.0 pt
\SetScale{1.0}
\SetWidth{0.1}      
\scriptsize    
\Text(20.0,65.0)[r]{$\nu _h$}
\Line(0.0,60.0)(28.0,60.0) 
\Text(50.0,70.0)[l]{$\nu _l$}
\Line(28.0,60.0)(49.0,70.0) 
\Text(50.0,50.0)[l]{$h_1$}
\DashLine(28.0,60.0)(49.0,50.0){3.0} 
\Text(70,20)[b] {$\displaystyle \frac{1}{2x}\left( -\sqrt{2}x y^\nu c_\alpha \cos{2\alpha _\nu}+m_{\nu _h} \sin{2\alpha _\nu}s_\alpha\right)$}
\end{picture} 
}%
\scalebox{1.33}{
\begin{picture}(170,79)(0,30)
\unitlength=1.0 pt
\SetScale{1.0}
\SetWidth{0.1}      
\scriptsize    
\Text(20.0,65.0)[r]{$\nu _h$}
\Line(0.0,60.0)(28.0,60.0) 
\Text(50.0,70.0)[l]{$\nu _l$}
\Line(28.0,60.0)(49.0,70.0) 
\Text(50.0,50.0)[l]{$h_2$}
\DashLine(28.0,60.0)(49.0,50.0){3.0} 
\Text(70,20)[b] {$\displaystyle \frac{1}{2x}\left( -\sqrt{2}x y^\nu s_\alpha \cos{2\alpha _\nu}-m_{\nu _h} \sin{2\alpha _\nu}c_\alpha\right)$}
\end{picture} 
}
\\\\
\scalebox{1.33}{
\begin{picture}(170,79)(0,30)
\unitlength=1.0 pt
\SetScale{1.0}
\SetWidth{0.1}      
\scriptsize    
\Text(20.0,65.0)[r]{$\nu _h$}
\Line(0.0,60.0)(28.0,60.0) 
\Text(50.0,70.0)[l]{$\nu _l$}
\Line(28.0,60.0)(49.0,70.0) 
\Text(50.0,50.0)[l]{$Z'_{B-L}$}
\Photon(28.0,60.0)(49.0,50.0){2.0}{4}
\Text(30,20)[b] {$\displaystyle g'_1\sin{2\alpha _\nu}$}
\end{picture} 
}
\scalebox{1.33}{
\begin{picture}(170,79)(0,30)
\unitlength=1.0 pt
\SetScale{1.0}
\SetWidth{0.1}      
\scriptsize    
\Text(20.0,65.0)[r]{$\nu _h$}
\Line(0.0,60.0)(28.0,60.0) 
\Text(50.0,70.0)[l]{$\nu _h$}
\Line(28.0,60.0)(49.0,70.0) 
\Text(50.0,50.0)[l]{$Z'_{B-L}$}
\Photon(28.0,60.0)(49.0,50.0){2.0}{4}
\Text(30,20)[b] {$\displaystyle g'_1\cos{2\alpha _\nu}$}
\end{picture} 
}
\\\\
\begin{displaymath}
\mbox{where}\ \ \ 
y^\nu = \frac{\sqrt{2 m_{\nu _l} m_{\nu _h}}}{v}\, ,\ \ \ 
\sin{2\alpha _\nu} = -2 \frac{y^\nu \frac{v}{\sqrt{2}}}{\sqrt{4 (y^\nu \frac{v}{\sqrt{2}})^2+m_{\nu _h}^2}}\, , \ \ \ 
\cos{2\alpha _\nu} = \frac{m_{\nu _h}}{\sqrt{4(y^\nu \frac{v}{\sqrt{2}})^2+m_{\nu _h}^2}}.
\end{displaymath}

\section{One-loop vertices}
For the correct analysis of the scalar sector of chapter~\ref{Ch:5}, one-loop vertices need to be considered. They couple the Higgs fields to massless particles, such as photon and gluon, for which no tree level interaction exists. In fact, such couplings are mediated by a loop of either (heavy) quarks or massive gauge bosons (or both). Despite the mass suppression, due to the integration of the particle in the loop, these interactions can still be sizable. In particular, the coupling to the gluons is responsible for the most effective production mechanism for the Higgs boson at the LHC, the so-called gluon-gluon fusion.

A detailed description of the implementation is as follows.

\begin{itemize}
\item The one-loop vertices $g-g-h_1(h_2)$, $\gamma -\gamma-h_1(h_2)$ and $\gamma -Z(Z')-h_1(h_2)$ via $W$ gauge bosons and heavy quarks (top, bottom, and charm) have been implemented, adapting the formulas in Ref~\cite{Gunion:1989we}.
\item Running masses for top, bottom, and charm quarks have been considered, evaluated at the Higgs boson mass: $Q= m_{h_1}(m_{h_2})$ (depending on which scalar boson is involved in the interaction).
\item Running of the QCD coupling constant has been included, at two loops with $5$ active flavours.
\end{itemize}
Finally, the NLO QCD $k$-factor for the gluon-gluon fusion process \cite{Graudenz:1992pv,Spira:1995rr,Djouadi:2005gi} \footnote{Notice 
that in Ref.~\cite{Spira:1995rr} (Ref.~\cite{Djouadi:2005gi}), $m_t=174$($178$) GeV, while we used $m_t=172.5$ GeV as top quark pole mass value.} has been used. Regarding the other processes, we decided to not implement their $k$-factors since they are much smaller in comparison.

It shall be noticed that the implementation of the formulas for the running quark masses and for the running QCD coupling constant, as well as the $k$-factor for the gluon-gluon fusion, have been done for the consistency of the NLO evaluation and to get in agreement with the cross sections in the literature (see, e.g., Ref.~\cite{Djouadi:2005gi}).

\vspace{0.5cm}

In the following subsection we list the complete set of tree level Feynman rules for the pure $B-L$ model, as given by \textsc{lanhep}.
Where possible, the couplings have been expanded upon the mass eigenstate basis.
The one-loop vertices are not included. We remand the reader to the formulas in Ref.~\cite{Gunion:1989we} for the SM, of which ours are a straightforward extension.

Concerning the notation we use, the following applies.
\begin{itemize}
\item $A$ is the photon, $G$ is the gluon.
\item $V_F$ is the Goldstone field of the vector $V$.
\item $C^V$ is the ghost field related to the vector $V$, and $\bar{C}^V$ is the corresponding conjugated field.
\item $p_n$ is the four-momentum of the $n^{th}$ vector field (as it appears in the left column).
\item $s_W = \sin_W$ is the sine of the EW angle (and $c_W$ the cosine).
\item $s_\alpha = \sin {\alpha}$ is the sine of the scalar mixing angle $\alpha$ (and $c_\alpha$ the cosine).
\item $s_{\alpha _\nu} = \sin {\alpha _\nu}$ is the sine of the neutrino mixing angle $\alpha _\nu$ (and $c_{\alpha _\nu}$ the cosine).
\end{itemize}

\section{Feynman rules for the pure $B-L$ model}\label{sect:feym_rules}
\begin{center}

\begin{tabular}{|l|l|} \hline
Fields in the vertex & Variational derivative of Lagrangian by fields \\ \hline
${A}_{\mu }$ \phantom{-} $W^+{}_{\nu }$ \phantom{-} $W^-{}_{\rho }$ \phantom{-}  &
	$- e\big(p_2^\rho g^{\mu \nu} -p_2^\mu g^{\nu \rho} -p_3^\nu g^{\mu \rho} +p_3^\mu g^{\nu \rho} +p_1^\nu g^{\mu \rho} -p_1^\rho g^{\mu \nu} \big)$\\[2mm]
${A}_{\mu }$ \phantom{-} $W^+{}_{\nu }$ \phantom{-} $W^-_F{}_{}$ \phantom{-}  &
	$ i \cdot e \cdot M_W\cdot g^{\mu \nu} $\\[2mm]
${A}_{\mu }$ \phantom{-} $W^+_F{}_{}$ \phantom{-} $W^-{}_{\nu }$ \phantom{-}  &
	$- i \cdot e \cdot M_W\cdot g^{\mu \nu} $\\[2mm]
${A}_{\mu }$ \phantom{-} $W^+_F{}_{}$ \phantom{-} $W^-_F{}_{}$ \phantom{-}  &
	$ e\big(p_3^\mu -p_2^\mu \big)$\\[2mm]
$\bar{C}^A{}_{}$ \phantom{-} $C^{W+}{}_{}$ \phantom{-} $W^-{}_{\mu }$ \phantom{-}  &
	$- e\cdot p_1^\mu $\\[2mm]
$\bar{C}^A{}_{}$ \phantom{-} $C^{W-}{}_{}$ \phantom{-} $W^+{}_{\mu }$ \phantom{-}  &
	$ e\cdot p_1^\mu $\\[2mm]
$\bar{b}{}_{a p }$ \phantom{-} $b{}_{b q }$ \phantom{-} ${A}_{\mu }$ \phantom{-}  &
	$\frac{1}{3} e\delta_{p q} \gamma_{a c}^\mu \cdot \delta_{c b} $\\
\hline
\end{tabular}

\begin{tabular}{|l|l|} \hline
Fields in the vertex & Variational derivative of Lagrangian by fields \\ \hline
$\bar{c}{}_{a p }$ \phantom{-} $c{}_{b q }$ \phantom{-} ${Z}_{\mu }$ \phantom{-}  &
	$-\frac{1}{12}\frac{ e}{ c_w \cdot s_w}\delta_{p q} \gamma_{a c}^\mu \big( (3-4 s_w {}^2)\cdot (1-\gamma^5)_{c b} -4 s_w{}^2 \cdot (1+\gamma^5)_{c b} \big)$\\[2mm]
$\bar{c}{}_{a p }$ \phantom{-} $c{}_{b q }$ \phantom{-} $Z_F{}_{}$ \phantom{-}  &
	$\frac{1}{2}\frac{ i \cdot e \cdot M_c}{ M_W \cdot s_w}\delta_{p q} \cdot \gamma_{a b}^5 $\\[2mm]
$\bar{c}{}_{a p }$ \phantom{-} $c{}_{b q }$ \phantom{-} $Z^\prime{}_{\mu }$ \phantom{-}  &
	$-\frac{1}{3} g^{\prime}_1\delta_{p q} \gamma_{a c}^\mu \cdot \delta_{c b} $\\[2mm]
$\bar{c}{}_{a p }$ \phantom{-} $d{}_{b q }$ \phantom{-} $W^+{}_{\mu }$ \phantom{-}  &
	$-\frac{1}{4}\frac{ e \cdot\sqrt{2} \cdot V_{cd}}{ s_w}\cdot (1-\gamma^5)_{c b} \delta_{p q} \gamma_{a c}^\mu $\\[2mm]
$\bar{c}{}_{a p }$ \phantom{-} $d{}_{b q }$ \phantom{-} $W^+_F{}_{}$ \phantom{-}  &
	$-\frac{1}{4}\frac{ i \cdot e \cdot M_c \cdot\sqrt{2} \cdot V_{cd}}{ M_W \cdot s_w}\cdot (1-\gamma^5)_{a b} \delta_{p q} $\\[2mm]
$\bar{c}{}_{a p }$ \phantom{-} $s{}_{b q }$ \phantom{-} $W^+{}_{\mu }$ \phantom{-}  &
	$-\frac{1}{4}\frac{ e \cdot\sqrt{2} \cdot V_{cs}}{ s_w}\cdot (1-\gamma^5)_{c b} \delta_{p q} \gamma_{a c}^\mu $\\[2mm]
$\bar{c}{}_{a p }$ \phantom{-} $s{}_{b q }$ \phantom{-} $W^+_F{}_{}$ \phantom{-}  &
	$\frac{1}{4}\frac{ i \cdot e \cdot\sqrt{2} \cdot V_{cs}}{ M_W \cdot s_w}\delta_{p q} \big( M_s\cdot (1+\gamma^5)_{a b} - M_c\cdot (1-\gamma^5)_{a b} \big)$\\
\hline
\end{tabular}

\begin{tabular}{|l|l|} \hline
Fields in the vertex & Variational derivative of Lagrangian by fields \\ \hline
$\bar{e}{}_{a }$ \phantom{-} $\nu _{h1}{}_{b }$ \phantom{-} $W^-_F{}_{}$ \phantom{-}  &
	$\frac{1}{4}\frac{ i}{ M_W   s_w}\big( s_{\alpha_\nu}   e   M_e  \sqrt{2}  (1-\gamma^5)_{a b} +2 s_w   c_{\alpha_\nu}   M_W   y^D_{n_1}  (1+\gamma^5)_{a b} \big)$\\[2mm]
${G}_{\mu p }$ \phantom{-} ${G}_{\nu q }$ \phantom{-} ${G}_{\rho r }$ \phantom{-}  &
	$ g_sf_{p q r} \big(p_3^\nu g^{\mu \rho} -p_3^\mu g^{\nu \rho} +p_1^\rho g^{\mu \nu} -p_1^\nu g^{\mu \rho} -p_2^\rho g^{\mu \nu} +p_2^\mu g^{\nu \rho} \big)$\\[2mm]
$\bar{C}^G{}_{p }$ \phantom{-} $C^G{}_{q }$ \phantom{-} ${G}_{\mu r }$ \phantom{-}  &
	$ g_s\cdot p_2^\mu f_{p q r} $\\[2mm]
$h_1{}_{}$ \phantom{-} $h_1{}_{}$ \phantom{-} $h_1{}_{}$ \phantom{-}  &
	$-\frac{3}{2}\frac{1}{ e \cdot v{}^2  \cdot x}\big(2 c_{\alpha}{}^2  s_w c_{\alpha} M_{h_2}{}^2  M_W x-2 c_{\alpha}{}^2  s_w c_{2\alpha} c_{\alpha} M_{h_2}{}^2  M_W x$ \\[2mm]
  & $+2 c_{\alpha}{}^2  s_w c_{\alpha} M_{h_1}{}^2  M_W x+2 c_{\alpha}{}^2  s_w c_{2\alpha} c_{\alpha} M_{h_1}{}^2  M_W x$ \\[2mm]
  & $- s_{\alpha}{}^3  e M_{h_1}{}^2  v{}^2 + s_{\alpha}{}^3  c_{2\alpha} e M_{h_1}{}^2  v{}^2 - s_{\alpha}{}^3  e M_{h_2}{}^2  v{}^2 - s_{\alpha}{}^3  c_{2\alpha} e M_{h_2}{}^2  v{}^2 $ \\[2mm]
  & $- c_{\alpha}{}^2  s_{\alpha} e M_{h_2}{}^2  s_{2\alpha} v x+2 s_w s_{\alpha}{}^2  c_{\alpha} M_{h_2}{}^2  M_W s_{2\alpha} v$ \\
\hline
\end{tabular}

\begin{tabular}{|l|l|} \hline
Fields in the vertex & Variational derivative of Lagrangian by fields \\ \hline
$h_1{}_{}$ \phantom{-} ${Z}_{\mu }$ \phantom{-} ${Z}_{\nu }$ \phantom{-}  &
	$\frac{ c_{\alpha} \cdot e \cdot M_W}{ c_w{}^2  \cdot s_w}g^{\mu \nu} $\\[2mm]
$h_1{}_{}$ \phantom{-} ${Z}_{\mu }$ \phantom{-} $Z_F{}_{}$ \phantom{-}  &
	$\frac{1}{2}\frac{ i \cdot c_{\alpha} \cdot e}{ c_w \cdot s_w}\big(p_3^\mu -p_1^\mu \big)$\\[2mm]
$h_1{}_{}$ \phantom{-} $Z_F{}_{}$ \phantom{-} $Z_F{}_{}$ \phantom{-}  &
	$-\frac{1}{2}\frac{1}{ e \cdot v{}^2 }\big(2 s_w c_{\alpha} M_{h_2}{}^2  M_W-2 s_w c_{2\alpha} c_{\alpha} M_{h_2}{}^2  M_W+2 s_w c_{\alpha} M_{h_1}{}^2  M_W$ \\[2mm]
  & $+2 s_w c_{2\alpha} c_{\alpha} M_{h_1}{}^2  M_W- s_{\alpha} e M_{h_2}{}^2  s_{2\alpha} v+ s_{\alpha} e M_{h_1}{}^2  s_{2\alpha} v\big)$\\[2mm]
$h_1{}_{}$ \phantom{-} $Z^\prime{}_{\mu }$ \phantom{-} $Z^\prime{}_{\nu }$ \phantom{-}  &
	$-8 s_{\alpha} \cdot g^{\prime}_1{}^2  \cdot x\cdot g^{\mu \nu} $\\[2mm]
$h_1{}_{}$ \phantom{-} $Z^\prime{}_{\mu }$ \phantom{-} $Z^\prime_F{}_{}$ \phantom{-}  &
	$-2 i \cdot s_{\alpha} \cdot g^{\prime}_1\big(p_1^\mu -p_3^\mu \big)$\\[2mm]
$h_1{}_{}$ \phantom{-} $Z^\prime_F{}_{}$ \phantom{-} $Z^\prime_F{}_{}$ \phantom{-}  &
	$\frac{1}{2}\frac{1}{ e \cdot v \cdot x}\big( s_{\alpha} e M_{h_1}{}^2  v- s_{\alpha} c_{2\alpha} e M_{h_1}{}^2  v+ s_{\alpha} e M_{h_2}{}^2  v$ \\
\hline
\end{tabular}

\begin{tabular}{|l|l|} \hline
Fields in the vertex & Variational derivative of Lagrangian by fields \\ \hline
 & $$ \\[2mm]
  & $+ c_{2\alpha} c_{\alpha} e M_{h_2}{}^2  v+2 s_w s_{\alpha} M_{h_2}{}^2  M_W s_{2\alpha}-2 s_w s_{\alpha} M_{h_1}{}^2  M_W s_{2\alpha}\big)$\\[2mm]
$\bar{\tau}{}_{a }$ \phantom{-} $\tau{}_{b }$ \phantom{-} ${A}_{\mu }$ \phantom{-}  &
	$ e\gamma_{a c}^\mu \cdot \delta_{c b} $\\[2mm]
$\bar{\tau}{}_{a }$ \phantom{-} $\tau{}_{b }$ \phantom{-} $h_1{}_{}$ \phantom{-}  &
	$-\frac{1}{2}\frac{ c_{\alpha} \cdot e \cdot M_\tau}{ M_W \cdot s_w}\cdot \delta_{a b} $\\[2mm]
$\bar{\tau}{}_{a }$ \phantom{-} $\tau{}_{b }$ \phantom{-} $h_2{}_{}$ \phantom{-}  &
	$-\frac{1}{2}\frac{ e \cdot M_\tau \cdot s_{\alpha}}{ M_W \cdot s_w}\cdot \delta_{a b} $\\[2mm]
$\bar{\tau}{}_{a }$ \phantom{-} $\tau{}_{b }$ \phantom{-} ${Z}_{\mu }$ \phantom{-}  &
	$\frac{1}{4}\frac{ e}{ c_w \cdot s_w}\gamma_{a c}^\mu \big( (1-2 s_w {}^2)\cdot (1-\gamma^5)_{c b} -2 s_w{}^2 \cdot (1+\gamma^5)_{c b} \big)$\\[2mm]
$\bar{\tau}{}_{a }$ \phantom{-} $\tau{}_{b }$ \phantom{-} $Z_F{}_{}$ \phantom{-}  &
	$-\frac{1}{2}\frac{ i \cdot e \cdot M_\tau}{ M_W \cdot s_w}\cdot \gamma_{a b}^5 $\\
\hline
\end{tabular}

\begin{tabular}{|l|l|} \hline
Fields in the vertex & Variational derivative of Lagrangian by fields \\ \hline
$\bar{s}{}_{a p }$ \phantom{-} $s{}_{b q }$ \phantom{-} $h_1{}_{}$ \phantom{-}  &
	$-\frac{1}{2}\frac{ c_{\alpha} \cdot e \cdot M_s}{ M_W \cdot s_w}\delta_{p q} \cdot \delta_{a b} $\\[2mm]
$\bar{s}{}_{a p }$ \phantom{-} $s{}_{b q }$ \phantom{-} $h_2{}_{}$ \phantom{-}  &
	$-\frac{1}{2}\frac{ e \cdot M_s \cdot s_{\alpha}}{ M_W \cdot s_w}\delta_{p q} \cdot \delta_{a b} $\\[2mm]
$\bar{s}{}_{a p }$ \phantom{-} $s{}_{b q }$ \phantom{-} ${Z}_{\mu }$ \phantom{-}  &
	$-\frac{1}{12}\frac{ e}{ c_w \cdot s_w}\delta_{p q} \gamma_{a c}^\mu \big(2 s_w{}^2 \cdot (1+\gamma^5)_{c b} - (3-2 s_w {}^2)\cdot (1-\gamma^5)_{c b} \big)$\\[2mm]
$\bar{s}{}_{a p }$ \phantom{-} $s{}_{b q }$ \phantom{-} $Z_F{}_{}$ \phantom{-}  &
	$-\frac{1}{2}\frac{ i \cdot e \cdot M_s}{ M_W \cdot s_w}\delta_{p q} \cdot \gamma_{a b}^5 $\\[2mm]
$\bar{s}{}_{a p }$ \phantom{-} $s{}_{b q }$ \phantom{-} $Z^\prime{}_{\mu }$ \phantom{-}  &
	$-\frac{1}{3} g^{\prime}_1\delta_{p q} \gamma_{a c}^\mu \cdot \delta_{c b} $\\[2mm]
$\bar{s}{}_{a p }$ \phantom{-} $t{}_{b q }$ \phantom{-} $W^-{}_{\mu }$ \phantom{-}  &
	$-\frac{1}{4}\frac{ e \cdot\sqrt{2} \cdot V_{ts}}{ s_w}\cdot (1-\gamma^5)_{c b} \delta_{p q} \gamma_{a c}^\mu $\\[2mm]
$\bar{s}{}_{a p }$ \phantom{-} $t{}_{b q }$ \phantom{-} $W^-_F{}_{}$ \phantom{-}  &
	$-\frac{1}{4}\frac{ i \cdot e \cdot\sqrt{2} \cdot V_{ts}}{ M_W \cdot s_w}\delta_{p q} \big( M_s\cdot (1-\gamma^5)_{a b} - M_t\cdot (1+\gamma^5)_{a b} \big)$\\
\hline
\end{tabular}

\begin{tabular}{|l|l|} \hline
Fields in the vertex & Variational derivative of Lagrangian by fields \\ \hline
$\bar{u}{}_{a p }$ \phantom{-} $s{}_{b q }$ \phantom{-} $W^+{}_{\mu }$ \phantom{-}  &
	$-\frac{1}{4}\frac{ e \cdot\sqrt{2} \cdot V_{us}}{ s_w}\cdot (1-\gamma^5)_{c b} \delta_{p q} \gamma_{a c}^\mu $\\[2mm]
$\bar{u}{}_{a p }$ \phantom{-} $s{}_{b q }$ \phantom{-} $W^+_F{}_{}$ \phantom{-}  &
	$\frac{1}{4}\frac{ i \cdot e \cdot M_s \cdot\sqrt{2} \cdot V_{us}}{ M_W \cdot s_w}\cdot (1+\gamma^5)_{a b} \delta_{p q} $\\[2mm]
$\bar{u}{}_{a p }$ \phantom{-} $u{}_{b q }$ \phantom{-} ${A}_{\mu }$ \phantom{-}  &
	$-\frac{2}{3} e\delta_{p q} \gamma_{a c}^\mu \cdot \delta_{c b} $\\[2mm]
$\bar{u}{}_{a p }$ \phantom{-} $u{}_{b q }$ \phantom{-} ${G}_{\mu r }$ \phantom{-}  &
	$ g_s\cdot \gamma_{a b}^\mu \lambda_{p q}^r $\\[2mm]
$\bar{u}{}_{a p }$ \phantom{-} $u{}_{b q }$ \phantom{-} ${Z}_{\mu }$ \phantom{-}  &
	$-\frac{1}{12}\frac{ e}{ c_w \cdot s_w}\delta_{p q} \gamma_{a c}^\mu \big( (3-4 s_w {}^2)\cdot (1-\gamma^5)_{c b} -4 s_w{}^2 \cdot (1+\gamma^5)_{c b} \big)$\\[2mm]
$\bar{u}{}_{a p }$ \phantom{-} $u{}_{b q }$ \phantom{-} $Z^\prime{}_{\mu }$ \phantom{-}  &
	$-\frac{1}{3} g^{\prime}_1\delta_{p q} \gamma_{a c}^\mu \cdot \delta_{c b} $\\[2mm]
$W^+{}_{\mu }$ \phantom{-} $W^-{}_{\nu }$ \phantom{-} ${Z}_{\rho }$ \phantom{-}  &
	$-\frac{ c_w \cdot e}{ s_w}\big(p_1^\nu g^{\mu \rho} -p_1^\rho g^{\mu \nu} -p_2^\mu g^{\nu \rho} +p_2^\rho g^{\mu \nu} +p_3^\mu g^{\nu \rho} -p_3^\nu g^{\mu \rho} \big)$\\
\hline
\end{tabular}

\begin{tabular}{|l|l|} \hline
Fields in the vertex & Variational derivative of Lagrangian by fields \\ \hline
$\bar{C}^{W-}{}_{}$ \phantom{-} $C^{W+}{}_{}$ \phantom{-} $h_2{}_{}$ \phantom{-}  &
	$-\frac{1}{2}\frac{ e \cdot M_W \cdot s_{\alpha}}{ s_w}$\\[2mm]
$\bar{C}^{W-}{}_{}$ \phantom{-} $C^{W+}{}_{}$ \phantom{-} ${Z}_{\mu }$ \phantom{-}  &
	$\frac{ c_w \cdot e}{ s_w}\cdot p_1^\mu $\\[2mm]
$\bar{C}^{W-}{}_{}$ \phantom{-} $C^{W+}{}_{}$ \phantom{-} $Z_F{}_{}$ \phantom{-}  &
	$-\frac{1}{2}\frac{ i \cdot e \cdot M_W}{ s_w}$\\[2mm]
$\bar{C}^{W-}{}_{}$ \phantom{-} $C^{Z}{}_{}$ \phantom{-} $W^+{}_{\mu }$ \phantom{-}  &
	$-\frac{ c_w \cdot e}{ s_w}\cdot p_1^\mu $\\[2mm]
$\bar{C}^{W-}{}_{}$ \phantom{-} $C^{Z}{}_{}$ \phantom{-} $W^+_F{}_{}$ \phantom{-}  &
	$\frac{1}{2}\frac{ i \cdot (1-2 s_w {}^2) \cdot e \cdot M_W}{ c_w \cdot s_w}$\\[2mm]
$\bar{C}^{Z}{}_{}$ \phantom{-} $C^{W+}{}_{}$ \phantom{-} $W^-{}_{\mu }$ \phantom{-}  &
	$-\frac{ c_w \cdot e}{ s_w}\cdot p_1^\mu $\\[2mm]
$\bar{C}^{Z}{}_{}$ \phantom{-} $C^{W+}{}_{}$ \phantom{-} $W^-_F{}_{}$ \phantom{-}  &
	$\frac{1}{2}\frac{ i \cdot e \cdot M_W}{ c_w \cdot s_w}$\\
\hline
\end{tabular}

\begin{tabular}{|l|l|} \hline
Fields in the vertex & Variational derivative of Lagrangian by fields \\ \hline
$\nu _1{}_{a }$ \phantom{-} $\nu _1{}_{b }$ \phantom{-} $h_1{}_{}$ \phantom{-}  &
	$-\frac{ s_{\alpha_\nu}}{ x}\big( c_{\alpha}\cdot  c_{\alpha_\nu}\cdot \sqrt{2}\cdot  x\cdot  y^D_{n_1}\cdot \delta_{a b} - s_{\alpha}\cdot  s_{\alpha_\nu}\cdot  m_{\nu h}\cdot \delta_{a b} \big)$\\[2mm]
$\nu _1{}_{a }$ \phantom{-} $\nu _1{}_{b }$ \phantom{-} $h_2{}_{}$ \phantom{-}  &
	$-\frac{ s_{\alpha_\nu}}{ x}\big( s_{\alpha}\cdot  c_{\alpha_\nu}\cdot \sqrt{2}\cdot  x\cdot  y^D_{n_1}\cdot \delta_{a b} + s_{\alpha_\nu}\cdot  c_{\alpha}\cdot  m_{\nu h}\cdot \delta_{a b} \big)$\\[2mm]
$\nu _1{}_{a }$ \phantom{-} $\nu _1{}_{b }$ \phantom{-} ${Z}_{\mu }$ \phantom{-}  &
	$\frac{1}{2}\frac{ c_{\alpha_\nu}{}^2  \cdot e}{ c_w \cdot s_w}\gamma_{a c}^\mu \gamma_{c b}^5 $\\[2mm]
$\nu _1{}_{a }$ \phantom{-} $\nu _1{}_{b }$ \phantom{-} $Z_F{}_{}$ \phantom{-}  &
	$ i \cdot c_{\alpha_\nu} \cdot s_{\alpha_\nu} \cdot\sqrt{2} \cdot y^D_{n_1}\cdot \gamma_{a b}^5 $\\[2mm]
$\nu _1{}_{a }$ \phantom{-} $\nu _1{}_{b }$ \phantom{-} $Z^\prime{}_{\mu }$ \phantom{-}  &
	$- g^{\prime}_1\gamma_{a c}^\mu \gamma_{c b}^5 \big( c_{\alpha_\nu}{}^2 - s_{\alpha_\nu}{}^2 \big)$\\[2mm]
$\nu _1{}_{a }$ \phantom{-} $\nu _1{}_{b }$ \phantom{-} $Z^\prime_F{}_{}$ \phantom{-}  &
	$-\frac{ i \cdot m_{\nu h} \cdot s_{\alpha_\nu}{}^2 }{ x}\cdot \gamma_{a b}^5 $\\[2mm]
$\nu _1{}_{a }$ \phantom{-} $\nu _{h1}{}_{b }$ \phantom{-} $h_1{}_{}$ \phantom{-}  &
	$\frac{1}{2}\frac{1}{ x}\big( s_{\alpha_\nu}{}^2   c_{\alpha} \sqrt{2}  x  y^D_{n_1} \delta_{a b} - c_{\alpha}  c_{\alpha_\nu}{}^2  \sqrt{2}  x  y^D_{n_1} \delta_{a b} +2 s_{\alpha}  s_{\alpha_\nu}  c_{\alpha_\nu}  m_{\nu h} \delta_{a b} \big)$\\
\hline
\end{tabular}

\begin{tabular}{|l|l|} \hline
Fields in the vertex & Variational derivative of Lagrangian by fields \\ \hline
$\nu _3{}_{a }$ \phantom{-} $\nu _3{}_{b }$ \phantom{-} $h_2{}_{}$ \phantom{-}  &
	$-\frac{ s_{\alpha_\nu}}{ x}\big( s_{\alpha}\cdot  c_{\alpha_\nu}\cdot \sqrt{2}\cdot  x\cdot  y^D_{n_3}\cdot \delta_{a b} + s_{\alpha_\nu}\cdot  c_{\alpha}\cdot  m_{\nu h}\cdot \delta_{a b} \big)$\\[2mm]
$\nu _3{}_{a }$ \phantom{-} $\nu _3{}_{b }$ \phantom{-} ${Z}_{\mu }$ \phantom{-}  &
	$\frac{1}{2}\frac{ c_{\alpha_\nu}{}^2  \cdot e}{ c_w \cdot s_w}\gamma_{a c}^\mu \gamma_{c b}^5 $\\[2mm]
$\nu _3{}_{a }$ \phantom{-} $\nu _3{}_{b }$ \phantom{-} $Z_F{}_{}$ \phantom{-}  &
	$ i \cdot c_{\alpha_\nu} \cdot s_{\alpha_\nu} \cdot\sqrt{2} \cdot y^D_{n_3}\cdot \gamma_{a b}^5 $\\[2mm]
$\nu _3{}_{a }$ \phantom{-} $\nu _3{}_{b }$ \phantom{-} $Z^\prime{}_{\mu }$ \phantom{-}  &
	$- g^{\prime}_1\gamma_{a c}^\mu \gamma_{c b}^5 \big( c_{\alpha_\nu}{}^2 - s_{\alpha_\nu}{}^2 \big)$\\[2mm]
$\nu _3{}_{a }$ \phantom{-} $\nu _3{}_{b }$ \phantom{-} $Z^\prime_F{}_{}$ \phantom{-}  &
	$-\frac{ i \cdot m_{\nu h} \cdot s_{\alpha_\nu}{}^2 }{ x}\cdot \gamma_{a b}^5 $\\[2mm]
$\nu _3{}_{a }$ \phantom{-} $\nu _{h3}{}_{b }$ \phantom{-} $h_1{}_{}$ \phantom{-}  &
	$\frac{1}{2}\frac{1}{ x}\big( s_{\alpha_\nu}{}^2   c_{\alpha} \sqrt{2}  x  y^D_{n_3} \delta_{a b} - c_{\alpha}  c_{\alpha_\nu}{}^2  \sqrt{2}  x  y^D_{n_3} \delta_{a b} +2 s_{\alpha}  s_{\alpha_\nu}  c_{\alpha_\nu}  m_{\nu h} \delta_{a b} \big)$\\[2mm]
$\nu _3{}_{a }$ \phantom{-} $\nu _{h3}{}_{b }$ \phantom{-} $h_2{}_{}$ \phantom{-}  &
	$\frac{1}{2}\frac{1}{ x}\big( s_{\alpha}  s_{\alpha_\nu}{}^2  \sqrt{2}  x  y^D_{n_3} \delta_{a b} - s_{\alpha}  c_{\alpha_\nu}{}^2  \sqrt{2}  x  y^D_{n_3} \delta_{a b} -2 s_{\alpha_\nu}  c_{\alpha}  c_{\alpha_\nu}  m_{\nu h} \delta_{a b} \big)$\\
\hline
\end{tabular}

\begin{tabular}{|l|l|} \hline
Fields in the vertex & Variational derivative of Lagrangian by fields \\ \hline
$\nu _{h3}{}_{a }$ \phantom{-} $\nu _{h3}{}_{b }$ \phantom{-} ${Z}_{\mu }$ \phantom{-}  &
	$\frac{1}{2}\frac{ e \cdot s_{\alpha_\nu}{}^2 }{ c_w \cdot s_w}\gamma_{a c}^\mu \gamma_{c b}^5 $\\[2mm]
$\nu _{h3}{}_{a }$ \phantom{-} $\nu _{h3}{}_{b }$ \phantom{-} $Z_F{}_{}$ \phantom{-}  &
	$- i \cdot c_{\alpha_\nu} \cdot s_{\alpha_\nu} \cdot\sqrt{2} \cdot y^D_{n_3}\cdot \gamma_{a b}^5 $\\[2mm]
$\nu _{h3}{}_{a }$ \phantom{-} $\nu _{h3}{}_{b }$ \phantom{-} $Z^\prime{}_{\mu }$ \phantom{-}  &
	$- g^{\prime}_1\gamma_{a c}^\mu \gamma_{c b}^5 \big( s_{\alpha_\nu}{}^2 - c_{\alpha_\nu}{}^2 \big)$\\[2mm]
$\nu _{h3}{}_{a }$ \phantom{-} $\nu _{h3}{}_{b }$ \phantom{-} $Z^\prime_F{}_{}$ \phantom{-}  &
	$-\frac{ i \cdot c_{\alpha_\nu}{}^2  \cdot m_{\nu h}}{ x}\cdot \gamma_{a b}^5 $\\[2mm]
${A}_{\mu }$ \phantom{-} ${A}_{\nu }$ \phantom{-} $W^+{}_{\rho }$ \phantom{-} $W^-{}_{\sigma }$ \phantom{-}  &
	$- e{}^2 \big(2g^{\mu \nu} g^{\rho \sigma} -g^{\mu \rho} g^{\nu \sigma} -g^{\mu \sigma} g^{\nu \rho} \big)$\\[2mm]
${A}_{\mu }$ \phantom{-} ${A}_{\nu }$ \phantom{-} $W^+_F{}_{}$ \phantom{-} $W^-_F{}_{}$ \phantom{-}  &
	$2 e{}^2 \cdot g^{\mu \nu} $\\[2mm]
${A}_{\mu }$ \phantom{-} $h_1{}_{}$ \phantom{-} $W^+{}_{\nu }$ \phantom{-} $W^-_F{}_{}$ \phantom{-}  &
	$\frac{1}{2}\frac{ i \cdot c_{\alpha} \cdot e{}^2 }{ s_w}\cdot g^{\mu \nu} $\\
\hline
\end{tabular}


\begin{tabular}{|l|l|} \hline
Fields in the vertex & Variational derivative of Lagrangian by fields \\ \hline
 & $$ \\[2mm]
  & $+ (1-6 c_{\alpha} {}^2   s_{\alpha} {}^2) M_{h_2}{}^2  s_{2\alpha} v x - (1-6 c_{\alpha} {}^2 \cdot  s_{\alpha} {}^2) M_{h_1}{}^2  s_{2\alpha} v x\big)$\\[2mm]
$h_1{}_{}$ \phantom{-} $h_1{}_{}$ \phantom{-} $W^+{}_{\mu }$ \phantom{-} $W^-{}_{\nu }$ \phantom{-}  &
	$\frac{1}{2}\frac{ c_{\alpha}{}^2  \cdot e{}^2 }{ s_w{}^2 }\cdot g^{\mu \nu} $\\[2mm]
$h_1{}_{}$ \phantom{-} $h_1{}_{}$ \phantom{-} $W^+_F{}_{}$ \phantom{-} $W^-_F{}_{}$ \phantom{-}  &
	$-\frac{1}{2}\frac{1}{ v{}^2  \cdot x}\big( c_{\alpha}{}^2  M_{h_2}{}^2  x- c_{\alpha}{}^2  c_{2\alpha} M_{h_2}{}^2  x+ c_{\alpha}{}^2  M_{h_1}{}^2  x$ \\[2mm]
  & $+ c_{\alpha}{}^2  c_{2\alpha} M_{h_1}{}^2  x + s_{\alpha}{}^2  M_{h_2}{}^2  s_{2\alpha} v- s_{\alpha}{}^2  M_{h_1}{}^2  s_{2\alpha} v\big)$\\[2mm]
$h_1{}_{}$ \phantom{-} $h_1{}_{}$ \phantom{-} ${Z}_{\mu }$ \phantom{-} ${Z}_{\nu }$ \phantom{-}  &
	$\frac{1}{2}\frac{ c_{\alpha}{}^2  \cdot e{}^2 }{ c_w{}^2  \cdot s_w{}^2 }g^{\mu \nu} $\\[2mm]
$h_1{}_{}$ \phantom{-} $h_1{}_{}$ \phantom{-} $Z_F{}_{}$ \phantom{-} $Z_F{}_{}$ \phantom{-}  &
	$-\frac{1}{2}\frac{1}{ v{}^2  \cdot x}\big( c_{\alpha}{}^2  M_{h_2}{}^2  x- c_{\alpha}{}^2  c_{2\alpha} M_{h_2}{}^2  x+ c_{\alpha}{}^2  M_{h_1}{}^2  x$ \\
\hline
\end{tabular}

\begin{tabular}{|l|l|} \hline
Fields in the vertex & Variational derivative of Lagrangian by fields \\ \hline
 & $$ \\[2mm]
  & $+ c_{\alpha}{}^4  \big( M_{h_2}{}^2 + M_{h_1}{}^2 \big)  v{}^2  + c_{\alpha}{}^4  c_{2\alpha} \big( M_{h_2}{}^2 - M_{h_1}{}^2 \big)  v{}^2 $ \\[2mm]
  & $+2 c_{\alpha}{}^2  s_{\alpha}{}^2  M_{h_2}{}^2  s_{2\alpha} v x -2 c_{\alpha}{}^2  s_{\alpha}{}^2  M_{h_1}{}^2  s_{2\alpha} v x\big)$\\[2mm]
$h_2{}_{}$ \phantom{-} $h_2{}_{}$ \phantom{-} $W^+{}_{\mu }$ \phantom{-} $W^-{}_{\nu }$ \phantom{-}  &
	$\frac{1}{2}\frac{ e{}^2  \cdot s_{\alpha}{}^2 }{ s_w{}^2 }\cdot g^{\mu \nu} $\\[2mm]
$h_2{}_{}$ \phantom{-} $h_2{}_{}$ \phantom{-} $W^+_F{}_{}$ \phantom{-} $W^-_F{}_{}$ \phantom{-}  &
	$-\frac{1}{2}\frac{1}{ v{}^2  \cdot x}\big( s_{\alpha}{}^2  M_{h_2}{}^2  x- s_{\alpha}{}^2  c_{2\alpha} M_{h_2}{}^2  x+ s_{\alpha}{}^2  M_{h_1}{}^2  x$ \\[2mm]
  & $+ s_{\alpha}{}^2  c_{2\alpha} M_{h_1}{}^2  x + c_{\alpha}{}^2  M_{h_2}{}^2  s_{2\alpha} v- c_{\alpha}{}^2  M_{h_1}{}^2  s_{2\alpha} v\big)$\\[2mm]
$h_2{}_{}$ \phantom{-} $h_2{}_{}$ \phantom{-} ${Z}_{\mu }$ \phantom{-} ${Z}_{\nu }$ \phantom{-}  &
	$\frac{1}{2}\frac{ e{}^2  \cdot s_{\alpha}{}^2 }{ c_w{}^2  \cdot s_w{}^2 }g^{\mu \nu} $\\
\end{tabular}
%
$Z_F{}_{}$ \phantom{-} $Z_F{}_{}$ \phantom{-} $Z^\prime_F{}_{}$ \phantom{-} $Z^\prime_F{}_{}$ \phantom{-}  &
	$-\frac{1}{2}\frac{ s_{2\alpha}}{ v \cdot x}\big( M_{h_2}{}^2 - M_{h_1}{}^2 \big)$\\[2mm]
$Z^\prime{}_{\mu }$ \phantom{-} $Z^\prime{}_{\nu }$ \phantom{-} $Z^\prime_F{}_{}$ \phantom{-} $Z^\prime_F{}_{}$ \phantom{-}  &
	$8 g^{\prime}_1{}^2 \cdot g^{\mu \nu} $\\[2mm]
$Z^\prime_F{}_{}$ \phantom{-} $Z^\prime_F{}_{}$ \phantom{-} $Z^\prime_F{}_{}$ \phantom{-} $Z^\prime_F{}_{}$ \phantom{-}  &
	$-\frac{3}{2}\frac{1}{ x{}^2 }\big( M_{h_1}{}^2 - c_{2\alpha} M_{h_1}{}^2 + M_{h_2}{}^2 + c_{2\alpha} M_{h_2}{}^2 \big)$\\ \hline
\end{tabular}

\end{center}



\chapter{Renormalisation group equations}\label{App:RGE}
In this appendix we present the complete set of one-loop RGEs for the minimal $U(1)_{B-L}$ extension of the SM. For some parameters, the equations will be equal to those of the SM, as no extra contribution arises at one loop.

\section{Gauge RGEs}
The RGEs for the $SU(3)_C$ and $SU(2)_L$ gauge couplings $g_S$ and $g$ are \cite{Arason:1991ic}
\begin{eqnarray}\label{rge_gs}
\frac{d}{dt}g_S &=& \frac{g_S^3}{16\pi ^2}\left[ -11+\frac{4}{3}n_g \right] = \frac{g_S^3}{16\pi ^2}\left( -7\right)\, ,\\ \label{rge_g}
\frac{d}{dt}g &=& \frac{g^3}{16\pi ^2}\left[ -\frac{22}{3}+\frac{4}{3}n_g+\frac{1}{6}\right] = \frac{g^3}{16\pi ^2}\left(
		-\frac{19}{6}\right) \, ,
\end{eqnarray}
where $n_g =3$ is the number of generations.

Following standard techniques, we obtain for the Abelian couplings \cite{Chankowski:2006jk,delAguila:1988jz}:
\begin{eqnarray}\label{RGE_g1}
\frac{d}{dt}g_1 &=& \frac{1}{16\pi ^2}\left[A^{YY}g_1^3 \right]\, , \\ \label{RGE_g2}
\frac{d}{dt}g_1' &=& \frac{1}{16\pi ^2}\left[A^{XX}g_1'^3+2A^{XY}g_1'^2\widetilde g+A^{YY}g_1'\widetilde g^2 \right] \, , \\ \label{RGE_g_tilde}
\frac{d}{dt}\widetilde g &=& \frac{1}{16\pi ^2}\left[A^{YY}\widetilde{g}\,(\widetilde g^2+2g_1^2)+2A^{XY}g_1'(\widetilde{g}^2+g_1^2)+A^{XX}g_1'^2\widetilde g \right]\, ,
\end{eqnarray}
with
\begin{equation}\label{A_charge}
A^{ab} = A^{ba} = \frac{2}{3} \sum _f Q_f^a Q_f^b + \frac{1}{3}\sum _s Q_s^a Q_s^b\, , \qquad (a,b=Y,X)\, ,
\end{equation}
where the first sum is over the left-handed two-component fermions and the second one is over the complex scalars. For the model we are discussing ($Y$ is the SM weak hypercharge, $X=B-L$ is the $B-L$ number), the coefficients of eq.~(\ref{A_charge}) are, respectively,      
\begin{equation}
A^{YY}=41/6\, ,\qquad A^{XX}=12\, ,\qquad A^{YX}=16/3.
\end{equation}

\section{Fermion RGEs}
From straightforward calculations we obtain:
\begin{equation}\label{RGE_yuk_top}
\frac{d}{dt}y_t = \frac{y_t}{16\pi ^2}\left( \frac{9}{2}y_t^2-8g_S^2-\frac{9}{4}g^2-\frac{17}{12}g_1^2-\frac{17}{12}\widetilde{g}^2 -\frac{2}{3}g_1^{'2}-\frac{5}{3}\widetilde{g}g'_1 \right)\, .
\end{equation}
For the right-handed neutrinos, it is not restrictive to consider the basis in which the Majorana matrix of couplings is real, diagonal and positive:
$y^M \equiv \mbox{diag}\, (y^M_1,y^M_2,y^M_3)$. Then we get \cite{Luo:2002iq,Iso:2009ss} \footnote{Notice the we get a difference of a factor $3$ in the third term in the RHS of the last expression in eq.~(14) contained in Ref.~\cite{Iso:2009ss}. The authors of Ref.~\cite{Iso:2009ss} acknowledged the difference and will correct their paper.}
\begin{equation}\label{RGE_nu_r_maj}
\frac{d}{dt}y^M_i = \frac{y^M_i}{16\pi ^2}\left( 4(y^M_i)^2+2Tr\big[ (y^M)^2\big] -6g_1^{'2} \right)\, , \qquad (i=1\dots 3)\, .
\end{equation}

\section{Scalar RGEs}
A very straightforward way to find the one-loop RGEs for the parameters of the scalar potential is to compute the one-loop
effective potential and to impose its independence from the renormalisation scale. To one-loop level, the scalar potential $V$ reads
\begin{equation}
V=V^{(0)}+\Delta V^{(1)}\, ,
\end{equation}
where $V^{(0)}$ is the tree-level potential and $\Delta V^{(1)}$ indicates the one-loop correction to it. To compute the latter it is useful to rewrite the tree-level potential
\begin{equation}\label{new-potential}
V^{(0)}(H,\chi ) = m^2H^{\dagger}H + \mu ^2\mid\chi\mid ^2 + \lambda _1 (H^{\dagger}H)^2 +\lambda _2 \mid\chi\mid ^4 + \lambda _3 H^{\dagger}H\mid\chi\mid ^2 
\end{equation}
in terms of the real scalar fields
\begin{equation}\label{H-CHI_degree}
H=\frac{1}{\sqrt{2}}\left( \begin{array}{c} \phi _1 +i\phi _2 \\ \phi _3 +i\phi _4\end{array}\right)\, , \hspace{2cm}
\chi =\frac{1}{\sqrt{2}}\left( \phi _5 +i\phi _6\right)\, .
\end{equation}
The only combinations of fields that are involved are $\phi ^2 =\phi _1^2+\phi _2^2+\phi _3^2+\phi _4^2 $ and
$\eta ^2 \equiv \phi _5^2+\phi _6^2$, so that eq.~(\ref{new-potential}) becomes
\begin{equation}\label{tree_lev_pot}
V^{(0)}(\phi,\eta )= \frac{1}{2}m^2\phi ^2 +\frac{1}{2}\mu ^2\eta ^2 +\frac{1}{4}\lambda _1 \phi ^4 +\frac{1}{4}\lambda _2 \eta ^4 +
			\frac{1}{4}\lambda _3 \phi ^2 \eta ^2\, .
\end{equation}
The one-loop correction to the tree-level potential (\ref{tree_lev_pot}) is,
in the Landau gauge,
\begin{equation}\label{1-loop_pot}
\Delta V^{(1)}(\phi ,\eta )=\frac{1}{64\pi ^2}\sum _i(-1)^{2s_i}(2s_i+1)M^4_i(\phi ^2,\eta ^2)\left[ \ln{\frac{M^2_i(\phi ^2,\eta ^2)}{\mu ^2}-c_i}\right]\, ,
\end{equation}
where $c_i$ are constants that depend on the renormalisation scheme (for example, in the $\overline{\rm MS}$ scheme, it is $c_i=3/2$
for scalars and fermions, $c_i=5/6$ for vectors). Expanding eq.~(\ref{1-loop_pot}) and keeping the contributions of the scalar fields (Higgs
and Goldstone bosons), of the top quark, of the gauge bosons, and of the RH neutrinos only, we obtain
\begin{eqnarray*}
\Delta V^{(1)}&=&\frac{1}{64\pi ^2}\left\{ 3G_1^2\left[ \ln{\frac{G_1}{\mu ^2}-\frac{3}{2}}\right] + G_2^2\left[ \ln{\frac{G_2}{\mu ^2}-\frac{3}{2}}\right]
		+ Tr\left( H^2\left[ \ln{\frac{H}{\mu ^2}-\frac{3}{2}}\right]\right) \right. \\
	&& -12T^2\left[ \ln{\frac{T}{\mu ^2}  -\frac{3}{2}}\right] + 
	6  M_W^2\left[ \ln{\frac{M_W}{\mu ^2}-\frac{5}{6}}\right] + 3Tr\left( M_G^2\left[ \ln{\frac{M_G}{\mu ^2}-\frac{5}{6}}\right]\right) \\
	&&\left. -2\sum _{i=1}^3 N_i^2\left[ \ln{\frac{N_i}{\mu ^2}-\frac{3}{2}}\right] \right\}\, ,
\end{eqnarray*}
where the field-dependent squared masses are, in a self-explanatory notation,
\begin{eqnarray}\label{3gold-field-dep}
G_1 (\phi ,\eta )&=& m^2+\lambda _1\phi ^2+\frac{\lambda _3}{2}\eta ^2\, ,\\ \label{1gold-field-dep}
G_2 (\phi ,\eta )&=& \mu ^2+\lambda _2\eta ^2+\frac{\lambda _3}{2}\phi ^2\, ,\\ \label{Higgs-field-dep}
H (\phi ,\eta )&=& \left( \begin{array}{cc} m^2+3\lambda _1\phi ^2+\frac{\lambda _3}{2}\eta ^2 & \lambda _3 \phi\eta\\ 
				\lambda _3\phi\eta & \mu ^2 +3\lambda _2\eta ^2+\frac{\lambda _3}{2}\phi ^2\end{array}\right)\, ,\\ \label{Top-field-dep}
T (\phi ,\eta )&=& \frac{1}{2}(y_t\phi )^2\, ,\\ \label{W-field-dep}
M_W (\phi ,\eta )&=& \frac{1}{4}(g\phi )^2\, ,\\ \label{RH-N-field-dep}
M_G (\phi ,\eta ) &=& \frac{1}{4}\left(
		\begin{array}{ccc}
		g_1^{\phantom{o}2}\phi ^2 & -gg_1\phi ^2 & g_1\widetilde{g}\phi ^2\\
		-gg_1\phi ^2 & g^2\phi ^2 & -g\widetilde{g}\phi ^2\\
		g_1\widetilde{g}\phi ^2 & -g\widetilde{g}\phi ^2 & \widetilde{g}^2\phi ^2 + 16\eta ^2 g_1^{'2}
		\end{array} \right) \, ,\\ \label{Gauge_bosonos-dep}
N_i (\phi ,\eta )&=&\frac{1}{2}(y^M_i \eta )^2\, .
\end{eqnarray}
As usual, we define the beta functions $\beta _i$ ($i=1\dots 3$) for the quartic couplings, the gamma
functions $\gamma _{m,\mu}$ for the scalar masses and the scalar anomalous dimensions $\gamma _{\phi,\,\eta}$ as follows ($t=\ln{Q}$):
\begin{eqnarray}\label{beta_i}
\frac{d\lambda _i}{dt} &=& \beta _i\, ,\\ \label{gamma _m}
\frac{dm^2}{dt} &=& \gamma _m m^2\, ,\\ \label{gamma _mu}
\frac{d\mu ^2}{dt} &=& \gamma _\mu \mu ^2\, ,\\ \label{gamma _phi}
\frac{d\phi ^2}{dt} &=& 2\gamma _\phi \phi^2\, ,\\ \label{gamma _eta}
\frac{d\eta ^2}{dt} &=& 2\gamma _\eta \eta ^2\, .
\end{eqnarray}

Now we can extract the RGEs for the parameters of the scalar potential just by requiring that the first derivative of the effective
potential with respect to the scale $t$ vanishes,
\begin{equation}\label{eff_pot_der_nulla}
\frac{d}{dt}V^{(1)} \equiv \frac{d}{dt}(V^{(0)}+\Delta V^{(1)})\equiv 0\, ,
\end{equation}
keeping only the one-loop terms. Reorganising it in a more convenient way, we see that eq.~(\ref{eff_pot_der_nulla}) implies the following
equations: 
\begin{eqnarray*}\label{RGE_1}
\frac{m^2\phi ^2}{2} \left[ \gamma _m +2\gamma _\phi  -\frac{1}{16\pi ^2}\left( 12\lambda _1+2\frac{\mu ^2}{m^2}\lambda _3\right) \right] &=& 0\, ,\\ \label{RGE_2}
\frac{\mu ^2 \eta ^2}{2} \left[\gamma _\mu +2\gamma _\eta -\frac{1}{16\pi ^2}\left( 8\lambda _2+4\frac{m^2}{\mu ^2}\lambda _3\right) \right] &=& 0\, ,\\ \label{RGE_3}
\frac{\phi ^4}{4}\left[\beta _1 +4\lambda _1\gamma _\phi -\frac{1}{16\pi ^2}\left( 24\lambda _1^2+\lambda _3^2
	-6y_t^4
	+\frac{9}{8}g^4+\frac{3}{8}g_1^4+\frac{3}{4}g^2g_1^2 \right.\right. \qquad && \\
	 \left.\left.
	+\frac{3}{4}g^2\widetilde{g}^2+\frac{3}{4}g_1^2\widetilde{g}^2 +\frac{3}{8}\widetilde{g}^4 \right)\right] &=&0\, ,\\ \label{RGE_4}
\frac{\eta ^4}{4}\left[\beta _2 +4\lambda _2\gamma _\eta -\frac{1}{8\pi ^2}\left( 10\lambda _2^2+\lambda _3^2
	-\frac{1}{2}Tr\big[ (y^M)^4\big] +48 g_1^{'4}\right)\right] &=&0\, ,\\ \label{RGE_5}
\frac{\phi ^2\eta ^2}{4} \left[\beta _3 +2\lambda _3(\gamma _\phi  +\gamma _\eta )-\frac{1}{8\pi ^2}\left( 6\lambda _1\lambda _3+4\lambda _2\lambda _3 +2\lambda _3^2 +6\widetilde{g}^2 g_1^{'2} \right)\right] &=&0\, .
\end{eqnarray*}

Imposing that each term between square brackets vanishes, we can obtain the RGEs for the parameters of the scalar potential after
inserting the explicit expressions of the scalar anomalous dimensions $\gamma _\phi$ and $\gamma _\eta$. The latter are easily
computed and read \cite{Luo:2002ey,Iso:2009ss,Sher:1988mj}
\begin{eqnarray}\label{gamma_phi}
\gamma _\phi &=& -\frac{1}{16 \pi ^2} \left( 3y_t^2 -\frac{9}{4}g^2-\frac{3}{4}g_1^2-\frac{3}{4}\widetilde{g}^2 \right)\, ,\\ \label{gamma_eta}
\gamma _\eta &=& -\frac{1}{16 \pi ^2} \left( 2Tr\left[ (y^M)^2\right] - 12 g_1^{'2}\right)\, .
\end{eqnarray}
Inserting eqs. (\ref{gamma_phi}) and (\ref{gamma_eta}) into the RGEs, we finally obtain the RGEs for the five parameters in the scalar potential:
\begin{eqnarray}\label{RGE_m}
\gamma _m \equiv \frac{1}{m^2}\frac{d m^2}{dt} &=& \frac{1}{16\pi ^2}\left( 12\lambda _1 +6y_t^2+2\frac{\mu ^2}{m^2}\lambda _3 -\frac{9}{2}g^2-\frac{3}{2}g_1^2-\frac{3}{2}\widetilde{g}^2\right)\, , \\ \label{RGE_mu}
\gamma _\mu \equiv \frac{1}{\mu ^2}\frac{d \mu ^2}{dt} &=& \frac{1}{16\pi ^2}\left( 8\lambda _2+4Tr\left[ (y^M)^2\right] +4\frac{m^2}{\mu ^2}\lambda _3 - 24 g_1^{'2}\right)\, ,\\ \nonumber
\beta _1 \equiv \frac{d \lambda _1}{dt} &=&  \frac{1}{16\pi ^2}\left( 24\lambda _1^2+\lambda _3^2
-6y_t^4 +\frac{9}{8}g^4 +\frac{3}{8}g_1^4 +\frac{3}{4}g^2g_1^2 +\frac{3}{4}g^2\widetilde{g}^2   \right. \\ \label{RGE_lamda1}
&& \left. +\frac{3}{4}g_1^2\widetilde{g}^2+\frac{3}{8}\widetilde{g}^4 + 12\lambda _1 y_t^2 -9\lambda _1 g^2-3\lambda _1 g_1^2-3\lambda _1 \widetilde{g}^2
	\right)\, ,\\ \nonumber
\beta _2 \equiv \frac{d \lambda _2}{dt} &=&  \frac{1}{8\pi ^2}\left( 10\lambda _2^2+\lambda _3^2-\frac{1}{2}Tr\left[ (y^M)^4\right] +48 g_1^{'4}+4\lambda _2Tr\left[ (y^M)^2\right] \right.\\ \label{RGE_lamda2}
&& \left. -24\lambda _2g_1^{'2} \right)\, ,\\ \nonumber
\beta _3 \equiv \frac{d \lambda _3}{dt} &=&  \frac{\lambda _3}{8\pi ^2}\left( 6\lambda _1+4\lambda _2+2\lambda _3+3y_t^2-\frac{9}{4}g^2-\frac{3}{4}g_1^2-\frac{3}{4}\widetilde{g}^2 \right.  \\ \label{RGE_lamda3}
&& \left. +2Tr\left[ (y^M)^2\right] - 12 g_1^{'2} + 6\frac{\widetilde{g}^2 g_1^{'2}}{\lambda _3}\right)\, .
\end{eqnarray}



\bibliographystyle{Classes/spiebib}
\renewcommand{\bibname}{References} 

\bibliography{References/references} 

\printindex

\end{document}